%% file: thesis_main.tex
\documentclass[11pt,twoside,openright]{book}

\usepackage{amsfonts}
\usepackage{amssymb,amsthm,amsmath}
\usepackage{fancyhdr}
\usepackage[dvips]{graphics}
\usepackage{graphicx}
\usepackage{floatflt}
\usepackage{multibox}
\include{bm}

\include{definitions}

\begin{document}

\frontmatter

\include{title}

\setlength{\oddsidemargin}{0.5cm}%
\setlength{\evensidemargin}{1.0cm}%


\setcounter{tocdepth}{1}\tableofcontents

\include{acknowlegments}


\mainmatter \setcounter{chapter}{0} \setcounter{page}{1}

\include{structure_thesis}

\include{preamble}


\renewcommand{\theequation}{\arabic{chapter}.\arabic{equation}}
\renewcommand{\thefigure}{\arabic{chapter}.\arabic{figure}}
\renewcommand{\thetable}{\arabic{chapter}.\arabic{table}}


\addcontentsline{toc}{part}{Part I. Conserved charges in
Lagrangian gauge theories \mbox{      } Application to the
mechanics of black holes}

\newpage
\thispagestyle{empty} \mbox{  }
\begin{center}
{\vspace{100pt} \bf \Huge \mbox{ }  Part I \\
\vspace{40pt}
Conserved charges in Lagrangian gauge theories \\
\vspace{40pt}
 Application to the mechanics \\ \vspace{10pt}
  of black holes}
\end{center}

\setcounter{chapter}{0} \setcounter{section}{0}
\setcounter{equation}{0} \setcounter{footnote}{0}

\include{general_formalism}

\include{charges_gravity_matter}

\include{blackholes}
\include{blackholes_examples}


\addcontentsline{toc}{part}{Part II. Asymptotically conserved
charges and their algebra. \mbox{      } Analyses in
three-dimensional gravity}

\cleardoublepage
\newpage
\thispagestyle{empty} \mbox{  }
\begin{center}
{\vspace{100pt} \bf \Huge \mbox{ }  Part II \\
\vspace{40pt} Asymptotically conserved charges and their algebra
\\ \vspace{40pt} Analyses in \\\vspace{10pt} three-dimensional gravity}
\end{center}

\setcounter{section}{0} \setcounter{equation}{0}
\setcounter{footnote}{0}

\include{theory_asympt_charges}

\include{asymptotic_analyses}

\include{conclusion}

\include{appendices}

\backmatter


\include{bibliothesis}

\end{document}

%% file: definitions.tex
\renewcommand{\chaptermark}[1]{         
  \markboth{#1}{}} %
\makeatletter
\def\cleardoublepage{\clearpage\if@twoside \ifodd\c@page\else%
    \hbox{}%
    \thispagestyle{empty}
    \newpage%
    \if@twocolumn\hbox{}\newpage\fi\fi\fi}
\makeatother

\renewcommand{\thechapter}{\arabic{chapter}}

\renewcommand{\theequation}{\arabic{chapter}.\arabic{equation}}
\renewcommand{\thefigure}{\arabic{chapter}.\arabic{figure}}
\renewcommand{\thetable}{\arabic{chapter}.\arabic{table}}

\newcommand{\dd}{\partial}
\renewcommand{\d}{\partial}

\renewcommand{\geq}{\,{\geqslant}\,}
\renewcommand{\leq}{\,{\leqslant}\,}

\newcommand{\binner}[2]{%
  {\langle}\kern-4.15pt{\langle}#1{,}\,#2{\rangle}\kern-4.15pt{\rangle}}

\newcommand{\half}{\mathchoice{%
    \ffrac{1}{2}}{\frac{1}{2}}{\frac{1}{2}}{\frac{1}{2}}}
\newcommand{\ffrac}[2]{\raisebox{.5pt}%
  {\footnotesize$\displaystyle\frac{#1}{#2}$}\kern1pt}

\newcommand{\ddl}[2]{\ffrac{\dd #1}{\dd #2}}

\newcommand{\vdl}[1]{\ffrac{{\delta}}{\delta #1}}

\newcommand{\vddl}[2]{{\ffrac{\delta #1}{\delta #2}}}

\def\bea{\begin{eqnarray}}
\def\eea{\end{eqnarray}}
\newcommand{\bref}[1]{\textbf{\ref{#1}}}
\def\5{\bar }
\def\6{\partial }
\def\7{\hat }
\def\4{\tilde }

\def\cA{\mathcal{A}}

\def\cC{\mathcal{C}}

\def\cE{\mathcal{E}}
\def\cF{\mathcal{F}}

\def\cH{\mathcal{H}}
\def\cI{\mathcal{I}}
\def\cJ{\mathcal{J}}
\def\cK{\mathcal{K}}
\def\cL{\mathcal{L}}
\def\cM{\mathcal{M}}
\def\cN{\mathcal{N}}

\def\cP{\mathcal{P}}
\def\cQ{\mathcal{Q}}

\def\cS{\mathcal{S}}
\def\cT{\mathcal{T}}

\def\cV{\mathcal{V}}

\def\Q#1#2{\frac{\partial #1}{\partial #2}}
\def\QS#1#2{\frac{\partial^S #1}{\partial #2}}
\def\varQ#1#2{\frac{\delta #1}{\delta #2}}

\def\eps{\epsilon}

\def\dx{\text{d}\hspace{-0.06em}x}

\def\dv{\text{d}_V}

\def\dH{\text{d}_H}
\newcommand{\eqalign}[1]{\begin{array}{rcl} #1 \end{array}}

\def\ndelta{\delta\hspace{-0.50em}\slash\hspace{-0.05em} }

\newcommand{\SL}{\mbox{SL}(2,\mathbb{R})}
\newcommand{\abs}[1]{\lvert#1\rvert}

\newtheorem{prop}{Proposition}
\newtheorem{lemma}[prop]{Lemma}

\newtheorem{corollary}[prop]{Corollary}
\newtheorem{theorem}[prop]{Theorem}

%% file: title.tex
\begin{titlepage}

\enlargethispage{5cm}

\vspace*{-2.5 cm}

\begin{center}
 \LARGE Universit\'{e} libre de Bruxelles\\
         Facult\'{e} des Sciences
\end{center}
\vspace{1cm}
\begin{center}
\end{center}
\hfill
\\ \\ \\ \\

\begin{center}
\textbf{ \huge{\mbox{Symmetries and conservation laws}\\
\vspace{0.4cm} in Lagrangian gauge theories
\\\vspace{0.8cm}} \Large{with applications to the} \\\vspace{0.8cm}
\huge{Mechanics of black holes\\\vspace{0.2cm} \Large{and to}
\\\vspace{0.4cm}\huge{Gravity in three dimensions} }} \vspace{20pt}
\bigskip
\vspace{100pt}
\end{center}

\begin{flushright}
\Large{Ph.D. thesis of Geoffrey Comp\`{e}re}
\vspace{20pt}\\
\Large{Supervised by Glenn Barnich}
\end{flushright}
\bigskip
\vspace{1cm}
\begin{center}
\Large{
Bruxelles                     \\
Academic year 2006-2007} \\

\end{center}

\cleardoublepage
\end{titlepage}

%% file: acknowlegments.tex
\chapter*{Acknowledgements}

Each person who contributed to this piece of work and who made me
a bit happier during this period is worth thanking. That includes a lot
of people. I will venture to give some names I am fond of. I
already apologize for the ones I forgot to mention.

First, I would like to show all my gratitude to Glenn Barnich
who has really been a wonderful advisor. Among his professional
qualities, maybe his determination to go to the right point,
develop the relevant mathematical structure around it and progress
towards understanding is the lesson for research that I
will remember the most. For sure, the very existence of this
thesis is mainly due to the constant attention and support, to the
unerring motivation, to the patience and to the ideas of Glenn.

I'm also first thinking of my parents, my brothers and sister for
all they gave me. It is always good to be back home. Thanks also
to all the members of this huge great family!

I would also like to pay tribute to Jan Govaerts, Vera Spiller and
Nazim who developed my taste and my passion for searching for what
we know (and know that we don't know) about nature. I would not be
doing a Ph.D. thesis without a marvellous first research
experience directed by Fran\c{c}ois Dupret.

I thank my flatmates Arnaud, Nazim, Pierre, Bastien and Mathieu and
also Anne-Fran\c{c}oise, Aude and Selma for the marvellous time
spent during these four years, between diners, music, parties,
dvds and all these little things that make life beautiful. I also
give a special mention to my ``chers brainois'' Maya and Nicolas.

I would like to thank sincerely all the other Ph.D. students I was
lucky to meet in Brussels: ``the old team'' (sorry but that's
unfortunately true) Claire, Elizabete, G\'erald, Laura, Paola,
Pierre, Marc, Nazim, Sandrine, Sophie and St\'ephane and ``the new
team'' Cyril, Daniel, my ``officemate'' Ella, Jean, Luca, Nathan,
Nassiba, Pierre and Vincent for all lunches and diners we enjoyed.
I also benefited from the jokes and the incredible stories of the
post-docs Carlo, Chethan, Jarah and Stanislav. I address also my
acknowledgments to Bruno Bertrand, Serge Leclercq, Mauricio
Leston, Vincent Mathieu, Domenico Orlando, Benoit Roland,
Alexander Wijns and all members of the classroom of general
relativity I really enjoyed to supervise this year.

Thanks also to everybody I had enriching scientific discussions
with. Thanks especially to the staff of the section ``Physique
math\'ematique des interactions fondamentales'': Riccardo Argurio,
Glenn Barnich, Laurent Houart, Frank Ferrari, Marc Henneaux and
Christiane Schomblond. Thanks also to J.~Gegenberg, G.~Giribet, D.
Klemm, C.~Mart\'{\i}nez, R.~Olea, F.~Schlenk, P.~Spindel,
R.~Troncoso and M.~Tytgat. I thank Beno\^{\i}t, Ella, Michael,
Nicolas, Nazim, St\'ephane for the nice discussions that we had
and will hopefully have again. I also thank a lot
my collaborators Maximo Ba\~nados, Glenn Barnich, St\'ephane
Detournay and Andres Gomberoff for the nice piece of work we did
together. I also thank the String Theory Group of the Milano I
University for its hospitality.

I would really like to express my thanks to Fabienne, Isabelle et
St\'ephanie for the wonderful job they do in eternal good mood.

Be blessed founders of the Modave Summer School: Xavier Bekaert,
Vincent Bouchard, Nicolas Boulanger, Sandrine Cnockaert, Sophie de
Buyl, St\'ephane Detournay, Alex Wijns and Stijn Nevens. Thanks
also to all the participants of these schools. It was definitely a
good idea.

I also greatly acknowledge the National Fund for Scientific
Research of Belgium for the research fellow that allowed me to
make this thesis. I thank M. Henneaux for all conferences and
schools he allowed me to attend. I was also supported partly by a
``P{\^o}le d'Attraction Interuniversitaire'' (Belgium), by
IISN-Belgium, convention 4.4505.86, by Proyectos FONDECYT 1970151
and 7960001 (Chile) and by the European Commission program
MRTN-CT-2004-005104, with which I am associated together with V.U.~Brussel.

\vspace{2pt}\emph{Et finalement, merci \`a toi Chantal d'\^etre
\`a mes c\^ot\'es ici et l\`a-bas\dots}

%% file: structure_thesis.tex

\chapter*{Overview of the thesis}
\addcontentsline{toc}{chapter}{Overview of the thesis}

This thesis may be summarized by the questions it
addresses.\vspace{15pt}

{\it What is energy in general relativity ? How can it be
described in general terms? Is there a concept of energy
independent from the spacetime asymptotic structure? Valid in any
dimension and for any solution? Are there unambiguous notions of
conserved quantities in general gauge and gravity theories?

\vspace{10pt}

Are the laws of black hole mechanics universal in any theory of
gravitation? Why? What can one tell about the geometry of
spacetimes with closed timelike curves? Has three dimensional
gravity specific symmetries? What can classical symmetries tell
about the semi-classical limit of quantum gravity?}\vspace{15pt}

In a preamble, a quick summary of the line of thought from
Noether's theorems to modern views on conserved charges in gauge
theories is attempted. Most of the background material needed for
the thesis is set out through a small survey of the literature.
Emphasis is put on the concepts more than on the formalism, which
is relegated to the appendices.

The treatment of exact conservation laws in Lagrangian gauge
theories constitutes the main axis of the first part of the
thesis. The formalism is developed as a self-consistent theory but
is inspired by earlier works, mainly by cohomological results,
covariant phase space methods and by the Hamiltonian formalism.
The thermodynamical properties of black holes, especially the
first law, are studied in a general geometrical setting and are
worked out for several black objects: black holes, strings and
rings. Also, the geometrical and thermodynamical properties of a
new family of black holes with closed timelike curves in three
dimensions are described.

The second part of the thesis is the natural generalization of the
first part to asymptotic analyses. We start with a general
construction of covariant phase spaces admitting asymptotically
conserved charges. The representation of the asymptotic symmetry
algebra by a covariant Poisson bracket among the conserved charges
is then defined and is shown to admit generically central
extensions. The asymptotic structures of three three-dimensional
spacetimes are then studied in detail and the consequences for
quantum gravity in three dimensions are discussed.

%% file: preamble.tex
\chapter*{Preamble}
\addcontentsline{toc}{chapter}{Preamble}
\chaptermark{Preamble}
\renewcommand{\theequation}{\arabic{equation}}
\renewcommand{\thefigure}{\arabic{figure}}
\renewcommand{\thetable}{\arabic{table}}


\section{Conservation laws and symmetries}
\label{first_sec}

\begin{floatingfigure}{0.45\textwidth}
\begin{center}
\resizebox{0.40\textwidth}{!}{\mbox{\includegraphics{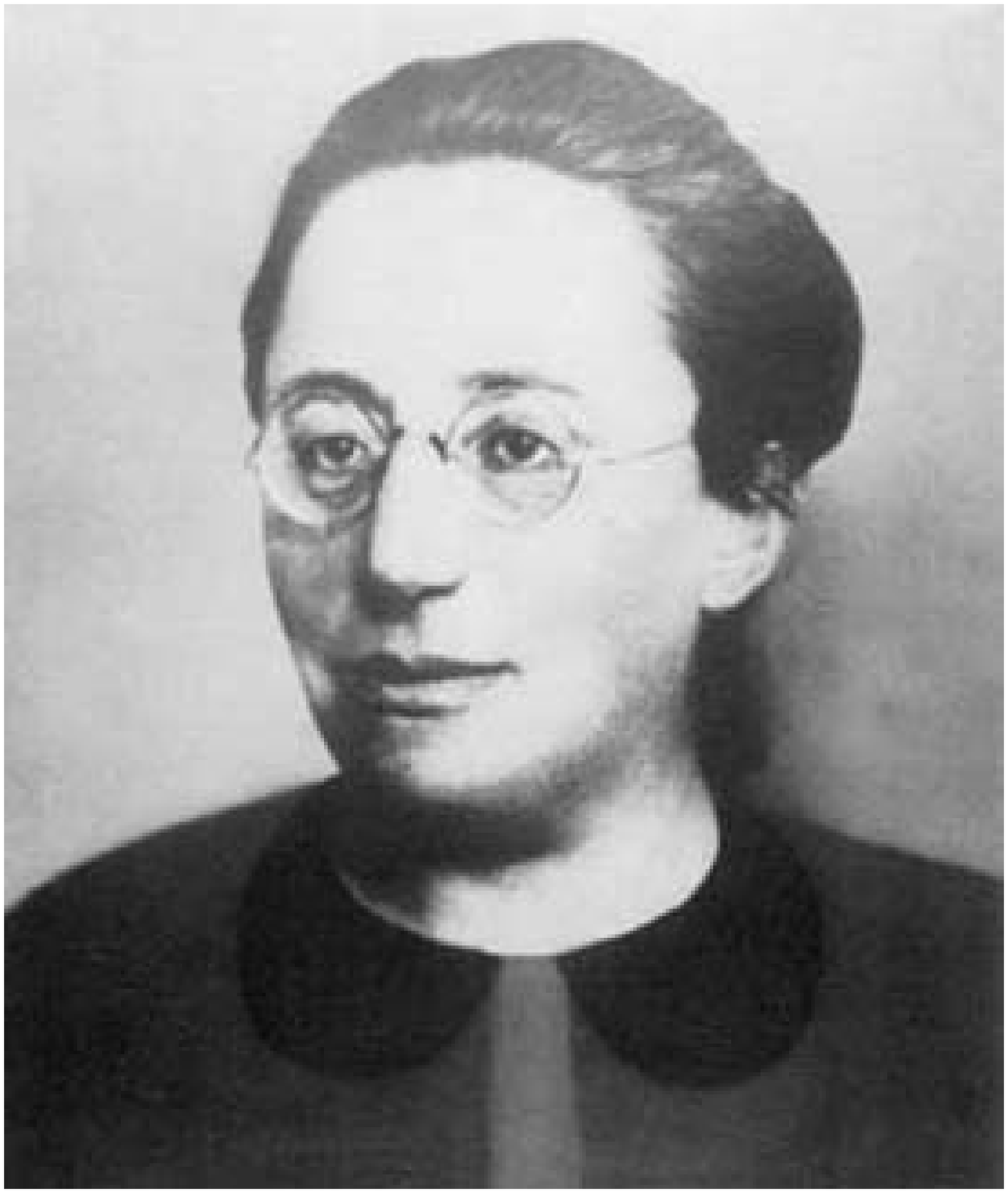}}}
\end{center}
~~~Emmy Noether [1882-1935]. \label{Noether}
\end{floatingfigure}
The concept of conservation law has a long and profound history in
physics. Whatever the physical laws considered: classical
mechanics, fluid mechanics, solid state physics, as well as
quantum mechanics, quantum field theory or general relativity,
whatever the constituents of the theory and the intricate
dynamical processes involved, quantities left dynamically
invariant have always been essential ingredients to describe
nature. The crowning conservation law, namely the constancy of the
total amount of energy of an isolated system, has been set up as
the first principle of thermodynamics and constitutes one of the
broadest-range physical law.

At the mathematical level, conservation laws are deeply connected
with the existence of a \emph{variational principle} which admits
\emph{symmetry transformations}. This crucial fact was fully
acknowledged by Emmy~Noether in 1918 \cite{Noether:1918aa}. Her
work, esteemed by F. Klein and D. Hilbert and remarked by Einstein
though hardly rewarded, provided a deep basis for the understanding
of global conservation laws in classical mechanics and in
classical field theories \cite{Taylor:1995aa,Byers:1998bd}. It
also prepared the ground to understanding the conservation laws in
Einstein gravity where the striking lack of local gravitational
stress-tensor called for further developments.

The essential ideas of linking symmetries and conservation laws
can be understood already in the classical description of a
mechanical system in the following way. Let $L[q^i,\dot q^i]$
denote the Lagrangian describing the motion of $n$ particles of
position $q^i$ and velocities $\dot q^i$. For a system
\emph{invariant under translations in time}, the total derivative
of the Lagrangian with respect to time $\frac{dL}{dt}$ contains
only the sum of implicit time variations $\Q{L}{q^i}\dot q^i$ and
$\Q{L}{\dot q^i}\ddot{q}^i$ for $i=1\dots n$. When the Lagrangian
equations hold, $\Q{L}{q^i} = \frac{d}{dt}\Q{L}{\dot q^i}$, the
time variation of the Lagrangian becomes $\frac{d}{dt}(\sum_i \dot
q^i \Q{L}{\dot q^i})$. The quantity $E \;\hat =\, \sum_i \dot q^i
\Q{L}{\dot q^i} - L$, the energy of the system, is then conserved
in time.

The same line of argument can be applied for an homogeneous and
isotropic in space action principle, which leads respectively to the conservation
of impulsion and angular momentum (see for
example~\cite{Landau:1982}). These arguments are applied equally to
non-relativistic or relativistic particles.

Similarly, conservation laws associated with global symmetries
appear in field theories. Let us consider the simple example of an
action principle depending at most on the first derivative of the
fields $I = \int d^nx\, L[\phi,\d_\mu\phi]$\footnote{All basic
definitions and conventions may be found in
Appendix~\ref{app:basicdef}.}. An infinitesimal transformation is
characterized by a transformation of the fields $\delta_X \phi^i =
X^i(x,[\phi])$\footnote{In this thesis, we consider infinitesimal
variations in characteristic form, see Appendix~\ref{app:basicdef}
for details.}. The transformation is called a global symmetry
if the Lagrangian is invariant under this transformation up to a
total derivative, $\delta_X L = \d_\mu k_X^\mu[\phi]$. Global
symmetries thus form a vector space.

As a main example, in relativistic field theories, the fields are
constrained to form a representation of the Poincar\'e group and
the Lagrangian has to be invariant (up to boundary terms) under
Poincar\'e transformations. The global symmetries for translations
and Lorentz transformations read respectively as
\begin{equation}\eqalign{
X^i[\d_\mu\phi] &=& -a^\mu \d_\mu \phi^i ,\\
X^i[x,\phi,\d_\mu\phi] &=& \half \omega_{\mu\nu}\left[ -(x^\mu
\eta^{\nu\alpha}- x^\nu \eta^{\mu\alpha} )\d_\alpha\phi^i + S^{i
\mu\nu}_j \phi^j \right],}\label{X_Poincare}
\end{equation}
where $a^\mu$, $\omega_{\mu\nu}= \omega_{[\mu\nu]}$ are the
constant parameters of the transformation $\delta x^\mu = a^\mu +
\omega^\mu_{\;\;\nu}x^\nu$, $\eta_{\mu\nu}$ is the Minkowski
metric used to raise and lower indices and $S^{i \mu\nu}_j $ are
the matrix elements of the representation of the Lorentz group to
which the fields $\phi^i$ belong. For a quick derivation see
e.g.~\cite{DiFrancesco:1997nk}.

Stated loosely, Noether's first theorem states that any global
symmetry corresponds to a conserved current. Indeed, by
definition, the variation $\delta_X L$ equals the sum of terms
$X^i \Q{L}{\phi^i}$ and $\d_\mu X^i \Q{L}{\d_\mu\phi^i}$. Using
the equations of motion, one then obtains that the current $j^\mu
\hat = X^i \Q{L}{\d_\mu\phi^i}-k^\mu_X$ is conserved on-shell,
$\d_\mu j^\mu \approx 0$. Using this current, one can define the
charge $\cQ = \int_\Sigma d^{n-1}x j^0$ on a spacelike surface
$\Sigma$ which is conserved, $\d_t \cQ = - \int_{\partial \Sigma}
d\sigma_i j^i = 0$ according to Stokes' theorem if the spatial
current vanishes at the boundary.

By way of example, associated with the translations and Lorentz
transformations ~\eqref{X_Poincare} is the current $j^\mu =
T^\mu_{\;\;\nu}a^\nu+\half j^{\mu\nu\rho}\omega_{\nu\rho}$ where
the canonical energy-momentum tensor $T^\mu_{\;\;\nu}$ and the
tensor $j^{\mu\nu\rho}$ are obtained as
\begin{equation}\eqalign{
T^\mu_{\;\;\nu} &=& \d_\nu \phi^i \Q{L}{\d_\mu \phi^i}- \delta^\mu_\nu L,\\
j^{\mu\nu\rho} &=& T^{\mu\nu}x^\rho - T^{\mu\rho}x^\nu + S^{i
\nu\rho}_j \phi^j \Q{L}{\d_\mu \phi^i}.}\label{j_Poincare}
\end{equation}
Remark that the energy $E = \int_\Sigma \d_0 \phi^i
\Q{L}{\d_0\phi^i} - L$ associated with $a^\mu = \delta^\mu_0$
correctly reduces to the mechanical expression in $0+1$ dimension.

In full generality, there is no bijective correspondence between
global symmetries and conserved currents. On the one hand, the
current $j^\mu$ is trivially zero in the case where the
characteristic of the transformation $X^i$ is a combination of the
equations of motion. On the other hand, one can associate with a
given symmetry the family of currents $j^\mu + \d_\nu
k^{[\mu\nu]}$ which are all conserved. It is nevertheless possible
to find quotient spaces where there is bijectivity. It is
necessary to first introduce the concept of gauge invariance.

A gauge theory is a Lagrangian theory such that its Euler-Lagrange
equations of motion admit non-trivial Noether identities, see
Appendix~\ref{app:fieldtheories} for definitions. Equivalently, as
a consequence of the second Noether theorem, a gauge theory is a
theory that admits non-trivial gauge transformations, i.e. linear
applications from the space of local functions to the vector space
of global symmetries of the Lagrangian. Vanishing on-shell
gauge transformations are defined as trivial gauge transformations.

Gauge transformations do not change the physics. It is therefore
natural to define equivalent global symmetries as symmetries of
the theory that differ by a gauge transformation. The resulting
quotient space is called the space of non-trivial global
symmetries.

On the other side, two currents $j^\mu$ and $j^{\prime \mu}$ will
be called equivalent if
\begin{equation}
j^\mu \sim j^{\prime \mu} + \partial_\nu k^{[\mu\nu]} +
t^\mu(\frac{\delta L}{\delta \phi}), \qquad t^\mu \approx
0,\label{ambiguity_Noether}
\end{equation}
where $t^\mu$ depends on the equations of motion. The complete
first Noether theorem can now be stated: \emph{There is an
isomorphism between equivalence classes of global symmetries and
equivalence classes of conserved currents (modulo constant
currents in dimension $n=1$)}. This theorem can be derived using
cohomological methods~\cite{Barnich:1995db,Barnich:1995ap}.

As a direct application of this theorem, one may consider tensors
equivalent to the energy-momentum tensor~\eqref{j_Poincare} which
differ by a divergence $\d_\rho B^{\rho\mu}_{\;\;\;\nu}$ with
$B^{\rho\mu}_{\;\;\;\nu} = B^{[\rho\mu]}_{\;\;\;\nu}$, and by a
tensor linear in the equations of motion and its derivatives
$t^{\mu}_{\;\;\nu}(\varQ{L}{\phi})$. This freedom may be used to
construct the so-called Belinfante stress-tensor $T_B^{\mu\nu}$
\cite{Belinfante:1940aa} which is symmetric in its two indices and
which satisfies $j^{\mu\nu\rho} = T_B^{\mu\nu}x^\rho -
T_B^{\mu\rho} x^\nu$, see discussions
in~\cite{Bak:1994gf,DiFrancesco:1997nk}.

Note also that there is a quantum counterpart to all these
classical considerations. However, we will not discuss these very
interesting issues in quantum field theory in this thesis.

\section{Puzzles in gauge theories}

In classical electromagnetism, besides the energy-momentum and the
angular momentum associated with global Poincar\'e symmetries
there is a conserved charge, the electric charge, associated with
the existence of a non-trivial Noether identity or, equivalently,
with the existence of a gauge freedom\footnote{See
Appendix~\ref{app:fieldtheories} for the background material used
in this section.}. Indeed, in arbitrary curvilinear coordinates,
the equations of motion read as $\d_\nu(\sqrt{-g}F^{\mu\nu}) =
4\pi \sqrt{-g}J^\mu $ where the charge-current vector $J^\mu$ has
to satisfy the continuity equation $\d_\mu (\sqrt{-g} J^\mu) = 0$
because of the Noether identity $\d_\mu
(\d_\nu(\sqrt{-g}F^{\mu\nu})) = 0$. The electric charge $\cQ$ can
be expressed as the integral over a Cauchy surface $\Sigma$
(usually of constant time),
\begin{equation}
\cQ = \int_\Sigma (d^{n-1}x)_\mu \sqrt{-g} J^\mu \approx
\frac{1}{4\pi} \int_{\partial \Sigma}
(d^{n-2}x)_{\mu\nu}\sqrt{-g}F^{\mu\nu},\label{elec_ch}
\end{equation}
where Stokes' theorem has been applied with $\partial \Sigma$ the
boundary of $\Sigma$, i.e. the $n-2$ sphere at spatial infinity.
Here we introduced the convenient notation
\begin{eqnarray*}
  (d^{n-p}x)_{\mu_1\dots\mu_p} \hat =
\frac 1{p!(n-p)!}\, \epsilon_{\mu_1\dots \mu_p \mu_{p+1}\cdots
\mu_n} dx^{\mu_{p+1}}\dots dx^{\mu_n},
\end{eqnarray*}
where $\epsilon_{\mu_1\dots\mu_n}$ is the numerically invariant
tensor with $\epsilon_{01\dots n-1} = 1$. Note that any current
$J^\mu$ can
 be reexpressed as a $n-1$ form $J = J^\mu (d^{n-1}x)_\mu$. A
conserved current $\d_\mu J^\mu = 0$ is equivalent to a closed
form $\text{d}J = 0$\footnote{In this section we will denote the
horizontal differential $\dH = dx^\mu \d_\mu$ simply as
$\text{d}$.}.

Noether's first theorem, however, cannot be used to describe this
conservation law. On the one hand, there is an
ambiguity~\eqref{ambiguity_Noether} in the choice of the conserved
current and, on the other hand, all gauge transformations are
thrown out of the quotient space of non-trivial global symmetries.
If one's derivation is based only on the first Noether theorem,
why would one choose the conserved current, $J^\mu$ in place of
$J^{\mu} = (\sqrt{-g})^{-1}\d_\nu k^{\mu\nu}$ with any $k^{\mu\nu}
= k^{[\mu\nu]}$, e.g. $k^{\mu\nu} = (4\pi)^{-1}\sqrt{-g}
F^{\mu\nu} + \text{const}\,
\sqrt{-g}F^{\alpha\beta}F_{\alpha\beta}F^{\mu\nu}$?

\begin{figure}[hbt]
\begin{center}
\resizebox{0.50\textwidth}{!}{\mbox{\includegraphics{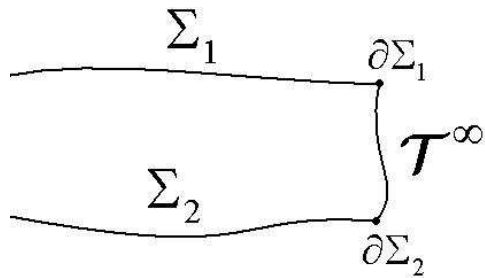}}}
\end{center}\caption{Two Cauchy surfaces $\Sigma_1$, $\Sigma_2$ and their intersection
with the spatial boundary $\cT^\infty$.} \label{CauchySurfaces}
\end{figure}

The problem can be cleared up by considering two Cauchy surfaces
$\Sigma_1$ and $\Sigma_2$ with boundaries $\d \Sigma_1$ and
$\d\Sigma_2$ and the $n-1$ surface at infinity $\cT^\infty$
joining $\d \Sigma_1$ and $\d\Sigma_2$, see
Fig.~\ref{CauchySurfaces}. Stokes' theorem implies the equality
\begin{equation*}
\int_{\d \Sigma_1} (d^{n-2}x)_{\mu\nu}\, k^{\mu\nu} - \int_{\d
\Sigma_2} (d^{n-2}x)_{\mu\nu}\, k^{\mu\nu} = \int_{\cT^\infty}
(d^{n-1}x)_{\mu} \sqrt{-g} \,J^{\mu}.
\end{equation*}
Now, for the integral of $k^{\mu\nu}$ to be a conserved
quantity, the right-hand side has to vanish on-shell. This is true
for $k^{\mu\nu} = (4\pi)^{-1}\sqrt{-g}F^{\mu\nu}$ because the
Noether current $J^\mu$ vanishes on-shell outside the sources but
it is not true for arbitrary $k^{\mu\nu}$. The point is that the
conservation of electric charge is a \emph{lower degree
conservation law}, i.e. not based on the conservation of a $n-1$
form $J = J^\mu (d^{n-1}x)_\mu$, but on the conservation of a
$n-2$ form $k = k^{\mu\nu}(d^{n-2}x)_{\mu\nu}$ with $dk = \d_\nu
k^{\mu\nu} (d^{n-1}x)_{\mu} \approx 0$.

The proof of uniqueness of the conserved $n-2$ form $k$ and its
relation to the gauge freedom of the theory goes beyond standard
Noether theorems even if they show part of the answer.

General relativity also admits gauge freedom, namely
diffeomorphism invariance. Infinitesimal transformations under
characteristic form $\delta_\xi g_{\mu\nu} = \cL_\xi g_{\mu\nu}$
are parameterized by arbitrary vector fields $\xi^\mu$. Here, a
straightforward application of the first Noether theorem fails to
provide even a proposal for a conserved quantity associated with
this gauge invariance.

More precisely, the variation of the Einstein-Hilbert Lagrangian
is given by $\delta L_{EH} = \delta g_{\mu\nu} \frac{\delta
L_{EH}}{\delta g_{\mu\nu}} + \partial_\mu\Theta^\mu(g,\delta g) $
with $\Theta^\mu(g,\delta g) = 2 \sqrt{-g}g^{\alpha[\beta}\delta
\Gamma^{\mu]}_{\alpha\beta}$\footnote{We use in this section units
such that $G = (16\pi)^{-1}$.} and $\frac{\delta L_{EH}}{\delta
g_{\mu\nu}} = -\sqrt{-g}G^{\mu\nu}$. For a diffeomorphism, one has
$\delta_\xi ( L_{EH} d^n x) = \text{d}i_\xi (L_{EH} d^n x )$ $=
\partial_\mu ( \xi^\mu L_{EH}) d^n x$ and $\delta_\xi
g_{\mu\nu}\frac{\delta L_{EH}}{\delta g_{\mu\nu}}  =
-2\partial_{\mu}(\sqrt{-g} G^{\mu\nu}\xi_\nu)$. The canonical
Noether current is then $J_\xi^\mu = \Theta^\mu(g,\mathcal L_\xi
g) - \xi^\mu L_{EH}$. By construction, it satisfies $\d_\mu
J_\xi^\mu = \d_\mu (2\sqrt{-g}G^{\mu\nu}\xi_\nu)$. Using the
algebraic Poincar\'e lemma, see Theorem~\ref{poincarelemma} on
page~\pageref{poincarelemma}, the Noether current can be written
as
\begin{equation*}
J_\xi^\mu =2\sqrt{-g}G^{\mu\nu}\xi_\nu + \d_\nu k^{[\mu\nu]},
\end{equation*}
for some skew-symmetric $k^{\mu\nu}$. An idea is to define the
charge associated with $\xi$ as $\int_{S^\infty}
(d^{n-2}x)_{\mu\nu}k_\xi^{\mu\nu}$ where $S^\infty$ is the sphere
at spatial infinity. However, this definition is completely
arbitrary. Indeed, since the Noether current is determined up to
the ambiguity~\eqref{ambiguity_Noether}, the Noether current
$J_\xi^\mu$ could also have been chosen to be zero.

A conserved surface charge can be defined from a conserved
superpotential $k_\xi^{\mu\nu}= k_\xi^{[\mu\nu]}$ such that
$\d_\nu k_\xi^{\mu\nu} \approx 0$. This superpotential would have
to be different from a total divergence $k^{\mu\nu} \approx
\d_\rho l^{[\mu\nu\rho]}$ for the charge to be non-trivial. The
point is that the Noether theorem is mute about the choice or, at
least the existence of a special choice, for this superpotential.

In the early relativity literature, conservation laws for
(four-dimensional) spacetimes which admit an expansion $g_{\mu\nu}
= \eta_{\mu\nu}+O(1/r)$ close to infinity\footnote{Some additional
conditions are required on the time-dependence and on the behavior
under parity of $g_{\mu\nu}$, see~\cite{Regge:1974zd} for detailed
boundary conditions for asymptotically flat spacetimes at spatial
infinity.} were given in terms of pseudo-tensors, i.e.
coordinate-dependent quantities $k^{\mu\nu}$ which are invariant
under diffeomorphisms vanishing fast enough at infinity and
which are covariant under Poincar\'e transformations at infinity.
A first pseudo-tensor was found by Einstein and many others where
built up afterwards, see for
example~\cite{Arnowitt:1961aa,Trautman::1962aa} for a synthesis,
see also~\cite{Brown:1993br} for a list of references. Note that
quasi-local methods present a modern point of view on
pseudo-tensors~\cite{Chang:1998wj}. This approach, which was quite
successful to describe the conserved momentum and angular momentum
of asymptotically flat spacetimes, unfortunately suffers from
serious drawbacks, e.g. the need for defining a rectangular
coordinate system at infinity, the profusion of alternative
definitions for $k^{\mu\nu}$, the lack of articulation with
respect to the gauge structure of the theory, the difficulties to
generalize and link the definition to other asymptotics, etc.

In Yang-Mills theory, similar problems as in general relativity
mainly arise because, as will be cleared later, the gauge
transformations involve the fields of the theory.

Nevertheless, a useful formula for the conserved quantity
associated with an exact Killing vector for a solution of
Einstein's equations in vacuum was given by Komar
\cite{Komar:1958wp}. This expression provided a sufficient tool to
unravel the thermodynamical properties of black holes
\cite{Bardeen:1973gs}. Unfortunately, this formula is only valid
in symmetric spacetimes without cosmological constant and one
needs to compare it with other definitions, e.g.
\cite{Arnowitt:1962aa}, in order to get the factors right.

All these puzzles called for further developments.

\section{Results in local cohomology}
\label{sec:localhomo}

A very convenient mathematical setting to deal with $(n-1)$ or
$(n-2)$-form conservation laws or more generally $p$-form
conservation laws ($0 \leq p < n $) is the study of local
cohomology in field theories. This subject was first developed in
the mathematical literature and research was proceeded by
physicists as well in the eighties and nineties
\cite{Vinogradov:1978,Vinogradov:1984,Tsujishita:1982aa,Wald:1990ic,Anderson1991,
Andersonbook,Bryant:1995,Barnich:1995db,Barnich:1995ap}. A
self-contained summary of important definitions and propositions
can be found in Appendix~\ref{app:basicdef}, see also
\cite{Torre:1997cd} for a pedagogical introduction to local
cohomology.

Two sets of conservations laws in field theories can be
distinguished: the so-called topological conservation laws and the
dynamical conservation laws. Topological conservation laws are
equivalence classes of $p$-forms $\omega$ which are identically
closed $\text{d}\omega = 0$ modulo exact forms $\omega =
\text{d}\omega^\prime$, irrespectively of the field equations of
the theory. These laws reflect topological properties of the
bundle of fields or of the base manifold itself. For example, if
the bundle of fields is a vector bundle, only the base manifold
can provide non-trivial cohomology and no interesting, i.e.
field-dependent, topological conservation laws appear
\cite{Wald:1990ic,Torre:1997cd}.

A famous example of topological conservation law is the ``kink
number'' first obtained by Finkelstein and Misner
\cite{Finkelstein:1959aa}. As described in~\cite{Torre:1994pf},
the bundle of Lorentzian signature metrics over a $n$-dimensional
manifold admits a cohomology isomorphic to the Rham cohomology of
$RP^{n-1}$. The only non-trivial cohomology is given by a $n-1$
form, a conserved current, in the case where $n$ is even. The kink
number is then defined as the integral of this form on a
$(n-1)$-dimensional surface. It can be shown to be an integer.
Remark that in the vielbein formulation of gravity, topological
charges are due to the constraint $\text{det}(e^\mu_a) > 0$ on the
vielbein manifold. The set of topological conserved p-forms is
then larger because it also contains non-invariant forms under
local Lorentz transformations of the
vielbein~\cite{Barnich:1995ap}.

Topological conservation laws mentioned here for completeness will
not be considered hereafter.

More fruitful are the dynamical conservations laws defined as the
conservation laws where the equations of motion are explicitly
used. The cohomology of closed forms on-shell $\text{d}\omega
\approx 0$ modulo exact forms on-shell $\omega \approx
\text{d}\omega^\prime$ is called the characteristic cohomology
$H^{n-p}_{char}(\text{d})$ on the stationary surface in form
degree $n-p$.

The cohomology $H^{n-1}_{char}(\text{d})$ is nothing but the
cohomology of non-trivial conserved currents which can be shown to
be equal to the cohomology of global symmetries of the theory.
This is in essence the first Noether theorem that was already
described in section~\ref{first_sec}. Note that for general
relativity, this cohomology is trivial as a consequence of the
nonexistence of non-trivial global
symmetries~\cite{Torre:1993jm,Anderson:1996eg}. In free theories,
this cohomology may be infinite-dimensional and can be difficult
to compute even for the Maxwell case
\cite{Lipkin,Morgan,Kibble,OConnell,Anco:2001}.

For a very large class of Lagrangians including Dirac,
Klein-Gordon, Chern-Simons, Yang-Mills or general relativity
theories which satisfy appropriate regularity conditions, the
cohomologies $H^{n-p}_{char}(\text{d})$ may be studied by tools
inspired from BRST methods~\cite{Barnich:1995ap,Barnich:2000zw}.
Each element of the cohomology $H^{n-2}_{char}(\text{d})$ can be
related to a non-trivial reducibility parameter of the theory,
i.e. a parameter of a gauge transformation vanishing on-shell such
that the parameter itself is non zero on-shell. For irreducible
gauge theories, this cohomology entirely specifies the
characteristic cohomology in degree $p < n-1$, in particular, in
Yang-Mills and Einstein theories.

For Maxwell's theory, a reducibility parameter $c$ for the gauge
field $A_\mu$ exists,
\begin{equation*}
\delta A_\mu = \d_\mu c \approx 0,
\end{equation*}
and is unique: $c = 1$ (up to a multiplicative constant that can
be absorbed in the choice of units). The associated conserved
$n-2$ form is obviously the electric charge~\eqref{elec_ch}. For
Einstein gravity or for Yang-Mills theory with a semi-simple gauge
group, no reducibility parameter exists, i.e.
\begin{eqnarray*}
\delta g_{\mu\nu} = D_\mu \xi_\nu +   D_\nu \xi_\mu \approx 0,\\
\delta A_\mu = \d_\mu \lambda^a + f^a_{bc}A^b_\mu \lambda^c
\approx 0,
\end{eqnarray*}
implies $\xi^\mu \approx 0$ and $\lambda^a \approx 0$. Stated
differently, no vector is a Killing vector of all solutions of
Einstein's equations, even when the Killing equation is only
imposed ``on-shell''. Because the reducibility equations for the
Yang-Mills case also depend on arbitrary fields, there are no
reducibility parameters either. As a consequence, there is no
general formula for a non-trivial conserved $n-2$ form locally
constructed from the fields in these theories.

This explains \emph{a posteriori} the insurmountable difficulties
people encountered when trying to define the analogue of the
energy-momentum tensor~\eqref{j_Poincare} for the gravitational
field. This impossibility was celebrated in Misner, Thorne and
Wheeler \cite{Misner:1970aa} in the quotation ``\textit{Anybody
who looks for a magic formula for ``local gravitational
energy-momentum'' is looking for the right answer to the wrong
question. Unhappily, enormous time and effort were devoted in the
past to trying to ``answer this question'' before investigators
realized the futility of the enterprise}''.

The lack of local energy-momentum tensor does not prevent,
however, the definition of conserved quantities for restricted
classes of spacetimes as the spacetimes admitting a Killing vector
(e.g. Komar integrals) or the spacetimes admitting a common
asymptotic structure (e.g. global energy-momentum for
asymptotically flat spacetimes) as we will explain below.

In the case of free or interacting $p$-form theories, the lower
degree cohomologies acquire importance because of the reducibility
of the gauge theory. In that case, the characteristic cohomologies
$H^{n-p}_{char}(\text{d})$ in form degree $p < n-1$ are generated
(in the exterior product) by the forms $\star \,H^a$ dual to the
field strengths $H^a$~\cite{Henneaux:1996ws}. More details on
conservations laws in $p$-form gauge theories will be given in
section~\ref{sec:pform} of Chapter~\ref{chap:matter}.

\section{Windows on the literature}

There is an impressive literature on conservations laws in
general relativity, see e.g. the review~\cite{Szabados:2004vb}.
Several lines of research have been followed, often with
intertwining and mutual progress. Some results such as the ADM
energy-momentum~\cite{Arnowitt:1962aa} for asymptotically flat
spacetimes or the Abbott-Deser charges for asymptotically anti-de
Sitter spacetimes~\cite{Abbott:1981ff} are seen as bench
marks that should be included within any viable theory of
conserved charges.

In the following paragraphs, the methods that are significant and
relevant for the thesis will be briefly set out. They will be
organized along the chronological order of their seminal work.
Certainly, this succinct presentation will be biased by personal
preferences and unintentional oversights.

A major progress towards the understanding of asymptotically
conserved quantities in general relativity was achieved by
Arnowitt, Deser and Misner~\cite{Arnowitt:1962aa}. These authors
reformulated general relativity in Hamiltonian terms and
identified the canonical generator conjugated to time displacement
at spatial infinity for asymptotically flat spacetimes.
In~\cite{Regge:1974zd}, Regge and Teitelboim provided a criteria,
namely the differentiability of the Hamiltonian, to uniquely
identify the surface terms to be added to the weakly vanishing
Hamiltonian associated with any asymptotic Poincar\'e
transformation at spatial infinity. Hamiltonian methods were later
successfully applied to asymptotically anti-de Sitter spacetimes
\cite{Henneaux:1985tv,Henneaux:1985ey}. The canonical theory of
representation of the Lie algebra of asymptotic symmetries by the
possibly centrally extended Poisson bracket of the canonical
generators was done in~\cite{Brown:1986ed,Brown:1986nw}. The
analysis of flat spacetimes was refined in later
works~\cite{Beig:1987aa,Szabados:2003yn} in which covariance was
kept manifest and boundary conditions were weakened.

An elegant construction to investigate the asymptotic structure of
spacetimes at null infinity was developed by
Penrose~\cite{Penrose:1962ij} inspired from the work of Bondi, van
der Burg and Metzner~\cite{Bondi:1962px}. It consisted in adding
to the physical spacetime a suitable conformal boundary. Conformal
methods were also developed for spatial
infinity~\cite{Geroch:1977aa} and the quantities constructed at
spatial and null infinity were
related~\cite{Ashtekar:1978aa,Christodoulou1993}. A review of
various constructions can be found in~\cite{Ashtekar1980}. An
alternative definition of spatial infinity was also given
in~\cite{Ashtekar:1991vb}. These methods were also successful to
describe conserved quantities in anti-de Sitter spacetimes by
using the electric part of the Weyl
tensor~\cite{Ashtekar:1984,Ashtekar:1999jx}.

A manifestly covariant approach was developed by Abbott and
Deser~\cite{Abbott:1981ff} by manipulating the linearized Einstein
equations. The method provided the first completely satisfactory
framework to study charges in anti-de Sitter spacetimes. A similar
line of argument led to the definition of charges in non-abelian gauge
theories~\cite{Abbott:1982jh}. Recently, higher curvature theories
were investigated~\cite{Deser:2002jk,Deser:2002rt,
Deruelle:2003ps,Deser:2007vs}.

A spinorial definition for energy was given
in~\cite{Nester:1982tr,Gibbons:1982jg,Gibbons:1983aq} following
the positive energy theorems proven
in~\cite{Schoen:1979aa,Witten:1981mf}. Positivity of energy in
locally asymptotically anti-de Sitter spacetimes was recently studied
in~\cite{Cheng:2005wk}.

Covariant phase space methods, also denoted as covariant
symplectic methods,~\cite{Crnkovic:1986be,
Crnkovic:1986ex,Ashtekar:1991aa} provided a powerful Hamiltonian
framework embedded in a covariant formalism. The study of local
symmetries~\cite{Lee:1990nz} in Lagrangian field theory led to
significant developments in general diffeomorphic invariant
theories~\cite{Wald:1993nt,Iyer:1994ys,Wald:1999wa}, see
also~\cite{Iyer:1995kg} for a comparison with Euclidean methods.
The representation of the Lie algebra of asymptotic symmetries
with a covariant Poisson bracket was developed
in~\cite{Koga:2001vq}. In first order theories, a prescription
depending only on the equations of motion was
given~\cite{Julia:1998ys,Silva:1998ii, Julia:2000er,Julia:2002df}
in order to define the integrated superpotential corresponding to
an arbitrary asymptotic symmetry. Fermionic charges were included
in the covariant phase space formalism recently
in~\cite{Hollands:2006zu}.

Quasi-local quantities, i.e. quantities defined with respect to a
bounded region of spacetime, may be defined by employing a
Hamilton-Jacobi analysis of the action
functional~\cite{Brown:1993br,Brown:2000dz}. These definitions are
in particular very suitable to perform numerical calculations for
realistic configurations. The covariant symplectic methods were
applied also for spatially bounded regions
in~\cite{Anco:2001gk,Anco:2001gm}.

Charges for flat and anti-de Sitter spacetimes have been defined
directly from the
action~\cite{Hawking:1995fd,Balasubramanian:1999re,Aros:1999id,Aros:1999kt}
after having prescribed the boundary terms to be added to the
Lagrangian.

Finally, cohomological techniques began with the observation
of~Anderson and Torre~\cite{Anderson:1996sc} that asymptotic
conservation laws can be understood as cohomology groups of the
variational bicomplex pulled back to the surface defined by the
equations of motion. Conservation laws and central extensions for
asymptotically linear configurations in irreducible Lagrangian
gauge theories were investigated in \cite{Barnich:2001jy} using
BRST techniques. Conserved charges associated with exact
symmetries were studied in~\cite{Barnich:2003xg}. A non-linear
theory for exact and asymptotic symmetries was developed
in~\cite{Barnich:2007bf}.

Different methods that apply to anti-de Sitter spacetimes have
been compared in detail in~\cite{Hollands:2005wt}. See
also~\cite{Papadimitriou:2005ii} for a link between counterterm
methods and covariant phase space techniques.

%
%

\section{The central idea: the linearized theory}

The Hamiltonian framework~\cite{Arnowitt:1962aa,Regge:1974zd} as
well as covariant
methods~\cite{Abbott:1981ff,Wald:1993nt,Anderson:1996sc} directly
or indirectly make use of the linearized theory around a reference
field. The linearized theory is either used as an approximation to
the full theory at the infinite distance boundary or as the first
order approximation when performing infinitesimal field
variations. This is the main theme underlying the present thesis.

In comparison to the full interacting theory, possibilities of
occurrence of conserved $n-2$ forms in the linearized theory are
greatly enhanced. Indeed, in that case, the characteristic
cohomology $H^{n-2}_{char}(\text{d})$ is determined by the
solutions of the reducibility equations of the linear theory which
may admit non-trivial solutions if the reference field is
symmetric, see
Appendix~\ref{app:fieldtheories}.\ref{app:linearized}. Moreover,
the conserved surface charges in regular gauge theories are
entirely classified by this cohomology.

For example, in Yang-Mills theory, the linearized theory around
the flat connection $A = g^{-1} \text{d} g$ admits $N$
reducibility parameters where $N$ is the number of generators of
the gauge group~\cite{Abbott:1982jh,Barnich:2001jy}. The
associated charges, however, are not very illuminating since they
vanish in interesting cases~\cite{Abbott:1982jh}.

In Einstein gravity, the reducibility equations of the linearized
theory around a reference solution $\bar g_{\mu\nu}$ admit as only
solutions the Killing vectors of the
background~\cite{Anderson:1996sc,Barnich:2004ts}. This completely determines
the characteristic cohomology in that case and,
therefore, provides unique expressions (up to trivialities) for
the conserved $n-2$ forms. Also, in higher spin fields theories,
$s > 2$, conserved $n-2$ forms are in one-to-one correspondence
with dynamical Killing tensors~\cite{Barnich:2005bn}.

In the full non-linear theory, the surface charges of the
linearized theory can be re-interpreted as one-forms in field
space, the appropriate mathematical framework being the
variational bicomplex associated with a set of Euler-Lagrange
equations, see Appendix~\ref{app:basicdef} for a summary. Two
different approaches make use of these charges one-forms.

An old successful method used in the asymptotic context, e.g.
\cite{Arnowitt:1962aa,Regge:1974zd}, consists in integrating
infinitesimal charge variations at infinity to get charge
differences between the background and the solutions of interest
by using boundary conditions on the fields so as to ensure
convergence, conservation and representation properties of the
charges. Besides the Hamiltonian framework, similar results have
been obtained in Lagrangian formalism, e.g.~\cite{Barnich:2001jy}
where detailed criteria for the applicability of the linearized
theory at the boundary have been studied.

Another approach, followed e.g. by Komar~\cite{Komar:1958wp}, consists in
considering a mini-superspace of solutions admitting a set of
Killing vectors of a reference solution~\cite{Barnich:2003xg}.
Finite charge differences generalizing Komar integrals can be
defined if a suitable integrability condition
hold~\cite{Wald:1999wa,Barnich:2003xg}. This allows one, e.g., to
derive more generally the first law of black holes
mechanics~\cite{Barnich:2004uw}.

We now turn to the formalism where these ideas will be developed
in more mathematical terms.

%% file: general_formalism.tex
\chapter{Classical theory of surface charges}
\label{chap:general_th}
\setcounter{equation}{0}\setcounter{figure}{0}\setcounter{table}{0}

We develop in this chapter a ``cohomological'' treatment of exact
symmetries in Lagrangian gauge theories. The extension to
asymptotic analyses is done in the second part of the thesis.

We begin by reviewing the construction of Noether charges for
global symmetries and we recall how central charges appear in that
context. We then fix our description of irreducible gauge theories
and recall that Noether currents associated with gauge symmetries
can be chosen to vanish on-shell. Surface one-forms, which are
$(n-2)$-forms in base space and one-forms in field space, are
constructed next from the weakly vanishing Noether currents. The
integrals of these surface one-forms on closed surfaces are the
surface charge one-forms which constitute the cornerstone in our
description of conservation laws in gauge theories.

In order to be self-contained, some results established in
\cite{Barnich:2001jy} are rederived, independently of BRST
cohomological methods: reducibility parameters, e.g. Killing
vectors in gravitation, form a Lie algebra and surface charge
one-forms associated with reducibility parameters are conserved
and represent the Lie algebra of reducibility parameters. A result
of \cite{Barnich:2001jy} is also recalled without
proof\footnote{As stated in Appendix~\ref{app:basicdef}, we assume
that the fiber bundle of fields is trivial.}: each equivalence
class of, local, closed $(n-2)$-forms modulo, local, exact
$(n-2)$-forms is associated with a reducibility parameter and
representatives for these conserved forms are given by the surface
one-forms.

The surface charge one-forms are constructed from the
Euler-Lagrange derivatives of the Lagrangian and thus do not
depend on total divergences added to the Lagrangian. So, from the
outset, our approach is free from the troublesome ambiguities of
covariant phase space
methods~\cite{Wald:1993nt,Iyer:1994ys,Wald:1999wa}. This property
may also be understood from the link between the surface charge
one-forms and what we call the invariant presymplectic $(n-1,2)$
form, distinguished from the usual covariant phase space
presymplectic $(n-1,2)$-form.

In another connection, the Hamiltonian
prescription~\cite{Regge:1974zd,Henneaux:1985tv,Henneaux:1985ey}
to define the surface charges is shown to be equivalent to our
definition. In that sense, our formalism provides the Lagrangian
counterpart of the Hamiltonian framework.

For first order actions, our definition of surface charges reduces
to the definition of~\cite{Julia:1998ys,Silva:1998ii,
Julia:2000er,Julia:2002df} which was motivated by the Hamiltonian
formalism. Because our formalism does not assume the action to be
of first order it therefore extends this proposal to Lagrangians
with higher order derivatives.

In the last section, we define the surface charges related to a
family of solutions admitting reducibility parameters by
integrating the surface charge one-forms along a path starting
from a reference solution. We explain how these charges are
well-defined if integrability conditions for the surface charge
one-forms are fulfilled. These conditions have been originally
discussed for surface charge one-forms associated with fixed
vector fields in the context of diffeomorphic invariant
theories~\cite{Wald:1999wa}. Here we point out that for a given
set of gauge fields and gauge parameters, the surface charge
one-forms should be considered as a Pfaff system and that
integrability is governed by Frobenius' theorem. This gives the
whole subject a thermodynamical flavor, which we emphasize by our
notation $\ndelta \cQ_f[\dv\phi]$ for the surface charge
one-forms. Eventually, we discuss some properties of the surface
charges and point out their relation to quantities defined at
infinity.

In Appendix~\ref{app:basicdef}, we give elementary definitions of
jet spaces, horizontal complex, variational bicomplex and homotopy
operators. We fix notations and conventions and recall the
relevant formulae. In particular, we prove crucial properties of
the invariant presymplectic $(n-1,2)$ form associated with the
Euler-Lagrange equations of motion. Some properties of classical
gauge theories are summarized in Appendix~\ref{app:fieldtheories}.

\section{Global symmetries and Noether currents}
\label{sec:glob-symm-charge}

In a Lagrangian field theory, the dynamics is generated from a
distinguished $n$-form, the Lagrangian $\cL=L\,d^nx$, through the
Euler-Lagrange equations of motion
\begin{equation}
  \label{eq:34}
  \vddl{L}{\phi^i}=0.
\end{equation}
A global symmetry $X$ is a vector field under characteristic form
(see~\eqref{char_def}) satisfying the condition $\delta_X \cL=\dH
k_X$. The Noether current $j_X$ is then defined through the
relation
\begin{eqnarray}
X^i\vddl{\cL}{\phi^i}=\dH j_X.\label{eq:31}
\end{eqnarray}
A particular solution is $j_X=k_X-I^n_X(\cL)$. Here, the operator
\[I^n_X(\cL)=(X^i\QS{L}{\phi^i_\mu}+\dots)(d^{n-1}x)_\mu,\]
is defined by equation \eqref{phihomotopy} for Lagrangians
depending on more than first order derivatives.
  Applying
$\delta_{X_1}$ to the definition of the Noether current for $X_2$
and using \eqref{eq:32} together with the facts that ${X_1}$ is a
global symmetry and that Euler Lagrange derivatives annihilate
$d_H$ exact $n$ forms, we get
\begin{eqnarray}
  \label{eq:alg_curr}
  \dH\Big(\delta_{X_1}j_{X_2}-j_{[X_1,X_2]}
-T_{X_1}[X_2,\vddl{\cL}{\phi}]\Big)=0,
\end{eqnarray}
with $T_{X_1}[X_2,\vddl{\cL}{\phi}]$ linear and homogeneous in the
Euler-Lagrange derivatives of the Lagrangian and defined in
\eqref{eq:27}. Under appropriate regularity conditions on the
Euler-Lagrange equations of
motion~\cite{Henneaux:1992ig,Barnich:2000zw}, which we always
assume to be fulfilled, two local functions are equal on-shell
$f\approx g$ if and only if $f$ and $g$ differ by terms that are
linear and homogeneous in $\vddl{L}{\phi^i}$ and their
derivatives. If the expression in parenthesis on the l.h.s of
\eqref{eq:alg_curr} is $\dH$ exact, we get the usual algebra of
currents on-shell
\begin{equation}
\delta_{X_1}j_{X_2} \approx j_{[X_1,X_2]} + \dH(\cdot
).\label{alg_cur}
\end{equation}

The origin of classical central charges in the context of Noether
charges associated with global symmetries are the obstructions for
the latter expression to be $\dH$ exact, i.e., the cohomology of
$\dH$ in the space of local forms of degree $n-1$. This cohomology
is isomorphic to the Rham cohomology in degree $n-1$ of the fiber
bundle of fields (local coordinates $\phi^i$) over the base space
$M$ (local coordinates $x^\mu$), see
e.g.~\cite{Andersonbook,Anderson1991}.

The case of classical Hamiltonian mechanics, $n=1$, $\cL=(p\dot
q-H)dt$ is discussed for instance in \cite{Arnoldbook}. Examples
in higher dimensions can be found in \cite{deAzcarraga:1989gm}.

\section{Gauge symmetries and vanishing Noether currents}
\label{sec:gauge-symmetries}

In order to describe gauge theories, one needs besides the fields
$\phi^i(x)$ the gauge parameters $f^\alpha(x)$. Instead of
considering the gauge parameters as additional arbitrary functions
of $x$, it is useful to extend the jet-bundle. Because we want to
consider commutation relations involving gauge symmetries, several
copies $f^\alpha_{a(\mu)}$, $a=1,2,3\dots$, of the jet-coordinates
associated with gauge parameters are
needed\footnote{Alternatively, one could make the coordinates
$f^\alpha_{(\mu)}$ Grassmann odd, but we will not do so here. For
expressions involving a single gauge parameter we will often omit
the index $a$ in order to simplify the notation.}. We will denote
the whole set of fields as $\Phi^\Delta_a=(\phi^i,f^\alpha_a)$ and
we will extend the variational bicomplex to this complete set,
e.g. $\dv^{\Phi}$ is defined in terms of $\Phi^\Delta_a$ and thus
also involve the $f^\alpha_a$. When $\dv^{\Phi}$ it is restricted
to act on the fields $\phi^i$ and their derivatives alone, we
denote it by $\dv$.

Let $\delta_{R_f}\phi^i=R^i_f$ be characteristics that depend
linearly and homogeneously on the new jet-coordinates
$f^\alpha_{(\mu)}$,
\begin{equation}
R^i_f=R^{i(\mu)}_\alpha f^\alpha_{(\mu)}.
\end{equation}
We assume that these characteristics define a generating set of
gauge symmetries of $\cL$\footnote{This means that they define
symmetries and that every other symmetry $Q_f$ that depends
linearly and homogeneously on an arbitrary gauge parameter $f$ is
given by $Q^i_f=R^{i(\mu)}_\alpha\partial_{(\mu)}
Z^\alpha_f+M^{+i}_f[\vddl{L}{\phi}]$ with $Z_f^\alpha =
Z^{\alpha(\nu)}f_{(\nu)}$ and
$M^{+i}_f[\vddl{L}{\phi}]=(-\partial)_{(\mu)}\Big(M_f^{[j(\nu)i(\mu)]}
\6_{(\nu)}\vddl{L}{\phi^j}\Big)$, see
e.g.~\cite{Henneaux:1992ig,Barnich:2000zw} for more details.}. For
simplicity, we assume the generating set in addition to be
irreducible\footnote{If $R^{i(\mu)}_\alpha\partial_{(\mu)}
Z^\alpha_f\approx 0$, where $\approx 0$ means zero for all
solutions of the Euler-Lagrange equations of motion, then
$Z^\alpha_f\approx 0$.}.

Because we have assumed that $\delta_{R_f} \phi^i=R^i_f$ provides
a generating set of non trivial gauge symmetries, the commutator
algebra of the non trivial gauge symmetries closes on-shell in the
sense that \bea \delta_{R_{f_1}}R^i_{f_2}-
\delta_{R_{f_2}}R^i_{f_1}
=-R^i_{[f_1,f_2]}+M^{+i}_{f_1,f_2}[\vddl{L}{\phi}] ,\label{1.18}
\eea with
$[f_1,f_2]^\gamma=C^{\gamma(\mu)(\nu)}_{\alpha\beta}f^\alpha_{1(\mu)}f_{2(\nu)}^\beta$
for some skew-symmetric functions
$C^{\gamma(\mu)(\nu)}_{\alpha\beta}$ and for some characteristic
$M^{+i}_{f_1,f_2}[\vddl{L}{\phi}]$. At any solution $\phi^s(x)$ to
the Euler-Lagrange equations of motion, the space of all gauge
parameters equipped with the bracket $[\cdot,\cdot]$ is a Lie
algebra\footnote{Proof: By applying $\delta_{R_{f_3}}$ to
\eqref{1.18} and taking cyclic permutations, one gets
$R_{[[f_1,f_2],f_3]}+{\rm cyclic}\ (1,2,3)\approx 0$ on account of
$\delta_{R_f}\vddl{L}{\phi^i}\approx 0$. Irreducibility then
implies the Jacobi identity
\begin{eqnarray*}
  [[f_1,f_2],f_3]^\gamma+{\rm cyclic}\
(1,2,3)\approx 0.
\end{eqnarray*}}.

For all collections of local functions $Q_i$ and $f^\alpha$, let
the functions $S^{\mu i}_\alpha(Q_i,f^\alpha)$ be defined by the
following integrations by part,
\begin{equation}
\forall Q_i,f^\alpha:\quad R^i_fQ_i = f^\alpha
R^{+i}_\alpha(Q_i)+\partial_\mu S^{\mu
  i}_\alpha(f^\alpha,Q_i),
\label{1.3}
\end{equation}
where $R^{+i}_\alpha$ is the adjoint of $R^{i}_\alpha$ defined by
$R^{+i}_\alpha \hat =\, (-\partial)_{(\nu)}[ R^{i}_\alpha \cdot
\;]$.

If $Q_i=\vddl{L}{\phi^i}$, we get on account of the Noether
identities $R^{+i}_\alpha(\vddl{L}{\phi^i})=0$ that the Noether
current for a gauge symmetry can be chosen to vanish weakly,
\begin{equation}
R^i_f\vddl{\cL}{\phi^i}= \dH S_f,\label{sec1cur}
\end{equation}
where $S_f =S^{\mu i}_\alpha(\vddl{L}{\phi^i},f^\alpha)
(d^{n-1}x)_\mu$. The algebra of currents~\eqref{alg_cur} is
totally trivial for gauge symmetries. In the simple case where the
gauge transformations depend at most on the first derivative of
the gauge parameter, $R_f^i = R_\alpha^i[\phi]f^\alpha +
R^{i\mu}_\alpha [\phi]\d_\mu f^\alpha$, the weakly vanishing
Noether current is given by
\begin{equation}
S_f = R^{i\mu}_\alpha[\phi]f^\alpha
\vddl{L}{\phi^i}(d^{n-1}x)_\mu.\label{eq:S_first}
\end{equation}

A relation similar to~\eqref{1.3} holds for trivial gauge
transformations,
\begin{equation}
M_f^{+i}[\vddl{L}{\phi}]Q_i=M_f^{[j(\nu)i(\mu)]}
\6_{(\nu)}\vddl{L}{\phi^j} \partial_{(\mu)}Q_i +\partial_\mu
M_f^{\mu
  ji}(\vddl{L}{\phi^j},Q_i).
\end{equation}
If $Q_i=\vddl{L}{\phi^i}$, one can use the skew-symmetry of
$M_f^{[j(\nu)i(\mu)]}$ to get
\begin{equation}
M_f^{+i}[\vddl{L}{\phi}]\vddl{\cL}{\phi^i}= \dH M_f,\label{defM}
\end{equation}
with $M_f=M_f^{\mu
ji}(\vddl{L}{\phi^j},\vddl{L}{\phi^i})(d^{n-1}x)_\mu$. Therefore,
the Noether current associated with a trivial gauge transformation
can be chosen to be quadratic in the equations of motion and its
derivatives.

\section{Surface charge one-forms and their algebra}
\label{sec:surface-charges}

Motivated by the cohomological results of \cite{Barnich:2001jy}
 introduced in the preamble, we define the $(n-2,1)$
 forms~\footnote{For convenience, these forms have
 been defined with an overall minus sign as compared to the definition used in~\cite{Barnich:2001jy}.}
\begin{eqnarray}
k_f[\dv\phi;\phi]=I^{n-1}_{\dv \phi} S_f, \label{def}
\end{eqnarray}
obtained by acting with the homotopy operator~\eqref{phihomotopy}
on the weakly vanishing Noether current $S_f$ associated with
$f^\alpha$. We will also call these forms the surface one-forms,
where the denomination ``surface'' refers to the horizontal degree
$n-2$. When the situation is not confusing, we will omit the
$\phi$ dependence and simply write $k_f[\dv\phi]$.

For first order theories and for gauge transformations depending
at most on the first derivative of gauge parameters, the surface
one-forms~\eqref{def} coincide with the proposal
of~\cite{Silva:1998ii,Julia:2002df}
\begin{equation}
k_f[\dv \phi] = \half \dv \phi^i \QS{}{\phi^i_\nu}\left(
\Q{}{dx^\nu}S_f \right),
\end{equation}
with $S_f$ given in~\eqref{eq:S_first}.

The surface charge one-forms are intimately related to the
invariant presymplectic $(n-1,2)$ form
$W_{{\delta\cL}/{\delta\phi}}$ discussed in more details in
Appendix~\ref{app:basicdef}.\ref{sec:prop-w_vddl-1} as follows

\begin{lemma}\label{lemma_ch} The surface one-forms satisfy
\begin{eqnarray}
\dH k_f[\dv\phi]=W_{{\delta\cL}/{\delta\phi}}[\dv\phi,R_{f}] -\dv
S_f +T_{R_f}[\dv\phi,\vddl{\cL}{\phi}] ,\label{eq:29a}
\end{eqnarray}
where $W_{{\delta\cL}/{\delta\phi}}[\dv\phi,R_{f}]\equiv -i_{R_f}
W_{{\delta\cL}/{\delta\phi}}$.
\end{lemma}
Indeed, it follows from \eqref{sec1cur} and \eqref{eq:36} that
\begin{eqnarray}
I^n_{\dv\phi} (\dH
S_{f})=W_{{\delta\cL}/{\delta\phi}}[\dv\phi,R_{f}]+
T_{R_{f}}[\dv\phi,\vddl{\cL}{\phi}]\label{eq:40}.
\end{eqnarray}
Combining~\eqref{eq:40} with equation \eqref{cc1a}, this gives the
Lemma~\ref{lemma_ch}.\qed

We will consider one-forms $\dv^s\phi$ that are tangent to the
space of solutions. These one-forms are to be contracted with
characteristics $Q_s$ such that
$\delta_{Q_s}\vddl{L}{\phi^i}\approx 0$.  In particular, they can
be contracted with characteristics $Q_s$ that define symmetries,
gauge or global, since $\delta_{Q_s}\cL=\dH(\cdot)$ implies
$\delta_{Q_s}\vddl{L}{\phi^i}\approx 0$ on account of \eqref{eq:3}
and \eqref{eA7}. For such one-forms, one has the on-shell relation
\begin{eqnarray}
\dH k_f[\dv^s\phi] \approx
W_{{\delta\cL}/{\delta\phi}}[\dv^s\phi,R_{f}].\label{eq:29abis}
\end{eqnarray}
Applying the homotopy operators $I_f^{n-1}$ defined
in~\eqref{eq:45} to \eqref{eq:29abis}, one gets
\begin{eqnarray}
k_f[\dv^s \phi]&\approx& I_f^{n-1}
W_{{\delta\cL}/{\delta\phi}}[\dv^s\phi,R_{f}] +\dH(\cdot).
\label{4.13}
\end{eqnarray}Remark that if the gauge theory satisfies the property
\begin{equation}\label{simplif}
I^{n-1}_f S_f = 0, \qquad I^{n-1}_f
T_{R_f}[\dv\phi,\vddl{\cL}{\phi}] = 0,
\end{equation}
then the relation~\eqref{4.13} holds off-shell,
\begin{eqnarray}
k_f[\dv\phi]&=& I_f^{n-1}
W_{{\delta\cL}/{\delta\phi}}[\dv\phi,R_{f}] +\dH(\cdot).
\label{4.13bis}
\end{eqnarray}
This condition holds for instance in the case of generators of
infinitesimal diffeomorphisms, see Chapter~\ref{chap:matter}, and
in the Hamiltonian framework, see next section.

It is easy to show that\footnote{Proof: Applying $i_{R_{f_1}}$ to
\eqref{eq:29a} in terms of $f_2$, and using $I^{n-1}_{f_1}$, we
also get $k_{f_2}[R_{f_1}] \approx - I^{n-1}_{f_1}
W_{{\delta\cL}/{\delta\phi}}[R_{f_1},R_{f_2}]+\dH(\cdot)$.
Comparing with $i_{R_{f_1}}$ applied to \eqref{4.13} in terms of
$f_2$, this implies~\eqref{eq:47}.}
\begin{eqnarray}
k_{f_2}[R_{f_1}]\approx -k_{f_1}[R_{f_2}]+\dH(\cdot).\label{eq:47}
\end{eqnarray}
We also show in Appendix~\ref{app:basicdef} that
\begin{eqnarray}
  \label{eq:79a}
 - W_{{\delta\cL}/{\delta\phi}}=\Omega_{\cL}+\dH
E_{\cL},\qquad\dv \Omega_{\cL}=0,
\end{eqnarray}
where $\Omega_{\cL}$ is the standard presymplectic $(n-1,2)$-form
used in covariant phase space methods, and $E_{\cL}$ is a suitably
defined $(n-2,2)$ form. Contracting~\eqref{eq:79a} with the gauge
transformation $R_f$, one gets
\begin{eqnarray}
  \label{eq:79abis}
 W_{{\delta\cL}/{\delta\phi}}[\dv\phi,R_f]=\Omega_{\cL}[R_f,\dv\phi]+\dH
E_{\cL}[\dv\phi,R_f].
\end{eqnarray}
Our expression for the surface charge one-form~\eqref{4.13} thus
differs on-shell from usual covariant phase space methods by the
term $E_{\cL}[\dv\phi,R_f]$.

For a given closed $n-2$ dimensional surface $S$, which we
typically take to be a sphere inside a hyperplane, the surface
charge one-forms are defined by integrating the surface one-forms
as
\begin{eqnarray}
\ndelta \cQ_f[\dv\phi]=\oint_{S} k_f[\dv\phi].
  \label{eq:17}
\end{eqnarray}
Equation \eqref{eq:47} then reads
\begin{eqnarray}
  \label{eq:48}
  \ndelta \cQ_{f_2}[R_{f_1}] \approx -\ndelta \cQ_{f_1}[R_{f_2}].
\end{eqnarray}

Let us denote by $\cE$ the space of solutions to the
Euler-Lagrange equations of motion. It is clear from
equation~\eqref{eq:29abis} that the surface one-form is
$\dH$-closed at a fixed solution $\phi_s \in \cE$, for one-forms
$\dv^s \phi$ tangent to the space of solutions and for gauge
parameters satisfying the so-called reducibility equations
\begin{equation}
R^i_{f^s}[\phi_s]=0.\label{eq:exactakv}
\end{equation}
In the case of general relativity, e.g., these equations are the
Killing equations for the solution $\phi_s$. The space $\mathfrak
e_{\phi_s}$ of non-vanishing gauge parameters $f^s$ that satisfy
the reducibility equations at $\phi_s$ are called the non-trivial
reducibility parameters at $\phi_s$. We will also call them
\emph{exact} reducibility parameters in distinction with
\emph{asymptotic} reducibility parameters that will be defined in
the asymptotic context in Chapter~\ref{chap:asymptcharges}. It
follows from \eqref{1.18} and from the Jacobi identity that
$\mathfrak e_{\phi_s}$ is a Lie algebra, the Lie algebra of exact
reducibility parameters at the particular solution $\phi_s$.

It then follows
\begin{prop}\label{prop1}
  The surface charge one-forms~$\ndelta \cQ_{f^s}[\dv^s
\phi]|_{\phi_s}$ associated with reducibility parameters
  only depend on the homology class of $S$.
\end{prop}

In particular, if $S$ is the sphere $t=constant,\,r=constant$ in
spherical coordinates, $\ndelta \cQ_{f^s}[\dv^s \phi]|_{\phi_s}$
is $r$ and $t$ independent and, therefore, is a constant.

Any trivial gauge transformation $\delta\phi^i =
M^{+i}_f[\vddl{L}{\phi}]$ can be associated with a $(n-2,1)$ form
$k_f = I_{\dv\phi}^{n-1}M_f$ in the same way as~\eqref{def} with
$M_f$ defined in~\eqref{defM}. Now, one has $k_f \approx 0$ since
the homotopy operator~\eqref{phihomotopy} can only ``destroy'' one
of the two equations of motion contained in $M_f$. Therefore,
trivial gauge transformations are associated with weakly vanishing
surface one-forms.

Up to here, we have constructed conserved surface charge one-forms
starting from reducibility parameters $f^s$. In fact, there is a
bijective correspondence between conserved charges and
reducibility parameters. More precisely, the following proposition
was demonstrated in \cite{Barnich:2001jy}
\begin{prop}\label{prop1bis} When restricted to solutions of the
equations of motion, equivalence classes of closed, local,
(n-2,1)-forms up to exact, local, (n-2,1)-forms correspond one to
one to non-trivial reducibility parameters. Representatives for
these (n-2,1)-forms are given by~\eqref{def}.
\end{prop}
This proposition provides the main justification of the
definition~\eqref{def} of the surface one-forms.

The following proposition is proved in
Appendix~\ref{app:proofs}.\ref{appb}:
\begin{prop}\label{lem11}
When evaluated at a solution $\phi_s$, for one-forms $\dv^s\phi$
tangent to the space of solutions and for reducibility parameters
$f^s$ at $\phi_s$, the surface one-forms $k_{f^s}[\dv^s \phi]$ are
covariant up to $\dH$ exact terms,
\begin{eqnarray}
  \label{eq:35}
  \delta_{R_{f_1}} k_{f^s_2}[\dv^s \phi] \approx -k_{[f_1,f^s_2]}[\dv^s
  \phi]+\dH (\cdot).
\end{eqnarray}
\end{prop}
If the Lie bracket of surface charge one-forms is defined by
\begin{eqnarray}
  \label{eq:60}
  [\ndelta \cQ_{f_1},\ndelta \cQ_{f_2}]=-\delta_{R_{f_1}} \ndelta \cQ_{f_2},
\end{eqnarray}
we thus have shown:
\begin{corollary}\label{cor3}
At a given solution $\phi_s$ and for one forms $\dv^s\phi$ tangent
to the space of solutions, the Lie algebra of surface charge
one-forms represents the Lie algebra of exact reducibility
parameters $\mathfrak e_{\phi_s}$,
\begin{eqnarray}
  \label{eq:47b}
  [\ndelta \cQ_{f^s_1} ,\ndelta \cQ_{f^s_2}][\dv^s\phi]|_{\phi_s} =
\ndelta \cQ_{[f^s_1,f^s_2]}[\dv^s\phi]|_{\phi_s}.
\end{eqnarray}
\end{corollary}

We finally consider one-forms $\dv^s\phi$ that are tangent to the
space of reducibility parameters at $\phi^s$. They are to be
contracted with gauge parameters $Q^s$ such that
\begin{eqnarray}
0=(\dv^s R_f)|_{\phi_s,f^s,Q_s}=
\delta_{Q_s}R_{f^s}|_{\phi_s}.\label{eq:90}
\end{eqnarray}
We recall that for $\cA$ a Lie algebra, the derived Lie algebra is
given by the Lie algebra of elements of $\cA$ that may be written
as a commutator. The derived Lie algebra is sometimes denoted as
$[\cA,\cA]$. It is an ideal of $\cA$. Definition \eqref{eq:60} and
Corollary \bref{cor3} imply
\begin{corollary}\label{cor4}
For field variations $\dv^s\phi$ preserving the reducibility
identities as~\eqref{eq:90}, the surface charge one-forms vanish
for elements of the derived Lie algebra ${\mathfrak
e}^\prime_{\phi_s}$  of exact reducibility parameters at $\phi_s$,
\begin{eqnarray}
  \label{eq:92}
\ndelta  \cQ_{[f^s_1,f^s_2]}[\dv^s\phi]|_{\phi_s}=0.
\end{eqnarray}
In this case, the Lie algebra of surface charge one-forms
represents non-trivially only the abelian Lie algebra ${\mathfrak
e}_{\phi_s}/{\mathfrak
  e}^\prime_{\phi_s}$.
\end{corollary}

\section{Hamiltonian formalism} \label{sec:hamilt-form}

In this section, we discuss the results obtained in the previous
section in the particular case of an action in Hamiltonian form
and for the surface $S$ being a closed surface inside the
space-like hyperplane $\Sigma_t$ defined at constant $t$.

We follow closely the conventions of \cite{Henneaux:1992ig} for
the Hamiltonian formalism. The Hamiltonian action is first order
in time derivatives and given by
\begin{equation}
S_H[z,\lambda] =\int \cL_H= \int dt d^{n-1}x \, (\dot z^A a_A - h
- \lambda^a \gamma_a )\,, \label{eqApp:1}
\end{equation}
where we assume that we have Darboux coordinates:
$z^A=(\phi^\alpha, \pi_\alpha)$ and $a_A = (\pi_\alpha,0)$. It
follows that $\sigma_{AB}=\partial_A a_B-\partial_B a_A$ is the
constant symplectic matrix with $\sigma^{AB} \sigma_{BC} =
\delta^A_C$ and $d^{n-1}x\equiv (d^{n-1}x)_0$. We assume for
simplicity that the constraints $\gamma_a$ are first class,
irreducible and time independent. In the following we shall use a
local ``Poisson'' bracket with spatial Euler-Lagrange derivatives
for spatial $n-1$ forms $\hat g= g\, d^{n-1}x$,
\begin{equation}
\{ \hat g_1,\hat g_2\} =
\varQ{g_1}{z^A}\sigma^{AB}\varQ{g_2}{z^B}\,  d^{n-1}x.
\end{equation}
If $\tilde \dH $ denotes the spatial exterior derivative, this
bracket defines a Lie bracket in the space $H^{n-1}(\tilde \dH )$,
i.e., in the space of equivalence classes of local functions
modulo spatial divergences, see e.g.~\cite{Barnich:1996mr}.

Similarly, the Hamiltonian vector fields associated with an $n-1$
form $\hat h = h \, d^{n-1}x$
\begin{eqnarray}
\stackrel{\leftarrow}{\delta_{\hat h}}\,(\cdot) =
\QS{}{z^A_{(i)}} (\cdot)\,
\sigma^{AB}\d_{(i)}\varQ{h}{z^B}=\{\cdot,\hat h\}_{alt},\\
\stackrel{\rightarrow}{\delta_{\hat h}}\,(\cdot) =
\d_{(i)}\varQ{h}{z^B}\sigma^{BA} \QS{}{z^A_{(i)}} (\cdot)\,
=\{\hat h,\cdot\}_{alt},
\end{eqnarray}
only depend on the class $[\hat h] \in H^{n-1}(\tilde\dH )$. Here
$(i)$ is a multi-index denoting the spatial derivatives, over
which we freely sum. The combinatorial factor needed to take the
symmetry properties of the derivatives into account is included in
$\frac{\partial^S}{\partial z^A_{(i)}}$. If we denote $\hat
\gamma_a = \gamma_a \, d^{n-1}x$ and $\hat h_E =\hat h +\lambda^a
\hat\gamma_a$, an irreducible generating set of gauge
transformations for \eqref{eqApp:1} is given by
\begin{eqnarray}
{\delta_{f}} z^A &=
  &
\{z^A,\hat \gamma_a f^a\}_{alt},\label{cang1}\\
{\delta_{f}} \lambda^a &=& \frac{D f^a}{Dt} +
\{f^a,\hat h_E\}_{alt}
+\cC_{bc}^a(f^b,\lambda^c)-\cV_b^a(f^b),\label{cang2}
\end{eqnarray}
where the arbitrary gauge parameters $f^a$ may depend on $x^\mu$,
the Lagrange multipliers and their derivatives as well as the
canonical variables and their spatial derivatives and
\begin{eqnarray}
 &&\frac{D}{Dt} = \Q{}{t} + \dot\lambda^a\Q{}{\lambda^a} + \ddot \lambda^a
 \Q{}{\dot\lambda^a} + \dots,\\
  && \{ \gamma_a, \hat \gamma_b \lambda^b\}_{alt} =
  \cC^{+c}_{ab}(\gamma_c,\lambda^b), \\
  &&\{\gamma_a,\hat h\}_{alt} = -\cV_a^{+b}(\gamma_b).
\end{eqnarray}

Let $d\sigma_i=2(d^{n-2}x)_{0i}$. For $S$ a closed surface inside
the hyperplane $\Sigma_t$ defined by constant $t$, the surface
charge one-forms are given by
\begin{eqnarray}
  \label{eq:101}
  \ndelta Q_f[\dv z,\dv \lambda]=\oint_S  k^{[0i]}_f[\dv z,\dv \lambda]d\sigma_i.
\end{eqnarray}
Therefore, only the $[0i]$ components of the surface one-forms are
relevant in order to construct the surface charges one-forms at
constant time. We prove in
Appendix~\ref{app:proofs}.\ref{proof_hamilt} the following result
first obtained in the Hamiltonian approach:

\begin{prop}\label{prop_hamilt}
  In the context of the Hamiltonian formalism, the surface
  one-forms at constant time do not depend on the Lagrange multipliers
  and are given by the opposite of boundary terms that arise when converting the
  variation of the constraints smeared with gauge parameters
  into an Euler-Lagrange derivative contracted with the
  undifferentiated variation of the canonical variables,
  \begin{eqnarray}
    \label{eq:102}
    \dv^z(\gamma_af^a)=\dv z^A\vddl{(\gamma_a
      f^a)}{z^A}-\d_ik^{[0i]}_f[\dv z;z].
  \end{eqnarray}
\end{prop}
Using this link between Hamiltonian and Lagrangian frameworks, one
can then use Propositions~\ref{prop1}, \ref{prop1bis}, \ref{lem11}
and their corollaries to study properties of the surface terms in
Hamiltonian formalism.

Note that, because of the simple way time derivatives enter into
the Hamiltonian action $\cL_H$, the
expressions~\eqref{eq:4}-\eqref{eq:27}-\eqref{eq:26} give for all
$Q^i_1,Q^i_2$,
\begin{eqnarray}
  \label{eq:42}
  W^0_{\vddl{\cL_H}{\phi}}[Q_1,Q_2]&=&-\sigma_{AB}Q_1^AQ_2^B\,,\\
T^{0}_{R_f}[\dv\phi, \vddl{\cL_H}{\phi}]&=&0,\qquad
 E^{0i}_{\cL_H}[\dv\phi, \dv\phi] =0\,.\label{eq:42c}
\end{eqnarray}
The last relation follows from our assumption that we are using
Darboux coordinates. As a consequence of the first relation, we
then also have
\begin{eqnarray}
  \label{eq:106}
  W^0_{\vddl{\cL_H}{\phi}}[\dv\phi,R_f]\, d^{n-1}x
&=&-\dv z^A\vddl{(\hat \gamma_a f^a)}{z^A}\,,\\
W^0_{\vddl{\cL_H}{\phi}}[R_{f_1},R_{f_2}]\, d^{n-1}x&=&
\{\hat\gamma_a f^a_1,\hat \gamma_b f^b_2\}\,,
\end{eqnarray}
which are useful in order to relate Hamiltonian and Lagrangian
frameworks.

\section{Exact solutions and symmetries}
\label{sec:exact-solut-symm}

Suppose one is given a family of exact solutions $\phi_s \in \cE$
admitting ($\phi_s$-dependent) reducibility parameters $f^s \in
\mathfrak e_{\phi_s}$. Let us denote by $\bar \phi$ an element of
this family that we single out as the reference solution with
reducibility parameter $\bar f \in \mathfrak e_{\bar \phi}$.

The surface charge $Q_{\gamma}$ of $\Phi_s=(\phi_s,f^s)$ with
respect to the reference $\bar\Phi=(\bar\phi,\bar f)$ is defined
as
\begin{eqnarray}
  \label{eq:82}
  \cQ_{\gamma} [\Phi,\bar \Phi] = \int_{\gamma}
\ndelta \cQ_{f_\gamma}[\dv^\gamma\phi]|_{\phi_\gamma} + N_{\bar f
}[\bar \phi],
\end{eqnarray}
where integration is done along a path $\gamma$ in the space of
exact solutions $\cE$ that joins $\bar \phi$ to $\phi_s$ for some
reducibility parameters that vary along the path from $\bar f$ to
$f^s$. Only charge differences between solutions are defined. The
normalization $N_{\bar f }[\bar \phi]$ of the reference solution
can be chosen arbitrarily. Note that these charges depend on $S$
only through its homology class because equation (\ref{eq:29a})
implies that $\dH k_{f^s}[\dv^s\phi]|_{\phi_s}=0$.


The natural question to ask for the charges $\cQ_{\gamma}$ is
whether they depend on the path $\gamma$ used in their definition.
If there is no de Rham cohomology in degree two in solution space,
the path independence of the charges $\cQ_{\gamma}$ is ensured if
the following integrability conditions
\begin{equation} \oint_S \dv^{\Phi,s}
k_{f^s}[\dv^s\phi]|_{\phi_s} = \oint_S \dv^s
k_{f^s}[\dv^s\phi]|_{\phi_s}+\oint_S k_{\dv
f^s}[\dv^s\phi]|_{\phi_s} =0\label{int_cond}
\end{equation}
are fulfilled. These conditions extend the conditions discussed
in~\cite{Wald:1999wa,Julia:2002df} to variable parameters $f^s$.

For one-forms $\dv^s \phi$ tangent to the family of solutions with
reducibility parameters $f^s$, one has
\begin{equation} \dv^{\Phi,s} R_{f^s}|_{\phi^s} = \dv^s R_{f^s}|_{\phi^s} + R_{\dv f^s}|_{\phi^s}=0.
\end{equation}
This implies together with equation \eqref{eq:29a} that $\dH
\dv^{\Phi,s} k_{f^s}[\dv^s\phi]|_{\phi_s}=0$, so that the
integrability conditions also only depend on the homology class of
$S$.

Now suppose that the solution space $\cE$ is entirely
characterized by $p$ parameters $a^A$, $A=1,\dots p$. In that
case, solutions $\phi_s(x;a)$ and reducibility parameters
$f^s(x;a)$ at $\phi_s(x;a)$ also depend on these parameters. Let
us denote by $e_{i}(x;a)$ a basis of the Lie algebra $\mathfrak
e_{\phi_s}$ with $i=1,\dots r$. For each basis element
$e_{i}(x;a)$, we consider the one-forms in parameter space
\[ \theta_i(a,da)=\oint_S k_{e_{ i}}[d^a\phi_s(x;a)],\] where $d^a$
is the pull-back of the vertical derivative to $\cE$, i.e. the
exterior derivative in parameter space. The integrability
conditions~\eqref{int_cond} are then a Pfaff system in parameter
space and the question of integrability can be addressed using
Frobenius' theorem, see e.g. \cite{Stephani:2003tm}:

\begin{theorem}(Frobenius' theorem) Let $\theta_i(a,da)$ be one-forms
linearly independent at a point $\phi_s \in \cE$. Suppose there
are one-forms $\tau^i_{\;\; j}(a,da)$, $i,j=1\dots r$, satisfying
\begin{equation}
d^a \theta^i = \tau^i_{\;\; j} \theta^j.
\end{equation}
Then, in a neighborhood of $\phi_s$ there are functions
$S^i_{\;j}(a)$ and $\cQ_j(a)$, such that $\theta^i = S^i_{\;j} d^a
\cQ_j$.
\end{theorem}

If the system is completely integrable, i.e. if there exists an
invertible matrix $S^i_{\;j}(a)$ and quantities $\cQ_j(a)$ such
that
\begin{eqnarray}
\theta_i(a,da) = S^j_{\;i}(a) d^a \cQ_j(a)\label{eq:69},
\end{eqnarray}
then there is a change of basis in the Lie algebra of reducibility
parameters $g_{j}(x;a)=(S^{-1})^{i}_{\;j}(a)e_{i}(x;a)$ such that
the integrability conditions~\eqref{int_cond} are satisfied in
that basis.

As a conclusion, in the absence of non-trivial topology in
solution space, the charges obtained by the resolution
of~\eqref{eq:69} provide path independent charges.

In the case where the action is the Hamiltonian
action~\eqref{eqApp:1} and where $S$ is the boundary of the $n-1$
dimensional surface $\Sigma_t$, $t=constant$, one can define the
functional associated with $\Phi = (\phi^s,f_s)$ as
\begin{equation}
\cH [\Phi,\bar \Phi] = \int_{\Sigma_t} \gamma_a f^a +
\int_{\d\Sigma_t}\cQ_\gamma[\Phi,\bar \Phi]
\end{equation}
As a direct consequence of Proposition~\bref{prop_hamilt}, $\cH
[\Phi,\bar \Phi]$ admits well-defined functional derivatives. This
completes the link with the Hamiltonian formalism.

The fact that the charge~\eqref{eq:82} depends on $S$ only through
its homology class for reducibility parameters $f$ has a nice
consequence. In the case where the surface $S$ surrounds several
sources that can be enclosed in smaller surfaces $S^i$, one gets
\begin{equation}
\oint_{S} \cQ_f[\Phi,\bar \Phi] = \sum_{i \in \text{
sources}}\oint_{S^i} \cQ_f[\Phi,\bar \Phi].\label{eq:propG}
\end{equation}
In electromagnetism, this properties reduces to the Gauss law for
static electric charges. For spacetimes in Einstein gravity with
vanishing cosmological constant, the Komar
formula~\cite{Komar:1958wp} obeys a property analogous
to~\eqref{eq:propG}. Here, we showed that the
property~\eqref{eq:propG} holds in a more general context when the
charges are defined as~\eqref{eq:82}.

Finally, let us consider the case where the surface charge is
evaluated at infinity. An interesting simplification occurs when
$\Phi$ approaches $\bar \Phi$ sufficiently fast at infinity in the
sense that the $(n-2,1)$-form can be reduced to
\begin{eqnarray}
k_{f}[\dv \phi;\phi]|_{S^\infty}= k_{\bar f}[\dv \phi;\bar
\phi]|_{S^\infty}.\label{AL}
\end{eqnarray}
We refer to this simplification as the \emph{asymptotically
linear} case because the charge~\eqref{eq:82} becomes manifestly
path-independent and reduces to the integral of the one-form
constructed in the linearized theory contracted with the deviation
$\phi - \bar \phi$ with respect to the background,
\begin{eqnarray}
  \label{eq:82bis}
  \cQ_{f} [\Phi,\bar \Phi] =
\oint_{S^\infty} k_{\bar f}[\phi-\bar \phi;\bar \phi] + N_{\bar f
}[\bar \phi],
\end{eqnarray}
This simplification allows one to compare the surface
charges~\eqref{eq:82} with definition at infinity, e.g. in general
relativity~\cite{Abbott:1982ff,Henneaux:1985tv,Henneaux:1985ey},
see section~\ref{sec:KerrAdS} of Chapter~\ref{chap:BHapp}. This
simplification is also relevant for particular boundary
conditions, see asymptotically anti-de Sitter and flat spacetimes
in three dimensions in Chapter~\ref{chap:asymptanalyses}.

%% file: charges_gravity_matter.tex
\chapter{Charges for gravity coupled to matter fields}
\label{chap:matter}
\setcounter{equation}{0}\setcounter{figure}{0}\setcounter{table}{0}

This part shows several applications to gravity of the general
theory developed in the preceding chapter. We begin in
section~\ref{sec:diffinv} by specializing the formalism to gauge
parameters which are infinitesimal diffeomorphism in generally
covariant theories of gravity. We then discuss in detail in
section~\ref{sec:general-relativity} the important case of
Einstein gravity in Lagrangian as well as in Hamiltonian
formalism. In sections \ref{sec:pform} and~\ref{sec:EM-chern}, we
extend the analysis to Einstein gravity coupled to matter fields
relevant in supergravity theories: scalars, $p$-form potentials
and Maxwell fields with or without a Chern Simons term.

Many of the expressions derived in this chapter were already known
in the literature. However, the unified way in which they are
derived allows us to highlight the differences and the
equivalences between different approaches as the covariant phase
space methods of~\cite{Wald:1993nt,Iyer:1994ys,Wald:1999wa}, the
covariant methods inspired from the Hamiltonian
prescription~\cite{Julia:1998ys,Silva:1998ii,
Julia:2000er,Julia:2002df}, Hamiltonian
methods~\cite{Regge:1974zd,Henneaux:1985tv,Henneaux:1985ey} and
methods based on the linearized Einstein
equations~\cite{Abbott:1981ff}. These comparisons complete the
picture given by earlier
works~\cite{Iyer:1995kg,Hollands:2005wt,Papadimitriou:2005ii}.

\section{Diffeomorphic invariant theories}
\label{sec:diffinv}

Gravities with higher curvature terms naturally appear in
effective theories describing semi-classical aspects of quantum
gravity~\cite{Birrell:1982ix} or in string
theories~\cite{Callan:1985ia,Callan:1986jb,Gross:1986mw}. The
minimal setting describing these general theories of gravity is an
action principle which is invariant under diffeomorphisms.

The definition of conserved quantities for arbitrary diffeomorphic
invariant theories has been addressed
in~\cite{Wald:1993nt,Iyer:1994ys,Brown:1995su} using covariant
phase space methods. More recent work includes, e.g., definitions
of energy for actions quadratic in the
curvature~\cite{Deser:2002jk,Deser:2007vs}.

In this section, we will derive the surface one-form associated
with an infinitesimal diffeomorphism for a diffeomorphic invariant
Lagrangian and we will study its properties. This surface one-form
will differ from the covariant phase space
result~\cite{Wald:1993nt,Iyer:1994ys} only by a term which
vanishes for a symmetry $\xi_s$ of the field configuration,
$\cL_{\xi_s}\phi^i = 0$.

Let us consider a Lagrangian $\cL[g_{\mu\nu},\psi^k]$ depending on
a metric $g_{\mu\nu}$, on the fields $\psi^k$ and on any finite
number of their derivatives which is invariant under
diffeomorphisms. The fields are collectively denoted by
$\phi^i=(g_{\mu\nu},\psi^k)$. An arbitrary $(p,s)$-form $\omega$
is invariant under diffeomorphism if it satisfies
\begin{equation}
\delta_{\cL_\xi \phi}\omega = \cL_\xi\omega,\label{prop_diff}
\end{equation}
where $\cL_\xi \omega = (i_\xi \dH + \dH i_\xi) \omega$ is the Lie
differential acting on $(p,s)$-forms,
see~\eqref{app:Liec}-\eqref{app:deltaL}, and $\cL_\xi \phi^i$ is
the usual Lie derivative of the field $\phi^i$. The invariance of
the lagrangian $n$-form $\cL$ implies
\begin{equation}
\delta_{\cL_\xi \phi}\cL = \dH i_\xi \cL.
\end{equation}
The variation formula~\eqref{cc2} in terms of $\cL$ reads as
\begin{equation}
\delta_{\cL_\xi \phi}\cL = \cL_\xi \phi^i \frac{\delta \cL}{\delta
\phi^i} + d_H I^n_{\cL_\xi \phi}\cL.\label{gauge}
\end{equation}
Results in the equivariant variational bicomplexes, see Theorem
5.3 of \cite{Anderson1991} and \cite{Anderson:1995} implies that a
choice for $I^{n}_{\cL_\xi \phi}\cL$ invariant under
diffeomorphisms can be made by suitably constructing the horizonal
homotopy operator. We refer the reader to \cite{Iyer:1994ys} for
such an explicit construction.

\paragraph{Surface one-form}

Using~\eqref{sec1cur}, the term $\cL_\xi \phi^i \frac{\delta
\cL}{\delta \phi^i}$ can be expressed as $\dH S_\xi$ where $S_\xi$
is the weakly vanishing Noether current which is linear in
$\xi^\mu$. We get
\begin{equation}
\dH( S_\xi +I^n_{\cL_\xi \phi}\cL - i_\xi \cL) = 0.
\end{equation}
 Acting on the latter expression with the contracting
homotopy $I_\xi^{n}$, the weakly vanishing current $S_\xi$ can be
expressed as
\begin{equation}
S_\xi = -I^{n}_{\cL_\xi \phi}\cL + i_\xi \cL - \dH
k^K_{\cL,\xi},\label{def_curr}
\end{equation}
where $k^K_{\cL,\xi} = -I_\xi^{n-1}I^{n}_{\cL_\xi \phi }\cL$ is a
representative for the Noether charge $n-2$ form
\cite{Wald:1993nt,Iyer:1994ys}. The pre-symplectic form
$\Omega_{\cL}[\cL_\xi\phi,\dv \phi] = i_{\cL_\xi\phi}\Omega_{\cL}$
reads here
\begin{equation}
\Omega_{\cL}[\cL_\xi \phi,\dv \phi] = \delta_{\cL_\xi
\phi}I^{n}_{\dv \phi}\cL - \dv I^{n}_{\cL_\xi \phi}\cL.
\end{equation}
Using then~\eqref{cc2a}, we get
\begin{equation}
\Omega_{\cL}[\cL_\xi\phi,\dv \phi] = \dv (i_\xi \cL -
I^{n}_{\cL_\xi \phi}\cL) + \dH i_\xi I^{n}_{\dv \phi}\cL-\dv
\phi^i i_\xi \frac{\delta \cL}{\delta \phi^i}.
\end{equation}
Replacing the expression between parenthesis
using~\eqref{def_curr}, we obtain
\begin{equation}
\Omega_{\cL}[\cL_\xi\phi,\dv \phi] = \dH (-\dv k^K_{\cL,\xi}
+i_\xi I^{n}_{\dv \phi}\cL)-\dv \phi^i i_\xi \frac{\delta
\cL}{\delta \phi^i} +\dv S_\xi.\label{eq:Omega_diff}
\end{equation}
Now, since we have $I_\xi^{n-1}T_{\cL_\xi \phi}[\dv\phi, \omega^n]
= I^{n-1}_\xi (\dv\phi^i i_\xi \varQ{\omega^n}{\phi^i}) = 0$ and
$I_\xi \dv S_\xi = \dv I_\xi S_\xi = 0$, the
property~\eqref{simplif} hold and we can use equations
\eqref{4.13bis} and \eqref{eq:79abis} to write the charge one-form
$k_\xi[\dv \phi]$ as
\begin{equation}
k_\xi[\dv \phi] = I_\xi^{n-1}\Omega_\cL[\cL_\xi
\phi,\dv\phi]-E_\cL[\cL_\xi \phi,\dv \phi]+
\dH(\cdot).\label{eq:def10}
\end{equation}
Finally, using~\eqref{eq:Omega_diff}, the surface one-form
$k_\xi[\dv \phi]$ reduces to
\begin{equation}
k_\xi[\dv \phi] =  -\dv k^K_{\cL,\xi}  +i_\xi I^{n}_{\dv
\phi}\cL-E_\cL[\cL_\xi \phi,\dv \phi] + \dH(\cdot).\label{k_diff}
\end{equation}
Note the relation~\eqref{inner_prod} useful to
express~\eqref{k_diff} in coordinates. Our definition of surface
one-form differs from the covariant phase space
methods~\cite{Iyer:1994ys,Iyer:1995kg} by the supplementary term
$E_\cL$. This supplementary term vanishes when $\xi_s$ is a
symmetry of the field configuration $\phi^i$, $\cL_{\xi_s}\phi^i=
0$.

\paragraph{Properties of the surface one-form}

By construction, the form~\eqref{k_diff} is independent on the
addition of boundary terms to the Lagrangian, which is not the
case for the expression obtained with covariant phase space
methods. Remark that these boundary terms should be diffeomorphic
invariant in order that the derivation of the previous paragraph
be valid.

This property can be explicitly checked by noting that for a
boundary term $\dH \mu$ in the Lagrangian, one has
\begin{eqnarray}
k^K_{\dH \mu,\xi} &=& -i_\xi \mu + I_{\cL_\xi
\phi}\mu+\dH(\cdot), \\
E_{\dH \mu}[\dv \phi,\cL_\xi \phi ] &=&  - \delta_{\cL_\xi
\phi}I_{\dv \phi}\mu +
\dv I_{\cL_\xi \phi} \mu+\dH(\cdot)\nonumber \\
&= & - i_\xi I_{\dv\phi}\dH \mu +\dv (I_{\cL_\xi \phi} \mu -i_\xi
\mu)+\dH(\cdot),
\end{eqnarray}
as implied by equations~\eqref{cc1a}-\eqref{cc1} and
\eqref{eq:26}.

Proposition~\bref{prop1} implies that the surface charge
one-forms~$\ndelta \cQ_{\xi^s}[\dv^s \phi]|_{\phi_s}$ associated
with reducibility parameters $\xi^s$ of a solution $\phi^s$, i.e.
$\cL_{\xi^s}\phi|_{\phi_s} = 0$, only depend on the homology class
of $S$.

Additional properties of the surface charge one-forms can be found
in Corollaries~\bref{cor3} and \bref{cor4}. For vectors $\xi$ that
are left invariant by the variation $\dv \xi = 0$, the
integrability condition reduces to the simple condition,
\begin{equation}
\oint_S  i_\xi
W_{{\delta\cL}/{\delta\phi}}[\dv\phi,\dv\phi]+\oint_S i_{\cL_\xi
\phi}\dv E_\cL[\dv\phi,\dv\phi]=0,\label{int_diff}
\end{equation}
after having used~\eqref{eq:comm2} and~\eqref{eq:79}. The first
term in \eqref{int_diff} vanishes for vector fields $\xi$ tangent
to the surface $S$. For a reducibility parameter $\xi^s$ of
$\phi$, the second term in the latter expression vanishes and the
integrability condition can be written equivalently as $\oint_S
i_{\xi^s} \Omega_{\cL}[\dv \phi,\dv \phi] = 0$, coinciding
with~\cite{Iyer:1994ys,Wald:1999wa}.

\section{General relativity}
\label{sec:general-relativity}

An introduction to the problem of defining conserved quantities in
general relativity was done in the preamble and we refer the
reader to this chapter for detailed discussions and references.

Here, we will first specialize the results obtained in
section~\ref{sec:diffinv} to Einstein gravity. Our expression for
the surface one-form will be shown to agree with the one found
in~\cite{Abbott:1981ff} in the context of anti-de Sitter
backgrounds. We will then apply the general method described in
Chapter~\ref{chap:general_th} to gravity in first order
Hamiltonian formalism and we will recover the surface terms
obtained by Hamiltonian
methods~\cite{Arnowitt:1962aa,Regge:1974zd,Henneaux:1985tv,Henneaux:1985ey}.
Finally, we will compare both approaches by reducing canonically
the covariant expression for the surface one-form using ADM
variables. The two expressions in ADM variables will be shown to
differ by terms that vanish for exact reducibility parameters
(i.e., here, Killing vectors).

\subsection{Lagrangian formalism}
\label{sec:cov_grav}

Pure Einstein gravity with cosmological constant $\Lambda$ is
described by the Einstein-Hilbert action
\begin{eqnarray}
  \label{eq:56}
  S[g]=\int\cL^{EH}= \int d^nx\, \frac{\sqrt{|g|}}{16\pi
  G}(R-2\Lambda).
\end{eqnarray}
A generating set of gauge transformations is given by
\begin{eqnarray}
  \label{eq:6}
  \delta_\xi g_{\mu\nu}=\cL_\xi g_{\mu\nu}=\xi^\rho\partial_\rho
  g_{\mu\nu}+\partial_\mu\xi^\rho g_{\rho\nu}+\partial_\nu\xi^\rho
  g_{\mu\rho}.
\end{eqnarray}
Reducibility parameters at $g$ are thus given by Killing vectors
of $g$. The weakly vanishing Noether current~\eqref{sec1cur} is
given by
\begin{eqnarray}
  \label{eq:15a}
  S^\mu_\xi[\varQ{L^{EH}}{g}]=2\varQ{L^{EH}}{g_{\mu\nu}}\xi_\nu=
\frac{\sqrt{|g|}}{8\pi G}(-G^{\mu\nu}-\Lambda g^{\mu\nu})\xi_\nu.
\end{eqnarray}
Note that from~\eqref{eq:3}, we have
\begin{equation}
\delta_{\cL_\xi g} \varQ{L^{EH}}{g_{\mu\nu}} = \d_\rho\left(
\xi^\rho \varQ{L^{EH}}{g_{\mu\nu}} \right) - \d_\rho \xi^\mu
\varQ{L^{EH}}{g_{\rho\nu}} - \d_\rho \xi^\nu
\varQ{L^{EH}}{g_{\mu\rho}}.
\end{equation}
It is convenient to define
\begin{eqnarray}
 \frac{\d^S L^{EH}}{\d g_{\gamma
\delta,\alpha\beta}}= G^{\alpha \beta\gamma\delta},\quad\quad
\frac{\d^S}{\d g_{\gamma \delta,\alpha\beta}} \left(
\varQ{L^{EH}}{g_{\mu\nu}} \right) = P^{\mu\nu\alpha\beta\gamma
\delta},
\end{eqnarray}
where
\begin{eqnarray}
G^{\alpha \beta\gamma\delta} &=&  \frac{\sqrt{-g}}{16\pi G} \big(
\half g^{\alpha\gamma} g^{\beta\delta} + \half g^{\alpha\delta}
g^{\beta\gamma} -  g^{\alpha\beta} g^{\gamma\delta} \big)\label{deWitt}\\
P^{\mu \nu \alpha\beta \gamma\delta} &= &\frac{\sqrt{-g}}{32\pi G}
\big( g^{\mu\nu} g^{\gamma(\alpha} g^{\beta)\delta} +
g^{\mu(\gamma}g^{\delta)\nu} g^{\alpha\beta} + g^{\mu(\alpha}
g^{\beta)\nu}g^{\gamma\delta}
\nonumber \\
&& - g^{\mu\nu} g^{\gamma\delta} g^{\alpha\beta}  -
g^{\mu(\gamma}g^{\delta)(\alpha}g^{\beta)\nu}  -g^{\mu(\alpha}
g^{\beta)(\gamma}g^{\delta)\nu} \big).
\end{eqnarray}
The tensor density $G^{\alpha \beta\gamma\delta} =
\frac{1}{n-2}g_{\mu\nu}P^{\mu\nu\alpha\beta\gamma \delta}$ called
the supermetric~\cite{Dewitt:1967yk} has the symmetries of the
Riemann tensor. The tensor density $P^{\mu\nu\alpha\beta\gamma
\delta}$ is symmetric in the pair of indices $\mu\nu$,
$\alpha\beta$ and $\gamma \delta$ and the total symmetrization of
any three indices is zero. The symmetries of these tensors are
thus summarized by the Young tableaux
\begin{equation}
G^{\alpha \beta\gamma\delta} \sim
\begin{picture}(28,25)(0,0)
\multiframe(0,5.75)(11.5,0){2}(11,11){$\alpha$}{$\beta$}
\multiframe(0,-5.75)(11.5,0){2}(11,11){$\gamma$}{$\delta$}
\end{picture},
\qquad P^{\mu \nu \alpha\beta\gamma\delta} \sim
\begin{picture}(28,25)(0,0)
\multiframe(0,11.5)(11.5,0){2}(11,11){$\mu$}{$\nu$}
\multiframe(0,0)(11.5,0){2}(11,11){$\alpha$}{$\beta$}
\multiframe(0,-11.5)(11.5,0){2}(11,11){$\gamma$}{$\delta$}
\end{picture}.\vspace{11pt}\label{pmunu}
\end{equation}
The explicit expression that one obtains for $k_\xi  =
I^{n-1}_{\dv g}S_\xi$ using~\eqref{phihomotopy} is
\begin{equation}
k_\xi[\dv g;g] = \frac{2}{3}(d^{n-2}x)_{\mu\nu}P^{\mu\delta\nu
\gamma \alpha\beta}(2 D_\gamma \dv g_{\alpha\beta}\xi_\delta - \dv
g_{\alpha\beta} D_\gamma \xi_\delta),\label{ch_grav1}
\end{equation}
or, more explicitly,
\begin{eqnarray}
k_{\xi}[\dv g ;g]=\frac{1}{16\pi G}(d^{n-2}x)_{\mu\nu} \sqrt{-
g}\Big( \xi^\nu   D^\mu h+  \xi^\mu   D_\sigma h^{\sigma\nu}+
\xi_\sigma
  D^\nu h^{\sigma\mu}\nonumber\\+\half h
D^\nu \xi^\mu+\half h^{\mu\sigma}  D_\sigma  \xi^\nu+\half
h^{\nu\sigma}   D^\mu  \xi_\sigma -(\mu\longleftrightarrow \nu)
\Big), \label{hom}
\end{eqnarray}
where indices are lowered and raised with the metric $g_{\mu\nu}$
and its inverse and where we introduced the notation $h_{\mu\nu}
\equiv \dv g_{\mu\nu}$ and $h \equiv g^{\alpha\beta}\dv
g_{\alpha\beta}$.

This expression can be shown to coincide with the one derived by
Abbott and Deser \cite{Abbott:1982ff} in the context of
asymptotically anti-de Sitter spacetimes:
\begin{eqnarray}
k^{\text{A-D}}_{\xi}[\dv g;g]=-\frac{1}{16\pi
G}(d^{n-2}x)_{\mu\nu} \sqrt{- g}\Big(  \xi_\rho   D_\sigma
H^{\rho\sigma\mu\nu} +\frac 12 H^{\rho\sigma\mu\nu}   D_\rho
\xi_\sigma \Big), \label{gsuperpot3}
\end{eqnarray}
where $H^{\rho\sigma\mu\nu}[\dv g;g]$ is defined by
\begin{eqnarray}
H^{\mu\alpha\nu\beta}[\dv g;g]&=& -\7h^{\alpha\beta}  g^{\mu\nu}
-\7h^{\mu\nu}  g^{\alpha\beta} +\7h^{\alpha\nu}  g^{\mu\beta}
+\7h^{\mu\beta}  g^{\alpha\nu},
\label{Hdef}\\
\7h_{\mu\nu}&=& h_{\mu\nu}-\frac{1}{2}   g_{\mu\nu} h.
\label{hath}
\end{eqnarray}
It can also be written as~\eqref{k_diff} where the first and
second term are expressed in the form derived with covariant phase
space methods~\cite{Iyer:1994ys,Wald:1999wa},
\begin{eqnarray}
k^K_{\cL^{EH},\xi} = \frac{\sqrt{-g}}{16\pi G} (D^\mu \xi^\nu -
D^\nu
\xi^\mu )(d^{n-2}x)_{\mu\nu},\label{Komar}\\
I_{\dv g}^{n}\cL^{EH}[\dv g] = \frac{\sqrt{-g}}{16\pi G}
(g^{\mu\alpha} D^\beta \dv g_{\alpha\beta} - g^{\alpha\beta} D^\mu
\dv g_{\alpha\beta} ) (d^{n-1}x)_\mu.\label{Theta_t}
\end{eqnarray}
Here, expression~\eqref{Komar} is called the Komar term. The
supplementary term
\begin{equation}
E_{\cL^{EH}}[\cL_\xi \phi,\dv g] = \frac{\sqrt{-g}}{16\pi G}
(\half g^{\mu\alpha}\dv g_{\alpha\beta} (D^\beta \xi^\nu + D^\nu
\xi^\beta) - (\mu \leftrightarrow \nu) )
(d^{n-2}x)_{\mu\nu},\label{suppl}
\end{equation}
vanishes for exact Killing vectors of $g$, but not necessarily for
asymptotic ones. In the case where $\xi$ may vary, it is
convenient to write~\eqref{k_diff} as
\begin{equation}
k_\xi[\dv \phi] =  -\dv^\Phi k^K_{\cL^{EH},\xi}
+k^K_{\cL^{EH},\dv\xi} +i_\xi I^{n}_{\dv
\phi}\cL^{EH}-E_{\cL^{EH}}[\cL_\xi \phi,\dv \phi],\label{k_diff2}
\end{equation}
where the extended vertical differential is defined
in~\eqref{vertdiff2} and where we omit the irrelevant exact
horizonal differential. The fundamental relation~\eqref{eq:29a}
reads in this case as
\begin{eqnarray}
  \dH k_\xi[\dv g;g]=W_{{\delta\cL^{EH}}/{\delta\phi}}[\dv
g,\cL_\xi g]-\dv^g S_\xi+T_{\cL_\xi g}[\dv g ,
\varQ{\cL^{EH}}{g}],\label{eq:19a}
\end{eqnarray}
where the invariant symplectic form $W$ and the weakly vanishing
form $T$ are given by
\begin{eqnarray}\begin{aligned}
W_{\varQ{\cL^{EH}}{\phi}}[\dv g,\cL_\xi g] &=
P^{\mu\delta\beta\gamma\varepsilon\zeta}\Big( \dv g
_{\beta\gamma}\nabla_\delta \cL_\xi g_{\varepsilon\zeta}- \cL_\xi
g_{\beta\gamma}\nabla_\delta \dv g_{\varepsilon\zeta}
 \Big) (d^{n-1}x)_\mu,\\
T_{\cL_\xi g}[\dv g , \varQ{\cL^{EH}}{g}]&=
 \dv g_{\alpha\beta}\varQ{L^{EH}}{g_{\alpha\beta}}\xi^\mu
(d^{n-1}x)_\mu.\end{aligned}\label{dHgrav}
\end{eqnarray}
The property~\eqref{simplif} is satisfied. The integrability
conditions for the surface one-forms are given
by~\eqref{int_diff}.

The covariant phase space expression~\cite{Iyer:1994ys} reads as
\begin{eqnarray}
k^{\text{I-W}}_\xi[\dv g;g] &=& \frac{\sqrt{-g}}{16\pi G} \Big[
\xi^\nu D^\mu h + \frac{1}{2} h D^\nu \xi^\mu +\xi^\mu D_\sigma
h^{\nu \sigma}+D^\nu
h^{\mu \sigma}\xi_\sigma \nonumber\\
&&\qquad  + h^{\mu\sigma}D_\sigma \xi^\nu - (\mu \leftrightarrow
\nu) \Big] (d^{n-2}x)_{\mu\nu}\label{kIW}
\end{eqnarray}
and differs from~\eqref{hom} by the term~\eqref{suppl} vanishing
for exact Killing vectors. As a consequence of~\eqref{eq:79abis}
and~\eqref{simplif}, we also have
\begin{eqnarray}\label{kIWand Omega}
k^{\text{I-W}}_\xi[\dv g;g] &=& I_\xi^{n-1}\Omega_{\cL^{EH}}
[\cL_\xi g,\dv g]
\end{eqnarray}
Remark that the expressions~\eqref{kIW} and \eqref{kIWand Omega}
lack in the beautiful symmetry properties of
expressions~\eqref{ch_grav1} and \eqref{dHgrav} where the tensor
$P^{\alpha\beta\gamma\delta \mu\nu}$ obeys~\eqref{pmunu}. This
provides an additional aesthetic argument in favor of
definition~\eqref{def}.

\subsection{General relativity in ADM form}

The surface terms that should be added to the Hamiltonian
generator of surface deformations in Einstein gravity are
well-known~\cite{Regge:1974zd,Henneaux:1985tv,Henneaux:1985ey}.
Although these surface terms were derived for deformations in the
asymptotic region, they can be used for infinitesimal surface
deformations inside the bulk. According to
Proposition~\bref{prop_hamilt}, the surface terms obtained by
varying the constraints smeared by the surface deformation
generators $\eps$ are given by the $[0a]$ component of the
$(n-2,1)$-form $k_\eps$ for Einstein gravity written in ADM
variables. These components are the only ones relevant in order to
compute the infinitesimal charges $\ndelta\cQ_\eps$~\eqref{eq:17}
associated with surface deformations $\eps$ on the surface $S$,
$t=constant$ and $r = constant$,
\begin{equation} \ndelta\cQ_\eps = \oint_S
\text{d}\sigma_a k_\eps^{[0a]},\label{eq:ch_0a}
\end{equation}
where $\text{d}\sigma_a \equiv 2(d^{n-2}x)_{0a}$. This section is
devoted to check that the surface terms obtained by our method
indeed reproduce the Hamiltonian surface terms.

The action for pure gravity in ADM variables
$(\gamma_{ab},\pi^{ab},N, N^a)$ in $n$ dimensions is the
straightforward generalization of the four dimensional
case~\cite{Arnowitt:1962aa},
\begin{equation}
S_{ADM} = \int dt d^{n-1}x \left[ \pi^{ab} \dot \gamma_{ab}  - N
\cH - N^a \cH_a \right].\label{lagr_ADM}
\end{equation}
It has the Hamiltonian form~\eqref{eqApp:1} with variables
$N^A=(N\equiv N^\perp,N^a)$ as Lagrange multipliers. The
constraints $\cH$ and $\cH_A$ are given by
\begin{equation}
\cH \equiv \frac{1}{\sqrt{\gamma}} (\pi^{ab}\pi_{ab} -
\frac{1}{n-2} \pi^2) - \sqrt{\gamma}\, {}^3\hspace{-2pt}R=0,
\qquad \cH_a \equiv -2{\pi_a^{\, \, \, b}}_{|b}=0.
\end{equation}
An arbitrary variation with gauge parameter $\eps^A$ leads to
\begin{eqnarray}
  \stackrel{\leftarrow}{\delta_\eps} \gamma_{ab} &=&
\stackrel{\leftarrow}{\delta_{\hat \cH_A\eps^A}}\gamma_{ab} =
\varQ{(\eps^A\cH_A)}{\pi^{ab}} \nonumber \\
&=& 2\eps^\perp\sqrt{\gamma}^{-1}
(\pi_{ab}- \frac{1}{n-2}\gamma_{ab}\pi) +\eps_{a|b}+\eps_{b|a} \label{eq:12re}\\
\stackrel{\leftarrow}{\delta_\eps} \pi^{ab} &=&
\stackrel{\leftarrow}{\delta_{\hat \cH_A\eps^A}} \pi^{ab} = -
\varQ{(\eps^A\cH_A)}{g_{ab}} =
-\eps^\perp\sqrt{\gamma}(R^{ab}-\frac{1}{2} \gamma^{ab}R) \nonumber\\
&+&\frac{1}{2}\eps^\perp\sqrt{\gamma}^{-1}\gamma^{ab}(\pi^{cd}\pi_{cd}
-\frac{1}{n-2}\pi^2)-2\eps^\perp\sqrt{\gamma}^{-1}(\pi^{ac}\pi_{c}^{\,\,\,
    b}-\frac{1}{n-2}\pi
  \pi^{ab})\nonumber \\
&+&\sqrt{\gamma}((\eps^\perp)^{|ab}-\gamma^{ab}(\eps^\perp)^{|c}_{\,\,\,
    |c}) +(\pi^{ab}\eps^c)_{|c}-\eps^a_{\,\, |c}\pi^{cb}-\eps^b_{\,\,
    |c}\pi^{ac}.\label{eq:12re2}
\end{eqnarray}

For arbitrary functions $\xi_{1,2}^A(x)$ vanishing sufficiently
fast at infinity, the Poisson brackets of the constraints are
explicitly given by \cite{Teitelboim:1972vw, Teitelboim:1973yj}
\begin{eqnarray}
&&  \{\int \hat \cH_A \xi_1^A, \int \hat \cH_B\xi_2^B \} =\int
\hat \cH_C C_{AB}^C(\xi_1^A,\xi_2^B)  ,\nonumber\\
&&C_{BC}^\perp(\xi_1^B,\xi_2^C)= \xi_1^a {\xi_2^\perp}_{,a}
 -\xi_2^a {\xi_1^\perp}_{,a} ,\\
&&C_{BC}^a(\xi_1^B,\xi_2^C) =
\gamma^{ab}(\xi_1^\perp{\xi_2^\perp}_{,b}-\xi_2^\perp
{\xi_1^\perp}_{,b}) +\xi_1^b {\xi_2^a}_{,b}-\xi_2^b
{\xi_1^a}_{,b}.\nonumber
\end{eqnarray}
The variation of the Lagrange multipliers is given by
\begin{equation} \delta_\eps N^A =
\partial_0 \eps^A + C^A_{BC}(\eps^B,N^C).\end{equation}
\paragraph{Surface charges} The weakly vanishing Noether current
$S^\mu_\eps=S^{\mu  I}_{B}(\varQ{L}{\phi^I},\eps^B)$ is obtained
by integration by
  parts,
\begin{eqnarray}
R^I_A (f^A) \varQ{L}{\phi^I}\hspace{-5pt} &=&\hspace{-5pt}
\stackrel{\leftarrow}{\delta_\eps}
\gamma_{ab}(-\d_0\pi^{ab}+\stackrel{\leftarrow}{\delta_N}
\pi^{ab})+ \stackrel{\leftarrow}{\delta_\eps}
\pi^{ab}(\d_0\gamma_{ab}-\stackrel{\leftarrow}{\delta_N}\gamma_{ab}
)+\delta_\eps N^A (-\cH_A)\nonumber\\
&=& \hspace{-5pt}\d_\mu S^{\mu I}_{A}(\varQ{L}{\phi^I},\eps^A).
\end{eqnarray}
Explicitly,
\begin{eqnarray}
S^0_\eps &=& - \eps^A \cH_A, \\
S^a_\eps &=&\eps^A\cH_A
N^a+\eps^\perp\cH^aN+2\eps_b(-\d_0\pi^{ab}+
\stackrel{\leftarrow}{\delta_N} \pi^{ab})+
\Big[\eps^a\pi^{cd}-\eps^c
\pi^{da}-\eps^d \pi^{ac}\nonumber\\
&&-  \eps^\perp \sqrt{\gamma}\gamma^{ad} D^b+ \eps^\perp
\sqrt{\gamma}\gamma^{bd} D^a +(\eps^\perp)^{|b}\sqrt{\gamma}
\gamma^{ad}
-(\eps^\perp)^{|a}\sqrt{\gamma} \gamma^{bd}\Big]\nonumber\\
&&\times \Big[\d_0\gamma_{cd}-
\stackrel{\leftarrow}{\delta_N}\gamma_{cd}\Big].\label{eq:S_frav}
\end{eqnarray}
Note that the factors explicitly depending on the dimension $n$ in
\eqref{eq:12re}-\eqref{eq:12re2} do not contribute to the current
because they do not involve derivatives of the parameters
$\eps^A$. The time-dependent terms in~\eqref{eq:S_frav} make up
the term $V^k_{B}[\dot z^B,\gamma_a f^a]$ in~\eqref{eqApp:5}.
Therefore, introducing the inverse De Witt
supermetric~\cite{Dewitt:1967yk} as in~\eqref{deWitt},
\begin{eqnarray}
G^{abcd} &=& \half \sqrt{\gamma}(\gamma^{ac}\gamma^{bd} +
\gamma^{ad}\gamma^{bc} - 2 \gamma^{ab}\gamma^{cd}),
\end{eqnarray}
we can straightforwardly write the expression~\eqref{eq:37} for
$k^{[0a]}_\eps$ as
\begin{eqnarray}
k_\eps^{\text{R-T}\; [0a]} = G^{abcd}(\eps^\perp D_b\dv
\gamma_{cd}-D_b\eps^\perp \dv \gamma_{cd}) + 2\eps^c\dv
\pi_c^{\,\,\, a} -\eps^a \dv \gamma_{cd}\pi^{cd},\label{k_Teit}
\end{eqnarray}
where $\dv \pi_c^{\,\,\,
a}=\gamma_{cd}\dv\pi^{da}+\dv\gamma_{cd}\pi^{da}$. This indeed
reproduces the Regge-Teitelboim expression \cite{Regge:1974zd} as
well as the expression used in anti-de Sitter
backgrounds~\cite{Henneaux:1985tv,Henneaux:1985ey}.

\subsection{Canonical reduction in Einstein gravity}
\label{sec:canonic}

In the last section, we showed that the Regge-Teitelboim
expression~\eqref{k_Teit} is the $[0a]$ component of the surface
one-form associated with the Lagrangian~\eqref{lagr_ADM}. In
section~\ref{sec:cov_grav}, we also showed that the
Einstein-Hilbert Lagrangian supplemented or not with boundary
terms leads to the surface
one-form~\eqref{ch_grav1}-\eqref{hom}-\eqref{gsuperpot3} that will
be referred to as the Abbott-Deser expression. Since both
computations use different homotopy formulas, one in terms of the
covariant metric $g_{\mu\nu}$ and the other in terms of the ADM
variables $(\gamma_{ab},\pi^{ab},N, N^a)$ the Regge-Teitelboim
expression~\eqref{k_Teit} and the $[0a]$ components of the
Abbott-Deser expression~\eqref{hom} might differ.

However, general results on the BRST
cohomology~\cite{Barnich:1994db} ensure the invariance of the
cohomology of reducibility parameters modulo trivial ones in the
transition from Lagrangian to Hamiltonian formalisms.
Proposition~\bref{prop1bis} on page~\pageref{prop1bis} then
guarantees the equivalence between the surface one-forms of both
formalisms up to boundary terms when the equations of motion hold
and when the reducibility equations hold. The Regge-Teitelboim and
the Abbott-Deser expressions may thus only differ by boundary
terms, by terms proportional to the equations of motion and their
derivatives and finally by terms proportional to the reducibility
equations and their derivatives. These terms are computed
hereafter.

We distinguish the indices $\mu = 0,i$, $i = 1,2,\dots n-1$ in the
coordinate basis and $A = \perp,a$, $a = 1,2, \dots n-1$ in the
Hamiltonian basis. In what follows, $\gamma_{ab}$ denote the
spatial metric $\gamma_{ab} = \delta_a^i \delta_b^j g_{ij}$.
Tensors are transformed under the change of basis according to the
following matrices
\begin{eqnarray}
B^\nu_{\;\;A} = \left( \begin{array}{cc} \frac{1}{N} & 0\\
-\frac{N^a}{N}\delta^{i}_{a} &  \delta^{i}_{a}
\end{array} \right), \qquad B^A_{\;\;\nu} = \left( \begin{array}{cc} N & 0\\
N^a &  \delta^{a}_{i}
\end{array} \right).
\end{eqnarray}
The connection one-form $\Gamma_{\rho}^\nu = \Gamma_{\mu\rho}^\nu
\text{d}x^\mu$ becomes in the new frame the connection one-form
$\omega^A_{B}$ given by
\begin{equation}
\omega^A_{\mu B} = B^{\nu}_{\;\;B,\mu}B^A_{\;\;\nu} +
\Gamma^\nu_{\mu\rho}B_{\;\;\nu}^A B^\rho_{\;\;B}.
\end{equation}
After a long but straightforward computation, one gets
\begin{eqnarray}
\omega^A_{\perp B} &= &\left(\begin{array}{cc}
\omega^\perp_{\perp\perp} & \omega^\perp_{\perp b}\\
\omega^a_{\perp \perp} & \omega^a_{\perp b}
\end{array} \right) = \left(\begin{array}{cc}
0 & N_{,b}/N\\
N^{,a}/N & -K^a_b+N^a_{\;,b}/N
\end{array} \right),\\
\omega^A_{a B} &= &\left(\begin{array}{cc}
\omega^\perp_{a\perp} & \omega^\perp_{a b}\\
\omega^c_{a \perp} & \omega^c_{a b}
\end{array} \right) = \left(\begin{array}{cc}
0 & -K_{ab}\\
-K^c_a & \mbox{}^{(3)}\Gamma^c_{ab}
\end{array} \right).
\end{eqnarray}
The gauge transformation $\cL_\xi g_{ab}$ reads as
\begin{equation}
\cL_\xi g_{ab} = D_b\xi_a + D_a \xi_b =
\xi_{a|b}+\xi_{b|a}+2K_{ab}\xi_\perp.
\end{equation}
Therefore, comparing the latter expression with~\eqref{eq:12re}
and using the equations of motion~$\pi^{ab} \approx -\sqrt{\gamma}
(K^{ab}-\gamma^{ab}K)$, one can identify on-shell the Hamiltonian
surface deformation $\eps$ with the Lagrangian infinitesimal
diffeomorphism generators, $\xi \approx \eps$.

Using $k_\xi^{[0i]} =B^{0}_{\;\; A}B^{i}_{\;\; B}k^{AB} =
\frac{1}{N} \delta^i_a k^{[\perp a]}$, one can write the
infinitesimal charge $\ndelta\cQ_\xi$~\eqref{eq:17} associated
with $\xi$ and adapted to the surface $S$, $t=constant$ and $r =
constant$ as
\begin{equation}
\ndelta\cQ_\xi = \oint_S \text{d}\sigma_a \frac{1}{N}
k_\xi^{[\perp a]}.\label{eq:int_kperp}
\end{equation}
Using
\begin{eqnarray} 2 \xi^c
\delta_h(\pi^a_{\,\,\,\, c})-\xi^a h_{cd}\pi^{cd} &=&
\sqrt{\gamma}\xi^a \big(h_{cd}K^{cd}+2\delta_h K \big) +
\nonumber\\
&&\hspace{-25pt} +\sqrt{\gamma}\xi_c \big(-2h^c_{\,\,\,
d}K^{ad}-K^{ac}\mbox{}^{(3)}h-2\delta_h K^{ac}\big),
\end{eqnarray}
and developing $\delta_h K^{ac}$ and $\delta_h K$ in terms of $\dv
g_{\mu\nu}=h_{\mu\nu}$, one can after some algebra relate the
$(n-2,1)$ forms~\eqref{gsuperpot3} and~\eqref{k_Teit} as
\begin{eqnarray}
(16\pi G)\frac{1}{N} k_\xi^{\text{A-D}\; [\perp a]} &\approx&
(16\pi G) \frac{1}{N} k_\eps^{\text{R-T}\;[\perp a]} +
\sqrt{\gamma}(h^{\perp
b}\xi^a-h^{\perp a }\xi^b)_{|b} \nonumber\\
&&\hspace{-40pt}- G^{abcd}h^\perp_{\;\; b}D_c\xi_{d} +
\frac{1}{2}\sqrt{\gamma}(h^{a}_{\;\; b}-h \delta^a_{\; b})(D^\perp
\xi^b + D^b \xi^\perp).\label{eq:rel1}
\end{eqnarray}
For exact Killing vectors, one recovers the result of the
reduction performed in~\cite{Anderson:1996sc}. The
Regge-Teitelboim~\eqref{k_Teit} and the Iyer-Wald~\eqref{kIW}
expressions are related by
\begin{eqnarray}
(16\pi G)\frac{1}{N} k_\xi^{\text{I-W}\;\perp a} &\approx& (16\pi
G)\frac{1}{N} k_\eps^{\text{Teit}\;[\perp a]} +
\sqrt{\gamma}(h^{\perp
b}\xi^a-h^{\perp a }\xi^b)_{|b} \nonumber \\
&&\hspace{-2.7cm}- \sqrt{\gamma}h^{a\perp}(D_\perp \xi^\perp-D_b
\xi^b)-
\frac{1}{2}\sqrt{\gamma}(\mbox{}^{(3)}h-h^\perp_{\;\;\perp})(D^\perp
\xi^a+D^a \xi^\perp)\label{eq:rel2}.
\end{eqnarray}
Besides a total divergence, the right-hand side
of~\eqref{eq:rel1}~\eqref{eq:rel2} contains terms proportional to
$D_\mu\xi_\nu+D_\nu\xi_\mu$. Therefore, we showed that the
one-forms obtained by integration of~\eqref{k_Teit},
\eqref{gsuperpot3} and \eqref{kIW} all agree on-shell for exact
Killing vectors, as expected. However, in the asymptotic context,
for vectors $\xi$ which are not Killing vectors, these expressions
might be different.

\section{Gravity coupled to a $p$-form potential and a scalar}
\label{sec:pform}

A great motivation to study classical conservation laws for
gravity coupled to $p$-forms with $p \geq 1$ and to scalar fields
is the natural occurrence of such theories in string theory and in
alternative theories of gravity.

A particular topic where such conservation laws are of interest is
the thermodynamics of black rings that will be studied in
Chapter~\ref{chap:BHapp}. The original black ring
solution~\cite{Emparan:2001wn} is a black hole solution to vacuum
Einstein gravity in five dimensions admitting a non-trivial
horizon topology. Once five-dimensional gravity is coupled to a
$2$-form potential, black rings may acquire a dipole
charge~\cite{Emparan:2004wy}.

Hamiltonian methods were developed in order to cover
conservation laws when $p$-form potential are
present~\cite{Sudarsky:1992ty,Copsey:2005se}. Covariant phase
space methods have also been
applied~\cite{Rogatko:2005aj,Rogatko:2006xh}. The main aim of this
section is to improve the covariant
analysis~\cite{Rogatko:2005aj,Rogatko:2006xh} by rederiving the
conserved charges using covariant cohomological
methods~\cite{Barnich:2001jy,Barnich:2003xg} in a notation taking
care of form factors. The conservations laws for gravity coupled
to a scalar field have been written in~\cite{Barnich:2002pi} and
will also be included here for completeness. The material
developed in this section was published in~\cite{Compere:2007vx}.

In what follows, we consider the action
\begin{eqnarray}
\hspace{-6pt} S[g,\mathbf A,\phi] \hspace{-6pt}&=& \hspace{-6pt}
\frac{1}{16 \pi G}\int  \left[\star 1\, (R - \frac{1}{2}\d_\mu\chi
\d^\mu \chi+ V(\chi) ) - \frac{1}{2} e^{-\alpha \chi} \mathbf H
\wedge \star \mathbf H \right],\label{act1}
\end{eqnarray}
where $\chi$ is a dilaton and $\mathbf H = d \mathbf A$ is the
field strength of a $p$-form $\mathbf A$, $p\geq 1
$~\footnote{Here, all forms are written with bold letters,
$\mathbf A = \frac{1}{p!}A_{\mu^1\cdots \mu^p}dx^{\mu^1}\wedge
\dots \wedge dx^{\mu^p}$.}. The fields of the theory are
collectively denoted by $\phi^i \equiv (g_{\mu\nu}, \mathbf
A,\chi)$. We will set $16\pi G = 1$ for convenience.

\subsection{Conservation laws}

In Minkowski spacetime $g_{\mu\nu}= \eta_{\mu\nu}$, $\chi = 0$ and
for a trivial bundle $\mathbf A$, all conservation laws are
classified by the characteristic cohomology of $p$-form gauge
theories~\cite{Henneaux:1996ws}. These laws are generated in the
exterior product by the forms $\star \mathbf H$ dual to the field
strength~\footnote{When magnetic charges are allowed, there are
additional conserved quantities as $\oint \mathbf H \neq 0$.
However, the field strength $\mathbf H$ cannot be written as the
derivative of a potential $\mathbf B$ and the action principle has
to be modified. This case will not be treated below.}. More
precisely, for odd $n-p-1$, one can construct the conserved
$n-p-1$-form $\star \mathbf H$. For even $n-p-1$, factors $\star
\mathbf H$ mutually commute and one may construct the conserved
forms $l(n-p-1)$ $\underbrace{\star \mathbf H \wedge \dots \wedge
\star\mathbf H}_l$ for any integer $l$ such that $l(n-p-1) < n-1$.

When gravity and the scalar field are present, the charges
\begin{eqnarray}
\mathbf Q^{(n-p-1)} &=& e^{-\alpha \chi} \star \mathbf H, \qquad
\qquad \qquad n-p-1\;\text{odd}\label{elc_ch}
\\\label{ch_other}\mathbf  Q^{l(n-p-1)}& =& e^{-l\alpha \chi}
\underbrace{\star \mathbf H \wedge \dots \wedge \star\mathbf
H}_l,\qquad n-p-1\;\text{even}
\end{eqnarray}
still enumerate the non-trivial conservation
laws~\cite{Barnich:1995ap,Henneaux:1996ws}~\footnote{The
conservations laws that we consider here are called dynamical
because they explicitly involve the equations of motion. There
exists also specific topological conservation laws, see
e.g.~\cite{Torre:1994pf}.}, see also discussions in the Preamble,
especially in section~\ref{sec:localhomo}.

In order to investigate the first law of thermodynamics, where
variations around a solution are involved, we now extend the
analysis to the linearized theory.

In linearized gravity, only $(n-2)$-form conservation laws are
allowed~\cite{Barnich:1995db,Barnich:2004ts}. The classification
of non-trivial conserved $(n-2)$-forms was described in
\cite{Barnich:2001jy} and is straightforward to specialize in our
case. The equivalence classes of conserved $(n-2)$-forms of the
linearized theory for the variables $\delta\phi^i$ around a fixed
reference solution $\phi^i$ are in correspondence with equivalence
classes of gauge parameters $\xi^\mu(x),\mathbf \Lambda(x)$
satisfying the reducibility equations $\delta_{\xi,\mathbf\Lambda}
\phi^i = 0$~\footnote{This correspondence is one-to-one for gauge
parameters that may depend on the linearized fields $\varphi^i$
and that
satisfy~$\delta_{\xi(x,\varphi^i),\mathbf\Lambda(x,\varphi^i)}
\phi^i \approx_{lin} 0$, i.e. zero for solutions $\varphi^i$ of
the linearized equations of motion. However, it has been proven
in~\cite{Barnich:2004ts} that this $\varphi$-dependence is not
relevant in the case of Einstein gravity. Such a dependence will
not be considered in this section anymore.}, i.e.
\begin{eqnarray}
\left\{\begin{array}{c}
  \cL_\xi  g_{\mu\nu} = 0,\\ \cL_\xi  {\mathbf A} + d
 \mathbf \Lambda = 0,\\
  \cL_\xi  \chi = 0.
\end{array}\right. \label{eq:red}
\end{eqnarray}

In the next section, we will compute the $(n-2,1)$-form $\mathbf
k_{\xi,\mathbf \Lambda}$ associated with gauge parameters
$(\xi,\mathbf \Lambda)$. For parameters satisfying the
reducibility equations~\eqref{eq:red}, the infinitesimal
charge~\eqref{eq:17} between solutions $\phi^i$ and $\phi^i+\delta
\phi^i$,
\begin{eqnarray}
  \ndelta \cQ_{ \xi, \mathbf\Lambda} \hat =
  \oint_S \mathbf k_{ \xi,\mathbf \Lambda}[ \delta \phi ;
  \phi],\label{ch_to}
\end{eqnarray}
will then only depend on the homology class of $S$.

\subsection{Conserved surface one-forms}

Following the lines of Chapter~\ref{chap:general_th}, one can
construct the weakly vanishing Noether currents associated with
the couple $(\xi,\mathbf \Lambda)$ by integrating by parts the
expression~$\delta_{\xi,\mathbf\Lambda} \phi^i \varQ{\mathbf
L}{\phi^i}$ and using the Noether identities. We obtain
\begin{eqnarray}
  \label{eq:4bis}
\mathbf S_{\xi,\mathbf\Lambda} &=& \star \bigg(
(-2G_\mu^{\;\,\nu}+ T_{\mathbf A\,\mu}^{\,\;\,\nu}+
T_{\chi\;\mu}^{\;\,\nu} )\xi_\nu dx^\mu \label{eq:S1}
\\
&-&  \frac{1}{(p-1)! }D_\beta(e^{-\alpha
\chi}H_\mu^{\;\,\beta\mu^1\cdots \mu^{p-1}})(\xi^\rho A_{\rho\mu^1
\cdots \mu^{p-1}}+\Lambda_{\mu^1 \cdots \mu^{p-1}})dx^\mu
\bigg),\nonumber
 \end{eqnarray}
where the stress tensors are given by
\begin{eqnarray}
T_{\mathbf A}^{\mu\nu} &=& e^{-\alpha \chi} \left( \frac{1}{p!}
H^{\mu}_{\;\;\mu^1 \cdots \mu^{p}}H^{\nu\mu^1 \cdots \mu^{p}} -
\frac{1}{2(p+1)!}g^{\mu\nu}H^2 \right) ,\\
T_\chi^{\mu\nu} &=& (\d^\mu \chi \d^\nu \chi -
\frac{1}{2}g^{\mu\nu}\d^\alpha \chi \d_\alpha\chi).
\end{eqnarray}
The conserved $(n-2,1)$ form~$\mathbf k_{\xi,\mathbf\Lambda}[\dv
\phi;\phi]=k_{\xi,\mathbf\Lambda}^{[\mu\nu]} (d^{n-2}x)_{\mu\nu}$
can be obtained as a result of a contracting homotopy $\mathbf
I^{n-1}_{\dv \phi}$ acting on the current $\mathbf
S_{\xi,\mathbf\Lambda}$, see~\eqref{def}. Using the
property~\eqref{cc1a} of the homotopy
operators~\footnote{In~\cite{Compere:2007vx}, the $(n-2)$-form
$\mathbf k_{\xi,\mathbf\Lambda}[\delta \phi;\phi]$ was computed
with $\delta\phi$ Grassmann even. Some sign factors have thus to
be adapted with respect to~\cite{Compere:2007vx}.},
\begin{equation}
-\dH \mathbf I_{\dv \phi}^{q-1}\mathbf \omega^{(q-1)}+ \mathbf
I_{\dv \phi}^{q} \dH \mathbf \omega^{(q-1)}= \dv\mathbf
\omega^{(q-1)},\qquad \forall \mathbf \omega^{(q-1)},\quad q \leq
n ,\label{prop_I}
\end{equation}
one has
\begin{equation}
\dH  \mathbf k_{\xi,\mathbf\Lambda} =- \dv \mathbf
S_{\xi,\mathbf\Lambda} + \mathbf I_{\dv
\phi}^{n-2}\left(\delta_{\xi,\mathbf\Lambda} \phi^i \varQ{\mathbf
L}{\phi^i} \right).
\end{equation}
The form $\mathbf k_{\xi,\mathbf\Lambda}[\dv \phi;\phi]$ is closed
 whenever $\phi^i$ satisfies the equations of motion, $\dv
\phi^i$ the linearized equations of motion and
$(\xi,\mathbf\Lambda)$ the system \eqref{eq:red}.

Let us now split the current into different contributions,
$\mathbf S_{\xi,\mathbf\Lambda} =\mathbf S^{g}_\xi + \mathbf
S^{\chi}_{\xi} + \mathbf S^{\mathbf A}_{\xi,\mathbf\Lambda}$ with
\begin{eqnarray}
\mathbf S^{g}_\xi &=&  \star (-2G_\mu^{\;\,\nu}\xi_\nu \,dx^\mu), \\
\mathbf S^{\chi}_{\xi} &=&\star ( T_{\chi \; \mu}^{\;\,\nu}
\xi_\nu \, dx^\mu),
\end{eqnarray}
and $\mathbf S^{\mathbf A}_{\xi,\mathbf\Lambda}$ being the
remaining expression. Since the homotopy $\mathbf I^{n-1}_{\dv
\phi}$ is linear in its argument, the conserved $n-2$ form can be
decomposed as $\mathbf k_{\xi, \mathbf\Lambda} = \mathbf
k^{g}_{\xi}+ \mathbf k^{\chi}_{\xi} + \mathbf k^{\mathbf A}_{\xi,
\mathbf\Lambda}$.

The gravitational contribution $\mathbf k^{g}_{\xi}$ which depends
only on the metric and its deviations was given in
section~\ref{sec:general-relativity}. This contribution can be
written in a form notation as\footnote{We recall that $\dv$ is
defined by~\eqref{vertdiff} and thus acts on the fields and not on
the gauge parameters.}
\begin{eqnarray}
\mathbf k^{g}_{ \xi}[\dv g] &=& -\dv \mathbf Q^g_{ \xi}
+i_{\xi}\mathbf \Theta -\mathbf E_\cL[\cL_\xi g, \dv
g],\label{grav_contrib}
\end{eqnarray}
where
\begin{eqnarray}
\mathbf Q^g_{\xi}&=& \star \Big( \half (D_\mu\xi_\nu-D_\nu\xi_\mu)
dx^\mu \wedge dx^\nu \Big),\label{Komar_term}
\end{eqnarray}
is the Komar $(n-2)$-form and
\begin{eqnarray}
\mathbf \Theta[\dv g]&=&\star \Big(  (D^\sigma \dv g_{\mu\sigma}-
g^{\alpha\beta} D_\mu \dv g_{\alpha\beta})\,dx^\mu\Big),\\
\mathbf E_\cL[\cL_\xi g, \dv g] &=& \star \Big( \half \dv
g_{\mu\alpha}(D^\alpha \xi_\nu + D_\nu \xi^\alpha) dx^\mu \wedge
dx^\nu \big).
\end{eqnarray}
The scalar contribution is easily found to be $\mathbf k^{\chi}_{
\xi}[\dv g,\dv\chi;g,\chi] = -i_\xi \mathbf
\Theta_\chi$~\cite{Barnich:2002pi} with
\begin{equation}
\mathbf \Theta_\chi = \star (\dv \chi  \, \dH \chi )
.\label{phicharge}
\end{equation}

Let us now compute the contribution $\mathbf k^{\mathbf A}_{\xi,
\mathbf\Lambda}$ from the $p$-form. After some algebra, one can
rewrite the current $\mathbf S^{\mathbf A}_{\xi,\mathbf\Lambda}$
as
\begin{eqnarray}
\mathbf  S^{\mathbf A }_{\xi,\mathbf\Lambda}&=&- \dH \mathbf
Q^{\mathbf A}_{\xi,\mathbf \Lambda} + e^{-\alpha \chi}(\cL_\xi
\mathbf A+d\mathbf \Lambda )\wedge \star \mathbf H - \half
e^{-\alpha \chi } i_\xi(\mathbf H \wedge \star \mathbf H)
\end{eqnarray}
with
\begin{equation}
\mathbf Q^{\mathbf A}_{ \xi,\mathbf\Lambda}= e^{-\alpha \chi} (
i_\xi \mathbf A +\mathbf \Lambda ) \wedge \star\mathbf
H.\label{def_QA}
\end{equation}
Using the property~\eqref{prop_I}, the $(n-2)$-form $\mathbf
k^{\mathbf A}_{ \xi,\mathbf\Lambda}$ reduces to
\begin{eqnarray}
\mathbf k^{\mathbf A}_{ \xi,\mathbf\Lambda} &=& -\dv\mathbf
Q^{\mathbf A}_{\xi,\mathbf\Lambda}  + \mathbf Q^{\mathbf
A}_{\dv\xi,\dv\mathbf\Lambda}- \dH  \mathbf
I^{n-2}_{\dv\phi}\mathbf Q^{\mathbf A}_{\xi,\mathbf\Lambda} \nonumber\\
&& + \mathbf I_{\dv \phi}^{n-1}\big( e^{-\alpha \chi}(\cL_\xi
\mathbf A+\dH\mathbf \Lambda )\wedge \star \mathbf H - \half
e^{-\alpha \chi } i_\xi(\mathbf H \wedge \star \mathbf H) \big),
\end{eqnarray}
where the exact term $\dH  \mathbf I^{n-2}_{\dv\phi}\mathbf
Q^{\mathbf A}_{\xi,\mathbf\Lambda}$ is trivial and can be dropped.
The last term can then be computed easily since it admits only
first derivatives of the gauge potential. The homotopy thus
reduces in that case to $I^{n-1}_{\dv \mathbf A} = \half \dv
\mathbf A \frac{\partial}{\partial \mathbf H}$. We eventually get
\begin{equation}
\mathbf k^{\mathbf A}_{ \xi,\mathbf\Lambda}[\dv g,\dv \mathbf
A,\dv \chi]=-\dv \mathbf Q^{\mathbf A}_{\xi,\mathbf\Lambda} +
\mathbf Q^{\mathbf A}_{\dv\xi,\dv\mathbf\Lambda} - i_\xi \mathbf
\Theta^{\mathbf A}-\mathbf E^{\mathbf A}_\cL[\cL_\xi \mathbf
A+\dH\mathbf\Lambda;\dv \mathbf A] \label{Bcharge}
\end{equation}
with
\begin{eqnarray}
\mathbf \Theta^{\mathbf A} &=& e^{-\alpha \chi} \dv \mathbf A
\wedge
\star \mathbf H,\label{ThetaA}\\
\mathbf E^{\mathbf A}_\cL[\cL_\xi \mathbf A+\dH\mathbf\Lambda;\dv
\mathbf A] &= & e^{-\alpha \chi} \star \big( \half
\frac{1}{(p-1)!}\dv \mathbf
A_{\mu\alpha_1\cdots \alpha_{p-1}} \nonumber \\
&&\hspace{-30pt}(\cL_\xi \mathbf A+\dH\mathbf
\Lambda)_\nu^{\;\,\,\alpha_1\cdots \alpha_{p-1}} dx^\mu\wedge
dx^\nu \big)
\end{eqnarray}
which has a very similar structure as the gravitational field
contribution~\eqref{grav_contrib}. For exact reducibility
parameters~\eqref{eq:red}, the term involving $\cL_{\xi} \mathbf
A+\dH \mathbf \Lambda$ will be zero. The form~\eqref{def_QA} will
be referred to as a Komar term, in analogy with the gravitational
Komar term~\eqref{Komar_term}.

\paragraph{Properties of the surface one-form.}

Let us suppose that $(\xi,\mathbf \Lambda)$ are exact reducibility
parameters. For $p=1$, the form~\eqref{Bcharge} reduces to
well-known expressions for Einstein-Maxwell theory, see e.g.
\cite{Gao:2001ut}. For $p$ arbitrary, expression~\eqref{Bcharge}
and the one derived in~\cite{Rogatko:2005aj,Rogatko:2006xh} have
the same structure but differ from form factors. More precisely,
both expressions agree when the right-hand side of equation~(10)
of~\cite{Rogatko:2005aj} and equation (4) of~\cite{Rogatko:2006xh}
are multiplied by $-\frac{p+1}{2}$.

As a consistency check, note that the form~\eqref{Bcharge}
satisfies the equality on-shell $\mathbf k^{\mathbf A}_{
\xi,\mathbf\Lambda}[\dv g=0,\dv \mathbf A =\dH \omega^{(p-1)},\dv
\chi=0;g] \approx \dH(\cdot)$ when~\eqref{eq:red} holds. The
charge difference~\eqref{ch_to} between two configurations
differing by a gauge transformation $\dv \mathbf A= \dH
\omega^{p-1}$, is thus zero on-shell.

Besides generalized Killing vectors $(\xi,\mathbf \Lambda)$ which
are also symmetries of the gauge field and of the scalar $\chi$,
there may be charges associated with non-trivial gauge parameters
$(\xi=0,\mathbf \Lambda \neq \dH (\cdot))$. For $p=1$, in
electromagnetism, $\mathbf \Lambda = constant \neq 0$ is such a
parameter and the associated charge is the electric
charge~\eqref{elc_ch}. For $p>1$, non-exact forms $\mathbf
\Lambda$ may exist if the topology of the manifold is non-trivial.
The charges with a non-trivial closed form $\mathbf \Lambda$ which
does not vary along solutions is given by
\begin{equation}
\cQ_{0,-\mathbf \Lambda} = \oint_S e^{-\alpha \chi} \mathbf
\Lambda \wedge \star \mathbf H = \oint_{T} e^{-\alpha \chi} \star
\mathbf H,\label{dipole_ch}
\end{equation}
where $S$ is a $n-2$ surface enclosing the non-trivial cycle $T$
dual to the form $\mathbf \Lambda$. It is simply the integral
of~\eqref{elc_ch} over the non-trivial cycle. The
charges~\eqref{dipole_ch} are thus the generalization for
$p$-forms of electric charges.

The properties of the form~\eqref{Bcharge} under transformations
of the potential $\mathbf A$ are worth mentioning. The
transformation $\mathbf A \rightarrow \mathbf A + \dH \epsilon$
preserves the reducibility equations~\eqref{eq:red} if $\dH
\cL_\xi \mathbf\epsilon = 0$. In that case, $\cL_\xi\mathbf
\epsilon$ can be written as the sum of an exact form and an
harmonic form that we denote as $f(\mathbf\epsilon,\xi) \mathbf
\Lambda^\prime$ with $\mathbf \Lambda^\prime$ not varying along
solutions, $\dv \mathbf \Lambda^\prime= 0$ and
$f(\mathbf\epsilon,\xi)$ constant. In Einstein-Maxwell theory, one
has $\mathbf \Lambda^\prime = 1$ and
$f(\mathbf\epsilon,\xi)=\cL_\xi\mathbf \epsilon$. Under the
transformation $\mathbf A \rightarrow \mathbf A + \dH \epsilon$,
the form~\eqref{Bcharge} changes according to
\begin{equation}
\mathbf k^{\mathbf A}_{ \xi,\mathbf\Lambda} \rightarrow \mathbf
k^{\mathbf A}_{ \xi,\mathbf\Lambda} -f(\epsilon,\xi)\mathbf
\dv(\mathbf \Lambda^\prime \wedge e^{-\alpha \chi}\star \mathbf H)
+ \dH (\cdot)+\mathbf t_{\xi},\qquad \mathbf t_{\xi} \approx 0.
\end{equation}
Defining the charge associated to $\mathbf \Lambda^\prime$
as~\eqref{dipole_ch}, one sees that the infinitesimal
charge~\eqref{ch_to} varies on-shell as
\begin{equation}
\ndelta\cQ_{ \xi,\mathbf\Lambda} \rightarrow \ndelta\cQ_{
\xi,\mathbf\Lambda} - f(\mathbf\epsilon,\xi) \delta\cQ_{0,-\mathbf
\Lambda^\prime}.\label{non-gauge_inv}
\end{equation}
As a consequence, a transformation $\mathbf A \rightarrow \mathbf
A + \dH \epsilon$ admitting a non-vanishing function
$f(\mathbf\epsilon,\xi)$ is not a proper gauge transformation
because such a transformation does not leave the conserved charges
of the solution invariant.

\section{Einstein-Maxwell with Chern-Simons term}
\label{sec:EM-chern}

Einstein-Maxwell theory for which a Chern-Simons term is present
can appear in general in the bosonic part of odd dimensional
supergravities~\cite{Cremmer:1980book}. This section provides the
necessary tools to define the conserved quantities in these
theories for general backgrounds. In section~\ref{sec:sugraGodel}
of Chapter~\ref{chap:BHapp}, we will use these tools to study some
particular solutions in five dimensions. A previous derivation of
conserved quantities using Komar integrals was done
in~\cite{Gauntlett:1998fz} (see also section~\ref{sec:firstlaw} of
Chapter~\ref{chap:geome} for comments on Komar integrals).

In odd space-time dimensions $n=2N+1$, the Einstein-Maxwell
Lagrangian with Chern-Simons term and cosmological constant reads
\begin{eqnarray}
L[g,A] && = \frac{\sqrt{-g}}{16\pi}[ R-2\Lambda -
F_{\mu\nu}F^{\mu\nu}]
\nonumber\\
&&-\frac{2\lambda}{16\pi(N+1)\sqrt{3}}
\epsilon^{\gamma\alpha\beta\cdots\mu\nu}A_\gamma
F_{\alpha\beta}\cdots F_{\mu\nu} .\label{eq:SUG1}
\end{eqnarray}
The bosonic part of $n=5$ minimal supergravity corresponds to
$\Lambda=0,\lambda=1$. The fields of the theory are collectively
denoted by $\phi^i\equiv(g_{\mu\nu}, A_\mu)$. Consider any fixed
background solution $\bar\phi^i$. Following
section~\ref{sec:pform}, the equivalence classes of conserved
$(n-2)$-forms of the linearized theory for the variables
$\varphi^i\equiv\phi^i-\bar\phi^i=(h_{\mu\nu}, a_\mu)$ can be
shown to be in one-to-one correspondence with equivalence classes
of field dependent gauge parameters
$\xi^\mu([\varphi],x),\epsilon([\varphi],x)$ satisfying
\begin{eqnarray}
\left\{\begin{array}{c} \label{eq:SUG2}
  \cL_\xi \bar g_{\mu\nu} = 0,\\ \cL_\xi \bar A_\mu + \partial_\mu
  \epsilon = 0,
\end{array}\right.
\end{eqnarray}
on-shell, i.e., when evaluated for solutions of the linearized
theory. Conserved $n-2$ forms are considered equivalent if they
differ on-shell from the exterior derivative of an $n-3$ form, while
field dependent gauge parameters are equivalent, if they agree
on-shell. If $n\geq 3$ and under reasonable assumptions on the
background $\bar g_{\mu\nu}$, the equivalence classes of solutions
to the first equation of \eqref{eq:SUG2} are classified by the
field independent Killing vectors $\bar \xi^\mu(x)$ of the
background $\bar g_{\mu\nu}$ \cite{Anderson:1996sc}. The second
equation then impose a further constraint on these Killing
vectors. It is straightforward to show that the system
\eqref{eq:SUG2} admits only one more equivalence class of
solutions characterized by $\xi^\mu=0,\epsilon=c\in\mathbb{R}$,
associated with the electric charge.

The weakly vanishing Noether currents are given by
\begin{eqnarray}
  \label{eq:SUG4bis}
  S^\mu_{\xi,\epsilon} =
\varQ{L}{g_{\mu\nu}}(2\xi_\nu) + \varQ{L}{A_\mu}(A_\rho
\xi^\rho)+\varQ{L}{A_\mu} \epsilon,
\end{eqnarray}
The $n-2$ form $k_{\xi,c}[\dv\phi]=k_{\xi,c}^{[\mu\nu]}
(d^{n-2}x)_{\mu\nu}$ is defined through~\eqref{def}.

For the parameters $(\xi,0)$, one can write $k_{ \xi,0}[\dv\phi] =
k^{g}_{\xi}[\dv g] + k^{A}_{ \xi}[\dv\phi] + \lambda
k^{CS}_{\xi}[\dv A ]$. Here, $k^{g}_{\xi}$ is the gravitational
contribution computed in~\eqref{ch_grav1} and whose convenient
equivalent form is given in~\eqref{k_diff2}
with~\eqref{Komar},\eqref{Theta_t} and \eqref{suppl}.
$k^{A}_{\xi}$ is the electromagnetic contribution computed
in~\eqref{Bcharge} with $\Lambda = 0$ and $\chi \equiv 0$. More
precisely, the Komar term~\eqref{def_QA} and the
term~\eqref{ThetaA} reduce in this case to
\begin{eqnarray}
Q^A_{\xi,0} &=& \frac{\sqrt{-g}}{16\pi
G} F^{\mu\nu}(\xi^\rho A_\rho)(d^{n-2}x)_{\mu\nu},\label{KomarQA}\\
\Theta^A[\dv A]& = &\frac{\sqrt{- g}}{16\pi G} F^{\mu\nu}\dv
A_{\nu}(d^{n-1}x)_{\mu},
\end{eqnarray}
where the factors $G$ have been restored. The Chern-Simons term
contributes as
\begin{equation}
k^{CS}_{\xi}[\dv A] = - \frac{N}
{4\sqrt{3}\pi}\eps^{\mu\nu\sigma\alpha\beta \cdots
\gamma\delta}\dv A_\sigma  F_{\alpha\beta}\cdots F_{\gamma\delta}(
A_\rho \xi^\rho)(d^{n-2}x)_{\mu\nu}.\label{eq:SUG6CS}
\end{equation}

For the $(n-2,1)$ form associated with the parameter $(\xi =
0,c=1)$, we get, up to a $\dH$ exact term,
\begin{eqnarray}
k_{0,1}[\dv A, \dv g] = -\delta (Q^{A}_{0,1}+ \lambda
J),\label{eq:SUG9}
\end{eqnarray}
where $Q^{A}_{0,1}$ is given in~\eqref{def_QA} and $J$ can be
written as
\begin{eqnarray}
J=\frac{1}{4\pi\sqrt{3}}\eps^{\mu\nu\sigma\alpha\beta\cdots
\gamma\delta}A_\sigma F_{\alpha\beta}\cdots
F_{\gamma\delta}(d^{n-2}x)_{\mu\nu}.\label{eq:SUG3bis}
\end{eqnarray}

%% file: blackholes.tex
\chapter{Geometric derivation of black hole mechanics}
\label{chap:geome}
\setcounter{equation}{0}\setcounter{figure}{0}\setcounter{table}{0}

In 3+1 dimensions, stationary axisymmetric black holes are
entirely characterized by their mass and their angular momentum.
This is part of the \emph{uniqueness theorems}, see
\cite{Heusler:1998ua} for a review. In higher dimensions, the
situation changes. First, the black hole may rotate in different
perpendicular planes. In 3+1 dimensions, the rotation group
$SO(3)$ has only one Casimir invariant, but in $n$ dimensions, it
has $D \equiv \lfloor (n-1)/2 \rfloor$ Casimirs. Therefore, one
expects that, as a general rule, a black hole will have $D$ conserved
angular momenta. This is what happens in the higher dimensional
Reissner-Nordstr{\o}m and Kerr black
holes~\cite{Myers:1986un,Gibbons:2004uw}.

More dramatically, higher dimensions allow for more exotic horizon
topologies than the sphere. For example, \emph{black ring}
solutions were recently found~\cite{Emparan:2001wn} in five
dimensions with horizon topology $S^1 \times S^2$. The initial
idea of the uniqueness theorems, namely that stationary
axisymmetric black holes are entirely characterized by a few
number of charges \emph{at infinity}, is thus challenged in higher
dimensions, see e.g. \cite{Bena:2004de,Horowitz:2004je} for two
contradictory points of view.

The laws of black hole mechanics were originally found for
asymptotically flat black holes with spherical topology in $3+1$
dimensions surrounded by a perfect fluid and possibly coupled to
an electromagnetic field~\cite{Bardeen:1973gs,Carter:1972}. Time
passing, these laws have been found to hold in far more general
cases.

Many derivations of the first law for higher dimensional black
holes explicitly assumed spherical topology or uniqueness results
which are not generally true, see discussion and references
in~\cite{Copsey:2005se}. Moreover, Komar integrals were used in
asymptotically flat spacetimes but are not suitable e.g. in
asymptotically anti-de Sitter spacetimes.

Bypassing these limitations, the first law of black hole mechanics
was demonstrated for arbitrary perturbations around a stationary
black hole with bifurcation Killing horizon in any diffeomorphism
invariant theory of gravity~\cite{Iyer:1994ys}. Also, this law has
been shown to hold when gravity is coupled to Maxwell or
Yang-Mills fields as a consequence of conservation laws and of
geometric properties of the
horizon~\cite{Sudarsky:1992ty,Gao:2003ys}.

Sections~\ref{sec:Event} and \ref{sec:equilstates} are a
brief review of the second and zeroth laws of black hole mechanics.
These laws will formally come out of the geometric properties of
event and Killing horizons, respectively. These sections are
mainly based on previous reviews on the thermodynamics of black
holes \cite{Carter1973,Carter:1986ca,Townsend:1997ku,Wald:1999vt}
and on a lecture given at the second edition of the Modave Summer
School in Mathematical Physics~\cite{Compere:2006my}.

In section~\ref{sec:firstlaw}, will be presented an unified geometric derivation of
the first law for Einstein gravity coupled to $p$-form fields and
to a scalar in $n$ dimensions. This derivation
will be independent on the asymptotic structure of the
gravitational field and on the topology of the Killing horizon.
Moreover, a generalized Smarr formula will be proven in general
relativity in any dimension.

Remark that the zero and first law of black hole mechanics may
also be generalized to black holes in non-stationary spacetimes.
This was done very recently in the framework of
``isolated horizons''~\cite{Ashtekar:2000sz,Ashtekar:2004cn}. However, in this
thesis, we limit the discussion to the original notion of Killing
horizon. Note also that we will not cover at all in this
thesis the quasi-local approach to the first
law~\cite{Brown:1993br}.

\section{Event horizons}
\label{sec:Event}

A black hole usually refers to a part of spacetime from which no
future directed timelike or null line can escape to arbitrarily
large distance in the outer asymptotic region. A white hole or
white fountain is the time reversed concept which is assumed not
to be physically relevant and will not be treated.

More precisely, if we denote by $\gimel^+$ the future asymptotic
region of a spacetime $(\mathcal M,g_{\mu\nu})$, e.g. null
infinity for asymptotically flat spacetimes and timelike infinity
for asymptotically anti-de Sitter spacetimes, the black hole
region $\mathcal B$ is defined as
\begin{equation}
\mathcal B \equiv \mathcal M - I^-(\gimel^+),
\end{equation}
where $I^-$ denotes the chronological past. The region
$I^-(\gimel^+)$ is what is usually referred to as the \emph{domain
of outer communication}, it is the set of points for which it is
possible to construct a future directed timelike line to arbitrary
large distance in the outer region.

\begin{figure}[hbtp]
\begin{center}
\resizebox{0.35\textwidth}{!}{\mbox{\includegraphics{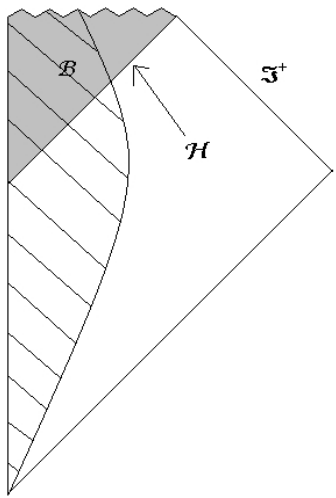}}}
\caption{\small{Penrose diagram of an asymptotically flat
spacetime with spherically symmetric collapsing star. Each point
is a $n-2$-dimensional sphere. Radial light rays propagate along
$45^\circ$ diagonals. The star region is hatched and the black
hole region is indicated in grey.}} \label{GEO:fig2}
\end{center}
\end{figure}

The \emph{event horizon} $\mathcal{H}$ of a black hole is then the
boundary of $\mathcal B$. Let us denote $J^-(U)$ the causal past
of a set of points $U \subset \mathcal M$ and $\bar J^-(U)$ the
topological closure of $J^-$. We have $I^-(U) \subset J^-(U)$. The
(future) event horizon of $\mathcal M$ can then equivalently be
defined as
\begin{equation}
\mathcal H \equiv \bar J^-(\gimel^+) - J^-(\gimel^+),
\end{equation}
i.e. the boundary of the closure of the causal past of $\gimel^+$.
See Fig.~\ref{GEO:fig2} for an example. The event horizon is a
concept defined with respect to the entire causal structure of
$\mathcal M$.

The event horizons are null hypersurfaces with peculiar
properties. The development of their properties will allow us to
sketch the proof of the area theorem~\cite{Hawking:1971vc} which
is concerned with the dynamical evolution of sections of the event
horizon at successive times. The area theorem is also called the
second law of black hole mechanics because it demonstrates that,
under reasonable conditions, the area of the event horizon always
increases as does the entropy in classical
thermodynamics~\cite{Bekenstein:1972tm}.

\subsection{Null hypersurfaces}

Let $S(x^\mu)$ be a smooth function and consider the $n-1$
dimensional null hypersurface $S(x^\mu) = 0$, which we denote by
$\mathcal{H}$. This surface will be the black hole horizon in the
subsequent sections. It is a null hypersurface, i.e. such that its
normal $\xi^\mu \sim g^{\mu\nu}\partial_\nu S$ is null,
\begin{equation}
\xi^\mu \xi_\mu \overset{\mathcal H}{=}0.\label{GEO:xi0}
\end{equation}
The vectors $\eta^\mu$ tangent to $\mathcal{H}$ obey $\eta_\mu
\xi^\mu |_\mathcal{H} = 0$ by definition. Since $\mathcal{H}$ is
null, $\xi^\mu$ itself is a tangent vector, i.e.
\begin{equation}
\xi^\mu = \frac{dx^\mu(t)}{dt}
\end{equation}
for some null curve $x^\mu(t)$ inside $\mathcal{H}$. One can then
prove that $x^\mu(t)$ are null geodesics\footnote{Proof: Let
$\xi_\mu = \tilde f S_{,\mu}$. We have
\begin{eqnarray}
\xi^\nu \xi_{\mu ; \nu} &=& \xi^\nu \partial_\nu \tilde f S_{,\mu}
+
\tilde f \xi^\nu S_{,\mu;\nu}\nonumber \\
&=& \xi^\nu \partial_\nu \ln \tilde f \xi_\mu + \tilde f \xi^\nu
S_{;\nu;\mu}\nonumber  \\
&=& \xi^\nu \partial_\nu \ln \tilde f \xi_\mu +\tilde f \xi^\nu (\tilde f^{-1}\xi_\nu )_{;\mu} \nonumber \\
&=& \xi^\nu \partial_\nu \ln \tilde f \xi_\mu +
\frac{1}{2}(\xi^2)_{,\mu} - \partial_\mu \ln \tilde f \xi^2.
\label{GEO:last1}
\end{eqnarray}
Now, as $\xi$ is null on the horizon, any tangent vector $\eta$ to
$\mathcal{H}$ satisfy $(\xi^2)_{;\mu}\eta^\mu=0$. Therefore,
$(\xi^2)_{;\mu} \sim \xi_\mu$ and the right-hand side of
\eqref{GEO:last1} is proportional to $\xi_\mu$ on the horizon. }
\begin{equation}
\xi^\nu \xi^\mu_{\,\,\, ; \nu} \overset{\mathcal H}{=} \kappa
\xi^\mu,  \label{GEO:geo}
\end{equation}
where $\kappa$ measures the extent to which the parameterization is
not affine.  If we denote by $l$ the normal to $\mathcal{H}$ which
corresponds to an affine parameterization $l^\nu l^\mu_{\,\,\, ;
\nu} = 0$ and $\xi = f(x)\, l$ for some function $f(x)$, then
$\kappa = \xi^\mu \partial_\mu \ln|f|$.

According to the Frobenius' theorem, a vector field $v$ is
hypersurface orthogonal if and only if it satisfies
$v_{[\mu}\partial_\nu v_{\rho]} =0$, see e.g. \cite{Wald:1984rg}.
Therefore, the vector $\xi$ satisfies the irrotationality
condition
\begin{equation}
\xi_{[\mu}\partial_\nu \xi_{\rho]}\overset{\mathcal H}{=}
0.\label{GEO:irr}
\end{equation}

A congruence is a family of curves such that precisely one curve
of the family passes through each point. In particular, any smooth
vector field defines a congruence. Indeed, a vector field defines at
each point a direction which can be uniquely ``integrated'' along
a curve starting from an arbitrary point.

\begin{figure}[hbtp]
\begin{center}
\resizebox{0.35\textwidth}{!}{\mbox{\includegraphics{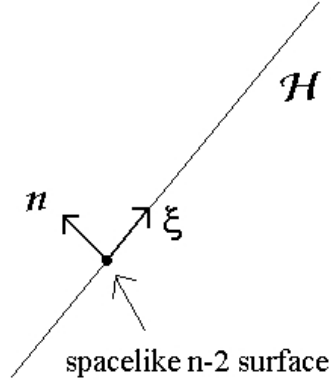}}}
\caption{\small{The null vector $n$ is defined with respect to
$\xi$.}} \label{GEO:fig3}
\end{center}
\end{figure}

Since $S(x)$ is also defined outside $\mathcal H$, the normal
$\xi$ defines a congruence only null
when restricted to $\mathcal{H}$. In order to study this
congruence outside $\mathcal{H}$, it is useful to define a
transverse null vector $n^\mu$ with
\begin{equation}
n^\mu n_\mu = 0, \qquad n_\mu \xi^\mu = -1.\label{GEO:norm}
\end{equation}
The normalization $-1$ is chosen so that if we consider $\xi$ to
be tangent to an outgoing radial null geodesic, then $n$ is
tangent to an ingoing one, see Fig.~\ref{GEO:fig3}. The
normalization conditions~\eqref{GEO:norm} (imposed everywhere,
$(n^2)_{;\nu} = 0 =(n \cdot \xi)_{;\nu}$) do not fix $n$ uniquely.
Let us choose arbitrarily one such $n$. The extent to which the
family of hypersurfaces $S(x) = const$ are not null is given by
\begin{equation}
\varsigma \equiv \frac{1}{2} (\xi^2)_{;\mu} n^\mu
.\label{GEO:defvarsig}
\end{equation}

The vectors $\eta$ orthogonal to both $\xi$ and $n$,
\begin{equation}
\eta^\mu\xi_\mu = 0 = \eta^\mu n_\mu,\label{GEO:tg_small}
\end{equation}
span a $n-2$ dimensional spacelike subspace of $\mathcal{H}$. The
metric can be written as
\begin{equation}
g_{\mu\nu} = - \xi_\mu n_\nu - \xi_\nu n_\mu +
\gamma_{\mu\nu}\label{GEO:metric}
\end{equation}
where $\gamma_{\mu\nu} = \gamma_{(\mu\nu)}$ is a positive definite
metric with $\gamma_{\mu\nu}\xi^\mu = 0 = \gamma_{\mu\nu}n^\mu$.
The tensor $\gamma^\mu_{\;\, \nu} =
g^{\mu\alpha}\gamma_{\alpha\nu}$ provides a projector onto the
$n-2$ spacelike tangent space to $\mathcal H$.

For future convenience, we also consider the hypersurface
orthogonal null congruence $l^\mu$ with affine parameter $\tau$
that is proportional to $\xi^\mu$ on $\mathcal H$\footnote{We
shall reserve the notation $\xi^\mu$ for vectors coinciding with
$l^\mu$ on the horizon but which are not null outside the
horizon.},
\begin{eqnarray}
l^\mu l_\mu = 0, \qquad l^\nu l^\mu_{\;\,;\nu}=0, \qquad l^\mu
\overset{\mathcal H}{\sim} \xi^\mu.
\end{eqnarray}
The vector field $l$ extends $\xi$ outside the horizon while
keeping the null property.

\subsection{The Raychaudhuri equation}

In this section, we shall closely follow the reference
\cite{Carter:1986ca}. We set out part of the material needed to
prove the area law.

Firstly, let us decompose the tensor $D_\mu \xi_\nu$ into the
tensorial products of $\xi$, $n$ and spacelike vectors $\eta$
tangent to $\mathcal H$ ,\footnote{Proof: Let us first decompose
$D_\mu \xi_\nu$ as
\begin{equation}
D_\mu \xi_\nu = v_{\mu\nu} + n_\mu (C_1 n_\nu + C_2 \xi_\nu + C_3
\eta_{\nu}) + \tilde \eta_\mu \xi_\nu+ \hat \eta_\mu n_\nu -
\xi_\mu \alpha_\nu,
\end{equation}
where $v_{\mu\nu} =
\gamma^\alpha_{\;\,\mu}\gamma^\beta_{\;\,\nu}v_{\alpha\beta}$ and
$\eta^\mu$, $\tilde \eta^\mu$, $\hat \eta^\mu$ are spacelike
tangents to $\mathcal H$. Contracting with $\xi^\mu$ and
using~\eqref{GEO:geo}, we find $C_1=0=C_3$, $C_2 = -\kappa$.
Contracting with $\gamma^\mu_{\;\,\alpha}n^\nu$, we find $\tilde
\eta_\mu = -\gamma^\alpha_{\;\,\mu}n^\beta D_\alpha \xi_\beta$.
Contracting with $\gamma^\mu_{\;\,\alpha}\xi^\nu$, we find finally
$\hat \eta_\mu = -1/2 \gamma^\alpha_{\;\,\mu}D_\alpha(\xi^2) = 0$
thanks to \eqref{GEO:xi0}.}
\begin{equation}
D_\mu \xi_\nu \overset{\mathcal H}{=} v_{\mu\nu} - \xi_\nu (\kappa
n_\mu + \gamma^\alpha_{\;\,\mu}n^\beta D_\alpha \xi_\beta) -
\xi_\mu n^\alpha D_\alpha \xi_\nu,\label{GEO:decomp}
\end{equation}
where the orthogonal projection $v_{\mu\nu} =
\gamma^\alpha_{\;\,\mu}\gamma^\beta_{\;\,\nu}D_\alpha \xi_\beta$
can itself be decomposed in symmetric and antisymmetric parts
\begin{equation}
v_{\mu\nu} = \theta_{\mu\nu} + \omega_{\mu\nu}, \qquad
\theta_{[\mu\nu]}=0,\qquad \omega_{(\mu\nu)}=0.
\end{equation}
The Frobenius irrotationality condition \eqref{GEO:irr} is
equivalent to $\omega_{\mu\nu}|_\mathcal{H} = 0$\footnote{Proof:
We have
\begin{equation}
\xi_{[\mu}\partial_\nu \xi_{\rho ]} = \xi_{[\mu}D_\nu \xi_{\rho]}
= \xi_{[\mu } v_{\nu\rho ]} = \xi_{[\mu} \omega_{\nu\rho]}.
\end{equation}As $\omega_{\mu\nu}$ is defined as a projection with $\gamma^\mu_{\;\,\nu}$, the equivalence is shown.}.
The tensor $\theta_{\mu\nu}$ is interpreted as the
expansion rate tensor of the congruence while its trace $\theta =
\theta_{\mu}^{\;\,\mu}$ is the divergence of the congruence. Any
smooth $n-2$ dimensional area element evolves according to
\begin{equation}
\frac{d}{d t}(d\mathcal A) = \theta \,d \mathcal
A.\label{GEO:incr}
\end{equation}
The shear rate is the trace free part of the strain rate tensor,
\begin{equation}
\sigma_{\mu\nu} = \theta_{\mu\nu} - \frac{1}{n-2}\theta
\gamma_{\mu\nu}.
\end{equation}
Defining the scalar $\sigma^2 =
(n-2)\sigma_{\mu\nu}\sigma^{\mu\nu}$, one has
\begin{equation}
\xi_{\mu;\nu}\xi^{\nu;\mu} \overset{\mathcal H}{=}
\frac{1}{n-2}(\theta^2+\sigma^2) + \kappa^2
+\varsigma^2,\label{GEO:eqxixi}
\end{equation}
where $\varsigma$ was defined in \eqref{GEO:defvarsig}. Note also
that the divergence of the vector field has three contributions,
\begin{equation}
\xi^\mu_{\;\,;\mu} \overset{\mathcal H}{=} \theta + \kappa -
\varsigma.\label{GEO:derContr}
\end{equation}
Now, the contraction of the Ricci identity
\begin{equation}
v^\alpha_{\; ;\mu ;\nu} - v^\alpha_{\; ;\nu;\mu} = -R^\alpha_{\;\,
\lambda \mu\nu }v^\lambda,
\end{equation}
implies the following identity
\begin{equation}
(v^\nu_{\;\, ;\nu})_{;\mu}v^\mu = (v^\nu v^\mu_{\; ;\nu})_{;\mu} -
v^{\nu;\mu}v_{\mu;\nu}-R_{\mu\nu}v^\mu v^\nu,\label{GEO:id}
\end{equation}
valid for any vector field $v$. The
formulae~\eqref{GEO:eqxixi}-\eqref{GEO:derContr} have their
equivalent for $l$ as
\begin{eqnarray}
l_{\mu;\nu}l^{\nu;\mu}=
\frac{1}{n-2}(\theta_{(0)}^2+\sigma_{(0)}^2), \qquad
l^\mu_{\;\,;\mu} = \theta_{(0)} ,
\end{eqnarray}
where the right hand side is expressed in terms of expansion
rate $\theta_{(0)} = \theta \frac{dt}{d\tau}$ and shear rate
$\sigma_{(0)} = \sigma \frac{dt}{d\tau}$ with respect to the
affine parameter $\tau$. The identity \eqref{GEO:id} becomes
\begin{equation}
\frac{d\theta_{(0)}}{d\tau} \hat  = \dot\theta_{(0)}
\overset{\mathcal H}{=}
-\frac{1}{n-2}(\theta_{(0)}^2+\sigma_{(0)}^2) - R_{\mu\nu}l^\mu
l^\nu,\label{GEO:Ray}
\end{equation}
where the dot indicates a derivation along the generator. It is the
final form of the Raychaudhuri equation for hypersurface
orthogonal null geodesic congruences in any dimension.

\subsection{Properties of event horizons}

As already mentioned, the main characteristic of event
horizons is them being null hypersurfaces. In the early
seventies, Penrose and Hawking further investigated the generic
properties of past boundaries whose event horizons are particular
representatives. We shall only enumerate these properties below
and refer the reader to the references
\cite{HawkingEllis,Townsend:1997ku} for explicit proofs. These
properties are crucial in order to prove the area theorem.

\begin{enumerate}
\item \emph{Achronicity property.} No two points of the horizon can be connected by a timelike
curve.

\item The null geodesic generators of $\mathcal H$
may have past end-points in the sense that the continuation of the
geodesic further into the past is no longer in $\mathcal H$.

\item The generators of $\mathcal H$ have no future end-points, i.e. no
generator may leave the horizon.

\end{enumerate}

The second property hold for example for collapsing stars where
the past continuation of all generators leave the horizon at the
time the horizon was formed. As a consequence of properties 2 and
3, null geodesics may enter $\mathcal H$ but not leave it.

\subsection{The area theorem}

The area theorem was initially demonstrated by Hawking
\cite{Hawking:1971vc}. We shall follow closely the reviews by
Carter \cite{Carter:1986ca} and Townsend \cite{Townsend:1997ku}.
The theorem reads as follows.
\begin{theorem}[Area law] If
\begin{itemize}
\item[(i)] Einstein's equations hold with a matter stress-tensor
satisfying the null energy condition, $T_{\mu\nu}k^\mu k^\nu \geq
0$, for all null $k^\mu$,
\item[(ii)] The spacetime is ``strongly asymptotically predictable''
\end{itemize}
then the surface area $\mathcal A$ of the event horizon can never
decrease with time.
\end{theorem}
The theorem was originally stated in $4$ dimensions but it is
actually valid in any dimension $n \geq 3$.

In order to understand the second requirement, let us recall
some definitions. The future domain of dependence $D^+(\Sigma)$ of
an hypersurface $\Sigma$ is the set of points $p$ in the manifold
for which every causal curve through $p$ that has no past
end-point intersects $\Sigma$. The significance of $D^+(\Sigma)$
is that the behavior of solutions of hyperbolic PDE's
\emph{outside} $D^+(\Sigma)$ is not determined by initial data on
$\Sigma$. If no causal curves have past end-points, then the
behavior of solutions inside $D^+(\Sigma)$ is entirely determined
in terms of data on $\Sigma$. The past domain of dependence
$D^-(\Sigma)$ is defined similarly.

A Cauchy surface is a spacelike hypersurface which every
non-spacelike curve intersects exactly once. It has as domain of
dependence $D^+(\Sigma) \cup D^- (\Sigma)$ the manifold itself. If
an open set $\mathcal N$ admits a Cauchy surface then the Cauchy
problem for any PDE with initial data on $\mathcal N$ is
well-defined. This is also equivalent to say that $\mathcal N$ is
globally hyperbolic.

The requirement (ii) means that there should be a globally
hyperbolic submanifold of spacetime containing both the exterior
spacetime \emph{and} the horizon. It is equivalent to say there
is a family of Cauchy hypersurfaces $\Sigma(\tau)$, such that
$\Sigma(\tau^\prime)$ is inside the domain of dependence of
$\Sigma(\tau)$ if $\tau^\prime > \tau$.

Now, the boundary of the black hole is the past event horizon
$\mathcal H$. It is a null hypersurface with generator $l^\mu$
(that is proportional to $\xi$ on $\mathcal H$). We can choose to
parameterize the Cauchy surfaces $\Sigma(\tau)$ using the affine
parameter $\tau$ of the null geodesic generator $l$.

The \emph{area of the horizon} $\mathcal A(\tau)$ is then the area
of the intersection of $\Sigma(\tau)$ with $\mathcal H$. We have
to prove that $\mathcal A(\tau^\prime) > \mathcal A(\tau)$ if
$\tau^\prime
> \tau$.

\vspace{5pt}\noindent {\bf Sketch of the proof:\vspace{5pt}}\\
The Raychaudhuri equation for the null generator $l$ reads
as~\eqref{GEO:Ray}. Therefore, wherever the energy condition
$R_{\mu\nu}l^\mu l^\nu \geq 0 $ holds, the null generator will
evolve subject to the inequality
\begin{equation}
\frac{d\theta_{(0)}}{d \tau} \leq -\frac{1}{n-2}\,\theta_{(0)}^2,
\end{equation}
except on possible singular points as caustics. It follows that if
$\theta_{(0)}$ becomes negative at any point $p$ on the horizon
(i.e. if there is a convergence) then the null generator can
continue in the horizon for at most a finite affine distance
before reaching a point $p$ at which $\theta_{(0)} \rightarrow
-\infty$, i.e. a point of infinite convergence representing a
caustic beyond which the generators intersect.

\begin{figure}[htbp]
\begin{center}
\resizebox{0.35\textwidth}{!}{\mbox{\includegraphics{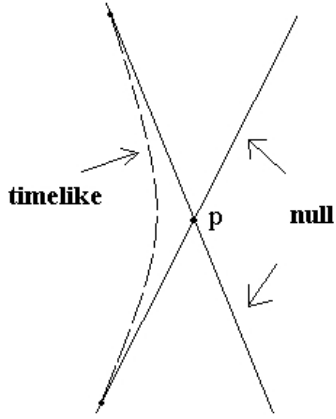}}}
\caption{\small{If two null generators of $\mathcal H$ cross, they
may be deformed to a timelike curve.}} \label{GEO:fig1}
\end{center}
\end{figure}

Now, from the third property of event horizons above, the
generators cannot leave the horizon. Therefore at least two
generators cross at $p$ inside $\mathcal H$ and, following Hawking
and Ellis (Prop 4.5.12 of \cite{HawkingEllis}), they may be
deformed to a timelike curve, see figure~\ref{GEO:fig1}. This is
however impossible because of the achronicity property of event
horizons. Therefore, in order to avoid the contradiction, the
point $p$ cannot exist and $\theta_{(0)}$ cannot be negative.

Since (at points where the horizon is not smooth) new null
generators may begin but old ones cannot end,
equation~\eqref{GEO:incr} implies that the total area $\mathcal
A(\tau)$ cannot decrease with increasing $\tau$,
\begin{equation}
\frac{d}{d\tau}\mathcal A \geq \oint \theta_{(0)} \,d\mathcal A
\geq 0.
\end{equation}
This completes the proof.

In particular, if two black holes with area $\mathcal A_1$ and
$\mathcal A_2$ merge then the area $\mathcal A_3$ of the combined
black hole have to satisfy
\begin{equation}
\mathcal{A}_3 > \mathcal{A}_1+\mathcal{A}_2.
\end{equation}
The area $A(\tau)$ do not change if $\theta = 0$ on the entire
horizon $\mathcal H$. The black hole is then stationary.

Note that this derivation implicitly assume regularity properties
of the horizon (as piecewise $C^2$) which may not be true for
generic black holes. Recently these gaps in the derivation have
been totally filled in~\cite{Chrusciel:2000cu}, see discussion
in~\cite{Wald:1999vt}.

\section{Equilibrium states}
\label{sec:equilstates}

\subsection{Killing horizons}

In any stationary and asymptotically flat spacetime with a black
hole, the event horizon is a Killing horizon \cite{HawkingEllis}.
This theorem firstly proven by Hawking~\cite{Hawking:1971vc} is
called the rigidity theorem. It provides an essential link between
event horizons and Killing horizons.\footnote{The theorem further
assumes the geometry is analytic around the horizon. Actually,
there exist a counter-example to the rigidity theorem as stated in
Hawking and Ellis \cite{HawkingEllis} but under additional
assumptions such as global hyperbolicity and simple connectedness
of the spacetime, the result is valid \cite{Chrusciel:1996bj}. See
also~\cite{Friedrich:1998wq} for a relaxation of the analyticity
hypotheses.}

A Killing horizon is a null hypersurface whose normal $\xi$ is a
Killing vector
\begin{equation}
\mathcal{L}_\xi g_{\mu\nu} = \xi_{\mu ; \nu} + \xi_{\nu ; \mu} =
0.\label{GEO:Kill}
\end{equation}
This additional property will allow us to explore many
characteristics of black holes.

The parameter $\kappa$ which we now call the surface gravity of
$\mathcal{H}$ is defined in~\eqref{GEO:geo}. In asymptotically
flat spacetimes, the normalization of $\kappa$ is fixed by
requiring $\xi^2 \rightarrow -1$ at infinity (similarly, we impose
$\xi^2 \rightarrow -\frac{r^2}{l^2}$ in asymptotically anti-de
Sitter spacetimes).

For Killing horizons, the expansion rate $\theta_{\mu\nu} =
\gamma_{(\mu}^{\;\,\alpha}\gamma_{\nu)}^{\;\,\beta}D_{\alpha}\xi_\beta
= 0$, so $\theta = \sigma = 0$. Moreover,
from~\eqref{GEO:derContr} and \eqref{GEO:Kill}, we deduce
$\varsigma = \kappa$. Equation~\eqref{GEO:eqxixi} then provides an
alternative definition for the surface gravity,
\begin{equation}
\kappa^2 = -\frac{1}{2}\xi_{\mu ; \nu} \xi^{\mu;\nu}|_{\mathcal
H}.
\end{equation}
Contracting~\eqref{GEO:geo} with the transverse null vector $n$ or
using the definition~\eqref{GEO:defvarsig}, one has also
\begin{equation}
\kappa = \xi_{\mu;\nu}\xi^\mu n^\nu |_\mathcal{H} = \frac{1}{2}
(\xi^2)_{,\mu} n^\mu |_\mathcal{H}.\label{GEO:kappa}
\end{equation}
The Raychaudhuri equation~\eqref{GEO:Ray} also states in this case
that
\begin{equation}
R_{\mu\nu}\xi^\mu \xi^\nu \overset{\mathcal H}{=}
0,\label{GEO:RayEQ}
\end{equation}
because $l$ is proportional to $\xi$ on the horizon.

From the decomposition~\eqref{GEO:decomp}, the irrotationality
condition~\eqref{GEO:irr} and the Killing property
$\xi_{[\mu;\nu]} = \xi_{\mu;\nu}$, it can be written
\begin{equation}
\xi_{\mu;\nu} \overset{\mathcal H}{=} \xi_\mu q_\nu - \xi_\nu
q_\mu,\label{GEO:defq}
\end{equation}
where the covector $q_\mu$ can be fixed uniquely by the
normalization $q_\mu n^\mu = 0$. Using \eqref{GEO:kappa}, the last equation can be decomposed
in terms of $(n,\xi,\{\eta\})$ as
\begin{equation}
\xi_{\mu;\nu} \overset{\mathcal H}{=} -\kappa (\xi_\mu n_\nu -
\xi_\nu n_\mu) + \xi_\mu \hat \eta_\nu - \hat \eta_\mu
\xi_\nu,\label{GEO:ortho}
\end{equation}
where $\hat \eta$ satisfy $\hat \eta \cdot \xi = 0 = \hat \eta
\cdot n$. In particular, it shows that for any spacelike tangent
vectors $\eta$, $\tilde \eta$ to $\mathcal H$, one has
$\xi_{\mu;\nu}\eta^\mu \tilde \eta^\nu \overset{\mathcal H}{=} 0$.

\subsection{Zero law}

We are now able to show that the surface gravity $\kappa$
is constant on the horizon under appropriate conditions. More
precisely,
\begin{theorem}[Zero law] \cite{Bardeen:1973gs}
If
\begin{itemize}
\item[(i)] The spacetime $(M,g)$ admits a Killing vector $\xi$
which is the generator of a Killing horizon $\mathcal H$,
\item[(ii)] Einstein's equations hold with matter satisfying the
dominant energy condition, i.e. $T_{\mu\nu}l^\nu$ is a
non-spacelike vector for all \mbox{$l^\mu l_\mu \leq 0$},
\end{itemize}
then the surface gravity $\kappa$ of the Killing horizon is
constant over $\mathcal H$.
\end{theorem}

Using the aforementioned properties of null hypersurfaces and
Killing horizons, together with
\begin{equation}
\xi_{\nu;\mu;\rho} = R_{\mu\nu\rho}^{\,\,\,\,\,\,\,\,\,\,
\tau}\xi_\tau,
\end{equation}
which is valid for Killing vectors, one obtains (see
\cite{Carter:1986ca} for a proof)
\begin{eqnarray}
\dot \kappa = \kappa_{,\mu}\xi^\mu &\overset{\mathcal H}{=}& 0,\\
\kappa_{,\mu}\eta^\mu &\overset{\mathcal H}{=}& -
R_{\mu\nu}\xi^\mu\eta^\nu,
\end{eqnarray}
for all spacelike tangent vectors $\eta$. Now, from the dominant
energy condition, $R_{\mu\nu}\xi^\mu$ is not spacelike. However,
the Raychaudhuri equation implies \eqref{GEO:RayEQ}. Therefore,
$R_{\mu\nu}\xi^\mu$ must be zero or proportional to $\xi_\nu$ and
$R_{\mu\nu}\xi^\mu\eta^\nu = 0$.

This theorem has an extension when gravity is coupled to
electromagnetism. If the Killing vector field $\xi$ is also a
symmetry of the electromagnetic field up to a gauge
transformation, $\mathcal L_\xi A_\mu +\partial_\mu \epsilon = 0$,
it can also be proven that the electric potential
\begin{equation}
\Phi = -(A_\mu \xi^\mu+\epsilon)|_{\mathcal H}\label{el_potential}
\end{equation}
is constant on the horizon. See the discussion
before~\eqref{eq:ntcycle} for a proof in the case of $p$-form
potentials, $p \geq 1$.

Note that the zeroth law can be derived from different
assumptions. As an example, constancy of surface gravity holds for
any black hole which is static or stationary-axisymmetric with the
$t-\phi$ reflection isometry (without assuming Einstein's
equations)~\cite{Carter1973,Racz:1995nh}.

\section{First law and Smarr formula}
\label{sec:firstlaw}

The geometric derivation of the Smarr relation and of the first
law of thermodynamics for four-dimensional asymptotically flat
black holes is usually based on Komar
integrals~\cite{Bardeen:1973gs,Carter:1972}. Komar integrals are
extremely useful since they allow one to easily express the
conserved quantities defined at infinity to properties associated
with the horizon of the black hole. However, they do not provide a
complete and systematic approach to conserved quantities.

Indeed, in order to give the correct definitions of energy and
angular momentum, the coefficients of the Komar integrals must be
fixed by comparison with the ADM
expressions~\cite{Arnowitt:1961aa,Arnowitt:1962aa}, see e.g.
discussions in~\cite{Iyer:1994ys,Townsend:1997ku}. Moreover,
although this approach can be extended to higher dimensional
asymptotically flat black holes
\cite{Myers:1986un,Gauntlett:1998fz}, it generally becomes
ambiguous for rotating asymptotically anti-de Sitter black
holes~\cite{Magnon:1985sc,Gibbons:2004ai}. Komar integrals are
also not applicable to more exotic black holes as the ones
immersed in G\"odel spacetimes~\cite{Gimon:2003ms,Banados:2005da}.

The aim of this section is to rederive the first law and the Smarr
formula using the Lagrangian charges defined in the preceding part
of the thesis, as sketched in~\cite{Barnich:2003xg}, without using
uniqueness results or assuming spherical horizon topology.

We will first derive the first law and the Smarr formula for
Einstein gravity and we will then extend the analysis to gravity
coupled to a $p$-form potential and to a scalar field. A proof of
the first law for an arbitrary theory of gravity and for
non-stationary perturbations will not be developed here. For that
analysis, we refer the reader to covariant phase space
methods~\cite{Wald:1993nt,Iyer:1994ys}.

Following section~\ref{sec:exact-solut-symm} of
Chapter~\ref{chap:general_th}, suppose that we have a family of
solutions $\cF$ with exact Killing vectors $\frac{\d}{\d t}$ and
$\frac{\d}{\d \varphi^a}$, $a=1 \dots \lfloor (n-1)/2 \rfloor$,
containing stationary and axisymmetric black holes with connected
Killing horizon. The generator of a Killing horizon is then a
combination of the Killing vectors,
\begin{equation}
\xi  = \frac{\partial}{\partial t} + \Omega^a
\frac{\partial}{\partial \varphi^a},\label{Kill_gen}
\end{equation}
where $\Omega^a$ are the angular velocities at the horizon. In
what follows, we will only consider one-forms $(\dv g,\dv\xi)$
contracted with stationary field variations~$(\delta g,\delta
\xi)$, i.e. satisfying~\eqref{eq:90},
\begin{equation}
\mathcal L_\xi \delta g_{\mu\nu} + \mathcal L_{\delta \xi}
g_{\mu\nu} = 0.\label{GEO:varKill}
\end{equation}

\subsection{The first law for Einstein gravity}

The differences of energy and angular momenta between two
configurations $g$ and $g+\delta g$ are defined by
\begin{equation}
\delta \mathcal E = \oint_{S^\infty} k_{\partial_t}[\delta
  g_{\mu\nu}],\qquad \delta
\mathcal J_a= -\oint_{S^\infty} k_{\partial_{\varphi^a}}[\delta
  g_{\mu\nu}].\label{GEO:def_EJ}
\end{equation}
Here, the relative sign difference between the definitions of
$\mathcal E$ and $\mathcal J^a$ traces its origin to the Lorentz
signature of the metric~\cite{Iyer:1994ys}. We assume that the
energy and angular momenta~\eqref{GEO:def_EJ} are integrable
in $\cF$, i.e. we require that the condition~\eqref{int_cond} or
more precisely~\eqref{int_diff} are satisfied for the Killing
vectors $\d_t$ and $\d_{\varphi^a}$, for $g \in \cF$ and for
one-forms $\delta g$ tangent to the family of solutions $\cF$.

Assuming that the de Rham cohomology in the solution space $\cF$
vanishes in (vertical) form degree two, the integrability
condition ensures that the charge one-forms~\eqref{eq:82} are
independent on the path $\gamma_s$ connecting $\bar g$ to $g$. The
energy and angular momenta are then obtained by integration
of~\eqref{GEO:def_EJ},
\begin{equation}
\cE = \int_{\gamma_s} \delta \cE + \bar\cE, \qquad \cJ_a =
\int_{\gamma_s} \delta \cJ_a + \bar\cJ_a.
\end{equation}

The \emph{equilibrium state version}~\footnote{There also exists a
physical process version, where an infinitesimal amount of matter
is send through the horizon from infinity.} of the first law for
the simple case of pure Einstein gravity can now be stated
as~\cite{Bardeen:1973gs,Iyer:1994ys,Jacobson:1993vj}
\begin{theorem}[First law]\label{th:fl}
Let $(\mathcal M,g)$ and $(\mathcal M+\delta \mathcal M,g+\delta
g)$ be two slightly different stationary black hole solutions of
Einstein's equations with Killing horizon. The difference of
energy $\delta\mathcal{E}$, angular momenta $\delta\mathcal{J}_a$
and area $\delta\mathcal A$ of the black hole are related by
\begin{equation}
\delta \mathcal{E} = \Omega^a \, \delta \mathcal{J}_a +
\frac{\kappa}{8\pi} \delta \mathcal A.\label{GEO:first09}
\end{equation}
\end{theorem}
Let us start with Proposition~\ref{prop1} stating the equality of
the charge associated with $\xi$ at a spacelike section $H$ of the
horizon and at infinity,
\begin{eqnarray}
  \label{GEO:eq_first}
  \oint_{S^\infty} k_\xi[\delta
  g]=\oint_{H} k_\xi[\delta g].
\end{eqnarray}
Using~\eqref{GEO:def_EJ}, the left-hand side of
\eqref{GEO:eq_first} is given by
\begin{eqnarray}
   \oint_{S^\infty} k_\xi[\delta
  g]=\delta \mathcal{E} - \Omega_a \,\delta
  \mathcal{J}^a.\label{GEO:leftside}
\end{eqnarray}
Using~\eqref{k_diff2}, the right-hand side of
\eqref{GEO:eq_first} may be rewritten as\footnote{The minus sign in front of
$I^{n}_{\delta g}\cL_{EH}$ comes from the fact that $i_\xi$ is
Grassman odd and that we use the Grassman even $\delta g$ in place
of $\dv g$.}
\begin{eqnarray}
\oint_{H} k_\xi[\delta g] = - \delta \oint_H k_{\cL^{EH},\xi}^K +
\oint_H k^K_{\cL^{EH},\delta \xi } -\oint_H i_\xi I^{n}_{\delta
g}\cL_{EH},\label{GEO:formula_charge}
\end{eqnarray}
where the integrands are given in~\eqref{Komar}-\eqref{Theta_t}.

On the horizon, the integration measure for $(n-2)$-forms is given
by
\begin{equation}
\sqrt{-g}(d^{n-2}x)_{\mu\nu} = \frac{1}{2} (\xi_\mu n_\nu - n_\mu
\xi_\nu) d\mathcal{A},\label{GEO:surf_form}
\end{equation}
where $d\mathcal{A}$ is the ``angular'' measure on $H$ and $n$ was
defined in~\eqref{GEO:norm}. Using the properties of Killing
horizons described in section~\ref{sec:equilstates}, the Komar
integral on the horizon can easily be computed as
\begin{equation}
\oint_H k_{\cL^{EH},\xi}^K = -\frac{\kappa \mathcal{A}}{8\pi
G},\label{Komar_hor}
\end{equation}
where $\mathcal{A}$ is the area of the horizon. Now, it turns out
that the local geometry around Killing horizons implies the
following property
\begin{prop}\label{prop:GEO}
\begin{eqnarray}
\oint_H k^K_{\cL^{EH},\delta \xi } -\oint_H i_\xi I^{n}_{\delta
g}\cL_{EH} & =& -\frac{\mathcal{A}}{8\pi G}\,\delta \kappa
.\label{GEO:to_prove}
\end{eqnarray}
\end{prop}
The computation which is straightforward but lengthly is
explicitly done in
Appendix~\ref{app:proofs}.\ref{GEO:app_firstlaw} \emph{without}
assuming specific coordinates as in the original
derivation~\cite{Bardeen:1973gs} and in some later
derivations~\cite{Carter1973,Wald:1993nt}. It would be interesting
to find a generalization of this proof for non-stationary
perturbations as well.

Using proposition~\ref{prop:GEO}, the right-hand side of
\eqref{GEO:eq_first} is finally given by
\begin{eqnarray}
\oint_{H} k_\xi[\delta
  g_{\mu\nu}] = \frac{\kappa}{8\pi G} \delta \mathcal{A},
\end{eqnarray}
as it should and the first law is proven.\qed

We can see in this derivation that the first law is a
\emph{geometrical} law in the sense that it relates the geometry
of Killing horizons to the geometric measure of energy and angular
momenta. Note that the derivation was done in arbitrary
dimensions, without hypotheses on the topology of the horizon and
for arbitrary stationary variations. The first law also applies in
particular for extremal black holes by taking $\kappa = 0$.

Finally remark that the first part of the derivation, especially
equation~\eqref{GEO:leftside}, is identical for any theory of
gravity, with appropriate definition of energy and angular
momenta~\eqref{GEO:def_EJ}. On the other hand, the surface terms
evaluated on the horizon~\eqref{GEO:formula_charge} will be
modified for other theories.

\subsection{The Smarr formula for Einstein gravity}
\label{sec:Smarr}

Let us now derive a formula relating the energy and angular
momenta of a black hole with Killing horizon to quantities defined
at the horizon. A general derivation can be found
in~\cite{Barnich:2004uw}.

Let us choose a path $g^{(s)}$, $s=0\dots 1$ in $\cF$
interpolating between the background $\bar g = g^{(0)} \in \cF$
and a black hole $g = g^{(1)} \in \cF$. It is not assumed that
there is a horizon defined for all metrics along the path.

Now, the conserved quantity associated with the Killing generator
$\xi$ of the target black hole solution $g =
g^{(1)}$~\eqref{Kill_gen} is
\begin{eqnarray}
\cQ_{\xi}[g,\bar g] = \cE - \Omega_a \cJ,
\end{eqnarray}
by linearity of $\cQ_{\xi}[g,\bar g] $ in $\xi$. Because this
quantity may be computed on any surface, we can write
\begin{eqnarray}
\cE - \Omega_a \cJ &=& - \oint_H k_{\cL^{EH},\xi}^K[g]+ \oint_S
k_{\cL^{EH},\xi}^K[\bar g] +\int_{\gamma_s}\oint_S i_\xi
I^{n}_{\delta g}\cL_{EH},
\end{eqnarray}
after having used~\eqref{GEO:formula_charge}. Here, $H$ is the
black hole horizon and $S$ is any surface that may be deformed to
the surface $S^\infty$ at infinity. The Komar integral $\oint_H
K^K_\xi[g]$ evaluated on the horizon is given
by~\eqref{Komar_hor}. We thus get
\begin{eqnarray}
  \label{eq:11}
  \cE-\Omega_a\cJ^a-\frac{\kappa}{8\pi}\cA= \oint_S
    k^K_{\cL^{EH},\xi}[\bar g] +\int_{\gamma_s}\oint_S i_\xi I^{n}_{\delta g}\cL_{EH}.
\label{PreSmarr}
\end{eqnarray}
The claim is that this relation gives the generalized Smarr
formula, which becomes the thermodynamical Euler relation, with
the standard identifications of temperature as
$\cT=\frac{\kappa}{2\pi}$ and entropy as $\cS=\frac{1}{4}\cA$.
This formula will be applied to Kerr-anti de Sitter black holes
and to their flat limit in Chapter~\ref{chap:BHapp}.

A generalization of the Smarr formula for Einstein gravity coupled
to a Maxwell field with Chern-Simons term will be given in
section~\ref{sec:sugraGodel} of Chapter~\ref{chap:BHapp}.

\subsection{First law for gravity coupled to a $p$-form potential and a dilaton}
\label{sec:pformlaw}

The first law of black hole mechanics was initially developed
taking into account dust as well as electromagnetic
fields~\cite{Bardeen:1973gs}. Also, the first law with Yang-Mills
fields were studied, e.g., in~\cite{Sudarsky:1992ty,Gao:2003ys}.

Hamiltonian~\cite{Copsey:2005se},
quasilocal~\cite{Astefanesei:2005ad} as well as covariant phase
space methods~\cite{Rogatko:2005aj,Rogatko:2006xh} have
investigated the role of $p$-form charges in the first law. The
aim of this section published in~\cite{Compere:2007vx} is to
continue the analysis started in section~\ref{sec:pform} of
Chapter~\ref{chap:matter} by deriving the first law using our
methods in a notation taking care of form factors. The first law
will be proven for gauge potentials that may be irregular on the
bifurcation surface, which is necessary in order to cover e.g. the
thermodynamics of black rings~\cite{Emparan:2004wy}. In that
respect, the covariant analysis
of~\cite{Rogatko:2005aj,Rogatko:2006xh} will be improved.

We will use the observation~\cite{Gao:2003ys} that a consistent
thermodynamics can be done on the future event horizon with
diverging potentials if, nevertheless, the potential is regular
when pulled-back on the future horizon. Our resulting expression
for the first law constitutes a generalization
of~\cite{Gao:2003ys} for $p$-form potentials (also coupled to a
scalar field). We will then show in section~\ref{sec:br} of
Chapter~\ref{chap:BHapp} that the potential for the black
rings~\cite{Emparan:2004wy} admits a regular pull-back on the
future event horizon and can thus be treated by this method. Note
that our analysis covers only electric-type charges and not
magnetic charges where the potential is necessarily singular on
the future event horizon.

We assume as in the previous section that the fields $\phi^i
\equiv (g,\mathbf A,\chi)$ and $\phi^i+\delta \phi^i$ are
stationary black hole solutions with Killing horizon $H$. The
variation of energy $\delta \cE$ and angular momenta $\delta
\cJ^a$ are defined as the charges associated with the Killing
vectors $\d_t$ and $-\d_{\varphi_a}$. We assume that $\xi$ is a
solution of~\eqref{eq:red} with $\mathbf \Lambda = 0$. We also
require that $\xi + \delta\xi$ is a symmetry of the perturbed
black hole $\phi^i+\delta\phi^i$.

The first law is then a consequence of the equality~\footnote{The
first law can be straightforwardly generalized to reducibility
parameters satisfying $\cL_\xi \mathbf A + \mathbf \Lambda = 0$
with $\mathbf\Lambda \neq d(\cdot)$. This simply amounts to add a
contribution at infinity and at the horizon.}
\begin{eqnarray}
  \oint_{S^\infty}\mathbf  k_{ \xi,0}[ \delta \phi ; \phi]
  =\oint_{H}\mathbf  k_{ \xi,0}[ \delta \phi; \phi],
\end{eqnarray}
where $S^\infty$ is a $(n-2)$-sphere at infinity. Using the
linearity of $\mathbf k_{ \xi,0}$ with respect to $\xi$, the
left-hand side is simply given by $\delta \cE - \Omega_a \delta
\cJ^a$. Splitting the right-hand side, we get
\begin{eqnarray}
\delta \cE - \Omega_a \delta \cJ^a  =\oint_{H}\mathbf  k^{g}_{
\xi,0}[ \delta \phi; \phi]+\oint_{H}\mathbf  k^{\chi}_{ \xi,0}[
\delta \phi; \phi]+\oint_{H}\mathbf  k^{\mathbf A}_{ \xi,0}[
\delta \phi; \phi].
\end{eqnarray}
We showed in the last section that geometric properties of the
Killing horizon allow one to express the gravitational
contribution into the form
\begin{eqnarray}
\oint_{H}\mathbf  k^{g}_{ \xi,0}[ \delta \phi; \phi]=
\frac{\kappa}{8\pi G} \delta \cA.\label{eq:geo_t}
\end{eqnarray}
Using~\eqref{phicharge}, the scalar contribution can be written as
\begin{equation}
\oint_{H}\mathbf  k^{\chi}_{ \xi,0}[ \delta \phi; \phi] = -
\oint_H d\cA\, \delta \chi (\cL_\xi \chi + \xi^2 \cL_n \chi) = 0,
\end{equation}
which vanishes thanks to the reducibility
equations~\eqref{eq:red}, assuming the regularity of the scalar
field on the horizon.

The contribution of the $p$-form can be computed using the
arguments of~\cite{Gauntlett:1998fz,Copsey:2005se}. The
Raychaudhuri equation gives $R_{\mu\nu}\xi^\mu \xi^\nu = 0$ on the
horizon. It follows by Einstein's equations and by the identity
$\cL_\xi \phi = 0$ that $i_\xi \mathbf{H}$ has vanishing norm on
the horizon. But as $i_\xi (i_\xi\mathbf H) = 0$, $i_\xi \mathbf
H$ is tangent to the horizon. $i_\xi\mathbf H$ has thus the form
$\mathbf\xi \wedge \dots \wedge \mathbf\xi$ by antisymmetry of
$\mathbf H$ and its pullback to the horizon vanishes. The equation
$\cL_\xi \mathbf A = 0$ can be written as $d i_\xi \mathbf A  =
-i_\xi \mathbf H$. Therefore, the pull-back of $i_\xi  \mathbf A$
on the horizon is a closed form.

For $p=1$, $-i_\xi \mathbf A = \Phi$ is simply the scalar electric
potential at the horizon~\eqref{el_potential}. When $p > 1$, the
quantity $-i_\xi \mathbf A $ pulled-back on the horizon is the sum
of an exact form $d \mathbf{e}$ and an harmonic form $\mathbf h$.
If the horizon has non-trivial $n-p-1$ cycles $T_a$, one can
define the harmonic forms dual to $T_a$ by duality between
homology and cohomology as
\begin{equation}
\int_{T_a} \mathbf\sigma = \int_H \mathbf\omega_a \wedge
\mathbf\sigma , \qquad \forall \mathbf\sigma.\label{eq:ntcycle}
\end{equation}
The harmonic form $\mathbf h$ is then a sum of terms $\mathbf h =
\Phi^a \mathbf\omega_a$ with $\Phi^a$ constant over the
non-trivial cycles.

The contribution from the potential contains three
terms~\eqref{Bcharge}. The Komar term~\eqref{def_QA} can be
written as
\begin{eqnarray}
\oint_H \mathbf Q^{\mathbf A}_{\xi,0} = -\Phi^a \oint_{T_a}
e^{-\alpha \chi}\star \mathbf H,
\end{eqnarray}
where the exact form $d \mathbf{e}$ do not contribute on-shell.
 We recognize on the right-hand side the
conserved form written in~\eqref{dipole_ch}. Let us denote by
$\cQ_a$ the integral $\oint_{T_a} e^{-\alpha \chi}\star \mathbf
H$.

Using~\eqref{GEO:surf_form}, the contribution~$\oint_H i_\xi
\mathbf \Theta_{\mathbf A}[\delta \phi,\phi]$ reads as
\begin{equation}
\oint_H i_\xi \mathbf \Theta_{\mathbf A}[\delta \phi,\phi] =
\oint_H e^{-\alpha\chi}(i_\xi \delta \mathbf A) \wedge \star
\mathbf H -\oint_H d\cA\, \xi^2 \star \Big( \delta \mathbf A
\wedge \star (i_n \mathbf H) \Big).\label{eq:k10}
\end{equation}
The first term of~\eqref{eq:k10} nicely combines with the second
term of~\eqref{Bcharge} into $-\oint_{T_a} \delta \Phi^a
e^{-\alpha \chi}\star \mathbf H = -\delta \Phi^a \cQ_a$ because
$\delta \Phi^a$ is constant as a consequence of the hypotheses on
the variation. In the second term of~\eqref{eq:k10}, one can
replace $\delta \mathbf A$ by its pull-back $\phi_* \delta \mathbf
A$ on the future horizon. Indeed, decomposing $\delta\mathbf A =
\mathbf n \wedge \omega^{(1)}+\phi_*\delta\mathbf A$, one sees
that the term involving $\mathbf n$ do not contribute because of
the antisymmetry of $\mathbf H$. Therefore, the second term
in~\eqref{eq:k10} will vanish if $\mathbf H$ is regular and if the
pull-back $\phi_* \delta \mathbf A$ on the future horizon is
regular.

Finally, the contribution from the potential on the horizon
reduces to
\begin{equation}
\oint_{H}\mathbf  k^{\mathbf A}_{ \xi,0}[ \delta \phi; \phi] =
\Phi^a \delta \cQ_a,
\end{equation}
as it should to give the first law
\begin{equation}
\delta \cE - \Omega_a \delta \cJ^a= \frac{\kappa}{8\pi G}\delta
\cA + \Phi^a \delta \cQ_a.\label{first_law_3}
\end{equation}

%% file: blackholes_examples.tex
\chapter{Black hole solutions and their thermodynamics}
\label{chap:BHapp}
\setcounter{equation}{0}\setcounter{figure}{0}\setcounter{table}{0}

General relativity provides a very elegant classical description
of the gravitational interaction. Remarkably, this theory predicts
the existence of black holes which satisfy laws analogous to the
laws of thermodynamics. In this chapter, we will try to get
further insights in the properties of black holes by finding new
solutions to gravity coupled to matter fields and by investigating
their thermodynamical properties.

In the first section, we will construct new G\"odel-type black
hole and particle solutions to Einstein-Maxwell theory in 2+1
dimensions with a negative cosmological constant and a
Chern-Simons term. These black holes can be seen as B(H)TZ
black holes~\cite{Banados:1992wn,Banados:1992gq} immersed into a
G\"odel background. We will show that a particular solution is
related to the original G\"odel universe. The solutions will also
be analyzed from the point of view of identifications. On-shell,
the electromagnetic stress-energy tensor will be seen to
effectively replace the cosmological constant by minus the square
of the topological mass and produce the stress-energy of a
pressure-free perfect fluid. Finally, the tools developed in the
preceding chapters will be used to compute the conserved charges
and work out the thermodynamics.

In section~\ref{sec:KerrAdS}, we will turn to higher dimensional
Kerr-anti-de Sitter black holes. The conserved charges will be
obtained by our methods and a generalized Smarr relation which is
valid both in flat and in anti-de Sitter backgrounds will be
derived. It will be also shown that the charges for higher
dimensional Kerr-adS black holes can be correctly computed from
the standard Hamiltonian or Lagrangian surface integrals at
infinity.

The definition of conserved quantities for G\"odel black holes was
an open problem in 2004~\cite{Gimon:2003ms,Klemm:2004wq} mainly
because the naive application of traditional approaches fails. In
section~\ref{sec:sugraGodel}, the mass, angular momenta and charge
of the G\"odel-type rotating black hole solution to five
dimensional minimal
supergravity~\cite{Gauntlett:2002nw,Gimon:2003ms} will be
computed, thereby providing a definition of charges in these
unconventional spacetimes. Moreover, a generalized Smarr formula
will be derived and the first law of thermodynamics will be
verified.

We conclude in sections~\ref{sec:br} and~\ref{sec:plan} with
applications of our formalism to black rings and with the
definition of energy in plane-waves geometries.

\section{Three-dimensional G\"odel black holes}
\label{sec:threedbh}

Exact solutions of higher dimensional gravity and supergravity
theories play a key role in the development of string theory.
Recently, a G\"odel-like exact solution of five-dimensional
minimal supergravity having the maximum number of supersymmetries
has been constructed~\cite{Gauntlett:2002nw}. As its four-dimensional
predecessor, discovered by G\"odel in
1949~\cite{Godel:1949ga}, this solution possesses a large number
of isometries. It can be lifted to higher dimensions and has
recently been extensively studied as a background for string and
M-theory, see e.g.~\cite{Boyda:2002ba,Harmark:2003ud}.

The G\"odel-like five-dimensional solution found in
\cite{Gauntlett:2002nw} is supported by an Abelian gauge field.
This gauge field has an additional Chern-Simons interaction and
produces the stress-energy tensor of a pressureless perfect fluid.
Since a Chern-Simons term can also be added in three dimensions,
it is a natural question to ask whether a G\"odel like solution
exists in three-dimensional gravity coupled to a
Maxwell-Chern-Simons field.

Actually, there is a stronger motivation to look for this kind of
solutions of three-dimensional gravity. The reason is that the
original four-dimensional G\"odel spacetime is already effectively
three dimensional, see e.g.~\cite{HawkingEllis}. In fact, the
metric has as direct product structure $ds^2_{(4)} = ds_{(3)}^2 +
dz^2$ where $ds_{(3)}^2$ satisfies a purely three-dimensional
Einstein equation.

The goal of this section, published as an article
in~\cite{Banados:2005da} with M.~Banados, G.~Barnich and
M.~Gomberoff, is twofold. On the one hand we will show that the
three-dimensional factor $ds_{(3)}^2$ of the G\"odel spacetime and
its generalizations \cite{Reboucas:1982hn} are exact solutions of
the three-dimensional Einstein-Maxwell-Chern-Simon theory
described by the action,
\begin{eqnarray}
I = \frac{1}{16\pi G} \int d^3x \left[\sqrt{-g}\left(R
+\frac{2}{l^2} - \frac{1}{4}
F_{\mu\nu}F^{\mu\nu}\right)-\frac{\alpha}{2}\epsilon^{\mu\nu\rho}A_\mu
F_{\nu\rho}\right]. \label{action}
\end{eqnarray}
The stress-energy tensor of the perfect fluid will be fully
generated by the gauge field $A_\mu$, in complete analogy with the
five-dimensional results reported in \cite{Gauntlett:2002nw}.

Our second goal deals with G\"odel particles and black holes.
Within the five-dimensional supergravity theory, rotating black
hole solutions on the G\"odel background have been investigated in
\cite{Herdeiro:2002ft,Gimon:2003ms,Brecher:2003wq,Gimon:2003xk,%
  Behrndt:2004pn,Klemm:2004wq,Gimon:2004if,Barnich:2005kq,Cvetic:2005zi}.
It is then natural to ask whether the three-dimensional G\"odel
spacetime $ds_{(3)}^2$ can be generalized to include horizons.
This is indeed the case and a general solution will be displayed.

The conserved charges - mass, angular momentum and electric charge
- will be computed for these solutions and the first law for the
three-dimensional black holes, adapted to an observer at rest with
respect to the electromagnetic fluid will be derived. We then show
how to adapt this first law in order to compare with the one for
adS black holes in the absence of the electromagnetic fluid.

In parallel to this work, three-dimensional black hole solutions
with naked closed time-like curves have also been obtained from
exact marginal deformations of the $SL(2,R)$ WZW
model~\cite{Detournay:2005fz}. G\"odel black hole solutions can
thus be promoted to exact string theory backgrounds. During the
writing of this thesis, the $\cN = 2$ supersymmetric extension of
the action~\eqref{action} has been constructed
in~\cite{Banados:2007sq}. It turns out that the three-dimensional
G\"odel solution preserves one half of the supersymmetries.

\subsection{Introduction}

Let us now briefly discuss the general structure of the
stress-energy tensor of Maxwell-Chern-Simons theory. The original
G\"odel geometry is a solution of the Einstein equations in the
presence of a pressureless fluid with energy density $\rho$ and a
negative cosmological constant $\Lambda$ such that $\Lambda=-4\pi
G\rho$. Equivalently, it can be viewed as a homogeneous spacetime
filled with a stiff fluid, that is, $p_{SF}=\rho_{SF}=\rho/2$ and
vanishing cosmological constant.

In (2+1)-spacetime dimensions, an electromagnetic field can be the
source of such a fluid. To see this it is convenient to write the
stress-energy tensor in terms of the dual field ${}^*\!F^{\mu}$,
\begin{equation}
16\pi G\,T^{\mu\nu} = {}^*\!F^{\mu} {}^*\!F^{\nu} - \frac{1}{2}
{}^*\!F^{\alpha} {}^*\!F_{\alpha}g^{\mu\nu} \ . \label{t}
\end{equation}
In any  region where the field ${}^*\!F^{\mu}$ is timelike, the
electromagnetic field behaves as a stiff fluid with
\begin{equation}
u^\mu = \frac{1}{\sqrt{-{}^*\!F^{\alpha}
    {}^*\!F_{\alpha}}}{}^*\!F^{\mu},  \ \ \ \
\rho_{SF}=p_{SF}=-{}^*\!F^{\alpha} {}^*\!F_{\alpha}/16\pi G.
\label{}
\end{equation}
If G\"odel's geometry is going to be a solution of the
Einstein-Maxwell system, then $\rho_{SF}=-{}^*\!F^{\alpha}
{}^*\!F_{\alpha}/2$ must be a constant. Moreover in comoving
coordinates, in which $g_{tt}=-1$, ${}^*\!F^{\mu}$ must be a
constant vector pointing along the time coordinate. One can easily
see that such a solution does not exist. In fact, the Maxwell
equations for this solution,
\begin{equation}
d{}^*\!F=0 , \label{max}
\end{equation}
imply in these coordinates that $g_{t[\varphi,r]}=0$ which cannot
be achieved for G\"odel. If the electromagnetic field acquires a
topological mass $\alpha$, however, Maxwell's equations
(\ref{max}) will be modified by the addition of the term $\alpha
F$. In that case, the timelike, constant, electromagnetic field
is, as we will see below, a solution of the coupled
Einstein-Maxwell-Chern-Simons system, and the geometry is
precisely that of G\"odel.

\subsection{Topologically massive
gravito-electrodynamics } \label{GODEL}

We start by reviewing the main properties, relevant to our
discussion, of the four-dimensional G\"odel spacetimes
\cite{Godel:1949ga,Reboucas:1982hn,Rooman:1998xf}. These metrics
have a direct product structure $ds_{(3)}^2+dz^2$ with
three-dimensional factor given by
\begin{eqnarray}
ds^2_{(3)} &=& -\left( d t +\frac{4\Omega}{\tilde
m^2}\sinh^2{\left( \frac{\tilde m
   \rho}{2}\right)}d\varphi\right)^2 +d\rho^2 \nonumber \\
 &&\quad + \frac{\sinh^2{(\tilde m \rho)}}{\tilde m^2}d\varphi^2.
\label{Godel}
\end{eqnarray}
The original solution discovered by G\"odel corresponds to $\tilde
m^2=2\Omega^2$. Furthermore, it was pointed out in
\cite{Reboucas:1982hn} that the property of homogeneity and the
causal structure of the G\"odel solution also hold for $\Omega$
and $\tilde m$ independent, provided that $0\leq \tilde m^2 <
4\Omega^2$, the limiting case $\tilde m^2= 4\Omega^2$
corresponding to anti-de Sitter space.

The three-dimensional metric (\ref{Godel}) has 4 independent
Killing vectors, two obvious ones, $\xi_{(1)}=\partial _t$ and
$\xi_{(2)}=\partial_\varphi$, and two additional ones,
\begin{eqnarray}
 \xi_{(3)} &=& \frac{2\Omega}{\tilde m^2} \tanh(\tilde m\rho/2)
\sin\varphi \frac{\partial}{\partial t}  - \frac{1}{\tilde m}
\cos\varphi \frac{\partial}{\partial \rho} + \nonumber \\ &&
\coth(\tilde m\rho) \sin\varphi\frac{\partial}{
\partial \varphi}\,,
 \label{xi3}\\
 \xi_{(4)} &=& \frac{2\Omega}{\tilde m^2} \tanh(\tilde m\rho/2) \cos\varphi
\frac{\partial}{\partial t}
 + \frac{1}{\tilde m} \sin\varphi  \frac{\partial}{\partial \rho} +
 \nonumber\\ && \coth(\tilde m\rho) \cos\varphi \frac{\partial}{
\partial \varphi}.
 \label{xi4}
\end{eqnarray}
which span the algebra $so(2,1) \times \mathbb R$. Finally, the
metric (\ref{Godel}) satisfies the three dimensional Einstein
equations,
\begin{equation}\label{EE}
G^{\mu\nu} - \Omega^2 g^{\mu\nu} = (4\Omega^2 - \tilde
m^2)\delta^\mu_{t} \delta^\nu_t,
\end{equation}
for all values of $\Omega,\tilde m$, and we see that $\Omega$
plays the role of a negative cosmological constant.

Note that a solution $ds^2_{(3)}$ to Einstein's equations in $3$
dimensions can be lifted to a solution in $4$ dimensions through
the addition of a flat direction $z$ if the additional components
of the stress-energy tensor are chosen as ${\cal T}^{\mu z}=0$ and
${\cal
  T}^{zz}=g_{\mu\nu}{\cal T}^{\mu\nu} +{\Omega^2}/{4\pi G}$. For the
solutions \eqref{Godel}, ${\cal T}^{zz}={(\tilde
m^2-2\Omega^2)}/{8\pi G}$ and vanishes, as it should, for the
original G\"odel solution.

Our first goal is to prove that (\ref{Godel}) can be regarded as
an exact solution to the equations of motion following from
(\ref{action}).

To this end, we need to supplement (\ref{Godel}) with a suitable
gauge field which will provide the stress-energy tensor (right
hand side of (\ref{EE})). Consider a spherically symmetric gauge
field in the gauge $A_r=0$,
\begin{equation}
A = A_t(\rho) dt  + A_\varphi(\rho) d\varphi.\label{9}
\end{equation}
Inserting this ansatz for the gauge field into the equations of
motion associated to the action (\ref{action}), and assuming that
the metric takes the form (\ref{Godel}), one indeed finds a
solution for $A_t$ and $A_\varphi$. Moreover, the two parameters
$\tilde m,\Omega$ entering in (\ref{Godel}) become related to the
coupling constants $\alpha$ and $1/l$ as
\begin{eqnarray}
 \Omega  &=& \alpha\,, \nonumber\\
 \tilde m^2 &=& 2 \left(\alpha^2 + \frac{1}{l^2} \right). \label{trans}
\end{eqnarray}
With this parameterization, the G\"odel sector is determined by
$\alpha^2l^2-1>0$, with $\alpha^2l^2=1$ corresponding to anti-de
Sitter space.  For future convenience, we shall write the solution
in terms of a new radial coordinate $r$ defined by
\begin{equation}
r= {2 \over \tilde m^2}\sinh^2\left( {\tilde m\rho\over 2}\right).
\label{r}
\end{equation}
Explicitly, the metric and gauge field are given by,
\begin{eqnarray}
ds^2 &=& -dt^2 - 4\alpha r dt d\varphi +
\left[2r-\left(\alpha^2l^2 - 1 \right){2r^2 \over l^2}
\right]d\varphi^2 \nonumber \\&&  +
\left(2r+(\alpha^2l^2+1){2r^2\over l^2} \right)^{-1} dr^2\,,  \label{Godel2}\\
A  &=&  \sqrt{\alpha^2l^2 -1}\, {2r \over l}\, d\varphi\,.
\label{gauge2}
\end{eqnarray}
{}From now on, we always write $\Omega$ and $\tilde m$ in terms of
$\alpha$ and $l$ using (\ref{trans}). The general solution for $A$
involves the addition of arbitrary constant terms along $dt$ and
$d\varphi$ in (\ref{gauge2}). At this stage, we choose the
constant in $A_t$ to be zero. We will come back to this issue when
we discuss black hole solutions below. A constant term in
$A_\varphi$ is not allowed, however, if one requires $A_\varphi
d\varphi$ to be regular everywhere. Indeed, near $r=0$, the
spacelike surfaces of (\ref{Godel2}) are $\mathbb R^2$ in polar
coordinates, the radial coordinate $r$ in (\ref{Godel2}) being the
square root of a standard radial coordinate over $\mathbb R^2$,
and thus $A_\varphi$ must vanish at $r=0$ because the 1-form
$d\varphi$ is not well defined there.

The gauge field (\ref{gauge2}) is also invariant under the
isometries of (\ref{Godel}), up to suitable gauge transformations:
for each Killing vector $\xi^\mu_{(a)}$ there exists a function
$\epsilon_{(a)}$ such that
\begin{equation}\label{} {\cal L}_{\xi_{(a)}} A_\mu
- \partial_\mu \epsilon_{(a)}=0.
\end{equation}
In this sense, the Killing vectors $\xi^\mu_{(a)}$ of
(\ref{Godel}) are lifted to gauge parameters
$(\xi^\mu_{(a)},\epsilon_{(a)})$ that leave the full gravity plus
gauge field solution invariant. The generalized G\"odel metric
(\ref{Godel2}) together with the gauge field (\ref{gauge2}) define
a background for the action (\ref{action}) with 4 linearly
independent symmetries of this type. We shall now use these
symmetries in order to find new solutions describing particles and
black holes.

\subsection{G\"odel particles: $\ {\alpha^2 l^2>1} $}
\label{PARTICLES}

We have proven in the previous section that the G\"odel metric can
be regarded as an exact solution to action (\ref{action}).  The
associated gauge field (\ref{gauge2}) is however real only in the
range $\alpha^2l^2\geq 1$. We consider in this section the case
$\alpha^2l^2>1$ and introduce particle-like objects on the
background (\ref{Godel2}) by means of spacetime identifications.

\subsubsection{G\"odel Cosmons }

Identifications in three-dimensional gravity were first introduced
by Deser, Jackiw and t'Hooft \cite{Deser:1984tn,Deser:1984dr} and
the resulting objects called ``cosmons". In the presence of a
topologically massive electromagnetic field, cosmons living in a
G\"odel background may also be constructed along these lines.

Take the metric (\ref{Godel2}) and make the following
identification along the Killing vectors $\partial _\varphi$ and
$\partial_t$
$$
(t, \varphi) \sim (t- 2  \pi j m,\varphi + 2 \pi m).
$$
where $m,j$ are real constants. If $m\neq 1$ this procedure will
turn the spatial plane into a cone. The cosmon lives in the tip of
this cone, and its mass is related to $m$ and $j$ (see below). The
time-helical structure given by $j$ will provide angular momentum.

To analyze the resulting geometry it is convenient to pass to a
different set of coordinates,
\begin{eqnarray}
\varphi &=& \varphi' m \nonumber \\
t &=& t' - j\varphi' m  \label{nc}\\
r &=& \frac{r'}{m} +\frac{j}{2\alpha} \nonumber .
\end{eqnarray}
where the above identification amounts to
\begin{equation}\label{}
\varphi' \sim \varphi' + 2\pi n, \ \ \ \ \ n\in Z.
\end{equation}
Also, the new time $t'$ flows ahead smoothly, that is, it does not
jump after encircling the particle.  Inserting these coordinates
into (\ref{Godel2}), and erasing the primes, we find the new
metric
\begin{eqnarray}
ds^2 &&= -dt^2 - 4\alpha r dt d\varphi\nonumber\\ && + \left[8G\nu
r-(\alpha^2l^2-1) {2r^2 \over l^2}  - \frac{4GJ}{\alpha} \right]
d\varphi^2 \nonumber \\ &&  + \left( (\alpha^2l^2+1){2r^2\over
l^2} + 8G\nu r - {{4GJ} \over \alpha}\right)^{-1} dr^2.
\label{particles}
\end{eqnarray}
For fixed  $m$ and $mj$, the new constants $\nu$ and $J$ are given
by
\begin{eqnarray}
 4G\nu=m\left(1+\frac{1+\alpha^2l^2}{\alpha l^2}j\right) ,  \label{const1}\\
 4GJ = -m^2 j\left(1+\frac{1+\alpha^2l^2}{2\alpha l^2}j\right).\label{const}
\end{eqnarray}
These constants will be shown to be related to the mass and
angular momentum respectively.

Since under (\ref{nc}) $\varphi$ scales with $m$ while $r$ with
$1/m$ we see that the $r$-dependent part of gauge field
(\ref{gauge2}) is invariant under (\ref{nc}).  However, the
manifold now has a non-trivial cycle, and it is not regular at the
point $r=r_0$ invariant under the action of the Killing vector
whose orbits are used for identifications. Explicitly, $r_0 =
-\frac{jm}{2\alpha}$ which corresponds to $r=0$ before the shift
of $r$ in \eqref{nc}. This means that one can now add a constant
piece to $A_\varphi$. The new gauge field becomes
\begin{equation}\label{}
A = (-\frac{4GQ}{\alpha} + \sqrt{\alpha^2l^2 -1}\, {2r \over
l})d\varphi.
\end{equation}
The constant $Q$ will be identified below as the electric charge
of the particle sitting at $r=0$.

The metrics \eqref{particles} only admit the 2 Killing vectors
$\partial_t$ and $\partial_\varphi$. Indeed, the other candidates
$\xi_{(3)}$ and $\xi_{(4)}$ do not survive as they do not commute
with the Killing vector along which the identifications are made
\cite{Banados:1993gq}.

So far we have only used the Killing vectors $\partial /\partial
\varphi$ and $\partial /\partial t$ of (\ref{Godel}) to make
identifications.  Besides these Killing vectors, the metric
(\ref{Godel}) has two other isometries defined by the vectors
(\ref{xi3}) and (\ref{xi4}), and one may consider identifications
along them. We shall not explore this possibility in this paper.

\subsubsection{Horizons, Singularities and Time Machines}

Distinguished places of the geometry (\ref{particles}) may appear
on those points where either $g_{\varphi\varphi}$ or $g^{rr}$
vanishes. The vanishing of $g_{\varphi\varphi}$ indicates that
$g_{\varphi\varphi}$ changes sign and hence closed timelike curves
(CTC) appear. On the other hand, the vanishing of $g^{rr}$
indicates the presence of horizons, as can readily be seen by
writting (\ref{particles}) in ADM form.

The function $g_{\varphi\varphi}$ in (\ref{particles}) is an
inverted parabola, and, it will have two zeros, say $r_1$ and
$r_2$ whenever
\begin{equation}
2G\nu^2 > \frac{J(\alpha^2 l^2-1)}{\alpha l^2}. \label{mu2}
\end{equation}
We must require this condition to be fulfilled in order to have a
``normal" region where $\partial_\varphi$ is spacelike. The
boundary of the normal region are two spacelike surfaces, the
velocity of light surfaces (VLS) at $r=r_1$ and $r=r_2$ (assume
$r_2>r_1$). These surfaces are perfectly regular as long as
$g_{t\varphi}\neq 0$ there, which is indeed the case for the
metric (\ref{particles}), when $\alpha\neq 0$.

On the other hand, it is direct to see from (\ref{particles}) that
\begin{equation}\label{grr}
g^{rr}=4\alpha^2 r^2+ g_{\varphi\varphi}.
\end{equation}
Since $g_{\varphi\varphi}$ is positive in the normal region, there
are no horizons there and $g^{rr}$ is positive in that region.
This means that, if any, both zeros of $g^{rr}$ are on the same
side of the normal region. The sides in which no zero of $g^{rr}$
are present are analog to the G\"odel time machine, an unbounded
region, free of singularities, where $\partial_\varphi$ is
timelike. If $\nu\geq 0$, the roots of $g^{rr}$ are smaller than
the roots of $g_{\varphi\varphi}$. Without loss of generality, we
can restrict ourselves to this case because the solutions
parametrized by $(\nu,J,Q)$ are related to those with
$(-\nu,J,-Q)$ by the change of coordinates $r\rightarrow -r$,
$\varphi\rightarrow -\varphi$.

The condition for ``would be horizons'' is
\begin{equation}
2G\nu^2 > \frac{J(\alpha^2 l^2+1)}{\alpha l^2}. \label{mu3}
\end{equation}
As depicted in Fig.~4.1, once one reaches the largest root
$r_+=r_0$ of $g^{rr}$, the manifold comes to an end. Indeed, the
signature of the metric changes as one passes $g^{rr}=0$.  This
can be seen by putting the metric in ADM form (see \eqref{ADM}
below). Note that in this case, given $(\nu,J)$, there is a unique
$(m,mj)$ satisfying \eqref{const1}-\eqref{const}.

Using then $r=r_++\kappa_0|\alpha r_+|\rho^2$, with $r_-$ the
smallest root of $g^{rr}$ and
$\kappa_0=\frac{(r_+-r_-)(\alpha^2l^2+1)}{2l^2|\alpha r_+|}$, one
finds near $r_+$,
\begin{eqnarray}
  ds^2\approx\kappa_0^2
  \rho^2dt^2+d\rho^2-4\alpha^2r_+^2(d\varphi+\frac{dt}{2\alpha
  r_+})^2.
\end{eqnarray}
This means that the spacetime has a naked singularity at $r_+$,
which is the analog of the one found in the spinning cosmon of
\cite{Deser:1984tn,Deser:1984dr}.

Alternatively, as proposed originally in \cite{Cvetic:2005zi} for
the case where the would be horizon is inside the time machine,
one can periodically identify time $t$ with period
$2\pi/\kappa_0$. This leads to having CTC's lying everywhere,
including the normal region.

\begin{figure}
\begin{center}
  \includegraphics[width=8cm]{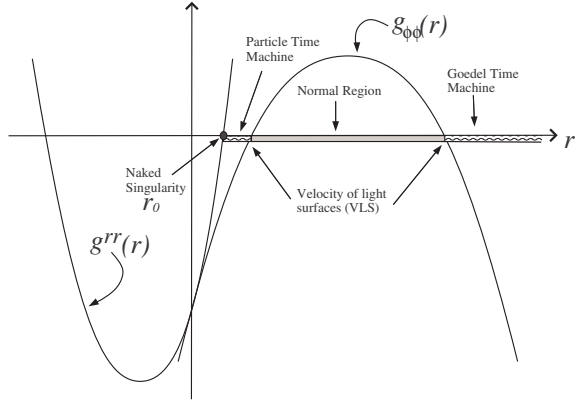}\label{pneg}
  \caption{G\"odel cosmons}
\end{center}
\end{figure}

\subsection{G\"odel black holes}
\label{BH}

\subsubsection{The $\alpha^2 l^2 <1 $ sector }

We have shown in Sec.~\ref{GODEL} that the metric (\ref{Godel})
can be embedded as an exact solution to the equations of motion
derived from (\ref{action}).  The necessary gauge field, given in
(\ref{gauge2}) is, however, real only in the range $\alpha^2l^2
\geq 1$.  As we mentioned in Sec.~\ref{GODEL}, the gauge field
(\ref{gauge2}) represents the most general static spherically
symmetric solution, given the metric (\ref{Godel}) (or, in the new
radial coordinate, (\ref{Godel2})).  This means that if we want to
find a real gauge field in the range $\alpha^2 l^2 <1$ we need to
start with a different metric. The goal of this section is to
explore the other sector, $\alpha^2l^2 < 1$, where black holes
will be constructed.

Starting from the metric (\ref{Godel2}) and gauge field
(\ref{gauge2}) it is easy to construct a new exact solution which
will be real in the range $\alpha^2 l^2<1$. Consider the following
(complex) coordinate changes \footnote{An equivalent way to do
this
  transformation without introducing the imaginary unit is by the
  following sequence of coordinate transformations (and analytic
  continuations) acting on (\ref{Godel2}):
$ t \rightarrow 2t^{1/2}$,  $t \rightarrow -t$, $ t \rightarrow {1
\over 4}t^2$, and the same for $\varphi$.} acting on
(\ref{Godel2}) and (\ref{gauge2}): $\varphi \rightarrow i\varphi$,
$t\rightarrow it$, and $r \rightarrow -r$. The new metric and
gauge field read,
\begin{eqnarray}
ds^2 &=& dt^2 - 4\alpha r dt d\varphi +
\left[2r-\left(1-\alpha^2l^2 \right) {2r^2\over l^2}\right]
d\varphi^2 \nonumber \\ &&  + \left( (\alpha^2 l^2+1){2r^2\over
l^2}
- 2r \right)^{-1} dr^2  \label{Godel3}\\
A &=& \sqrt{1-\alpha^2l^2}\, {2r \over l}d\varphi. \label{gauge3}
\end{eqnarray}

Several comments are in order here. First of all, the intermediate
step of making some coordinates complex is only a way to find a
new solution. From now on, all coordinates $t,r,\varphi$ are
defined real, and, in that sense, the fields (\ref{Godel3}) and
(\ref{gauge3}) provide a new exact solution to the action
(\ref{action}) which is real in the range $\alpha^2l^2 <1$.

Second, in the original metric (\ref{Godel2}), the coordinate
$\varphi$ was constrained by the geometry to have the range $0\leq
\varphi < 2\pi$.  This is no longer the case in the metric
(\ref{Godel3}). The 2-dimensional sub-manifold described by the
coordinates $r,\varphi$ does not have the geometry of $\mathbb
R^2$ near  $r\rightarrow 0$ anymore; the coordinate $\varphi$ is
thus not constrained to be compact, and in principle it should
have the full range
\begin{equation}
-\infty < \varphi <  \infty.
\end{equation}
The reason that $\varphi$ in (\ref{Godel3}) is not constrained by
the geometry is that the $g^{rr}$ component of the metric
(\ref{Godel3}) changes sign as we approach $r=0$. This is an
indication of the presence of a horizon, although this surface is
not yet compact.

Finally, it is worth mentioning that the metrics (\ref{Godel3})
and (\ref{Godel2}) are real and are related by a coordinate
transformation, so that all local invariants involving the metric
alone have the same values. However, as solutions to the
Einstein-Maxwell equations, they are inequivalent. Indeed, the
diffeomorphism and gauge invariant quantity
$({}^*F)^2={4}(1-\alpha^2l^2)/l^2$ changes sign when going from
(\ref{Godel2})-(\ref{gauge2}) to (\ref{Godel3})-(\ref{gauge3}).
This is different from the pure anti-de Sitter case where
particles and black holes are obtained by identifications
performed on the same background.

\subsubsection{The G\"odel black
  hole}

Let us go back to (\ref{Godel3}) and note that the function
$g^{rr}$ vanishes at $r_+>0$. In order to make the $r=r_+$ surface
a regular, finite area, horizon we shall use the Killing vector
$\partial_\varphi$ of (\ref{Godel3}) to identify points along the
$\varphi$ coordinate. In this case, $\partial _\varphi$ has a
non-compact orbit and identifications along it does not produce a
conical singularity, but a ``cylinder".  More generically, we may
proceed in analogy with the cosmon case and identify along a
combination of both $\partial_\varphi$ and $\partial_t $ so that
$$
(t, \varphi) \sim (t- 2  \pi j m,\varphi + 2 \pi m).
$$
so that the resulting geometry will also carry angular momentum.
We again pass to a different set of coordinates,
\begin{eqnarray}
\varphi &=& \varphi' m \\
t &=& t' - j\varphi' m \\
r &=& \frac{r'}{m} - \frac{j}{2\alpha} ,
\end{eqnarray}
so that the new angular coordinate $\varphi'$ is identified in
$2\pi$, and the time $t'$ flows ahead smoothly.

The new metric reads (after erasing the primes),
\begin{eqnarray}
ds^2 \hspace{-4pt} &=& \hspace{-3pt} dt^2 \hspace{-2pt}- 4\alpha r
dt d\varphi \hspace{-2pt}+\hspace{-2pt} \left( 8G\nu
r-(1-\alpha^2l^2) {2r^2
\over l^2}  - \frac{4GJ}{\alpha} \right) d\varphi^2 \nonumber \\
&& + \left( (\alpha^2l^2+1){2r^2\over l^2} - 8G\nu r +
  \frac{4GJ}{\alpha}\right)^{-1} dr^2.  \label{blackhole}
\end{eqnarray}
As for the particles analyzed in the previous section, for given
$(m,mj)$, we define new constants $\mu$ and $J$ according to
\begin{eqnarray}
 4G\nu=m\left(1+\frac{1+\alpha^2l^2}{\alpha l^2}j\right) ,\label{const5} \\
 4GJ = m^2 j\left(1+\frac{1+\alpha^2l^2}{2\alpha l^2}j\right).\label{const6}
\end{eqnarray}
Again, these constants will be related below to the mass and the
angular momentum and without loss of generality, we can limit
ourselves to the case $\nu\geq 0$.

In the new coordinates, the electromagnetic potential takes the
form $A=A_\varphi d\varphi$, where
\begin{equation}\label{aphi}
A_\varphi(r) = -\frac{4GQ}{\alpha} + \sqrt{1-\alpha^2l^2}\, {2r
\over l}.
\end{equation}
The constant $Q$ is arbitrary because, once again, the nontrivial
topology allows the addition of an arbitrary constant in
$A_\varphi$. It is worth stressing that if $\varphi$ was not
compact, then $m$ and $Q$ would be trivial constants. It also
follows that the Killing vectors of (\ref{Godel3}) have the same
form as those of (\ref{Godel2}), but with the trigonometric
functions $\cos(\varphi)$ and $\sin(\varphi)$ replaced by
hyperbolic ones. Again, these vectors do not survive after the
identifications.

\subsubsection{Horizons, Singularities and Time Machines}

\begin{figure}
\begin{center}
\includegraphics[width=9cm]{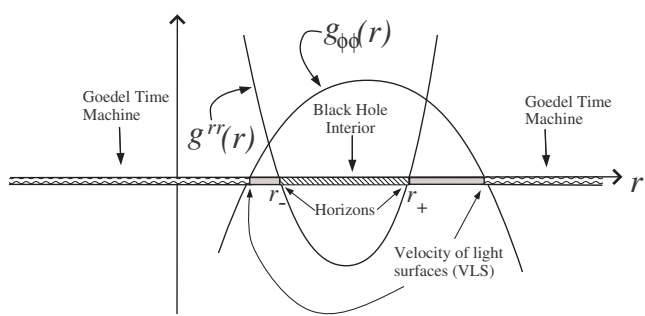}\caption{\label{posp}
G\"odel black holes}
\end{center}
\end{figure}

We now proceed to analyze the metric in the same way we did in the
preceding section.  Again we have a condition for having a normal
region, which, in this case reads
\begin{equation}
2G\nu^2 > \frac{J(1-\alpha^2 l^2)}{\alpha l^2}.
\end{equation}
The functions $g^{rr}$ and $g_{\varphi\varphi}$ now behave as in
Fig.~\ref{posp}.

Note that~\begin{equation}\label{} g^{rr} = -
g_{\varphi\varphi}+4\alpha^2 r^2,
\end{equation}
and therefore horizons may only exist in the normal region of
positive $g_{\varphi\varphi}$. Note, however, that for horizons to
exist we must require
\begin{equation}
2G\nu^2 \ge \frac{J(1+\alpha^2l^2)}{\alpha l^2} . \label{reqhor}
\end{equation}
If this requirement is fulfilled, we get two horizons inside the
normal region, $r_-=mj/(2\alpha)$ and $r_+$, which coincide in the
extremal case. The whole normal region is in fact an ergoregion
because $\partial/\partial t$ is spacelike everywhere. Again, for
given $(\nu,J)$, one can then find a unique solution $(m,mj)$
satisfying \eqref{const5}-\eqref{const6}.

Following Carter \cite{Carter1973}, the metric and the gauge field
can be made regular at both horizons by a combined coordinate and
gauge transformation. Indeed, if
$$\Delta(r)=(\alpha^2l^2+1){2r^2\over l^2} - 8G\nu r +
\frac{4GJ}{\alpha},$$ the black hole metric can be written as
\begin{eqnarray}
  ds^2 = (dt - 2\alpha r d\varphi)^2 -\Delta d\varphi^2 +
\frac{dr^2}{\Delta}.  \label{blackhole1}
\end{eqnarray}
The analog of ingoing Eddington-Finkelstein coordinates are the
angle $\varphi^{\hspace{-6pt}\leftharpoonup}$ and the time
${t}^{\hspace{-6pt}\leftharpoonup}$ defined by
$d\varphi=d\varphi^{\hspace{-6pt}\leftharpoonup}-\frac{1}{\Delta}dr$,
$dt=dt^{\hspace{-6pt}\leftharpoonup}-\frac{2\alpha r}{\Delta}dr$,
giving the regular metric
\begin{eqnarray}
  ds^2 = (dt^{\hspace{-6pt}\leftharpoonup} - 2\alpha r
  d\varphi^{\hspace{-6pt}\leftharpoonup})^2
  -\Delta d{\varphi^{\hspace{-6pt}\leftharpoonup}}^2
  +2d\varphi^{\hspace{-6pt}\leftharpoonup} dr.  \label{blackhole2}
\end{eqnarray}
With $A_\varphi(r)$ given by \eqref{aphi}, the $r$ dependent gauge
transformation $A^{\hspace{-6pt}\leftharpoonup}=A+d\epsilon$,
where $\epsilon=\int dr \frac{A_\varphi(r)}{\Delta}$ gives the
regular potential $A^{\hspace{-6pt}\leftharpoonup}=A_\varphi(r)d
\varphi^{\hspace{-6pt}\leftharpoonup}$ whose norm
${A^{\hspace{-6pt}\leftharpoonup}}^2$ is zero.

Outgoing Eddington-Finkelstein coordinates are defined by
$d\varphi=-d\varphi^{\hspace{-6pt}\rightharpoonup}+\frac{1}{\Delta}dr$,
$dt=-dt^{\hspace{-6pt}\rightharpoonup}+\frac{2\alpha
r}{\Delta}dr$. The metric then takes also the
form~\eqref{blackhole2} with $t^{\hspace{-6pt}\leftharpoonup}$ and
$\varphi^{\hspace{-6pt}\leftharpoonup}$ replaced by
$-t^{\hspace{-6pt}\rightharpoonup}$ and
$-\varphi^{\hspace{-6pt}\rightharpoonup}$ and the potential can be
regularized by $A^{\hspace{-6pt}\rightharpoonup}=A-d\epsilon$.

The null generators of the horizons are $\frac{\partial}{\partial
t}+\frac{1}{2\alpha r_\pm}\frac{\partial}{\partial \varphi}$. The
associated ignorable coordinates which are constant on these null
generators are then given by
\begin{equation}
dt^{\pm} = dt - 2\alpha r_{\pm} d\varphi.
\end{equation}
Kruskal type coordinates $(t^{\pm},U^\pm,V^\pm)$ are obtained by
defining
\begin{eqnarray}
k_\pm \frac{dV^{\pm}}{V^\pm}
&= & d\varphi^{\hspace{-6pt}\leftharpoonup} = d\varphi + \frac{dr}{\Delta},\\
k_\pm \frac{dU^{\pm}}{U^\pm}& = &
d\varphi^{\hspace{-6pt}\rightharpoonup} = -d\varphi +
\frac{dr}{\Delta },
\end{eqnarray}
where
$$k_\pm = \frac{l^2}{1+\alpha^2l^2}\frac{1}{r_\pm - r_\mp}.$$
In these coordinates, the metric is manifestly regular at the
bifurcation surfaces,
\begin{eqnarray}
ds^2 &= &\big[dt^\pm-\alpha k_\pm (r-r_\mp) (U^\pm dV^\pm-V^\pm
dU^\pm)\big]^2\nonumber\\
&+&\frac{2k_\pm (r-r_\mp)^2}{r_\pm-r_\mp}dU^\pm dV^\pm,
\end{eqnarray}
with $r$ given implicitly by
\begin{equation}
U^\pm V^\pm = \left( \frac{r-r_+}{r-r_-}\right)^{\pm 1}.
\end{equation}

In Kruskal coordinates, the gauge field \eqref{aphi} becomes
\begin{eqnarray}
  A&=&\frac{k_\pm}{2}
  \left(\frac{A_\varphi(r_\pm)}{U^\pm
  V^\pm}+\frac{\sqrt{1-\alpha^2l^2}}{l}(r-r_\mp) \right)\nonumber\\
  && \times (U^\pm dV^{\pm} -V^\pm dU^\pm).  \label{eq:10}
\end{eqnarray}
The potential can be regularized at $r=r_\pm$ by the
transformations
\begin{eqnarray}
\hspace*{-8pt}  \tilde
A^\pm&=&A-d[A_\varphi(r_\pm)\frac{k_\pm}{2}\ln{
  \frac{V^\pm}{U^\pm}}]\nonumber\\
&=& \frac{k_\pm\sqrt{1-\alpha^2l^2}}{2l}(r-r_\mp)(U^\pm
dV^{\pm}-V^\pm dU^{\pm}).
\end{eqnarray}
In the original coordinates, however, the parameters of these
transformations explicitly involve the angle $\varphi$, $\tilde
A^\pm=A-d[A_\varphi(r_\pm)\varphi]$ and, as explicitly shown
below, they change the electric charge. In order to avoid this,
one can add a constant piece proportional to $dt^\pm$, so that
\begin{eqnarray}
  A^\pm&=& \tilde A^\pm - d(\frac{A_\varphi(r_\pm)}{2\alpha
r_\pm}t^\pm).
\end{eqnarray}
In the original coordinates, the gauge parameter is now a linear
function of $t$ alone,
\begin{eqnarray}
A^\pm &=& A - d(\frac{A_\varphi(r_\pm)}{2\alpha
r_\pm}t).\label{time_gauge}
\end{eqnarray}
According to the definition given below, such a transformation
does not change the charges.

In the published paper~\cite{Banados:2005da}, a naive
Carter-Penrose diagram for these black holes was drawn. This
diagram, however, is premature in view of the two following issues
that have still to be addressed: namely, the clarification of the
global topology of these spacetimes, and the existence of a
conformal completion. These considerations are left for further
work.

%

\subsection{Vacuum solutions $\alpha^2l^2=1$}

In the case $\alpha^2l^2=1$ the gauge field vanishes and the
G\"odel metric (\ref{Godel2}) reduces to the three-dimensional
anti-de Sitter space (to see this, do the coordinate
transformations $\varphi \rightarrow \varphi+ \alpha t$ and $2r
\rightarrow r^2$).  This means that the identifications in this
case yield the usual three-dimensional black holes, and conical
singularities.

\subsection{The general solution}
\label{section6}

\subsubsection{Reduced equations of motion}

We have seen in previous sections that the G\"odel metrics
(\ref{Godel}) and (\ref{Godel3}), as well as the corresponding
quotient spaces describing particles and black holes, can be
regarded as exact solutions to the action (\ref{action}).

We have distinguished three cases according to the values of the
dimensionless quantity $\alpha^2l^2$. Our purpose in this section
is to write a general solution which will be valid for all values
of $\alpha^2 l^2$. We shall now construct the solution by looking
directly at the equations of motion.  It is useful to write a
general spherically symmetric static ansatz in the form
\cite{Clement:1993kc,Andrade:2005ur}
\begin{eqnarray}
ds^2 &=& \frac{dr^2}{h^2 - p\, q} + p\, dt^2 + 2h\, dt d\varphi +
q\, d\varphi^2,\label{ds1}
\end{eqnarray}
where $p,q,h$ are functions of $r$ only. This ansatz can also be
written in the ``ADM form",
\begin{equation}\label{ADM}
ds^2   = - \frac{h^2 - pq}{q} dt^2 + \frac{dr^2}{h^2-pq} + q
\left( d\varphi + \frac{h}{q} dt\right)^2.
\end{equation}
This confirms that the function $g^{rr}$
\begin{equation}\label{}
f(r) = h^2(r) - p(r)q(r),
\end{equation}
controls the existence of horizons. Note that for all $p,q,h$, the
determinant of this metric is $\det(-g)=1.$ For the gauge field,
we use the radial gauge $A_r=0$, and assume that $A_t$ and
$A_\varphi$ depend only on the radial coordinate,
\begin{equation}
A = A_t(r)\, dt + A_\varphi(r)\, d\varphi.
\end{equation}
In this parametrization, Einstein's equations take the remarkably
simple form,
\begin{eqnarray}
 h'' &=& -A_t'\, A_\varphi' \nonumber \\
 p'' &=& -A_t'^{\, 2} \nonumber\\
 q'' &=& -A_\varphi'^{\, 2} \label{3}\\
 (h^2-pq)''&=&h'^2-p'q' + \frac{4}{l^2},\nonumber
\end{eqnarray}
where primes denote radial derivatives. Maxwell's equations reduce
to
\begin{eqnarray}
 (h A_t' - p A_\varphi' - 2\alpha A_t)'=0, \nonumber\\
 (q A_t'-h A_\varphi' - 2\alpha A_\varphi)'=0 . \label{5}
\end{eqnarray}

Before we write the solution to these equations, we make some
general remarks on the structure of the stress-energy tensor
associated to topologically massive electrodynamics. As we pointed
out in the introduction, we will seek for solutions with a
constant electromagnetic field ${}^*\!F$. Hence, we will only
consider potentials $A$ which are linear in $r$. In this case,
Eqs.~(\ref{5}) are
\begin{eqnarray}
 h' A_t' - p' A_\varphi' &=& 2\alpha A_t', \nonumber\\
 q' A_t'-h' A_\varphi' &=& 2\alpha A'_\varphi . \label{dd}
\end{eqnarray}
We now multiply the first by $h'$ and the second by  $p'$, then we
subtract them to obtain
$$
(h'^2-p'q') A_t' = 2\alpha (h'A_t' - p'A'_\varphi) = 4\alpha^2
A_t'.
$$
In the last step we have used Eq.~(\ref{dd}). This implies that,
if $A_t'\neq 0$ then $(h'^2-p'q')=4\alpha^2$. By properly
manipulating Eqs.~(\ref{dd}) we see that this result is also valid
if $A_t'=0$ but $A_\varphi'\neq 0$, and therefore is it true as
long as the electromagnetic field does not vanish.  Now we insert
this in the last equation in (\ref{3}), and obtain,
\begin{eqnarray}
{}^*\!F^\mu {}^*\!F_\mu &=& q(A_t')^2 + p(A_\varphi')^2 -2hA_t'\,
A_\varphi' \nonumber \\ &=& {4\over l^2}\left( 1 -\alpha^2l^2
\right). \label{f2}
\end{eqnarray}

This equation tells us that when the topological mass $\alpha^2$
is greater (smaller) than the negative cosmological constant
$1/l^2$, the theory only supports timelike (spacelike) constants
fields. Therefore, for the generalized G\"odel spacetimes
(\ref{Godel}), we will need a topological mass $\alpha^2>1/l^2$.
In the other region, the constant electromagnetic field will
describe a tachyonic perfect fluid. Anyway, as we will see below,
it is this region in which black hole solutions are going to
exist.

\subsubsection{The solution}

By direct computation one can check that equations
(\ref{3})-(\ref{5}) are satisfied by the field
\begin{eqnarray}
p(r)    &=& 8G\mu   \nonumber\\
q(r)   &=& -\frac{4G J}{\alpha}+2r - 2\frac{\gamma^2}{l^2} r^2  \nonumber\\
h(r)   &=& - 2\alpha r   \label{Sol2}\\
A_{t}(r)   &=&  \frac{\alpha^2l^2-1}{\gamma \alpha l}
+ \zeta \nonumber\\
A_{ \varphi}(r)   &=& -\frac{4G}{\alpha} Q + 2\frac{\gamma}{l} r,
\nonumber
\end{eqnarray}
where
\begin{equation}
\gamma = \sqrt{\frac{1-\alpha^2 l^2}{8G\mu}}.
\end{equation}
The parameters $\mu$, $J$ and $Q$ are integration constants with a
physical interpretation as they will be identified with mass,
angular momentum and electric charge below. The arbitrary constant
$\zeta$ on the other hand will be shown to be pure gauge. For
later convenience, it is however useful to keep it along and not
restrict ourselves to a particular gauge at this stage.  This will
be discussed in details Sec.~\ref{conschar}.

In the sector $\alpha^2 l^2 > 1$, the solution is real only for
$\mu$ negative. These are the G\"odel particles, i.e., the conical
singularities, discussed in Sec.~\ref{PARTICLES}. The
metric~\eqref{particles} is recovered when $\mu=-2G\nu^2$ and the
change of variables $t \rightarrow t/\sqrt{-8G\mu}$, $r
\rightarrow \sqrt{-8G\mu}\, r$ is performed. For the special
values $\mu = -1/8G$ and $J=0$,  which correspond to the trivial
identification $j=0$, $m=1$ in Sec.~\ref{PARTICLES}, the conical
singularities disappear and we are left with the G\"odel
universes~\eqref{Godel2}, used for the identifications producing
the cosmons.

When $\alpha^2 l^2 < 1$, $\mu$ has to be positive. The black hole
metrics \eqref{blackhole} of Sec.~\ref{BH} are recovered when
$\mu=2G\nu^2$ and $t \rightarrow t/\sqrt{8G\mu}$, $r \rightarrow
\sqrt{8G\mu}\, r$. For $\mu=1/8G$ and $J=0$, they reduce to the
solution \eqref{Godel3} from which the black holes have been
obtained from non-trivial identifications.

By construction, the electromagnetic stress-energy tensor for the
solutions~\eqref{Sol2} takes the form
\begin{eqnarray}
 \label{eq:TEM}
8\pi G T_{EM}^{\mu\nu}&=& (\alpha^2-\frac{1}{l^2}) g^{\mu\nu}+
8\pi G{\mathcal
 T}^{\mu\nu},\\ {\mathcal
 T}^{\mu\nu}&=&\frac{|1-\alpha^2 l^2|}{4\pi G l^2}u^\mu u^\nu,
\label{eq:fluid}
\end{eqnarray}
where the unit tangent vector of the fluid is $u =
\frac{1}{\sqrt{8G|\mu|}}\frac{\partial}{\partial t}$. For
$\alpha^2l^2\neq 1$, the effect of the electromagnetic field can
be taken into account by replacing the original cosmological
constant $-\frac{1}{l^2}$ by the effective cosmological constant
$-\alpha^2$ and introducing a pressure-free perfect, ordinary or
tachyonic, fluid with energy density $\frac{|1-\alpha^2l^2|}{4\pi
G l^2}$. {}From this point of view, \emph{the Chern-Simons
coupling transmutes into a cosmological constant}. For
$1-\alpha^2l^2 < 0$, the fluid flows along timelike curves while
for $1-\alpha^2l^2 > 0$, the fluid is tachyonic.

When $\alpha^2l^2=1$, the fluid disappears, the stress-energy
tensor vanishes and the solution is real for $\mu \in \mathbb R$.
The metric ~\eqref{Sol2} reduces to the BTZ metric
\cite{Banados:1992wn}, as can be explicitly seen by transforming
to the standard frame that is non-rotating at infinity with
respect to anti-de Sitter space,
\begin{equation}
\varphi \rightarrow \varphi + \alpha t, \qquad r \rightarrow
\frac{r^2}{2}+\frac{2GJ}{\alpha}.\label{frame}
\end{equation}
As will be explained in more details below, in the rotating frame
that we have used, the energy and angular momentum are $\mu$ and
$J$ respectively, while they become $M\equiv \mu-\alpha J$ and $J$
in the standard non-rotating frame.

Regular black holes have the range (see Fig.~\ref{diag2})
 \begin{equation}
 \mu \geq 0,\qquad \mu \geq 2\alpha J.
 \end{equation}
Note that the solution still possess a topological charge $Q$. It
has been discussed in more details in \cite{Andrade:2005ur}.

When $\alpha^2l^2 \neq 1$, the limit $\mu \rightarrow 0$ can be
taken smoothly in the coordinates $\hat r =\gamma r$, $\hat t =
t/\gamma$ in which the solution becomes
\begin{eqnarray}
p(\hat r)    &=& 1-\alpha^2l^2   \nonumber\\
q(\hat r)   &=& -\frac{4G J}{\alpha}+\frac{2}{\gamma}\hat r -
\frac{2}{l^2} \hat r^2  \nonumber\\
h(\hat r)   &=& - 2\alpha \hat r   \label{Sol3}\\
A_{\hat t}(\hat r)   &=&  \frac{\alpha^2l^2-1}{\alpha l}
+ \hat \zeta \nonumber\\
A_{ \varphi}(\hat r)   &=& -\frac{4G}{\alpha} Q + \frac{2}{l} \hat
r, \nonumber
\end{eqnarray}
where $\hat \zeta = \gamma \zeta$.

\begin{figure}
\begin{center}
  \includegraphics[width=6cm]{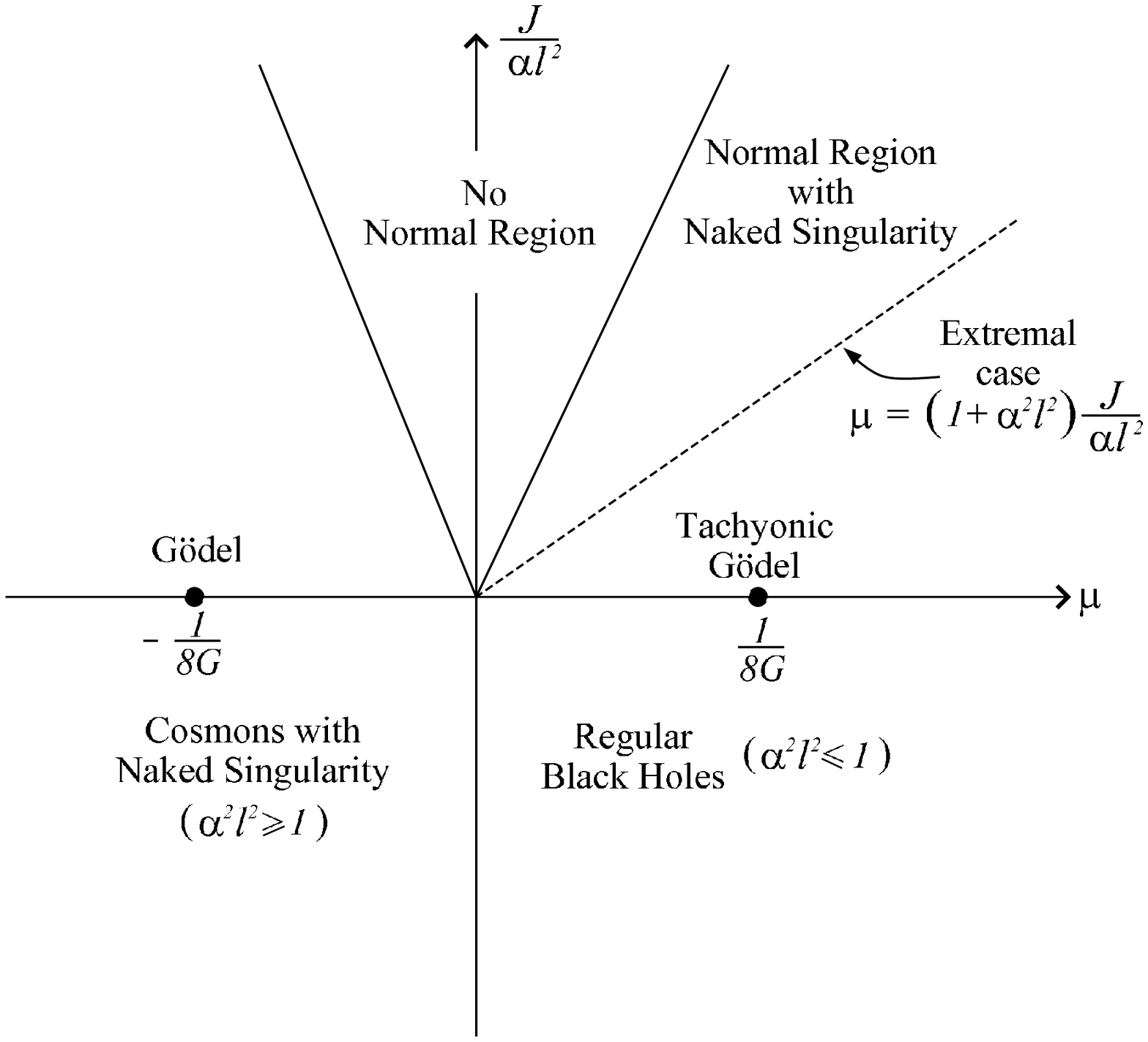}
\caption{\label{diag1} Sectors of the general solution.}
\end{center}
\end{figure}

\begin{figure}
\begin{center}
  \includegraphics[width=6cm]{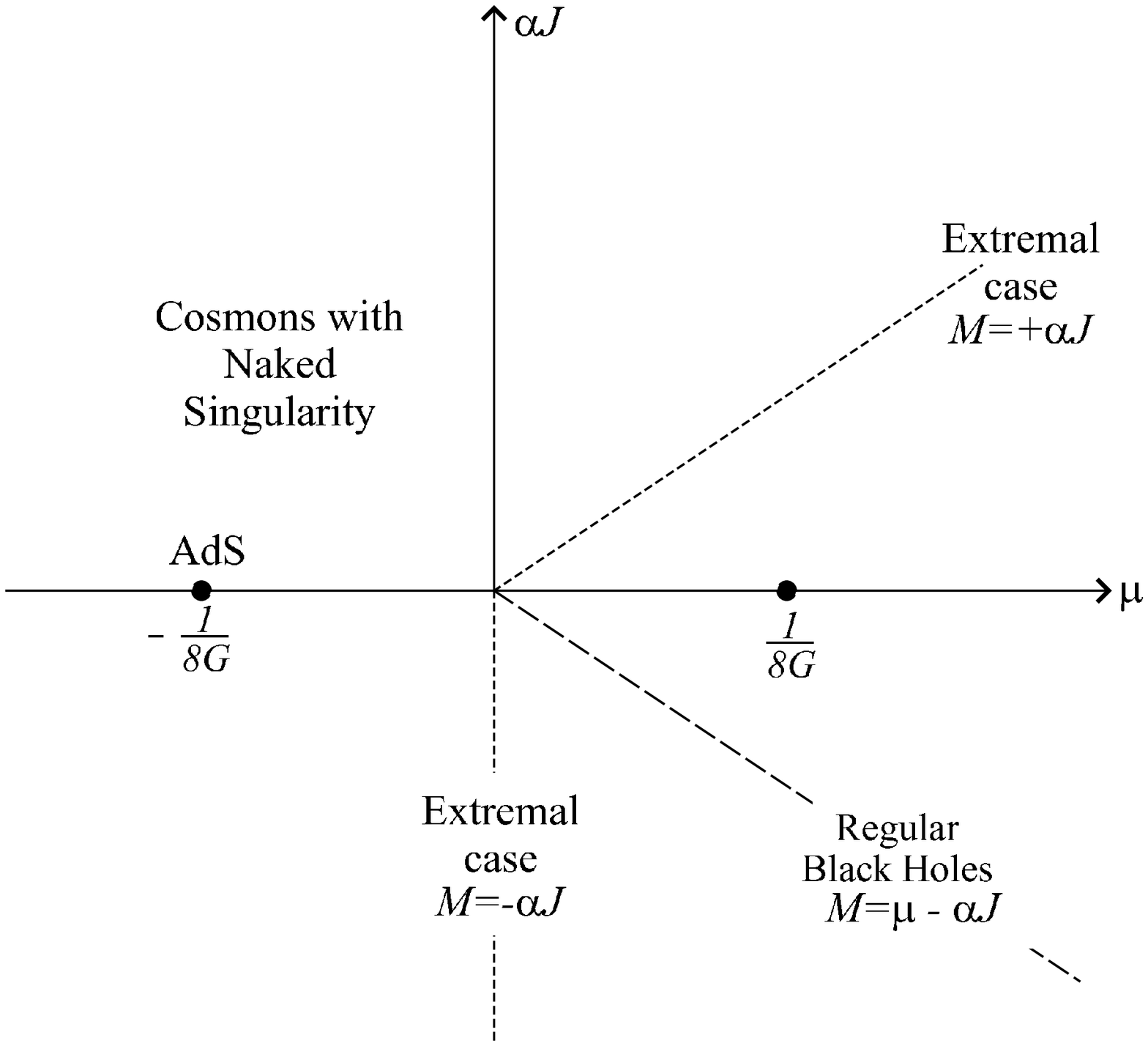}
\caption{\label{diag2} Sectors of the $\alpha^2l^2 = 1$ solution.
The BTZ mass axis $M=\mu-\alpha J$ and the extremal solutions are
explicitly indicated.}
\end{center}
\end{figure}

\subsection{Conserved charges}
\label{conschar}

\subsubsection{Angular momentum, electric charge and energies}

The charge differences between a given solution $(g_{\mu\nu},
A_\mu)$ and an infinitesimally close one $(g_{\mu\nu}+\delta
g_{\mu\nu}, A_\mu+\delta A_\mu)$ were computed in
section~\ref{sec:EM-chern} of Chapter~\ref{chap:matter}.

Particularizing to three dimensions and contracting the vertical
one-forms $(\dv g, \dv \xi, \dv A,\dv \epsilon)$ with variations
$(\delta g, \delta \xi, \delta A,\delta \epsilon)$ satisfying the
reducibility equations
\begin{eqnarray}
\left\{\begin{array}{c} \label{eq:SUG22}
  \cL_\xi g_{\mu\nu} = 0,\\ \cL_\xi A_\mu + \partial_\mu
  \epsilon = 0,
\end{array}\right.
\end{eqnarray}
the (1,1)-forms $k_{\xi,\epsilon}$ can be simplified as
\begin{eqnarray}\label{kCS2}
k_{\xi,\epsilon}[\delta g,\delta A] = k^{g}_{\xi} + k^{A}_{
 \xi,\epsilon} +  k^{CS}_{\xi,\epsilon},
\end{eqnarray}
with\footnote{The minus sign of in front of $i_\xi I^n_{\delta
g}\cL^{EH}$ as compared to~\eqref{k_diff} comes from the fact that
$\delta g$ is Grassmann even ($[\delta g,i_\xi] = 0$) while $\dv
g$ is Grassmann odd ($\{\dv g,i_\xi\} = 0$).}
\begin{eqnarray}
k^{g}_{ \xi} &=& -\delta k^K_{\cL^{EH},\xi} + k^K_{\cL^{EH},\delta
\xi} -i_\xi I^n_{\delta g}\cL^{EH},\label{eq:6grav}
\end{eqnarray}
where
\begin{eqnarray}
k^K_{\cL^{EH},\xi}&=& dx^\rho \frac{\sqrt{-g}}{16\pi G }
\epsilon_{\rho\mu\nu} D^\mu\xi^\nu,\label{eq:7}
\end{eqnarray}
is the Komar 1-form and
\begin{eqnarray}
i_\xi I^n_{\delta g}\cL^{EH}&=&dx^\rho \frac{\sqrt{-
 g}}{16\pi G} \epsilon_{\rho\nu\mu}
\xi^{\mu} (g^{\nu\alpha}D^\beta \delta g_{\alpha\beta}-
g^{\alpha\beta}D^\nu \delta g_{\alpha\beta} ).
\label{eq:Theta}\nonumber
\end{eqnarray}
The electromagnetic contribution is\footnote{The same remark as
the preceding footnote applies to the term $i_\xi \Theta^A$ as
compared to \eqref{Bcharge}.}
\begin{eqnarray}
 k^{A}_{ \xi,\epsilon} = -\delta Q^{A}_{
   \xi,\epsilon}+Q^{A}_{\delta \xi,\delta\epsilon} + i_\xi \Theta^{A},
\label{eq:6em}
\end{eqnarray}
where
\begin{eqnarray}
 Q^{A}_{\xi,\epsilon} &=&
 dx^\rho \epsilon_{\rho\mu\nu} \frac{\sqrt{-g}}{32\pi G} \left(
   F^{\mu\nu}(\xi^\rho
   A_\rho +  \epsilon)\right),\label{eq:8bis}\\
i_\xi \Theta^{A}&=&dx^\rho \epsilon_{\rho\mu\nu}\frac{\sqrt{-
     g}}{16\pi G} \xi^\nu  F^{\mu\alpha} \delta A_\alpha .\label{eq:Thetaem}
\end{eqnarray}
The Chern-Simons term contributes as
\begin{equation}
k^{CS}_{ \xi,\epsilon} = dx^\rho \alpha \frac{\sqrt{-g}} {8\pi G}
\delta A_\rho (A_\sigma \xi^\sigma+\epsilon). \label{eq:6CS}
\end{equation}

For generic metrics and gauge fields of the form (\ref{Sol2}), the
general solution $(\xi,\epsilon)$ of (\ref{eq:SUG22}) is a linear
combination of $(0,-1)$, $(-\frac{\partial}{\partial \varphi},0)$
and $(\frac{\partial}{\partial t},0)$.  These basis elements are
associated to infinitesimal charges as follows,
\begin{eqnarray}
 \label{eq:BH1}
 \oint_S k_{ 0,-1}&=&\delta Q,\
\oint_S k_{-\frac{\partial}{\partial \varphi },0 }=\delta
(J-\frac{2G}{\alpha}Q^2),\ \nonumber\\\oint_S
k_{\frac{\partial}{\partial t},0 }&=&\delta \mu -\zeta\delta Q,
\end{eqnarray}
where the contribution proportional to $\delta Q$ in $\oint_S
k_{-\frac{\partial}{\partial \varphi },0 }$ and $\oint_S
k_{\frac{\partial}{\partial t},0 }$ originate from the
Chern-Simons term through \eqref{eq:6CS}.  The conserved charges
associated with $(0,-1)$, $(-\frac{\partial}{\partial \varphi},0)$
are thus manifestly integrable.  We choose to associate the
angular momentum to $(-\frac{\partial}{\partial
\varphi},-\frac{4GQ}{\alpha})$ so that its value be algebraically
independent of $Q$. If one takes as basis element
$(\frac{\partial}{\partial t},-\zeta)$ instead of
$(\frac{\partial}{\partial t},0)$, one gets a third integrable
conserved charge equal to $\delta \mu$.

The integrated charges computed with respect to the background
$\mu=0=J=Q$ and associated to $(\frac{\partial}{\partial
t},-\zeta)$, $(-\frac{\partial}{\partial
  \varphi},-\frac{4GQ}{\alpha})$ and $(0,-1)$ are the mass, the
angular momentum and the total electric charge respectively,
\begin{eqnarray}
{\mathcal E}=\mu,\ \ {\mathcal
 J}=J,\ {\ \mathcal Q}=Q.
\end{eqnarray}
Note that even though the metric and gauge fields in \eqref{Sol2}
become singular at the background $\mu=0=J=Q$, we can see from the
form \eqref{Sol3}  that this is just a coordinate singularity.

The parameter $\zeta$ is pure gauge because the variation $\delta
\zeta$ is not present in the infinitesimal charges~\eqref{eq:BH1}.
Note however that $\zeta$ appears explicitly in the definition of
the mass by associating it with the basis element
$(\frac{\partial}{\partial
  t},-\zeta)$. It is only in the gauge $\zeta= 0$, that the mass is
associated with the time-like Killing vector
$(\frac{\partial}{\partial t},0)$. This definition ensures in
particular that the mass of the black hole does not depend on the
gauge transformations (\ref{time_gauge}) needed to regularize the
potential on the bifurcation surfaces.

In order to compare with standard adS black holes, one has to
compute the mass in the frame \eqref{frame} instead of using the
rest frame for the fluid. The conserved charge $\mathcal E^\prime$
associated with $(\partial/\partial t-\alpha
\partial /\partial \varphi,-\zeta+4GQ)$ is now given by
\begin{equation}
\mathcal E^\prime = \mathcal E-\alpha \mathcal J = \mu - \alpha
J=M,
\end{equation}
which coincides with the conventional definition of the mass for
the BTZ black holes.

\subsubsection{Horizon and first law - General derivation}
\label{sec:geometry-2}

\label{sec:general-formulas}

When it exists, the outer horizon $H$ is located at $r_+$, the
largest positive root of $f(r)$. In the following, a subscript $+$
on a function means that it is evaluated at $r_+$. The generator
of the horizon is given by $\xi = \frac{\partial}{\partial t} +
\Omega \frac{\partial}{\partial \varphi}$, where the angular
velocity $\Omega$ of the horizon has the value
 \begin{eqnarray}
\Omega=-\varepsilon_{h_+}\varepsilon_{q_+}\sqrt{\frac{p_+}{q_+}}
=-\frac{h_+}{q_+}, \label{eq:9}
\end{eqnarray}
where $\varepsilon_{h_+}$ denotes the sign of $h_+$. The first law
can be derived by starting from
\begin{eqnarray}
\delta{\mathcal E}&=&\oint_{S} k_{\frac{\partial}{\partial
 t},- \zeta} \nonumber\\&=& \oint_{S}k_{ \xi,0}+
 \Omega \oint_{S} k_{-\frac{\partial}{\partial
 \varphi},-\frac{4GQ}{\alpha}}+\oint_{S}k_{- \zeta+\frac{4GQ}{\alpha}\Omega,0}\nonumber\\
&=&\oint_{H}k_{ \xi,0}+  \Omega \delta \mathcal{J} + ( \zeta
-\frac{4GQ}{\alpha}\Omega )\delta {\mathcal Q}.
\end{eqnarray}
The first term on the right-hand side was computed in
section~\ref{sec:firstlaw} of Chapter~\ref{chap:geome} with as
final result
\begin{equation}
\delta \mathcal{E} = \frac{\kappa}{8\pi G} \delta {\mathcal A} +
\Omega
 \delta \mathcal{J}+  \Phi_H^{tot} \delta Q,\label{firstL}
\end{equation}
where the total electric potential is given by
\begin{eqnarray}
 \Phi_H^{tot} =  \Phi_H +  \zeta
-\frac{4GQ}{\alpha}\Omega,\qquad
 \Phi_H = - (i_\xi  A)_+.
\end{eqnarray}
The surface gravity is given by
\begin{eqnarray}
 \kappa=\left.\sqrt{|-\frac{1}{2}(D^\mu \xi^\nu)
     (D_\mu \xi_\nu)|}\right|_H = \frac{|f^\prime_+|}{2\sqrt{|
     q_+|}}, \label{eq:BH11}
\end{eqnarray}
and the proper area by
\begin{eqnarray}
\ {\cal A}=2 \pi \sqrt{|q_+|}.\label{s}
\end{eqnarray}
Note that the choice of signs in the definition of electric charge
and angular momentum were made so that the first laws appear in
the conventional form~\eqref{firstL}.

\subsubsection{Horizon and first law - Explicit values and discussion}

\label{sec:numerical-values}

We have
\begin{eqnarray}
f(r)= 2\frac{(1+\alpha^2 l^2)}{l^2}r^2-16G\mu\left(r -
\frac{2GJ}{\alpha}\right)
\end{eqnarray}
so that
\begin{eqnarray}
 r_+=\frac{4l^2G\mu}{1+\alpha^2l^2}\Big[ 1+\sqrt{1-
\frac{J(1+\alpha^2l^2)}{\alpha l^2\mu}}\Big]\label{rplus}
\end{eqnarray}
In order to explicitly verify the first law \eqref{firstL}, we
start by showing that $\Phi^{tot} = 0$. We need to verify that
\begin{eqnarray}
 -A_{ t}( r_+)-
 \Omega A_{ \varphi}( r_+)+ \zeta - \Omega\frac{4GQ}{\alpha}=0.
\end{eqnarray}
Using the explicit expressions for the components of $A$, this
equation reduces to
\begin{eqnarray}
  \label{eq:15}
\Omega=\frac{4G\mu}{\alpha r_+}.
\end{eqnarray}
Taking into account $ \Omega=-{h_+}/{q_+}$ together with
$q_+=h^2_+/p_+$, this equality can then easily be checked using
$h_+=-2\alpha r_+$, $p_+=8G\mu$, implying
$q_+=\alpha^2r_+^2/(2G\mu)$. Since $f^\prime_+=4(1+\alpha^2
l^2)r_+/l^2-16G\mu$, the first law reduces to
\begin{eqnarray}
  \label{eq:16a}
  \delta\mu-\frac{4G\mu}{\alpha r_+}\delta J=
[\frac{\alpha^2l^2+1}{4Gl^2}r_+-\mu][\frac{2\delta
r_+}{r_+}-\frac{\delta\mu}{\mu}],
\end{eqnarray}
which can be explicitly checked using \eqref{rplus}.

In particular, the first law~\eqref{firstL} can be evaluated in
the gauge where the potential is regular on the horizon $r_+$.
Because the two forms (\ref{blackhole}) and (\ref{Sol2}) of the
black hole solution are related by the change of coordinates $t
\rightarrow t\sqrt{-8G\mu}$, $r \rightarrow r/\sqrt{-8G\mu}$, the
gauge~\eqref{time_gauge} now corresponds to
\begin{equation}
A_t =A_t^+= -\Omega A_\varphi^+. \label{value_zeta}
\end{equation}
This amounts to the choice $ \zeta = \frac{4G Q}{\alpha}\Omega$ in
\eqref{Sol2}. It follows that $\Phi^{tot}=\Phi=0$ and that the
vector associated to $A$ is proportional to $\xi$ on the horizon.

The first law adapted to the energy $\mathcal E^\prime=\mathcal
E-\alpha \mathcal J$ is obtained by changing $\Omega$ to
$\Omega^\prime=\Omega-\alpha$ in \eqref{firstL}. This form of the
first law reduces to the standard form for 3 dimensional adS black
holes (with or without topological charge) when $\alpha=\pm 1/l$.

Finally, we note that the first law \eqref{firstL} applies both to
the outer event horizon of a black hole in the normal region and
to the horizon at $r_0$ of a cosmon, when time is identified with
real period $2\pi/|\kappa|$.

\section{Kerr-anti-de Sitter black holes}
\label{sec:KerrAdS}

The general Kerr-anti-de Sitter metrics in arbitrary spacetime
dimensions $n\geq 4$ were found recently in~\cite{Gibbons:2004uw},
generalizing the results of Myers and Perry~\cite{Myers:1986un} to
non-vanishing cosmological constant.

As has been emphasized in~\cite{Gibbons:2004ai}, not even in four
dimensions do all authors obtain the same expression for the
energy of Kerr-adS black holes. Much worse, some of these
expressions are in disagreement with the first law.
In~\cite{Gibbons:2004ai}, Gibbons et al.~computed the energy of
such black holes indirectly by integrating the first law. In
\cite{Deruelle:2004mv}, the mass and energy have been computed
directly by using the BKL superpotentials~\cite{Katz:1996nr}. In a
completed version of their paper, Gibbons et al.~then have also
computed the energy directly by using the Ashtekar-Magnon-Das
definition~\cite{Ashtekar:1984,Ashtekar:1999jx}.

In this section, published in~\cite{Barnich:2004uw}, we compute
the conserved charges - mass and angular momenta - for the
Kerr-adS black holes by using the surface integrals developed in
the preceding chapters and we find agreement with the results of
\cite{Gibbons:2004ai,Deruelle:2004mv}.  We also show explicitly
that, in this case, the surface integrals integrated along a path
of solutions reduce to the standard
Lagrangian~\cite{Abbott:1982ff} or
Hamiltonian~\cite{Henneaux:1985tv,Henneaux:1985ey} surface
integrals at infinity.

Finally, we give a detailed and geometric derivation of the
generalized Smarr relation for the higher dimensional Kerr-adS
black holes, in the continuation of section~\ref{sec:Smarr} of
Chapter~\ref{chap:geome}. The derivation can also be applied
straightforwardly to asymptotically flat black holes in the limit
of vanishing cosmological constant.

\subsection[Description of the solutions]{Description of the
  solutions}
\label{sec:descr-solut}

The general Kerr anti-de Sitter metrics in $n = 2 N +1 + \eps$
dimensions\footnote{In this section we shall use the notations of
  \cite{Gibbons:2004ai} except the spacetime dimension denoted by $n$
  and the indices $a,b$, which run from 1 to $N$, while $i,j$
  run from 1 to $N+\eps$. When $\eps= 1$, $a_{N+\eps} \equiv
  0$. We also use $G = 1$} where $\eps \equiv n-1 \text{ mod } 2$ were obtained in
\cite{Gibbons:2004uw,Gibbons:2004js}.  They have $N$ independent
rotation parameters $a_a$ in $N$ orthogonal 2-planes.  Gibbons et
al.~start from the $n$ dimensional anti-de Sitter metric in static
coordinates,
\begin{equation}
\bar{ds}^2 = -(1+y^2l^{-2})dt^2 +
\frac{dy^2}{1+y^2l^{-2}}+y^2\sum_{a=1}^N \hat \mu^2_a d\phi_a^2
+y^2\sum_{i=1}^{N+\eps} d\hat \mu^2_i,\label{adsy}
\end{equation}
with $\sum_{i=1}^{N+\eps}\hat \mu_i^2 = 1$. They then consider the
change of variables to Boyer-Linquist spheroidal coordinates
$(\tau,r,\varphi_a,\mu_i)$. These coordinates depend on $N$
arbitrary parameters $a_a$ and are defined by
\begin{equation}
y^2 \hat \mu_i^2 = \frac{(r^2+a_i^2)}{\Xi_i}\mu_i^2,\qquad
\varphi_a = \phi_a,\qquad \tau = t.\label{coordtra}
\end{equation}
Note that for later convenience, we have renamed the variables
$t,\phi^a$ as $\tau,\varphi^a$ already at this stage. The anti-de
Sitter metric then becomes
\begin{eqnarray}
\bar{ds}^2 &=& - W(1+r^2l^{-2}) d\tau^2 + \frac{U}{V}\, dr^2 +
\sum_{a=1}^N
\frac{r^2+a^2_a}{\Xi_a}\mu_a^2 d{\varphi}_a^2 \label{adstilde}\\
&& \nonumber + \sum_{i=1}^{N+\eps}\frac{r^2+a^2_i}{\Xi_i}d\mu_i^2
- \frac{l^{-2}}{W(1+r^2l^{-2})}\big( \sum_{i=1}^{N+\eps}
\frac{r^2+a_i^2}{\Xi_i}\mu_i d\mu_i \big)^2,
\end{eqnarray}
where
\begin{eqnarray}
W &\equiv& \sum_{i=1}^{N+\eps} \frac{\mu_i^2}{\Xi_i},\qquad U
\equiv r^\eps \sum_{i=1}^{N+\eps}
\frac{\mu_i^2}{r^2+a^2_i}\prod_{a=1}^N(r^2+a_a^2),\quad
\sum_{i=1}^{N+\eps} \mu_i^2 = 1,\\
V&\equiv& r^{\eps-2}(1+r^2 l^{-2})\prod_{a=1}^N (r^2+a^2_a),\qquad
\Xi_i \equiv 1-a_i^2l^{-2}.
\end{eqnarray}
In the coordinates $(\tau,r,\varphi^a,\mu^i)$, the Kerr-adS
solutions $g_{\mu\nu}$, depending on $N+1$ parameters $M,a_a$, are
related to the AdS metric $\bar g_{\mu\nu}$ as follows:
\begin{eqnarray}
{ds}^2 = \bar{ds}^2|_{\eqref{adstilde}} + \frac{2M}{U}\big( W\,
d\tau - \sum_{a=1}^N \frac{a_a \mu_a^2}{\Xi_a}\, d{\varphi}_a
\big)^2 + \frac{2M U}{V(V-2M)}\,dr^2.\label{KerrAdsmetric}
\end{eqnarray}
as can be directly verified by comparing with equation (4.2) of
\cite{Gibbons:2004ai}. In these coordinates, defining the metric
deviations $h_{\mu\nu}$ through
\begin{eqnarray}
ds^2= \bar
{ds}^2|_{\eqref{adstilde}}+h_{\mu\nu}dx^\mu\dx^\nu\label{eq:KerrAdS2}
\end{eqnarray}
and using $U =r^{n-3}+o(r^{n-3})$, $V = r^{n-1}l^{-2}+o(r^{n-1})$,
it is straightforward to see that
\begin{eqnarray}
h_{AB}\sim O(r^{-n+3}), \quad h_{rr} \sim
O(r^{-n-1}),\label{falloff}
\end{eqnarray}
with $A = (\tau,\varphi_a)$, while all other components of
$h_{\mu\nu}$ vanish.

The Killing vectors of the Kerr metric are given in coordinates
$(t,y,\phi_a,\hat\mu_i)$ and $(\tau,r,\varphi_a,\mu_i)$ by
\begin{equation}
k \equiv \Q{}{t} = \frac{\d}{\d \tau},\qquad m^a \equiv
\Q{}{\phi_a} = \Q{}{\varphi_a}. \label{eq:KerrAdSkill2}
\end{equation}

\subsection[Mass and angular momenta]{Mass and angular momenta}
\label{sec:mass-angular-momenta}

\paragraph{Surface charges}

The $(n-2,1)$-forms for general relativity were computed in
section~\ref{sec:general-relativity} of Chapter~\ref{chap:matter}.
For exact Killing vectors $\xi$ of the metric $g$ and for
variations $(\dv g_{\mu\nu},\dv \xi^\mu)$ contracted with $(\delta
g_{\mu\nu} \equiv h_{\mu\nu},0)$, one can simplify the
expression~(\ref{k_diff}) with \eqref{Komar}, \eqref{Theta_t} and
\eqref{suppl} to
\begin{eqnarray}
  \label{eq:wald2}
  k_{\xi}[\delta g ; g]=- \delta k^K_{\cL^{EH},\xi} - i_\xi \Theta[\delta g;g]
\end{eqnarray}
where $\Theta[\delta g ;g]=(d^{n-1}x)_\mu\frac{\sqrt{-g}}{16\pi
G}\Big( D_\sigma h^{\mu\sigma}-D^\mu h\Big)$.

The conserved charges for the family of
solutions~(\ref{KerrAdsmetric}) are then obtained as outlined in
section~\ref{sec:exact-solut-symm} of
Chapter~\ref{chap:general_th}. Let $g^{(s)}_{\mu\nu}$ with $s\in
[0,1]$ denote a one parameter family of solutions to Einstein's
equations interpolating between the anti-de Sitter background
$\bar g_{\mu\nu}=g^{(0)}_{\mu\nu}$ and the Kerr-adS solution
$g_{\mu\nu}=g^{(1)}_{\mu\nu}$. For $s\in [0,1]$, $g^s_{\mu\nu}$
can be obtained by replacing $M$ by $sM$ in (\ref{KerrAdsmetric}).
Let $\xi$ be a Killing vector field for this family\footnote{We
consider only vectors $\xi$ that do not vary along the path
$\xi^{(s)}\equiv \xi$.}, $\cL_{\xi}g^{(s)}_{\mu\nu}=0$, and
$h^{(s)}_{\mu\nu}=\frac{d}{ds}g^{(s)}_{\mu\nu}$ be the tangent
vector to $g^{(s)}_{\mu\nu}$ in solution space. The charge
associated with $\xi$ is then defined as
\begin{eqnarray}
  \label{eq:7bis}
 \cQ_{\xi}[g;\bar g] = \oint_S \int_0^1 ds\, k_{\xi}[h^{(s)};g^{(s)}],
\end{eqnarray}
and depend only on the homology class of $S$. We will check below
that the charge associated with the Killing
vectors~\eqref{eq:KerrAdSkill2} are integrable. Because the space
of parameters of the solutions~\eqref{KerrAdsmetric} has trivial
topology, it will imply that the charges~\eqref{eq:7bis} do not
depend on the particular path chosen. More explicitly, we have
\begin{eqnarray}
  \label{eq:3bis}
  Q_\xi[g;\bar g]= -\oint_S k^K_{\cL^{EH},\xi}[g]+\oint_S
  k^K_{\cL^{EH},\xi}[\bar g]+\oint_S\cC_{\xi;\gamma},\\
\cC_{\xi;\gamma} =- \int_0^1 ds\,
 i_\xi \Theta[h^{(s)};g^{(s)}] , \label{intcCbis}
\end{eqnarray}
The total energy of spacetime is defined to be $\cE \equiv \cQ_k$,
while the total angular momenta are $\cJ_a \equiv-\cQ_{m^a}$.
Because the charges $Q_\xi$ only depend on the homology class of
$S$, one can evaluate them on the sphere at infinity $S^\infty$ in
order to allow their comparison with the usual Lagrangian and
Hamiltonian surface charges at
infinity~\cite{Abbott:1982ff,Henneaux:1985tv,Henneaux:1985ey}.

\paragraph{Useful integrals}

Let us define the spheroid $S^\infty$ in coordinates
$(\tau,r,\varphi_a,\mu_i)$ by $r=cst \longrightarrow \infty$,
$\tau=cst$. Using $\sqrt{-g}=\sqrt{-\bar g}$ given explicitly in
equation (A.9) of \cite{Gibbons:2004ai} and expressing
$\mu_{N+\eps}$ as a function of the remaining $\mu_\alpha$, $1\leq
\alpha \leq N+\eps - 1$, it is straightforward to show that
\begin{eqnarray}
\cA^{sphoid}&\equiv&\int_{S^\infty} \prod_{\alpha=1}^{N+\eps-1}
d\mu_\alpha \prod_{a=1}^{N} d\varphi_a\ \frac{\sqrt{-\bar
g}}{r^{n-2}} = \frac{\cA_{n-2}}{\prod_{a=1}^N \Xi_a},
\end{eqnarray}
where $\cA_{n-2}$ is the volume of the unit $n-2$ sphere, given
explicitly for instance in (4.9) of \cite{Gibbons:2004ai}.

Similarly,
\begin{eqnarray}
\hspace{-17pt}\cI \hspace{-7pt}&\equiv &\hspace{-7pt}
\int_{S^\infty} \prod_{\alpha=1}^{N+\eps-1} d\mu_\alpha
\prod_{a=1}^{N} d\varphi_a\ \frac{\sqrt{-\bar g}}{r^{n-2}} \, W =
\frac{2}{n-1}\big( \sum_{a=1}^N \frac{1}{\Xi_a}+ \frac{\eps}{2}
\,\big) \cA^{sphoid}.
\end{eqnarray}
This identity has been verified using {\it Mathematica} up to
$n=8$. We suppose it holds for higher $n$.

\paragraph{Angular momenta}

Because $m^a=\frac{\partial}{\partial \varphi_a}$ is tangent to
$S^\infty$, the charge~\eqref{eq:3bis} reduces to the standard
expression for the angular momenta in terms of Komar integrals:
\begin{eqnarray}
  \cJ_a=\oint_{S^\infty}  k^K_{\cL^{EH},m^a}[g]-\oint_{S^\infty}
  k^K_{\cL^{EH},m^a}[\bar g],\label{angul1}
\end{eqnarray}
and is path independent. Explicitly, one gets
\begin{eqnarray}
\cJ_a  &=& \int_{S^\infty} \prod_{\alpha=1}^{N+\eps-1} d\mu_\alpha
\prod_{a=1}^{N} d\varphi_a\ \frac{\sqrt{-\bar g}}{16\pi} \left(
g^{\tau\alpha}g^{rr} g_{\alpha \varphi^a,r}-\bar
g^{\tau\alpha}\bar g^{rr}
\bar g_{\alpha \varphi^a,r}\right)\nonumber\\
&=&\frac{M a_a}{8\pi} (n-1)
\int_{S^\infty}\prod_{\alpha=1}^{N+\eps-1} d\mu_\alpha
\prod_{a=1}^{N} d\varphi_a\ \frac{\sqrt{-g}}{r^{n-2}}
\frac{\mu_a^2}{\Xi_a}\nonumber \\
&=& \frac{M a_a }{4\pi\Xi_a}\,\cA^{sphoid}.
\end{eqnarray}
Here, the Komar integral evaluated for the background does not
contribute because $\bar g_{\tau \varphi_a}=\bar g^{\tau
\varphi_a} = 0$. The result agrees with the one given in
\cite{Gibbons:2004ai}.

\paragraph{Mass}

In order to compute the mass, we evaluate \eqref{eq:3bis} with
$\xi=k=\frac{\partial}{\partial \tau}$ on $S^\infty$. We have
\begin{eqnarray}
\int_{S^\infty}( -K^K_{k}[g]+K^K_{k}[\bar g])&=&\nonumber \\
\mbox{}\hspace{2cm}&&\hspace{-3.5cm}\int_{S^\infty}
\prod_{\alpha=1}^{N+\eps-1} d\mu_\alpha \prod_{a=1}^{N}
d\varphi_a\ \frac{\sqrt{-\bar g}}{16\pi} \left(
g^{\tau\alpha}g^{rr} g_{\alpha \tau,r} - {\bar
g}^{\tau\alpha}{\bar g}^{rr} {\bar g}_{\alpha
\tau,r}\right).\label{eq:KerrAdSint1}
\end{eqnarray}
Let decompose the metric as $g_{\mu\nu} = \bar g_{\mu\nu} +
h_{\mu\nu}$.  The asymptotic behavior~\eqref{falloff} of
$h_{\mu\nu}$ implies that $h^\mu_{\,\,\nu} = \bar
g^{\mu\alpha}h_{\alpha\nu} = O(r^{-n+1})$.  Hence, in the
expansion of the inverse metric $g^{\mu\nu}$
\begin{equation} g^{\mu\nu} = \bar
  g^{\mu\alpha}(\delta^{\,\,\nu}_{\alpha}
-h^{\,\,\nu}_{\alpha} + h^{\,\,\beta}_{\alpha}h^{\,\,\nu}_{\beta}
-h^{\,\,\gamma}_{\alpha}
 h^{\,\,\beta}_{\gamma} h^{\,\,\nu}_{\beta} +\cdots).\label{decomp}
\end{equation}
only the first two terms will contribute to
integral~\eqref{eq:KerrAdSint1}, since the following terms fall
off faster and keeping only the first two terms will give finite
contributions, as we will show.  Injecting this expansion into
\eqref{eq:KerrAdSint1}, one gets terms that are at most quadratic
in $h_{\mu\nu}$. The terms of order $0$ will cancel, while the
terms quadratic in $h_{\mu\nu}$ can directly be shown not to
contribute. Hence, only terms linear in $h_{\mu\nu}$ will
contribute to \eqref{eq:KerrAdSint1} with the result
\begin{eqnarray}
\int_{S^\infty} (-K^K_{k}[g]+K^K_{k}[\bar g]) &=& \frac{M}{8\pi}
\int_{S^\infty}\nonumber \prod_{\alpha=1}^{N+\eps-1} d\mu_\alpha
\prod_{a=1}^{N} d\varphi_a\
\frac{\sqrt{-\bar g}}{r^{n-2}} \Big[ (n-1)W-2 \Big] \\
&=&  \frac{M \cA_{n-2}}{4 \pi (\prod_{a} \Xi_a)}\left(
\sum_{b=1}^N \frac{1}{\Xi_b} + \frac{\eps}{2}-1 \right).
\end{eqnarray}

The integral $\oint_{S^{\infty}} \cC_{k;\gamma}$ defined
in~(\ref{intcCbis}) reduces to
\begin{equation}
\oint_{S^{\infty}} \cC_{k;\gamma} = \int_0^1 ds\, \int_{S^\infty}
\prod_{\alpha=1}^{N+\eps-1} d\mu_\alpha \prod_{a=1}^{N}
d\varphi_a\ \frac{\sqrt{-\bar g}}{16\pi} (D^{(s)}_\sigma
h_{(s)}^{r\sigma} -
\partial^r h^{(s)}),\label{intcCKerr}
\end{equation}
where $h^{(s)}_{\mu\nu}=\frac{d
  g^{(s)}_{\mu\nu}}{ds}$ (and indices are lowered and raised with
$g^{(s)}_{\mu\nu}$ and its inverse). Note that the equality
$\sqrt{-g^{(s)}} = \sqrt{-\bar g}$ implies $h^{(s)} \equiv
g^{\mu\nu}_{(s)}h^{(s)}_{\mu\nu} = 0$. From the definition of the
metric~\eqref{KerrAdsmetric}, one can see that
\begin{equation}
h^{(s)}_{\mu\nu}=h_{\mu\nu}+ o (h_{\mu\nu}),\quad
g^{(s)}_{\mu\nu}=\bar g_{\mu\nu}+ s h_{\mu\nu} + o
(h_{\mu\nu}),\label{fall_s}
\end{equation}
where $\bar g_{\mu\nu}$ is the adS metric and $h_{\mu\nu}$ is
defined in \eqref{eq:KerrAdS2}. Now, as the leading terms in
expression~\eqref{fall_s} give finite contributions to the
integral~\eqref{intcCKerr}, as we will show below, the sub-leading
terms $o(h_{\mu\nu})$ will not contribute. Expanding
$g^{\mu\nu}_{(s)}$ as in \eqref{decomp}, we get
\begin{equation} g_{(s)}^{\mu\nu} \sim \bar
  g^{\mu\alpha}(\delta^{\,\,\nu}_{\alpha}
-s h^{\,\,\nu}_{\alpha} + s^2
h^{\,\,\beta}_{\alpha}h^{\,\,\nu}_{\beta} -
\cdots),\label{decomp2}
\end{equation}
where the indices are raised with $\bar g^{\mu\nu}$ and where
$\sim$ indicates that the sub-leading terms in
equation~\eqref{fall_s} have been dropped. Again, we will show
below that the first two terms of \eqref{decomp2} give finite
contributions to the integral~\eqref{intcCKerr}. As the following
terms in~\eqref{decomp2} fall off faster, we can safely ignore
them in the computation. If we now expand the expressions
$g_{\mu\nu}^{(s)}$, $g_{(s)}^{\mu\nu}$ and $h_{\mu\nu}^{(s)}$ in
the integrand $\sqrt{-\bar g} D^{(s)}_\sigma h_{(s)}^{r\sigma}$ in
terms of $\bar g_{\mu\nu}$ and of $h_{\mu\nu}$, we obtain after
some work that
\begin{equation}
\sqrt{-\bar g} D^{(s)}_\sigma h_{(s)}^{r\sigma} = \sqrt{-\bar g}
{\bar D}_\sigma h^{r\sigma} + O(r^{-n+1})\label{eq:dem1}
\end{equation}
where all the dependence in $s$ appear only in the vanishing term
$O(r^{-n+1})$. As a consequence, the integral~\eqref{intcCKerr}
does not depend on the path.

Explicitly, one shows after some computations that $D_\sigma
h^{r\sigma}$ reduces to $r^{-1}h^{rr} + o(r^{-n+2})$.  Therefore,
$\oint_{S^\infty} \cC_{k;\gamma}$ becomes
\begin{eqnarray}
\oint_{S^\infty} \cC_{k;\gamma} &=& \frac{M}{8\pi} \cA^{sphoid}=
\frac{M}{8\pi}\frac{\cA_{n-2}}{(\prod_{a}
\Xi_a)}.\label{eq:KerrAdScC}
\end{eqnarray}
Finally, the energy is obtained by summing the two contributions
$\oint(-K^K_k[g]+K^K_k[\bar g])$ and $\oint \cC_{k;\gamma}$, which
gives explicitly
\begin{equation}
\cE  = \frac{M \cA_{n-2}}{4 \pi (\prod_{a} \Xi_a)}\left(
\sum_{b=1}^N \frac{1}{\Xi_b} - \frac{(1-\eps)}{2}
\right),\label{eq:KerrAdSee}
\end{equation}
in agreement with \cite{Gibbons:2004ai,Deruelle:2004mv}.

\paragraph{Comparison with alternative surface charges}

Actually, in~\eqref{eq:dem1} and because $h^{(s)}=0$, we showed
that
\begin{equation}
\oint_{S^{\infty}} \cC_{k;\gamma} = -\oint_{S^\infty} i_k
\Theta[h, \bar g],
\end{equation}
with $h_{\mu\nu} = g_{\mu\nu}-\bar g_{\mu\nu}$ because all terms
of which are of order $1$ or higher in an expansion according to
$s$ of $\oint_{S^{\infty}} \cC_{k;\gamma}$ vanish when one
approaches the boundary at infinity. Hence, we have shown that at
$S^\infty$, the mass can be computed using
\begin{eqnarray}
\cQ_{\bar \xi}[g,\bar g] = \oint_{S^\infty} k_{\bar \xi}[g-\bar
g,\bar g].\label{eq:22bis}
\end{eqnarray}
with $\bar \xi =k$ and where the $(n-2)$-form is given
in~\eqref{eq:wald2}. Moreover, as shown in~\eqref{angul1}, the
same relation~\eqref{eq:22bis} hold for $\bar \xi$ replaced by the
Killing vectors $m^a$.

Now, because of the equivalence of expression~\eqref{eq:wald2}
with~\eqref{gsuperpot3} proven in section~\ref{sec:cov_grav} of
Chapter~\ref{chap:matter}, the conserved charge~\eqref{eq:22bis}
for $\bar \xi = k,\; m^a$ is exactly the Abbott-Deser surface
charge~\cite{Abbott:1982ff} associated with the Killing vector
$\bar \xi$ of the anti-de Sitter background $\bar g$.

Moreover, using the results of section~\ref{sec:canonic} of
Chapter~\ref{chap:matter}, one can write the conserved
charge~\eqref{eq:22bis} related to the Killing vector $\bar
\xi^\mu$, $\mu = 0,i$ of the background in the hamiltonian form
derived in~\cite{Henneaux:1985tv,Henneaux:1985ey},
\begin{eqnarray}
\oint_{S^\infty} k_{\bar\xi}[\delta\gamma,\delta
\pi;\bar\gamma,\bar\pi] = \oint_{S^\infty} \frac{1}{16\pi G}
(d\sigma)_{a} \Big( \bar G^{abcd}\big[\bar \nabla_b
\delta\gamma_{cd} \bar\xi^\perp
-\bar \nabla_b\bar\xi^\perp \delta\gamma_{cd}\big]\nonumber\\
 + 2\bar\xi_b \delta
\pi^{ab} - \bar\xi^a \delta\gamma_{cd} \bar \pi^{cd} \Big),
\label{eq:k0iHamilt}\\
\bar G^{abcd} =\half \sqrt{\bar \gamma}(\bar \gamma^{ac}\bar
\gamma^{bd} + \bar \gamma^{ad}\bar \gamma^{bc} - 2 \bar
\gamma^{ab}\bar \gamma^{cd}),
\end{eqnarray}
with $\delta\gamma_{cd} = \gamma_{cd}-\bar \gamma_{cd}$ and
$\delta \pi^{ab} =  \pi^{ab}- \bar  \pi^{ab}$. In this expression,
$a=1,\dots ,n-1$, $\bar \gamma_{ab}$ denotes the spatial
background three metric, which is used, together with its inverse
$\bar \gamma^{bc}$ to lower and raise indices, $\bar \nabla_a$ is
the associated covariant derivative, $\bar \pi^{ab}$ are the
conjugate momenta, $\bar \xi_a=\delta_a^i\bar \xi_i$, with
$i=1,\dots n-1$ and $\bar \xi^\perp=N\bar \xi^0$, with $N$ the
lapse function.

Finally, the charge derived in~\cite{Katz:1996nr} is defined on
the sphere at infinity $S^\infty$ as
\begin{eqnarray}
  \cQ^{BKL}_{\xi}[g;\bar
  g]&=&-\oint_{S^\infty} k^K_{\cL^{EH},\xi}[g]+\oint_{S^\infty}k^K_{\cL^{EH},\xi}[\bar g]\nonumber \\
  &&-\oint_{S^\infty}(d^{n-2}x)_{\mu\nu}\frac{\sqrt{-g}}{16\pi}\Big(\xi^\mu
k^\nu[g,\bar g]-(\mu\leftrightarrow \nu)\Big),\label{eq:BKL}
\end{eqnarray}
with
\begin{eqnarray}
  \label{eq:BKLbis}
k^\nu[g,\bar g]=g^{\nu\rho}(\Gamma^\sigma_{\rho\sigma}-\bar
\Gamma^\sigma_{\rho\sigma})-g^{\rho\sigma}(\Gamma^\nu_{\rho\sigma}-
\bar\Gamma^\nu_{\rho\sigma}).
\end{eqnarray}
This expression coincides to first order in $h_{\mu\nu}$ with
$\oint_{S^\infty} k_\xi[g_{\mu\nu}- \bar g_{\mu\nu} ; \bar
g_{\mu\nu}]$ where the one-form $k_\xi$ is given
in~\eqref{eq:wald2} as can easily be seen by using $\delta
\Gamma^\nu_{\rho\sigma}=\frac 12 ( D_\sigma h^\nu_\rho+ D_\rho
h^\nu_\sigma- D^\nu h_{\rho\sigma})$. Hence, the BKL expression
gives the same results as the charge $\cQ_\xi[g,\bar g]$ because
the  boundary conditions are such that the terms quadratic and
higher in $h_{\mu\nu}$ vanish asymptotically.

\subsection[Generalized Smarr relation]{Generalized Smarr relation}
\label{sec:gener-smarr-relat}

The Smarr relation is given in general relativity by the
expression~\eqref{PreSmarr}. Let us now evaluate the terms on its
right-hand side.

The integral $\oint_H k_{\cL^{EH},\xi}^K[\bar g]$ evaluated on the
surface $r = r_+$, where the horizon radius $r_+$ is the largest
root of $V(r)-2m = 0$, is given by
\begin{eqnarray}
\oint_H k_{\cL^{EH},\xi}^K[\bar g] = - \frac{\cA_{n-2}}{8\pi
l^2(\prod_a \Xi_a)} r_+^\eps\prod_{a=1}^N (r_+^2+a^2_a).
\end{eqnarray}
Note that this integral vanishes in Minkowski space $(l\rightarrow
\infty)$.

In Kerr-adS spacetimes, the Komar integrand $k_{\cL^{EH},\xi}^K$
of a Killing vector $\xi$ is not closed. Indeed, using the
equations of motion $R_{\mu\nu} = - (n-1)l^{-2}g_{\mu\nu}$, we
have
\begin{eqnarray}
\dH
k_{\cL^{EH},\xi}^K[g]&=&\frac{1}{16\pi}(d^{n-1}x)_{\nu}\sqrt{-g}
\big(D_\mu
D^\mu\xi^\nu- D_\mu D^\nu\xi^\mu \big)\\
&=& -\frac{n-1}{8\pi l^2}(d^{n-1}x)_{\nu}\sqrt{-g} \xi^\nu.
\end{eqnarray}
Because $\sqrt{-g}=\sqrt{-\bar g}$, we have
$\dH(-k_{\cL^{EH},\xi}^K[g]+k_{\cL^{EH},\xi}^K[\bar g])=0$. It
then follows from the definition of
$\cC_{\xi;\gamma}$~\eqref{intcCbis} and from the identity $\dH
k_{\xi}=0$ that $\dH\,\cC_{\xi;\gamma} = 0$. We thus can move the
integral on the horizon back out to infinity,
\begin{eqnarray}
\oint_H\cC_{\xi;\gamma} =\oint_{S^\infty}\cC_{k;\gamma}
+\Omega_a\oint_{S^\infty}\cC_{m^a;\gamma}.
\end{eqnarray}
The first term on the right hand side has already been computed in
\eqref{eq:KerrAdScC}, while the second term vanishes because
$m^a=\frac{\partial}{\partial \varphi^a}$ does not vary along the
path and is tangent to $S^\infty$.

We can now write the Smarr formula (\ref{PreSmarr}) as
\begin{eqnarray}
\cE - \Omega^a \cJ_a &=& \frac{\kappa \cA^{sphoid}}{8\pi} +
\frac{\cA^{sphoid}}{8\pi}\big( M -
\frac{r_+^\eps}{l^2}\prod_{b=1}^N (r_+^2+a^2_b)
\big),\label{eq:KerrAdSsmarrfin}
\end{eqnarray}
in complete agreement with the results obtained by Euclidean
methods in \cite{Gibbons:2004ai}.

In the limit $l \rightarrow \infty$, we recover the Smarr formula
for Kerr black holes in flat backgrounds since then $\cA^{sphoid}
= \cA_{n-2}$, $\oint_H k^K_{\cL^{EH},\xi}[\bar g]= 0$. Combining
\eqref{eq:KerrAdScC} with \eqref{eq:KerrAdSee} then gives $\oint
\cC_\xi = (n-2)^{-1}\cE$. Injected into \eqref{PreSmarr}, we
finally have
\begin{eqnarray}
  \frac{n-3}{n-2}\cE - \Omega^a \cJ_a &=& \frac{\kappa \cA_{n-2}}{8\pi}.
\end{eqnarray}

The first law for these black holes holds as a consequence of
Theorem~\ref{th:fl} on page~\pageref{th:fl}.

\section{G\"odel black holes in supergravity}
\label{sec:sugraGodel}

Black hole solutions in supergravity theories have attracted a lot
of interest recently for two main reasons. On the one hand, higher
dimensional supersymmetric theories play a prominent role in the
effort of unifying gravity with the three microscopic forces and
on the other hand, black hole solutions are preferred laboratories
to study effects of quantum gravity.

Among the supersymmetric solutions of five dimensional minimal
supergravity \cite{Gauntlett:2002nw}, a maximally supersymmetric
analogue of the G\"odel universe \cite{Godel:1949ga} has been
found. This solution can be lifted to 10 or 11 dimensions (see
also \cite{Tseytlin:1996as}) and has been intensively studied as a
background for string and M-theory, see
e.g.~\cite{Boyda:2002ba,Harmark:2003ud}.

Black holes in G\"odel-type backgrounds have been proposed in
\cite{Herdeiro:2002ft,Gimon:2003ms,Herdeiro:2003un,Brecher:2003wq,Behrndt:2004pn}.
Usually, given new black hole solutions, the conserved charges are
among the first properties to be studied, see e.g.
\cite{Myers:1986un,Gauntlett:1998fz,Gibbons:2004ai}. Indeed, they
are needed in order to check whether these solutions satisfy the
same remarkable laws of thermodynamics as their four dimensional
cousins \cite{Bardeen:1973gs,Carter:1972}.

The computation of the mass, angular momenta and electric charge
of the G\"odel black holes was an open problem in 2004, mentioned
explicitly in \cite{Gimon:2003ms} with partial results obtained in
\cite{Klemm:2004wq} because the naive application of traditional
approaches fails. The aim of the computation below, published as a
paper in~\cite{Barnich:2005kq}, is to solve this problem for the
five dimensional spinning G\"odel-type black
hole~\cite{Gimon:2003ms} and to derive both the generalized Smarr
formula and the first law.

In what follows, we consider the bosonic part of minimal
supergravity in $n=5$ dimensions described by the
Lagrangian~\eqref{eq:SUG1} with $\Lambda=0$ and $\lambda=1$.

The G\"odel-type solution~\cite{Tseytlin:1996as,Gauntlett:2002nw}
to the field equations is given by
\begin{eqnarray}
\bar{ds}^2 &&= - (dt + j\,r^2\sigma_3)^2 + dr^2 +\nonumber\\&&+
\frac{r^2}{4}(d\theta^2+d\psi^2+d\phi^2+2\cos{\theta}d\psi d\phi),\\
\bar A &&= \frac{\sqrt{3}}{2}j\,
r^2\sigma_3,\nonumber\label{GodelSol}
\end{eqnarray}
where the Euler angles $(\theta,\phi,\psi)$ belong to the
intervals $0\leq\theta \leq\pi$, $0\leq \phi \leq 2\pi$, $0\leq
\psi < 4\pi$ and where $\sigma_3 = d\phi+\cos{\theta}d\psi$. It is
the reference solution with respect to which we will measure the
charges of the black hole solutions of \cite{Gimon:2003ms} that we
are interested in. These latter solutions can be written as
\begin{eqnarray}
&&ds^2 = \bar{ds}^2 + \frac{2m}{r^2}(dt-\frac{l}{2}\sigma_3)^2
\nonumber -2m j^2r^2\sigma_3^2 \\ && \,\,\qquad + (k(r)-1)dr^2,
\qquad \quad A = \bar A,\label{eq:SUGKerrGodel} \\
 \vspace*{-0.7cm} && k^{-1}(r) = 1 -\frac{2m}{r^2}+\frac{16j^2m^2}{r^2}
+ \frac{8jml}{r^2} +\frac{2ml^2}{r^4}.\nonumber
\end{eqnarray}
They reduce to the Schwarzschild-G\"odel black hole when $l = 0$,
whereas the five dimensional Kerr black hole with equal rotation
parameters is recovered when $j = 0$.

The $(n-2,1)$-forms constructed from the
Lagrangian~\eqref{eq:SUG1} were described in
section~\ref{sec:EM-chern} of Chapter~\ref{chap:matter}.

Consider a path $\gamma$ in solution space joining the solution
$\phi$ to the background $\bar \phi$. Whenever two $n-2$
dimensional closed hypersurfaces $S$ and $S^\prime$ can be chosen
as the only boundaries of an $n-1$ dimensional hypersurface
$\Sigma$, the charges defined by
\begin{eqnarray}
Q_{\bar\xi,c}=\oint_S \int_{\gamma} k_{\bar\xi,c}[d_V\phi]
\label{eq:SUG3}
\end{eqnarray}
where $(\bar \xi,c)$ are reducibility parameters~\eqref{eq:SUG2}
do not depend on the hypersurfaces $S$ or $S^\prime$ used for
their evaluation. Furthermore, the integrability conditions
satisfied by $k_{\bar\xi,c}[d_V\phi]$, as shown in the
computations below, and the absence of topological obstructions
imply that these charges do not depend on the path, but only on
the initial and the final solutions.

We choose to integrate over the surface $S$ defined by
$t=constant=r$, while the path $\gamma:(g^{(s)},A^{(s)})$
interpolating between the background G\"odel-type universe $(\bar
g,\bar A)$ and the black hole $(g,A)$ is obtained by substituting
$(m,l)$ by $(sm,sl)$ in (\ref{eq:SUGKerrGodel}), with $s\in
[0,1]$. Because $A^{(s)}_\mu=\bar A_\mu$ for all $s$, the mass
\begin{eqnarray}
\cE\equiv\oint_S \int_{\gamma} k_{\frac{\partial}{\partial
t},0}[d_V\phi]
\end{eqnarray}
of the black hole comes from the gravitational part~\eqref{k_diff}
only
\begin{eqnarray}
  \label{eq:SUG6}
  \cE &=&- \left[ \oint_S k^K_{\cL^{EH},\d_t}  \right]^{g}_{\bar g} +\int_0^1 ds\,
\oint_S i_{\d_t}I^{n}_{\dv g}\cL^{EH}  \nonumber\\
&=& \frac{3\pi}{4} m - 8\pi j^2\, m^2 -\pi j\ m\ l.
\end{eqnarray}
Unlike the five dimensional Kerr black hole
\cite{Myers:1986un,Gauntlett:1998fz}, the mass of which is
recovered for $j=0$, we also see that the rotation parameter $l$
brings a new contribution to the mass with respect to the
Schwarzschild-G\"{o}del black hole.

Note that the integral over the path is really needed here in
order to obtain meaningful results, because the naive application
of the Abbott-Deser~\eqref{gsuperpot3}, Iyer-Wald~\eqref{k_diff}
or Regge-Teitelboim~\eqref{k_Teit} expressions evaluated on $\bar
g  $ gives as a result
\begin{eqnarray}
  \label{eq:SUG17}
  \cE^{naive}= \oint_S k_{\frac{\partial}{\partial t},0}
[g-\bar g] = 8 \pi m^2 j^4 r^2 + O(1),
\end{eqnarray}
which, as pointed out in \cite{Klemm:2004wq}, diverges for large
$r$. A correct application consists in using these expressions to
compare the masses of infinitesimally close black holes, i.e.,
black holes with $m+\delta m,l+\delta l$ as compared to black
holes $(g,A)$ with $m,l$. Indeed, $\oint_S
k_{\frac{\partial}{\partial t},0} [\dv g]=\delta\cE$, with $\cE$
given by the r.h.s of \eqref{eq:SUG6}, which is finite and $r$
independent as it should since $\dH k_{\frac{\partial}{\partial
t},0} [\dv g]=0$. Finite mass differences can then be obtained by
adding up the infinitesimal results. This procedure is for
instance also needed if one wishes to compute in this way the
masses of the conical deficit solutions  \cite{Deser:1983tn} in
asymptotically flat 2+1 dimensional gravity.

Because our computation of the mass does not depend on the radius
$r$ at which one computes, one can consider, if one so wishes,
that one computes inside the velocity of light surface. Similarly,
if one uses this method to compute the mass of de Sitter black
holes, one can compute inside the cosmological horizon, and
problems of interpretation, due to the fact that the Killing
vector becomes space-like, are avoided.

The expression for the angular momentum
\begin{eqnarray}
\cJ^\phi\equiv-\oint_S \int_{\gamma} k_{\frac{\partial}{\partial
\phi},0}[d_V\phi] \label{eq:SUG10}
\end{eqnarray}
reduces to
\begin{eqnarray}
\cJ^\phi &=& \left[ \oint k^K_{\cL^{EH},\d_\phi}\right]^g_{\bar g}
+ \left[ \oint Q^{A}_{\d_\phi,0} \right]^{A,g}_{\bar A,\bar
g}.
\end{eqnarray}
Using~\eqref{Komar} and~\eqref{KomarQA}, we get
\begin{eqnarray}
\cJ^\phi  &=&\frac{1}{2}\pi m l -\pi j m l^2  -4 \pi j^2 m^2 l
,\label{eq:SUGangul3}
\end{eqnarray}
while the angular momenta for the other 3 rotational Killing
vectors \cite{Gimon:2003ms} vanish.

The electric charge picks up a contribution from the Chern-Simons
term and is explicitly given by~\eqref{eq:SUG9},
\begin{eqnarray}
  \label{eq:SUG11}
  \cQ\equiv-\oint_S \int_\gamma k_{0,1}
=\left[ Q^{A}_{0,1}+{\lambda }J\right]^{g,\bar A}_{\bar
  g,\bar A} =
2\sqrt{3}\pi\, j m l.
\end{eqnarray}
In particular, it vanishes for the Schwarzschild-G\"odel black
hole.

\paragraph{Generalized Smarr formula and first law.}
Consider a stationary black hole with Killing horizon determined
by $\xi_H=k+\Omega^H_a m^a$, where $k$ denotes the time-like
Killing vector, $\Omega^H_a$ the angular velocities of the horizon
and $m^a$ the axial Killing vectors and let $\cE=\oint_S
\int_\gamma k_{k,0}$, $\cJ^a=-\oint_S \int_\gamma k_{m^a,0}$. As
discussed in section~\ref{sec:Smarr} of Chapter~\ref{chap:geome},
the definition of $\xi_H$ and the charges imply
\begin{eqnarray}
  \label{eq:SUG12}
  \cE - \Omega^H_a \cJ^a = \oint_H \int_\gamma
  k_{\xi_H,0}[\dv\phi],
\end{eqnarray}
where $H$ is a $n-2$ dimensional surface on the horizon. Because
$A^{(s)}_\mu =\bar A_\mu$, the r.h.s becomes
\begin{eqnarray} \label{eq:SUGchargesmarr}
\oint_H \int_\gamma
  k_{\xi_H,0}[\dv\phi] &=& -\left[\oint_H k^K_{\cL^{EH},\xi_H}\right]^{g}_{\bar
g} - \left[ \oint_H Q_{\xi_H,0}^{A} \right]_{\bar g,\bar
A}^{g,\bar
  A} +
\oint_H\cC_{\xi_H;\gamma},\nonumber\\
\cC_{\xi_H;\gamma}&=& \int_0^1 ds\, i_{\xi_H} I_{\dv g}^n
\cL^{EH}. \label{intcC}
\end{eqnarray}
Now, $-\oint_H k^K_{\cL^{EH},\xi_H} = \frac{\kappa \cA}{8\pi}$,
where $\kappa$ is the surface gravity and $\cA$ the area of the
horizon, while $- \left[ \oint_H Q_{\xi_H,0}^{A}\right]_{\bar
g,\bar A}^{g,\bar A} = \Phi_H \cQ$, where $\Phi_H = -i_{\xi_H} A$
is the co-rotating electric potential, which is constant on the
horizon \cite{Carter:1972,Gauntlett:1998fz}. We thus get
\begin{eqnarray}
  \label{eq:SUG13}
\hspace{-11pt}  \cE - \Omega^H_a \cJ^a \hspace{-4pt}&=&
\hspace{-4pt}\frac{\kappa \cA}{8\pi} +\Phi_H \cQ
 + \oint_H  k^K_{\cL^{EH},\xi_H}[\bar g] + \oint_H\cC_{\xi_H;\gamma}
\end{eqnarray}
which generalizes~\eqref{PreSmarr}.

In order to apply this formula in the case of the black hole
(\ref{eq:SUGKerrGodel}), we have to compute the remaining
quantities. The radius $r_H$ and the angular velocities
$\Omega_\phi^H$ and $\Omega_\psi^H$ are solutions of
\begin{eqnarray}
\left[ \Q{\xi^2}{\Omega^\phi}\right]_{r_H,\Omega^H_a} = 0, \left[
\Q{\xi^2}{\Omega^\psi}\right]_{r_H,\Omega^H_a} = 0,
\left[\xi^2\right]_{r_H,\Omega^H_a} = 0.
\end{eqnarray}
Defining for convenience $\alpha  =
(1-8j^2m)(1-8j^2m-8jl-2m^{-1}l^2)$ and $\beta =
1-8j^2m-4r_H^2j^2+2ml^2r_H^{-4}$, we find
\begin{eqnarray*}
r_H^2 &=& m-4jml-8j^2m^2+ m \sqrt{\alpha}\\
\Omega^H_\phi& =& 4\frac{j +m lr_H^{-4} }{\beta},\qquad
\Omega^H_\psi = 0.
\end{eqnarray*}
The electric potential is given by $\Phi_H = -i_{\xi_H} \bar A =
-\frac{\sqrt 3}{2}j r_H^2\Omega^\phi_H$. The area and surface
gravity of the horizon are
\begin{equation}
\cA = 2\pi^2r_H^3 \sqrt{\beta},\qquad \kappa = \frac{2m
\sqrt{\alpha} }{r^3_H\sqrt{\beta} }.
\end{equation}
For the G\"odel-Schwarzschild black hole, we recover the results
of \cite{Gimon:2003ms,Klemm:2004wq}:
\begin{eqnarray}
  \label{eq:SUG16}
  r^2_H=2m(1-8j^2m),\ \cA =
  2\pi^2\sqrt{8m^3(1-8j^2m)^5},\nonumber\\
  \Omega^H_\phi=\frac{4j}{(1-8j^2m)^2},\
  \kappa = \frac{1}{\sqrt{2m(1-8j^2m)^3}}.\nonumber
\end{eqnarray}
Using
\begin{eqnarray}
\oint_H k^K_{\cL^{EH},\xi_H}[\bar g]= -\pi j^2 r_H^4 - \pi
j^3r_H^6\Omega_\phi^H,\\
\oint_H\cC_{\xi_H;\gamma} = \frac{\pi m}{4} -4\pi j^2m^2-\pi j m
l+2\pi j^2mr_H^2,
\end{eqnarray}
together with the explicit expressions for all the other
quantities, one can verify that the generalized Smarr
formula~\eqref{eq:SUG13} reduces indeed to an identity.

We can also compare with the generalized Smarr formula derived for
asymptotically flat black holes in five-dimensional supergravity
\cite{Gauntlett:1998fz}: for the G\"odel type black hole
(\ref{eq:SUGKerrGodel}) we get
\begin{equation}
\frac{2}{3} \, \cE - \Omega_a \cJ^a -\frac{\kappa \cA}{8\pi} -
\frac{2}{3}\,\Phi_H \cQ =- \frac{2\pi}{3}jm(2jm+l).
\end{equation}
The right hand side, which vanishes when $j=0$, describes the
breaking of the Smarr formula for asymptotically flat black holes
due to the presence of the additional dimensionful parameter $j$.
This is somewhat reminiscent to what happens for Kerr-adS black
holes \cite{Gibbons:2004ai}, see
equation~\eqref{eq:KerrAdSsmarrfin}. In the latter case, different
values of the cosmological constant $\Lambda$ describe different
theories because $\Lambda$ appears explicitly in the action. Even
though this is not the case for $j$, we have also taken $j$ here
as a parameter specifying the background because all charges have
been computed with respect to the G\"odel background.

As for Kerr-adS black holes, the spinning G\"odel black hole
satisfies a standard form of the first law. Indeed, using the
explicit expressions for the quantities involved, one can now
explicitly check that the first law
\begin{equation}
\delta \cE = \Omega_a \delta \cJ^a + \Phi_H \delta \cQ +
\frac{\kappa}{8\pi}\delta \cA \label{first_law}
\end{equation}
holds. As pointed out in \cite{Gibbons:2004ai}, the validity of
the first law provides a strong support for our definitions of
total energy and angular momentum. Furthermore, in the limit of
vanishing $j$, we recover the usual expressions for 5 dimensional
asymptotically flat black holes.

\paragraph{Discussion.}

In the case of the non-rotating G\"odel black hole, $l=0=
\cJ^\phi=\cQ$, the parameterization $M^* = 2m-16j^2m^2$, $\beta^*
= \frac{2j}{1-8j^2m}$ suggested by the analysis of
\cite{Gimon:2003xk} allows to write a non anomalously broken
Smarr formula of the form $\frac{2}{3} \, \cE^* =\frac{\kappa
\cA}{8\pi}$, where $\cE^*=\frac{3\pi}{8}M^*$, with $\kappa$ and
$\cA$ unchanged. With $\cE^*$ being the energy and $\beta^*$ the
fixed parameter characterizing the G\"odel background, the first
law is however not satisfied.

A way out, in the case $l=0$, is to consider the Killing vector
\begin{equation}
k^\prime=(1+{\beta^*}^2M^*)^{-2/3}\frac{\partial}{\partial
t}.\label{varyingk}
\end{equation}
which is a particular example of a variable reducibility parameter
($\dv k^\prime \neq 0$).

The associated energy is
\[\cE^\prime\equiv\oint_S
\int_\gamma
k_{k^\prime}=\frac{3\pi}{8}M^*(1+{\beta^*}^2M^*)^{-2/3}.
\] The first law now holds and in addition, with $\kappa^\prime$
defined with respect to $k^\prime$, so does the non anomalously
broken Smarr formula $\frac{2}{3} \, \cE^\prime
=\frac{\kappa^\prime
  \cA}{8\pi}$. Furthermore, it turns out that the
prefactor acts as an integrating factor and the first law is
verified for variations of both $M^*$ and $\beta^*$.

\section{Application to black rings}
\label{sec:br}

Let us consider the black ring with dipole charge described
in~\cite{Emparan:2004wy}. This black ring is a solution to the
action~\eqref{act1} in five dimensions for a two-form $\mathbf A$.
The solution admits three independent parameters: the mass, the
angular momentum and a dipole charge $\oint_{S^2} e^{-\alpha \chi}
\star \mathbf H$ where $S^2$ is a two-sphere section of the black
ring whose topology is $S^2 \times S^1$.

The thermodynamics of this solution was worked out in the original
paper~\cite{Emparan:2004wy}. As shown in~\cite{Copsey:2005se}, the
computations of~\cite{Sudarsky:1992ty,Iyer:1994ys} are not
directly applicable to these black rings. The role of dipole
charges in the formalism of Sudarsky and Wald
\cite{Sudarsky:1992ty} was elucidated in~\cite{Copsey:2005se}.

The metric, the scalar field and the gauge potential are written
in equations~(3.2)-(3.3)-(3.4) of~\cite{Copsey:2005se}. There, the
gauge potential
\begin{equation}
\mathbf A = B_{t\psi} dt \wedge d\psi,
\end{equation}
was shown to be singular on the bifurcation surface in order to
avoid a delta function in the field strength on the black ring
axis. Here, we point out that this singularity in the potential
does not prevent from studying thermodynamics on the future event
horizon along the lines of section~\ref{sec:pformlaw} of
Chapter~\ref{chap:geome} since the pull-back of the potential is
regular there.

Indeed, following~\cite{Emparan:2006mm}, one can introduce ingoing
Eddington-Finkelstein coordinates near the horizon of the black
ring as
\begin{eqnarray}
d\psi &=& d\psi^\prime + \frac{dy}{G(y)}\sqrt{-F(y)H^N(y)},\\
dt &=& dv - C D R \frac{(1+y)\sqrt{-F(y)H^N(y)} }{F(y) G(y)}dy.
\end{eqnarray}
The metric is regular in these coordinates and the gauge potential
can be written as
\begin{equation}
\mathbf A = B_{t\psi}dv \wedge d\psi^\prime + dy \wedge
\omega^{(1)},
\end{equation}
for some $\omega^{(1)}$. The pull-back of the gauge potential to
the future horizon $y = -1/\nu$ is explicitly regular because
$B_{t\psi}$ is finite and $v$ and $\psi^\prime$ are good
coordinates.

The first law for black rings may then be seen as a consequence
of~\eqref{first_law_3}.

\section{Application to black strings in plane waves}
\label{sec:plan}

Conservations laws have been defined in asymptotically flat and
anti-de Sitter backgrounds, see e.g. the seminal
works~\cite{Arnowitt:1962aa,Regge:1974zd,Abbott:1982ff}. A natural
question, raised
in~\cite{Gimon:2003xk,Hubeny:2003ug,Hashimoto:2004ve}, is how mass
can be defined in asymptotic plane wave geometries.

We show in this section that the conserved charges defined
in~Chapter~\ref{chap:general_th} can be used in this context and
lead to the correct first law. More precisely, we show that the
integration of the $(n-2,1)$-form $\mathbf k_{\d_t,0}[\dv
\phi,\phi]$ along a path $\gamma$ in solution
space~\cite{Wald:1999wa,Barnich:2003xg},
\begin{equation}
\cE = \int_\gamma \oint_{S^\infty} \mathbf k_{\d_t,0}[\dv
\phi,\phi]
\end{equation}
provides a natural definition of mass, satisfying the first law of
thermodynamics.

The action of the NS-NS sector of bosonic supergravity in
$n$-dimensions in string frame reads
\begin{eqnarray} S[G,B,\phi_s] &=&  \frac{1}{16\pi G} \int d^{n}
x \sqrt{-G} e^{-2\phi_s} \left[ R_G +4\d_\mu\phi_s \d^\mu \phi_s
\nonumber - \frac{1}{12}H^2 \right],
\end{eqnarray}
when all fields in the $D-n$ compactified dimensions vanish. In
Einstein frame, $g_{\mu\nu} = e^{-4\tilde\phi/(n-2)} G_{\mu\nu}$,
$\phi = \alpha \phi_s$, the action can be written as~\eqref{act1}
with $\alpha = \sqrt{8/(n-2)}$ and $\mathbf A = B$.

Neutral black string in the $n$-dimensional maximally symmetric
plane wave background $\cP_n$, with $n> 4$, are given by
\cite{Gimon:2003xk,Hubeny:2003ug,Hashimoto:2004ve}
\begin{eqnarray}
ds_s^2 &=& -\frac{f_n(r)(1+\beta^2
r^2)}{k_n(r)}dt^2-\frac{2\beta^2 r^2 f_n(r)}{k_n(r)}dt dy +r^2
d\Omega^2_{n-3} \nonumber\\\nonumber &+& \left( 1-\frac{\beta^2
r^2}{k_n(r)}\right) dy^2 + \frac{dr^2}{f_n(r)}-\frac{r^4
\beta^2(1-f_n(r))}{4k_n(r)}\sigma^2_n,\\
e^{\phi_s} &=& \frac{1}{\sqrt{k_n(r)}}, \qquad B= \frac{\beta
r^2}{2k_n(r)}(f_n(r)dt+dy)\wedge \sigma_n
\end{eqnarray}
where
\begin{equation}
f_n(r) = 1-\frac{M}{r^{n-4}}, \qquad k_n(r) = 1+\frac{\beta^2
M}{r^{n-6}}.
\end{equation}
The black strings have horizon area per unit length given by $\cA
= M^{\frac{n-3}{n-4}}A_{n-3}$ where
\begin{equation}
A_{n-3} = \frac{2\pi^{\frac{n-2}{2}}}
{\Gamma\left(\frac{n-2}{2}\right)},
\end{equation}
is the area of the $n-3$ sphere. Choosing the normalization of the
horizon generator as $\xi = \d_t$, the surface gravity is given by
$\kappa = \sqrt{-1/2 (D_\mu \xi_\mu D^\mu \xi^\nu)} =
\frac{n-4}{2}M^{-\frac{1}{n-4}}$.

Using the $(n-2,1)$-forms defined above, the charge difference
 associated with $\Q{}{t}$ between two infinitesimally close black string solutions $\phi$,
$\phi+\delta \phi$ is given by
\begin{equation}
\delta \cQ_{\d_t} = \oint k_{\d_t,0}[\delta \phi,\phi] =
\frac{n-3}{16\pi G } A_{n-3} \delta M,
\end{equation}
which reproduces the expectations of
\cite{Gimon:2003xk,Hubeny:2003ug,Hashimoto:2004ve}. This quantity
is integrable and allows one to define $\cQ_{\d_t} =
\frac{n-3}{16\pi G } A_{n-3} M$ where the normalization of the
background has been set to zero. It is easy to check
that the first law is satisfied.

Note that one freely can choose a different normalization for the
generator $\xi^\prime = N \d_t$. In that case, the surface gravity
changes according to $\kappa^\prime = N \kappa$, the charge
associated to $\xi^\prime$ becomes $\delta \cQ_{\xi^\prime} =
\frac{n-3}{16\pi G } A_{n-3} \,N\,\delta M$ and the first law is
also satisfied. However, $N$ cannot be a function of $\beta$.
Otherwise, the charge $\cQ_{\xi^\prime}$ would not be defined.

%% file: theory_asympt_charges.tex
\chapter{Classical theory of asymptotic charges}
\label{chap:asymptcharges}
\setcounter{equation}{0}\setcounter{figure}{0}\setcounter{table}{0}

In this chapter, we first provide general conditions in order to
define a phase space of fields and gauge transformations for
manifolds admitting a particular closed surface $S$ in an
asymptotic region. Asymptotic symmetries at $S$ are defined as the
quotient space of gauge transformations by gauge transformations
admitting vanishing charges, i.e. proper gauge transformations. We
prove that that asymptotic symmetries form a Lie subalgebra of the
Lie algebra of gauge transformations. We then show that the
representation of this algebra by a covariant Poisson bracket
among the associated conserved charges can be centrally extended.
The representation theorem that we obtain is the Lagrangian
analogue of the theorem proven in Hamiltonian
formalism~\cite{Brown:1986ed,Brown:1986nw}. It was obtained in
covariant phase space methods as well~\cite{Koga:2001vq}. We also
discuss the consequences of the existence of a variational
principle admitting $S$ as a boundary. Finally, we describe two
algorithms allowing one to construct consistent phase spaces and
gauge transformations. Applications for diffeomorphic invariant
theories and Einstein gravity are mentioned. We refer the reader
to~\cite{Barnich:2007bf} for a synthesis.

\section{Phase space of fields and gauge parameters}
\label{sec:phasespace}

Let us start our asymptotic analysis with a particular fixed
closed surface of a $n$-dimensional manifold which we take for
definiteness to be the limit $S^{\infty,t}$ of the sphere
$S^{r,t}$ for $t$ constant and $r$ going to infinity. Here,
$S^{r,t}$ is the intersection of the hyperplane $\Sigma_t$ defined
by constant $t$ and the (usually timelike or null) hyperplane
$\cT^r$ defined by constant $r$. Note that all considerations
below only concern the region of the manifold close to
$S^{\infty,t}$.

We now define a space of allowable field configurations $\cF$ and
for each $\phi^i \in \cF$ a space of allowable gauge parameters
$f \in \cA_\phi$ such that $\delta_{R_f}\phi^i$ are gauge
transformations. The intersection of the configuration space $\cF$
with the stationary surface $\cE$ (where the equations of motion
hold) will be denoted as $\cF^s$. The space $\cF^s$ is the set of
asymptotic solutions that fulfils the required boundary
conditions.

Besides standard smoothness properties we impose the following
requirements on the fields $\phi^i \in \cF$, the tangent one-forms
$\dv\phi^i$ to $\cF$ and the gauge parameters $f^\alpha$:
\begin{itemize}
\item Finiteness of the surface charges,
\begin{eqnarray}
  \label{eq:105}
  \oint_{S^{r,t}} \cL_{\d_r}   k_{f}[\dv\phi]=o(r^{-1})\,.
\end{eqnarray}
This condition compels any surface charge~\eqref{eq:17} for
$S = S^{r,t}$ to be finite in the limit $r\rightarrow \infty$. It may
be understood equivalently as the independence of the surface
charges on smooth deformations of $S^{\infty,t}$ on the hyperplane
$\Sigma_t$ in the asymptotic region $r \rightarrow \infty$.
\item Integrability of the surface charges,
\begin{eqnarray}
  \oint_{S^{r,t}} \dv  k_{f}[\dv \phi]=o(r^0),\qquad \oint_{S^{r,t}} k_{\dv f}[\dv \phi]=o(r^0)\,.\label{eq:92a}
\end{eqnarray}
These conditions guarantee that the surface charges~\eqref{eq:82}
are independent on the path $\gamma \in \cF$ given that no global
obstruction in $\cF$ occurs, which is also asked. The second
condition expresses that $\dv f$ is irrelevant to satisfy
the integrability condition. The last condition will be used to
prove Proposition~\bref{lem4}.

\item Conservation in time of the surface charges for solutions $\phi^s \in \cF^s$
and tangent one-forms $\dv^s\phi$ to $\cF^s$,
\begin{eqnarray}\label{conserveda}
  \oint_{S^{r,t}} \cL_{\d_t} k_{f}[\dv^s\phi]|_{\phi^s}=\oint_{S^{r,t}}i_{\d_t}
W_{\delta\cL/\delta\phi}[\dv^s\phi,R_{f}]|_{\phi^s}=o(r^0)\,,
\end{eqnarray}
where the equality follows from~(\ref{eq:29a}) and from Stokes'
theorem.
\item Closure of the form $E_\cL$
\begin{eqnarray}
  \label{eq:109}
   \oint_{S^{r,t}}i_{R_f} \dv  E_{\cL}[\dv\phi,\dv\phi]=o(r^0),\; \oint_{S^{r,t}}\delta_{R_f} \dv  E_{\cL}[\dv\phi,\dv\phi]=o(r^0).
\end{eqnarray}
These quite technical assumptions are used to prove
Proposition~\bref{lem4} and to prove that the asymptotic
symmetries form an algebra. There are two motivations for them. On
the one hand, these conditions are satisfied for exact
reducibility parameters, $R_f = 0$. On the other hand, it is
argued in section~\ref{sec:varprincip} that
$\oint_{S^{\infty,t}}\dv E_{\cL} = 0$ is a consequence of the
existence of a variational principle with boundary $S^{\infty,t}$.

\item By consistency, the gauge transformations should transform fields
$\phi^i \in \cF$ into other allowable configurations,
\begin{eqnarray}\label{eq:condRf}
  \delta_{R_f} \phi^i = R_f^i\, \text{ should be tangent to $\cF$}.
\end{eqnarray}
It implies that all the other relations are valid for $\dv\phi^i$
contracted with $R^i_f$.
\end{itemize}

For diffeomorphisms, the integrability condition~\eqref{eq:92a}
and the condition on the closure of $E_\cL$~\eqref{eq:109} become
\begin{eqnarray}
  \oint_{S^{r,t}} i_\xi W[\dv\phi,\dv\phi]=o(r^0),&\qquad &\oint_{S^{r,t}} k_{\dv \xi}[\dv \phi]=o(r^0)\,,  \label{eq:92a_diff2}\\
&&\hspace{-3.5cm}\oint_{S^{r,t}} i_{\cL_\xi \phi}\dv
E_\cL[\dv\phi,\dv\phi]=o(r^0), \label{eq:92a_diff}
\end{eqnarray}
as a consequence of~\eqref{int_diff} and~\eqref{eq:79}. As a
consequence of~\eqref{4.13bis}, if $\xi = \d_t, \, \d_r$ are
allowable gauge transformations, the first equation
of~\eqref{eq:92a_diff2} implies together with~\eqref{eq:condRf}
finiteness and conservation of the charges~\eqref{eq:105},
\eqref{conserveda}.

Note that the additional condition~\eqref{eq:109} is automatically
fulfilled in the Hamiltonian formalism in Darboux coordinates
because of \eqref{eq:42c}.

\section{Asymptotic symmetry algebra}
\label{sec:asympt-symm-algebra}

The set of allowable gauge parameters, $f \in \cA_\phi$,
satisfying
\begin{eqnarray}
  \label{eq:107}
  \oint_{S^{r,t}} k_{f}[\dv\phi] = o(r^0),
\end{eqnarray}
for all $\dv\phi$ tangent to $\cF$ will be called proper gauge
parameters of the field $\phi$. The associated transformations
$\delta\phi^i=R^i_{f}$ will be called proper gauge
transformations. On the contrary, gauge parameters (resp.
transformations) related to non-identically vanishing surface
charges will be called improper gauge parameters (resp.
transformations). Improper gauge transformations send field
configurations into inequivalent field configurations because they
change their conserved charges, as will be cleared in
section~\ref{sec:poissonbracket}.

Using the properties~\eqref{eq:92a}, \eqref{eq:109} and
\eqref{eq:condRf} of the phase space, one can prove the following
proposition, see Appendix~\bref{app:proofs}.\ref{appc}.,
\begin{prop}\label{lem4}
  For any field $\phi^s \in \cF^s$, one-form $\dv^s\phi$ tangent to $\cF^s$ at $\phi^s$ and for
  allowable gauge parameters $f_a \in \cA_{\phi^s}$, the identity
\begin{eqnarray}
  \label{eq:87}
   \oint_{S_{\infty,t}}k_{[f_a,f_b]}[\dv^s\phi]|_{\phi^s}  =\oint_{S_{\infty,t}}\dv^s k_{f_a}[R_{f_b}]|_{\phi^s}
\end{eqnarray}
holds.
\end{prop}
The gauge parameters at a solution $\phi^s \in \cF^s$, $f_a \in
\cA_{\phi^s}$, may then be characterized by the following
corollary
\begin{corollary}\label{cor:14}
  The space of allowable gauge parameters $\cA_{\phi^s}$ at $\phi^s \in \cF^s$ form a Lie algebra.
\end{corollary}
The proof of Corollary~\ref{cor:14} goes as follows. Applying
$\cL_{\d_\mu}$ with $\mu = t,\, r$ to \eqref{eq:87} and
using~\eqref{eq:105}, \eqref{conserveda} and~\eqref{eq:condRf}, we
get that $[f_a,f_b]$ corresponds to finite and conserved charges
for fields belonging to $\cF^s$ and for one-forms tangent to
$\cF^s$. As a consequence of~\eqref{eq:92a} we have
$\oint_{S_{r,t}} k_{[\dv
f_a,f_b]}[\dv^s\phi]=-\oint_{S_{r,t}}\dv^s k_{\dv
f_a}[R_{f_b}]=o(r^0) $. Applying $\dv$ to \eqref{eq:87}, the
integrability conditions~\eqref{eq:92a} for $[f_a,f_b]$ are
fulfilled. Using
$[\delta_{R_{f_a}},\delta_{R_{f_b}}]=\delta_{[R_{f_a},R_{f_b}]}$,
\eqref{1.18} and \eqref{eq:comm2}, it is easy to check
that~\eqref{eq:109} and \eqref{eq:condRf} are satisfied for
$[f_a,f_b]$ as well in $\cF^s$.\qed

Note that this derivation shows the consistency of our definition
of phase space. Proposition~\bref{lem4} also trivially involves
the corollary
\begin{corollary}\label{cor17}
The proper gauge transformations at $\phi^s \in \cF^s$ form an
ideal $\cN_{\phi^s}$ of $\cA_{\phi^s}$.
\end{corollary}

The quotient space $\cA_{\phi^s}/\cN_{\phi^s}$ is therefore a Lie
algebra which we call the asymptotic symmetry algebra $\mathfrak
e^{as}_{\phi^s}$ at $\phi^s \in \cF^s$. The asymptotic symmetry
algebra at $\phi^s$ consists in equivalence classes of improper
gauge transformations at $\phi^s$ modulo proper gauge
transformations.

The exact reducibility parameters $f^s \in \mathfrak e_{\phi^s}$
which are associated with (off-shell) finite and integrable
surface one-forms are allowable gauge parameters, i.e. $f^s \in
\cA_{\phi^s}$. If, for any reducibility parameter $f^s$ the phase
space contains at least one solution $\phi^s$ and a tangent
one-form $\dv\phi$ such that $\ndelta \cQ_{f^s}[\dv
\phi]|_{\phi^s} \neq 0$, the space $\mathfrak e^{as}_{\phi^s}$
will hold in representatives of the exact reducibility parameters
$\mathfrak e_{\phi^s}$.

If the gauge theory satisfies~\eqref{simplif} and if the algebra
of gauge parameters closes off-shell, i.e. if~\eqref{1.18} hold
with $M^{+i}_{f_1,f_2}[\varQ{L}{\phi}]=0$, then the proof carried
out in Appendix~\bref{app:proofs}.\ref{appc} can be repeated
off-shell and the following corollary occurs
\begin{corollary}
  If condition~\eqref{simplif} hold and if the bracket
of gauge parameters closes off-shell, the proper gauge
transformations at $\phi \in \cF$ form an ideal $\cN_\phi$ of
$\cA_\phi$. The space of asymptotic symmetries $\mathfrak
e^{as}_{\phi} \equiv \cA_{\phi}/\cN_{\phi}$ at any $\phi \in \cF$
then forms a Lie algebra.
\end{corollary}

\section{Representation by a Poisson bracket}
\label{sec:poissonbracket}

Let us turn to the representation of the Lie algebra of asymptotic
symmetries by a possibly centrally extended Poisson bracket
defined on the associated charges. In this section we derive the
Lagrangian analogue of the theorem of canonical representation of
the Lie algebra of asymptotic symmetries proven in Hamiltonian
formalism in~\cite{Brown:1986ed,Brown:1986nw}. The alternative
analysis achieved in covariant phase space
methods~\cite{Koga:2001vq} is also compared with our results.

Let us define the quantity
\begin{eqnarray}\label{eq:16}
\cK_{f_a,f_b}[\phi^s]=\oint_{S_{\infty,t}}
k_{f_a}[R_{f_b}]|_{\phi^s}= \oint_{S_{\infty,t}} I^{n-1}_{f_b}
W_{{\delta\cL}/{\delta\phi}} [R_{f_a},R_{f_b}]|_{\phi^s}.
\end{eqnarray}
Applying consecutively $i_{R_{f_b}}$ and $i_{R_{f_c}}$
to~\eqref{eq:92a}, the integrability conditions imply
\begin{eqnarray}
\oint_{S^{\infty,t}} k_{f_a}[R_{[f_b,f_c]}] &=&
\oint_{S^{\infty,t}} \big( \delta_{R_{f_c}}k_{f_a}[R_{f_b}]- (b
\leftrightarrow c)\big).
\end{eqnarray}
Using~\eqref{eq:87} on the two terms on the r.h.s. and the
antisymmetry~\eqref{eq:47}, we get
\begin{corollary}\label{lem2}
$\cK_{f_a,f_b}[\phi^s]$ defines a Chevalley-Eilenberg 2-cocycle on
the Lie algebra $\mathfrak e^{as}_{\phi^s}$,
\begin{equation}\eqalign{
  \cK_{f_a,f_b}[\phi^s]+\cK_{f_b,f_a}[\phi^s]&=0,\vspace{3pt}\\
 \cK_{[f_a,f_b],f_c}[\phi^s]+\, {\rm cyclic}\ (a,b,c)&=0.}\label{eq:95}
\end{equation}
\end{corollary}

The surface charges $\cQ[\Phi,\bar \Phi]$ of $\Phi=(\phi,f)$,
$\phi \in \cF$, $f \in \cA_{\phi}$ with respect to the reference
$\bar \Phi=(\bar \phi ,\bar f)$, $\bar\phi \in \cF$, $\bar f \in
\cA_{\bar \phi}$ are defined as
\begin{equation}
  \cQ[\Phi,\bar \Phi] \;\hat = \oint_{S^{\infty,t}} \int_{\gamma}
  k_{f_\gamma}[\dv^{\gamma} \phi]|_{\phi_\gamma} + N_{\bar f}[\bar \phi],
\label{def_Q}
\end{equation}
where the integration is done along a path $\gamma$ in $\cF$
joining $\bar \Phi$ to $\Phi$. We have assumed that there are no
global obstruction in $\cF$ for the integrability
conditions~\eqref{eq:92a} to guarantee that the surface charges
$\cQ[\Phi,\bar \Phi]$ are independent on the path $\gamma \in
\cF$. Note that if asymptotic linearity holds~\eqref{AL}, the
charges~\eqref{def_Q} simplify as~\eqref{eq:82bis}.


We denote $\cQ_a \equiv \cQ[\Phi_a,\bar \Phi_a]$ the charge
related to $\Phi_a =(\phi,f_a)$. The covariant Poisson bracket of
these surface charges is defined by
\begin{eqnarray}
  \label{eq:18}
  \{\cQ_{a} ,\cQ_{b} \}_c \;\hat = -\delta_{R_{f_a}}
  \cQ_{b}=-\oint_{S^{\infty,t}} k_{f_b}[R_{f_a}].
\end{eqnarray}
This covariant Poisson bracket coincides on solutions $\phi^s \in
\cF^s$ with $\cK_{f_a,f_b}[\phi^s]$.

For an arbitrary path $\gamma\in\cF^s$, the
definition~(\ref{eq:16}) leads to
\begin{eqnarray}
  \cK_{f_a,f_b}[\phi^s]- \cK_{\bar f_a,\bar f_b}[\bar \phi^s] &=&
  \int_{\gamma} \oint_{S^{\infty,t}}  \dv^{\gamma}
  \big(k_{f_{a,\gamma}}[R_{f_{b,\gamma}}]|_{\phi_\gamma} \big)\\ &=&
\int_{\gamma}\oint_{S^{\infty,t}}
   k_{[f_{a,\gamma},f_{b,\gamma}]}[\dv^{\gamma}\phi]|_{\phi_\gamma},
\label{equal1}
\end{eqnarray}
where Proposition \bref{lem4} has been used in the last line.
 Using~\eqref{eq:18} and denoting as $\cQ_{[a,b]}$ the charge associated with $[f_a,f_b]$, the
equality~\eqref{equal1} implies
\begin{theorem}\label{lem5}
In $\cF^s$, the charge algebra between a fixed reference solution
$\bar\phi^s$ and a final solution $\phi^s$ is determined by
\begin{eqnarray}
  \label{eq:71}
  \{ \cQ_{a} , \cQ_{b} \}_c=\cQ_{[a,b]}+\cK_{\bar f_a,\bar f_b}[\bar \phi^s]-N_{[\bar f_a,\bar f_b]}[\bar \phi^s],
\end{eqnarray}
where the central charge $\cK_{\bar f_a,\bar f_b}[\bar \phi^s]$ is
a two-cocycle on the Lie algebra of asymptotic symmetries
$\mathfrak e^{as}_{\bar \phi^s}$.
\end{theorem}
The central extension is trivial if it can be reabsorbed in the
normalization of the charges. On the contrary, a central charge
$\cK_{\bar f_a,\bar f_b}[\bar \phi^s]$ is non-trivial if it cannot
be written as a linear function of the bracket $[\bar f_a,\bar
f_b]$ only. Observe that the central charge involving an exact
reducibility parameter of the reference field automatically
vanishes. Also, for a semi-simple algebra $\mathfrak e^{as}_{\bar
\phi^s}$, the property $H^2(\mathfrak e^{as}_{\bar \phi^s})=0$
 guarantees that the
central charge can be absorbed by a suitable normalization of the
background. The property $H^1(\mathfrak e^{as}_{\bar \phi^s})=0$
implies that this completely fixes the normalization.

As a consequence of the theorem together with
Corollary~\bref{cor17}, the proper gauge transformations are
characterized by
\begin{corollary}Any proper gauge transformation $f^{prop}$ acts trivially on
the charges
\begin{equation}
\delta_{R_{f^{prop}}}\cQ_a = 0,
\end{equation}
once we assume that the normalizations associated with the proper
gauge transformations all vanish.
\end{corollary}

\paragraph{Note on general relativity} For Einstein gravity, the explicit formula for the central charge
follows from~\eqref{hom} and is given by\footnote{This
  expression differs from the one given in \cite{Barnich:2001jy} by an
  overall sign because we have changed the sign convention for the
  charges and also by the fact that we use here the
  Misner-Thorne-Wheeler convention for the Riemann tensor.}
\begin{eqnarray}
\cK_{\xi^\prime,\xi}[g]& =& \frac{1}{16\pi G }\oint_{S}
(d^{n-2}x)_{\mu\nu}\sqrt{- g} \Big( - 2  D_\sigma \xi^\sigma D^\nu
\xi^{\prime\mu} + 2  D_\sigma \xi^{\prime\sigma} D^\nu \xi^\mu
\nonumber
 \\
&& + 4  D_\sigma \xi^\nu D^\sigma \xi^{\prime \mu}
+\frac{8\Lambda}{2-n}\xi^\nu \xi^{\prime\mu} - 2 R^{
\mu\nu\rho\sigma}\xi_\rho \xi_\sigma^\prime \nonumber\\
&&+ (D^\sigma \xi^{\prime \nu}+ D^\nu \xi^{\prime \sigma})(D^\mu
\xi_\sigma + D_\sigma \xi^\mu) \Big).\label{G_cbis}
\end{eqnarray}
Note that this expression vanishes if either $\xi$ or $\xi^\prime$
is a Killing vector of $g$. The last term is due to the
contribution from \eqref{suppl} and again vanishes for exact
Killing vectors of $g$ but not necessarily for asymptotic ones.

The application of covariant phase space
methods~\cite{Iyer:1994ys} leads to the surface
charges~\eqref{kIW} and then to a central charge
$\cK^{IW}_{\xi^\prime,\xi}$ equal to
$\cK_{\xi^\prime,\xi}$~\eqref{G_cbis} where the last term is
dropped~\cite{Koga:2001vq,Silva:2002jq}. See also
\cite{Koga:2001vq,Silva:2002jq} for a discussion on the
deficiencies of the expressions derived in
\cite{Carlip:1998wz,Carlip:1999cy,Dreyer:2001py} in the context of
asymptotic symmetries close to horizons. Following the reasoning
of Chapters~\ref{chap:general_th} and~\ref{chap:asymptcharges}, it
can be shown that Corollary~\bref{lem2} and the
Theorem~\bref{lem5} also hold for the surface charges~\eqref{kIW}
and the associated central charge $\cK^{IW}_{\xi^\prime,\xi}$, see
also~\cite{Koga:2001vq}. In that case, the
hypothesis~\eqref{eq:109} is \emph{not} required to prove these
propositions. However, as explained in section~\ref{sec:cov_grav}
of Chapter~\ref{chap:matter} these surface charges depend on
boundary terms that may be added to the Lagrangian, which is not
the case with our definitions.

\section{Existence of a variational principle}
\label{sec:varprincip}

In this section we study conditions for the existence of a
variational principle for spacetimes $\cM$ containing as a
boundary $\cT^\infty$, which is the limit of the null or timelike
hyperplane $\cT^r$ for $r \rightarrow \infty$. We will follow
closely the references~\cite{Iyer:1995kg,Julia:2002df}. We then
analyze how~$\oint_{S^{\infty,t}} \dv E_\cL = 0$ is a consequence
of these conditions.

The boundary term $-\oint_{\cT^\infty} I_{\dv\phi}^n \cL$ obtained
by varying the action, see~\eqref{cc2a}, may not vanish and may
thus prevent the action from being extremal for arbitrary variations.
Let us define a subset $\cF_{X}$ of the phase space $\cF$ where a
$(n-1)$-form $B_{X}$ satisfying
\begin{equation}
\int_{\cT^\infty} (- I_{\dv\phi}^n \cL + \dv B_{X}) =
0,\label{vanish_Th}
\end{equation}
for any field of $\cF_X$ and any tangent vector $\dv\phi$ to
$\cF_{X}$ is defined on $\cT^\infty$. Adding the boundary term
$\int_{\cT^\infty} B_{X}$ to the action will then provide a
correct variational problem in $\cF_X$. Here, the label $X$ refers
to the additional constraints imposed on $\cF$ in order to define
the restricted phase space $\cF_X$. If one can find a $(n-1)$-form
$B$ such that~\eqref{vanish_Th} hold for all variations tangent to
$\cF$, the entire phase space admits a variational principle and
no constraint $X$ is needed.

The boundary term $B_X$ may be constructed if one can find
furthermore a $(n-2)$-form $\mu_{X}[\dv \phi]$ defined on
$\cT^\infty$ such that
\begin{equation}
\dv B_{X} = I^n_{\dv\phi}\cL|_{\cT^\infty} + \dH
\mu_{X}[\dv\phi],\qquad
\mu_{X}[\dv\phi]|_{\partial\cT^\infty}=\dH(\cdot),\label{def_B}
\end{equation}
for any variation $\dv\phi$ tangent to the phase space $\cF_X$.
Note that there is the following ambiguity in the definition of
$B_{X}$ and $\mu_{X}$,
\begin{eqnarray}
B_{X} \rightarrow B_{X}-\dH C_{X},\qquad \mu_{X}[\dv\phi]
\rightarrow \mu_{X}[\dv\phi] + \dv C_{X} +
\dH(\cdot),\label{ambiguity_Bmu}
\end{eqnarray}
for any $(n-2)$-form $C_{X}$ vanishing at $\partial \cT^{\infty}$.
The relation~\eqref{def_B} implies that the symplectic form
$\Omega_\cL$~\eqref{eq:66a} obeys
\begin{equation}
\Omega_\cL[\dv\phi,\dv\phi]|_{\cT^\infty} = \dH\dv
\mu_{X}[\dv\phi].
\end{equation}
The $E_\cL$ form~\eqref{eq:26} is obtained as a result of the
horizontal homotopy $\half I^{n-1}_{\dv \phi}$ applied to
$I^{n}_{\dv \phi}\cL$. If the boundary conditions are such that
this homotopy can be equally applied to $I^{n}_{\dv
\phi}\cL|_{\cT^\infty}$~\footnote{In the very similar computation
of~\cite{Julia:2002df}, such an argument was proven for a
particular set of boundary conditions. Unfortunately, we do not
know a proof for general boundary conditions.}, one gets
\begin{equation}\label{app:argm}
E_{\cL}[\dv\phi,\dv\phi]|_{S^{\infty,t}} = \half \dv
(I^{n-1}_{\dv\phi}B_{X} -\mu_{X} )+\dH(\cdot),
\end{equation}
which leads to the equality
\begin{equation}
\oint_{S^{\infty,t}} \dv E_{\cL}[\dv\phi,\dv\phi] =
0,\label{vanish_E}
\end{equation}
for all tangent vector to $\cF_X$. Note that this latter equation
is independent on boundary terms added to the Lagrangian, as shown
in~\eqref{app:invE}. For first order theories, this condition
reads as
\begin{equation}
\oint_{S^{\infty,t}} \dv \phi^k \wedge \dv \phi^j \wedge \dv\phi^i
\QS{}{\phi^k_\nu}\QS{}{\phi^j_\mu}\varQ{L}{\phi^i}(d^{n-2}x)_{\mu\nu}
 = 0.\label{suff_cond_varprin}
\end{equation}

As a conclusion, the existence of a variational principle on $\cF$
leads under the aforementioned hypotheses to the
equality~\eqref{vanish_E} for all tangent vectors to $\cF$, which
implies~\eqref{eq:109} because of the condition~\eqref{eq:condRf}.
The proof, however, is incomplete and one should still answer the
following questions: (i) which extent conditions~\eqref{def_B} are
necessary for the variational problem to be well-defined, (ii)
under which precise boundary conditions the argument
before~\eqref{app:argm} is valid. These considerations are left
for further work.

\paragraph{Integrated charge for diffeomorphisms.}

In the case of diffeomorphism invariant theories, one can work out
the consequences of assuming the existence of a covariant
$(n-1)$-form $B$ and a covariant $(n-2)$-form $\mu$~\eqref{def_B}.
The charge one-form~\eqref{k_diff} associated with infinitesimal
diffeomorphisms reduces to
\begin{eqnarray}
k_\xi[\dv \phi]|_{S^{\infty,t}} &=& - \dv \left( k^K_{\cL,\xi}
+i_\xi B_{X}+\half \mu_{X}[\cL_\xi \phi]-\half I_{\cL_\xi
\phi}B_{X} \right)\nonumber\\
&& -\half \delta_{\cL_\xi \phi} \left( I_{\dv\phi}B_{X} +
\mu_{X}[\dv\phi]\right) + \dH(\cdot),
\end{eqnarray}
and is independent on the ambiguity~\eqref{ambiguity_Bmu} for
covariant $C_X$. The second term in the latter expression does not
explicitly satisfy the integrability condition~\eqref{eq:92a}.
However, in the integrable case, one can try to find a covariant
$n-2$ form $D_{X}$ defined at the boundary $\cT^\infty$ such that
\begin{equation}
I_{\dv\phi}B_{X} + \mu_{X}[\dv\phi]   = 2\dv D_{X}+
\dH(\cdot).\label{strong_int}
\end{equation}
When there exists forms $B_{X}$, $\mu_{X}[\dv\phi]$ and $D_{X}$
satisfying~\eqref{def_B}, \eqref{strong_int}, the phase space
$\cF_X$ will be called \emph{strongly integrable}. The charge
one-form $k_\xi[\dv\phi]$ will then be the exact variation of the
charge
\begin{equation}
\cQ_{X,\xi}[\phi] = -\oint_{S^{\infty,t}} \left( k^K_{\cL,\xi}
+i_\xi B_{X}+ \mu_{X}[\cL_\xi \phi]
\right),\label{back_indept_charge}
\end{equation}
which is also independent on the ambiguity~\eqref{ambiguity_Bmu}
for covariant $C_X$. Remark that the last term vanishes for exact
symmetries. The surface charge~\eqref{def_Q} then equals to
$\cQ_{X,\xi}[\phi] -\cQ_{X,\xi}[\bar \phi]+N_\xi[\bar \phi]$. It
provides an integrated formula for the surface charge in the phase
space $\cF_X$. Remark that while the surface charge~\eqref{def_Q}
is finite for asymptotic Killing vectors, the
expression~\eqref{back_indept_charge} may be infinite. It is
therefore unappropriate to interpret $\cQ_{X,\xi}[\bar \phi]$ as
the natural normalization of the background $N_\xi[\bar \phi]$.

\paragraph{Note about general relativity.}
In the Palatini formulation of Einstein gravity in four
dimensions, a variational principle for asymptotically flat
spacetimes was defined~\cite{Ashtekar:2000hw}. In the metric
formalism, it is well-known that the Einstein-Hilbert action
supplemented by the Gibbons-Hawking term,
\begin{equation}
S_{EH+GH} = \frac{1}{16\pi G}\oint_\cM \sqrt{-g}R + \frac{1}{8\pi
G} \oint_{\partial \cM}\sqrt{-h}K,
\end{equation}
does \emph{not} provide a satisfactory variational principle for
asymptotically flat spacetimes because~\eqref{vanish_Th} is not
satisfied with $B_X = (8\pi G)^{-1}\sqrt{-h}K$. However, this
variational principle is well-defined when Dirichlet boundary
conditions $X$ are laid down on the induced metric at $\partial
\cM$. For recent progress in obtaining boundary forms $B_{X}$,
$\mu_{X}$ solving~\eqref{def_B} for general asymptotically flat
spacetimes, see the proposals of \cite{Mann:2005yr,Mann:2006bd}.
For the construction of a variational principle for anti-de Sitter
spacetimes, see for example
\cite{Papadimitriou:2005ii,Olea:2006vd}.

\section{Algorithms}

We discussed in the previous section the general conditions one
can impose on the fields and on the gauge parameters in order to
obtain a well-defined theory of asymptotic charges. However, we
have not yet discussed how to fullfil these conditions and actually
find the asymptotic form of the allowable fields and gauge
parameters. This is the aim of this section. We will discuss two
algorithms that allow one to define a phase space $\cF$ and spaces
of gauge parameters $\cA_\phi$.

\subsection{Starting from particular solutions}
\label{sec:algoH}

One can start by constructing a small phase space $\bar \cF$
containing solutions of interest with, in particular, a background
solution $\bar \phi$ admitting a non-trivial set of exact
reducibility parameters $e_{\bar \phi}$. One then imposes that the
asymptotic symmetry algebra $e^{as}_{\phi}$ contains as a
subalgebra $e_{\bar \phi}$ for all fields $\phi$. Acting on the
phase space $\bar \cF$ with the exact reducibility parameters, one
then generates a set of fields $\cF$ that are then constrained to
admit finite, integrable and conserved charges. The algebra of
gauge transformations that leaves invariant this phase space and
that admits non-identically vanishing charges is then defined as
the asymptotic symmetry algebra, which includes the exact
reducibility algebra $e^{as}_{\phi}$.

The hereby presented method was successfully used in the context of
asymptotically anti-de Sitter spacetimes in general
relativity~\cite{Henneaux:1985tv,Henneaux:1985ey}. In three
dimensions, the asymptotic symmetry algebra was found to be the
conformal algebra containing two copies of the Virasoro algebra,
see section~\ref{sec:adS} of Chapter~\ref{chap:asymptanalyses}.

A great advantage of this method is the simplicity of the argument
and the rapidity of the computation. However, allowable
configurations not generated by the exact reducibility parameters
may exist, see e.g. section~\ref{sec:adS} of
Chapter~\ref{chap:asymptanalyses}, and asymptotic symmetries explicitly
depending on $\phi$ may also be relevant, see e.g.
G\"odel spacetimes in section~\ref{sec:Godel} of
Chapter~\ref{chap:asymptanalyses}.

It is therefore of interest to find alternative points of
departure for defining $\cF$ and $\cA_\phi$ in order to check the
generality of the boundary conditions and of the asymptotic
symmetries. An alternative method, applied in
Chapter~\ref{chap:asymptanalyses}, goes as follows.

\subsection{Starting from the reducibility equations}
\label{sec:algo}

One considers a particular background solution $\bar\phi$ to the
Euler-Lagrange equations of motion. The idea is to define the Lie
algebra $\cA_{\bar \phi}$ of allowable gauge transformations at
$\bar \phi$ before defining the space of asymptotic fields. One
then constructs fields $\cF$ admitting an isomorphic Lie algebra
of gauge transformations. Eventually, one restricts the phase
space so that all conditions described in
section~\ref{sec:phasespace} hold.

\paragraph{(A) Determination of the algebra $\cA_{\bar \phi}$.} We
proceed in three steps. First, (A1) the reducibility equations are
solved to leading order at the background $\bar \phi$. Next, (A2)
one requires the expression~\eqref{eq:16} to be a finite constant
and (A3) one finally imposes that the Lie bracket of two such
parameters also fulfils the latter conditions.

The first condition is an adaptation of the exact symmetry
equations~\eqref{eq:exactakv} in the asymptotic context. Likewise,
in pure gravity, asymptotic Killing vectors can be defined as
vectors fields obeying the Killing equations to ``as good an
approximation as possible'' as one approaches the
boundary~\cite{Wald:1984rg}. The second condition expresses
finiteness~\eqref{eq:105} and conservation~\eqref{conserveda} of
the one-forms evaluated on the background in the particular case
where $\dv\phi$ is $\bar R_{f^\prime}$. In fact, this condition
expresses the only constraints on finiteness and conservation that
one can impose at this stage. The third condition simply ensures
that the gauge parameters form a Lie algebra.

More precisely, we first expand the gauge parameters $f^\alpha(x)$
in $r$ as
\begin{eqnarray}
f^\alpha(x) = {\chi^\alpha(r)} \tilde f^\alpha(y^a)+o({
\chi^\alpha}),\label{eq:3a}
\end{eqnarray}
for some undetermined $\chi^\alpha(r)$, typically of the form
$r^{-m_\alpha}$, with $m_\alpha$ allowed to be $+\infty$ and with
$\tilde f^\alpha(y^a)$ not identically vanishing. If $\bar R^i_f$
denote the gauge transformations at $\bar \phi^i(x)$, $\bar R^i_f
= O(\rho^i)$, where $\rho^i(r)$ depend on the still undetermined
$\chi^\alpha$. One then solves
\begin{equation}
\bar R^i_f= o(\rho^i),\label{def_akV2}
\end{equation}
with the slowest decreasing $\chi^\alpha$ or, in other words, the
highest order in $r$. In general, the slowest decreasing
$\chi^\alpha$ are not uniquely defined and some choice may be
necessary. This choice can and should be done in such a way that
all exact reducibility parameters at $\bar \phi$ are also gauge
parameters. The first step of the procedure thus determines the
fall-offs $\rho^i(r)$, and restricts the form of the leading order
components $\tilde f^\alpha(y^a)$ of the gauge parameters at $\bar
\phi$.

Further constraints are then set by equations~\eqref{eq:105}
and~\eqref{conserveda} evaluated on the background $\bar \phi$ and
for $\dv\phi$ contracted with $\bar R_{f^\prime}$. For
the algebra of gauge parameters to be well-defined, i.e.
$[f_1,f_2]|_{\bar\phi} \in \cA_{\bar\phi}$ for all $f_1$, $f_2 \in
\cA_{\bar\phi}$, one has in general to specify subleading terms,
\begin{equation}
f^\alpha= \tilde f^\alpha(y^a) \chi^\alpha(r) +
f^\alpha_{alg}(r,y^a) + o(f^\alpha_{alg}) .\label{asK_final}
\end{equation}
These subleading terms will as a general law functionally depend on the
leading functions $\tilde f^\alpha$ but additional functions
independent of $\tilde f^\alpha$ may also appear.

\paragraph{(B) First determination of $\cF$ and $\cA_\phi \approx \cA_{\bar
\phi}$.} The other aspect in defining the asymptotic structure is
the definition of the boundary conditions on the fields. We start
the construction of $\cF$ by imposing that the algorithm described in (A) applied on $\phi$ in place of $\bar
\phi$ leads to the same constraints on the gauge parameters. As a
result of this construction, we will have an isomorphism $\cA_\phi
\approx \cA_{\bar \phi}$.

More precisely, we define the following three steps. First, for
gauge parameters of the form~\eqref{eq:3a} with $\chi^\alpha$ and
$\tilde f^\alpha$ arbitrary, we select fields $\phi^i$ such that
\begin{eqnarray}
  \label{eq:18a}
  R^i_f[\phi]=O(\rho^i).
\end{eqnarray}
with $\rho^i$ determined at $\bar\phi$ and such that the only
solutions $\tilde f^\alpha$ to
\begin{eqnarray}
  \label{eq:18b}
  R^i_f[\phi]=o(\rho^i),
\end{eqnarray}
be given by the solutions $\tilde f^\alpha(y^a)$ determined in the
previous paragraph.

Now, if one starts the procedure of the previous paragraph with
any of the $\phi$'s just found, one might find fall-offs
$\chi^\alpha$ which decrease more slowly than those determined at
$\bar\phi$. Let us therefore, as a second step, select the
fields which lead exactly to the same fall-offs $\chi^\alpha$ as initially
obtained. Since these fields also satisfy
\eqref{eq:18a}-\eqref{eq:18b}, they lead to the previously obtained solutions~\eqref{eq:3a}.

As a third step, one imposes equations~\eqref{eq:105}
and~\eqref{conserveda} evaluated on $\phi$ and for $\dv\phi$
contracted with $R_{f^\prime}$. Finally, as the constraints on the
Lie algebra do not depend on $\bar \phi$, they are imposed in the
same way for $\phi$ and we have constructed the phase space $\cF$
such that $\cA_\phi \approx \cA_{\bar \phi}$.

\paragraph{(C) Restrictions on $\cF$ and $\cA_\phi$.}

As final step, we impose all conditions~\eqref{eq:condRf},
\eqref{eq:105}, \eqref{eq:92a}, \eqref{conserveda} and
\eqref{eq:109} on both $\cA_\phi$ and $\cF$. We choose to
implement these constraints in such a way as to keep all elements
of $\cA_\phi$ that are associated with non-identically vanishing
charges. We choose to restrict the subleading terms in $f$ that
lead to vanishing charges~\eqref{eq:107} prior to restrictions on
the fields.

For a solution $\phi^s \in \cF^s$, the asymptotic symmetry algebra
is obtained as the quotient
$e^{as}_{\phi^s}=\cA_{\phi^s}/\cN_{\phi^s}$ of gauge
transformations at $\phi^s \in \cF^s$ by the ideal $\cN_{\phi^s}$
of proper gauge transformations.

\paragraph{Discussion. }
A distinctive feature of this algorithm is that it does not
require exact solutions of the equations of motion (except the
starting point $\bar \phi$) in order to construct the phase space.
An other one is that starting from any field $\phi \in \cF$, one
will recover exactly the same phase space $\cF$ in the end. It is
not necessary to start the procedure with a highly symmetric
background $\bar \phi$ since the exact reducibility equations are
never used.

This approach however has a major shortcoming which is the
non-geometrical nature of the first condition (A1). This condition
may depend on the way to approach the boundary, i.e. on the
coordinates near the boundary. Moreover, condition (A1) is not
necessary to define the phase space which is truly defined by the
conditions of section~\ref{sec:phasespace}. Nevertheless, in
practice, the method is very powerful. We will show in
Chapter~\ref{chap:asymptanalyses} how the algorithm allows
 to study asymptotically anti-de Sitter spacetimes,
asymptotically flat spacetimes at null infinity and G\"odel
spacetimes. It is noteworthy that all these asymptotic structures
in general relativity may be handled by this unifying method.

%% file: asymptotic_analyses.tex
\chapter{Asymptotic analyses in three dimensional gravity}
\label{chap:asymptanalyses}
\setcounter{equation}{0}\setcounter{figure}{0}\setcounter{table}{0}

A successful approach to certain aspects of quantum gravity has
been the study of lower-dimensional gravity, see
e.g.~\cite{Carlip:1998uc} for a review. Three-dimensional gravity
was first classically analyzed in the eighties by Deser, Jackiw
and 't Hooft~\cite{Deser:1983tn,Deser:1984dr}. In the nineties, a
black hole solution, the so-called B(H)TZ black hole, was found in
gravity with negative cosmological
constant~\cite{Banados:1992wn,Banados:1992gq}. It was therefore
understood that three-dimensional gravity may be used as a simpler
setting to investigate intricate issues such as black hole
entropy, see e.g. the
reviews~\cite{Carlip:2005zn,Carlip:1998qw,Banados:1998sm}.

In particular, Strominger's derivation of BTZ black hole entropy
exactly reproduces the geometrical Bekenstein-Hawking entropy
\cite{Strominger:1997eq}. This semi-classical computation
essentially relies on two earlier works: one by Brown and Henneaux
\cite{Brown:1986nw}, who showed that the canonical realization of
asymptotic symmetries of $adS_3$ is represented by two Virasoro
algebras with non-vanishing central charge, and another by Cardy
et al. \cite{Bloete:1986qm,Cardy:1986ie} who derived the so-called
Cardy formula which allows to count in the semi-classical limit
the asymptotic density of states of a conformal field theory, even
if the full details of the theory are not known. It turns out that
application of the Cardy formula with the anti-de Sitter central
charge yields the expected number of states of the BTZ black hole
even if a precise description of the microscopic states of these
black holes is still missing so far~\cite{Carlip:2005zn}.

In this chapter, we will try to broader the scope of Strominger's
reasoning by a deeper analysis of the asymptotic structure of
three-dimensional spacetimes.

First, we re-analyze asymptotically anti-de Sitter spacetimes
along the lines of the algorithm developed in
section~\ref{sec:algo} of Chapter~\ref{chap:asymptcharges}. The
charge algebra consisting in two copies of the Virasoro algebra
will be recovered but more general metrics than
in~\cite{Brown:1986nw} will be found. The link with the Chern
Simons formalism will be shown.

Second, we will derive the symmetry algebra of asymptotically flat
spacetimes at null infinity. In three dimensions, this algebra is
the semi-direct sum of the infinitesimal diffeomorphisms on the
circle with an abelian ideal of supertranslations. The associated
charge algebra will be shown to admit a non trivial classical
central extension of Virasoro type closely related to that of the
anti-de Sitter case.

We will finally consider Einstein-Maxwell theory with Chern-Simons
term in (2+1) dimensions. We will define an asymptotic symmetry
algebra for the G\"odel spacetimes discussed in
section~\ref{sec:threedbh} of Chapter~\ref{chap:BHapp} which will
turn out to be the semi-direct sum of the diffeomorphisms on the
circle with two loop algebras. A class of fields admitting this
asymptotic symmetry algebra and leading to well-defined conserved
charges will be found. The covariant Poisson bracket of the
conserved charges will then be shown to be centrally extended to
the semi-direct sum of a Virasoro algebra and two affine algebras.
The subsequent analysis of three-dimensional G\"odel black holes
indicates that the Virasoro central charge is negative.

All analytical expressions relevant for Einstein gravity can be
found in section~\ref{sec:general-relativity} of
Chapter~\ref{chap:matter}. The expressions for the charges
specialized to three dimensions were also stated in
section~\ref{conschar} of Chapter~\ref{chap:BHapp}.

\section{Asymptotically anti-de Sitter spacetimes} \label{sec:adS}

The anti-de Sitter asymptotic symmetry groups in 3, 4 and $n$
dimensions were extensively studied in
\cite{Brown:1986nw,Henneaux:1985tv,Henneaux:1985ey,Barnich:2003qn}.
For dimensions $n > 3$, non-trivial asymptotic Killing vectors are
in one-to-one correspondence with the exact Killing vectors of the
anti-de Sitter metric and the asymptotic symmetry algebra is
$so(2,n-1)$. In three dimensions, the exact algebra is enhanced in
the asymptotic context to the infinite-dimensional conformal
algebra containing two copies of the Virasoro algebra. This fact
is relevant in the context of the ${\rm adS}_3/{\rm CFT}_2$
correspondence \cite{Aharony:1999ti} and was used to give a
microscopical derivation of the Bekenstein-Hawking entropy for
black holes with near horizon geometry that is locally ${\rm
adS}_3$ \cite{Strominger:1997eq}. The analysis of the asymptotic
charge algebra was subsequently performed in the context of
asymptotically de Sitter spacetimes at timelike infinity
\cite{Strominger:2001pn} with results very similar to those
of the anti-de Sitter case.

In what follows, the algorithm developed in section~\ref{sec:algo}
of Chapter~\ref{chap:asymptcharges} is applied to derive the
asymptotic algebra and the space of asymptotic fields in the
three-dimensional case. As a result, the conformal algebra will be
recovered but more general metrics than developed in
\cite{Brown:1986nw} will be found. Our boundary conditions will
also be expressed in the Chern Simons formalism.

\subsection{Phase space, diffeomorphisms and asymptotic symmetry
algebra.}

In global coordinates, the background three dimensional anti-de
Sitter metric is written as
\begin{equation}
\bar {ds}^2 =
-(1+\frac{r^2}{l^2})dt^2+\frac{1}{(1+\frac{r^2}{l^2})} dr^2+r^2 d
\theta^2 ,\label{adS_metric}
\end{equation}
and the boundary is located at $r=constant \rightarrow +\infty$.
The first step (A1) of the algorithm described in
section~\ref{sec:algo} of Chapter~\ref{chap:asymptcharges} leads
easily to the vectors~\cite{Barnich:2003qn}
\begin{eqnarray}
\xi = (l T(t,\theta)+o(r^{0}))\Q{}{t}
-(r\Theta_{,\theta}(t,\theta)+o(r))\Q{}{r}+(\Theta(t,\theta)
+o(r^{0}))\Q{}{\theta},\label{akV_ads3}
\end{eqnarray}
where $lT_{,t} = \Theta_{,\theta}$ and $l\Theta_{,t}=
T_{,\theta}$. As step (A2), the central charge~\eqref{G_cbis} is
found to be finite for all vectors of the form,
\begin{eqnarray}
\xi = (l T(t,\theta)\hspace{-1pt}+\hspace{-1pt}O(r^{-1}))\Q{}{t}
-(r\Theta_{,\theta}(t,\theta)\hspace{-1pt}+\hspace{-1pt}O(r^0))\Q{}{r}+(\Theta(t,\theta)
\hspace{-1pt}+\hspace{-1pt}O(r^{-1}))\Q{}{\theta}.\label{akV2_ads3}
\end{eqnarray}
The central charge then becomes
\begin{eqnarray}
\cK_{\xi,\xi^\prime} = \frac{l}{8\pi G} \int_{0}^{2\pi}d\theta
\left( T_{,\theta} \Theta^\prime_{,\theta\theta} -
T^\prime_{,\theta} \Theta_{,\theta\theta}\right),\label{cc_ads}
\end{eqnarray}
which is the covariant analogue of~\cite{Brown:1986nw} found
in~\cite{Terashima:2001gb,Barnich:2001jy}. The step (A3) is
trivial because the algebra of these vectors is well-defined.
Therefore, the general form of admissible infinitesimal
diffeomorphisms $\xi \in \cA_{\bar g}$ is given
by~\eqref{akV2_ads3}.

The space of asymptotic metrics is firstly defined by condition
(B). The largest class of metrics satisfying these conditions is
given by
\begin{equation}\eqalign{
&g_{tt} = -C_{tt}\frac{r^2}{l^2}+o(r^2), \qquad g_{tr} =
O(r^{-1}), \qquad
g_{t\theta}=o(r^2),\\
&g_{rr} = C_{rr}\frac{l^2}{r^2}+o(r^{-2}), \qquad
g_{r\theta}=O(r^{-1}),\qquad
g_{\theta\theta}=C_{tt}r^2+o(r^2),}\label{cond_ads_g}
\end{equation}
where $C_{tt}$, $C_{rr}$ are constants. The gauge transformations
$\xi \in \cA_{g}$ are defined as~\eqref{akV2_ads3}.

Let us turn to step (C). We first impose that the surface charges
be finite off-shell. Boundary conditions compatible with the
equations of motion $C_{rr} \approx 1$, $\dv^s C_{tt} \approx 0$,
$g_{t\theta} \approx O(r^0)$, $r^4 g_{rr}+l^2
g_{\theta\theta}-2l^2 r^2 \approx O(r^0)$ and with the adS
background are given by
\begin{eqnarray}
g_{tt} &=& -\frac{r^2}{l^2}-\frac{r}{l^2}g_1(t,\theta) + O(r^0),
\qquad g_{tr} = O(r^{-1}),\nonumber\\
g_{t\theta}&=&O(r^0), \qquad g_{rr} =
\frac{l^2}{r^2}-\frac{l^2}{r^3}g_1(t,\theta)+O(r^{-4}),\label{cond_ads_gfinal} \\
g_{r\theta}&=&O(r^{-1}),\qquad g_{\theta\theta}=r^2+g_1(t,\theta)
r +O(r^0).\nonumber
\end{eqnarray}
The gauge transformations~\eqref{akV2_ads3} are tangent to the
phase space determined by \eqref{cond_ads_gfinal} if one further
restricts the subleadings of the gauge transformations as
\begin{eqnarray}
\xi = (l T(t,\theta)\hspace{-1pt}+\hspace{-1pt}O(r^{-2}))\Q{}{t}
-(r\Theta_{,\theta}(t,\theta)\hspace{-1pt}+\hspace{-1pt}O(r^{0}))\Q{}{r}+(\Theta(t,\theta)
\hspace{-1pt}+\hspace{-1pt}O(r^{-2}))\Q{}{\theta}.\label{akV3_ads3bis}
\end{eqnarray}
This is the final form of the gauge transformations $\xi\in
\cA_g$. With the boundary conditions~\eqref{cond_ads_gfinal}, the
charge one-forms are also integrable off-shell. The surface
charges~\eqref{def_Q} can then be written as
\begin{eqnarray}
\cQ_{\xi}[g,\bar g] &=& \oint_{S^{\infty,t}} k_\xi [g-\bar g,\bar
g]+
N_{\xi}[\bar g]\nonumber\\
&& + \oint_{S^{\infty,t}}\frac{1}{16\pi G l^3} \left[ r^2 l^2
g_{tr}^2 -r^2 g_{r\theta}^2 - \frac{5}{4}l^2 g_1(\theta)^2 \right]
T(t,\theta) ,\label{ch_offshellads}
\end{eqnarray}
where the first terms on the r.h.s would be obtained by a naive
calculation from the linear analysis and the last term is a
non-linear contribution. The surface charge is given on-shell by
the expression
\begin{eqnarray}
\cQ_{\xi}[g,\bar g] &=& \cQ_{\xi}[g] - \cQ_{\xi}[\bar g] +
N_{\xi}[\bar g],\\
\cQ_{\xi}[g] &\approx &\oint_{S^{\infty,t}}\frac{1}{16\pi G}
\bigg[ T\left( \frac{1}{l}g_{\theta\theta} + l g_{tt} +
\frac{r}{l}\d_\theta
g_{r\theta}+ l r \d_t g_{tr} \right) \nonumber\\
&&+ \Theta \left( r \d_\theta g_{tr} + r \d_t g_{r\theta} + 2
g_{t\theta} \right) \bigg].\label{charges_ads_onshell}
\end{eqnarray}
We have $\cQ_{\xi}[\bar g] = 0$ for all $\xi$ except
$\cQ_{\d_t}[\bar g] = -\frac{1}{8G}$. Incidently, these values
correspond to the normalization of the anti-de Sitter background
obtained by supersymmetry arguments~\cite{Coussaert:1993jp} and
which are relevant for the microscopic explanation of the entropy
of BTZ black holes~\cite{Strominger:1998eq}.

The conservation in time of these charges follows from $W^r[\dv^s
\phi,\dv^s \phi]|_{\phi^s} = o(r^{0})$. The phase
space~\eqref{cond_ads_gfinal} also satisfies
\begin{equation}
\oint_{S^{r,t}} E_{\cL^{EH}}[\dv \phi,\dv\phi] = O(r^{-1}),
\end{equation}
which implies~\eqref{eq:109}. As can be seen
in~\eqref{charges_ads_onshell}, proper coordinate transformations
or proper diffeomorphisms $\xi^{prop} \in \cN_{g}$ consist in all
infinitesimal gauge transformations $\xi \in \cA_{g}$ admitting
vanishing functions $T(t,\theta)$ and $\Theta(t,\theta)$. The
asymptotic Killing vectors are defined by the quotient
$\cA_{g}/\cN_{g}$. The asymptotic Killing vectors are generated by
the two sets $\{\xi^{(1)}_n\}$ and $\{\xi^{(2)}_n\}$ given by
\begin{equation}
T^{(1)}_n = \Theta^{(1)}_n =
\frac{1}{2i}e^{in(\frac{t}{l}+\theta)},\qquad T^{(1)}_n = -
\Theta^{(2)}_n
 = \frac{1}{2i}e^{in(\frac{t}{l}-\theta)}.\label{genads}
\end{equation}
They define two independent Witt algebras\footnote{The two sets of
generators correspond to $T^{\pm}$ in \cite{Strominger:1998eq}
where the normalization factor $\frac{1}{2i}$ should be added to
obtain the correct normalization of the Witt algebra.}
\begin{equation}
[\xi^{(a)}_m,\xi^{(b)}_n] = (m-n)\xi^{(a)}_{n+m}\delta^{(a)(b)},
\qquad \forall \,a, b = 1, 2.
\end{equation}
According to Theorem~\ref{lem5}, the two copies of the Witt
algebra are represented at the level of conserved charges by two
copies of the Virasoro algebra with central charge $c= 3l/2G$ as
can be checked by plugging~\eqref{genads} into~\eqref{cc_ads}.

\subsection{Link with previous boundary conditions and with the
Chern-Simons formulation}

The fall-off conditions defining asymptotically adS metrics were
found in \cite{Brown:1986nw} by acting on the conic geometry
representing a spinning particle in adS with the exact anti-de
Sitter symmetry group as described in section~\ref{sec:algoH} of
Chapter~\ref{chap:asymptcharges}. The result was given by
\begin{equation}\eqalign{
g_{tt} &= -\frac{r^2}{l^2}+O(r^0), \qquad g_{tr} = O(r^{-3}),
\qquad
g_{t\theta}=O(r^0),\\
g_{rr} &= \frac{l^2}{r^2}+O(r^{-4}), \qquad
g_{r\theta}=O(r^{-3}),\qquad
g_{\theta\theta}=r^2+O(r^0),}\label{cond_ads_gold}
\end{equation}
Here, we found that the anti-de Sitter phase space can be rather
defined by~\eqref{cond_ads_gfinal} where boundary conditions are
less restrictive. The metric~\eqref{cond_ads_gold} can be obtained
via a gauge fixing of the coordinates close to the boundary by
using the proper gauge transformations generated by $\xi^{prop} =
O(r^{-2})\d_t + O(r^0)\d_r+ O(r^{-2})\d_\theta$. The infinitesimal
diffeomorphisms leaving the metric~\eqref{cond_ads_gold} invariant
are then given by
\begin{eqnarray}
\xi^t & =& l T(t,\theta)+\frac{l^4}{2r^2}\d_\theta\d_t
\Theta(t,\theta)+O(r^{-4}),\\
\xi^r & =& -r\d_\theta \Theta(t,\theta)
+O(r^{-1})\\
\xi^\theta &=&
\Theta(t,\theta)-\frac{l^2}{2r^2}\d_\theta\d_\theta\Theta(t,\theta)
+ O(r^{-4}).
\end{eqnarray}
With the fall-off conditions~\eqref{cond_ads_gold}, the non-linear
terms in~\eqref{ch_offshellads} do not appear. In fact, with the
boundary conditions~\eqref{cond_ads_gold}, the surface one-forms
become asymptotically linear in the sense of~\eqref{AL} and the
surface charges indeed reduce to expression~\eqref{eq:82bis} which
is linear in the metric deviation $h_{\mu\nu} = g_{\mu\nu}-\bar
g_{\mu\nu}$.

Remark that non-linear terms in the charges were also shown to
occur in the context of gravity coupled to scalar
fields~\cite{Henneaux:2004zi,Henneaux:2006hk}.

Three-dimensional gravity with negative cosmological constant can
be reformulated as a Chern-Simons theory with gauge group
$SL(2,\mathbb R) \times SL(2,\mathbb R)$. The boundary
conditions~\eqref{cond_ads_gfinal} can be translated in terms of
the connections $A$ and $\tilde A$\footnote{The connections $A$
and $\tilde A$ are related to the triad $e$ and spin connection
$\omega$ through $A=e+\omega$, $\tilde A = - e + \omega$ with
$\omega^a = -\half \eps^{abc}\omega_{bc}$. We use $\eps^{012}=+1$
and the generators of $\mathfrak{sl}(2,\mathbb R)$ are given by
$T_- = \left(
\begin{array}{cc} 0&0\\1&0\end{array}\right)$, $T_+ = \left(
\begin{array}{cc} 0&1\\0&0\end{array}\right)$ and $T_2 = \left(
\begin{array}{cc} 1/2&0\\0&-1/2\end{array}\right)$ with $x^\pm = t
\pm \theta$.} as
\begin{eqnarray}
\hspace{-20pt}A &\hspace{-5pt}=\hspace{-5pt} &\left(
\begin{array}{cc} \frac{dr}{2r}+O(1)dx^+ & O(r^{-1})dx^+ \\ r dx^+
+\frac{g_1}{2}dx^+ +O(r^{-2})dr & -\frac{dr}{2r}+O(1)dx^+
\end{array}\right) + O(r^{-1})dx^-,\\
\hspace{-20pt}\tilde A &\hspace{-5pt}=\hspace{-5pt}& \left(
\begin{array}{cc} -\frac{dr}{2r}+O(1)dx^- & r dx^-
+\frac{g_1}{2}dx^- +O(r^{-2})dr
\\ O(r^{-1})dx^- & \frac{dr}{2r}+O(1)dx^-
\end{array}\right) + O(r^{-1})dx^+.
\end{eqnarray}
According to \cite{Coussaert:1995zp}, the boundary conditions
imposing that the lightlike components $A_-$ of $A$ and $\tilde
A_+$ of $\tilde A$ are set to zero imply that the Chern-Simons
theory reduces asymptotically to the $SL(2,\mathbb R)$ non-chiral
Wess-Zumino-Witten model. We also have that $A_+^{(-)}$ and
$\tilde A^{(+)}_-$ are independent of $t$ and $\theta$ at leading
order in $r$ but contrary to the boundary conditions imposed in
\cite{Brown:1986ed,Coussaert:1995zp}, the components $A_+^{(3)}$
and $\tilde A_-^{(3)}$ are \emph{not} vanishing at infinity. In
fact, this is due entirely to the slower fall-off conditions on
$g_{tr}$ and $g_{r\theta}$. These boundary conditions probably
allow for a boundary theory more general than the Liouville theory
on a flat background.

Using the Fefferman-Graham-Lee theorems, the boundary conditions
of~\cite{Coussaert:1995zp} were already improved
in~\cite{Rooman:1999km,Rooman:2000ei}. The resulting boundary
theory was found to be Liouville theory on a two-dimensional
\emph{curved} background. Although we have not compared in detail
our results with theirs, we expect that our boundary conditions
are mainly a reformulation of the conditions derived
in~\cite{Rooman:1999km,Rooman:2000ei}\footnote{We thank M. Banados
for his judicious comments.}.

\section{Asymptotically flat spacetimes at null infinity}
\label{sec:nullinfinity}

For asymptotically flat-spacetimes, the appropriate boundary from
a conformal point of view is null infinity \cite{Witten:2001kn}.
The asymptotic symmetry algebra has been derived a long time ago
in four dimensions \cite{Bondi:1962px,Sachs:1962wk,Sachs:1962aa}
and more recently by conformal methods \cite{Penrose:1962ij} also
in three dimensions \cite{Ashtekar:1996cd}.

The purpose of this section is to complete the picture for
classical central charges in three dimensions. We begin by
computing the symmetry algebra $\mathfrak{bms_n}$ of
asymptotically flat spacetimes at null infinity in $n$ dimensions,
i.e., the $n$-dimensional analog of the four dimensional
Bondi-Metzner-Sachs algebra, by solving the Killing equations to
leading order according to the procedure outlined in
section~\ref{sec:algo} of Chapter~\ref{chap:asymptcharges}.

In four dimensions, we make the obvious observation that the
asymptotic symmetry algebra can be larger than the one originally
discussed in \cite{Sachs:1962aa} if the conformal transformations
of the $2$-sphere are not required to be globally well-defined.
In three dimensions, we recover the known
results~\cite{Ashtekar:1996cd}: $\mathfrak{bms_3}$ is the
semi-direct sum of the infinitesimal diffeomorphisms on the circle
with the abelian ideal of supertranslations.

In three dimensions, we then derive the space of allowed metrics
by following the algorithm presented in section~\ref{sec:algo} of
Chapter~\ref{chap:asymptcharges}, namely by requiring (i) that
$\mathfrak{bms_3}$ be the symmetry algebra for all allowed
metrics, (ii) that the asymptotic symmetries leave the space of
allowed metrics invariant, (iii) that the associated charges be
finite, integrable and conserved on-shell. As a new result, the
associated Poisson algebra of charges is shown to be centrally
extended. A non trivial central charge of Virasoro type with value
$c=\frac{3}{G}$ appears between the Poisson brackets of the
charges of the two summands. To conclude our analysis we point out
that the centrally extended asymptotic charge algebras in flat and
anti-de Sitter spacetimes are related in the same way than their
exact counterparts~\cite{Witten:1988hc}.

Most of the material here was published in~\cite{Barnich:2006av}
but, here, the assumption of asymptotic linearity~\eqref{AL} is
relaxed and more general boundary conditions are computed. Related
recent work on holography in asymptotically flat spacetimes can be
found for example
in~\cite{deBoer:2003vf,Arcioni:2003xx,Arcioni:2003td,
Dappiaggi:2005ci,Astefanesei:2005ad,Mann:2005yr,Astefanesei:2006zd}.
We stress, however, that none of these references mentioned the
central extension occurring in the representation of the
$\mathfrak{bms_3}$ algebra.

\subsection{The $\mathfrak{bms_n}$ algebra} \label{sec:bmsalg}

Introducing the retarded time $u=t-r$, the luminosity distance $r$
and angles $\theta^A$ on the $n-2$ sphere by $x^1= r\cos
\theta^1$, $x^A = r\sin \theta^1\dots\sin \theta^{A-1}\cos
\theta^A$, for $A=2,\dots,n-2$, and $x^{n-1}=r\sin \theta^1\dots
\sin \theta^{n-2}$, the Minkowski metric is given by
\begin{equation}
d\bar s^2=-du^2-2dudr
+r^2\sum_{A=1}^{n-2}s_A(d\theta^A)^2,\label{metric_flat}
\end{equation}
where $s_1 = 1$, $s_A = \sin^2 \theta^1\dots \sin^2\theta^{A-1}$
for $2\leq A\leq n-2$. The (future) null boundary is defined by $r
=constant\rightarrow \infty$ with $u,\theta^A$ held fixed.

We require infinitesimal diffeomorphisms to satisfy the Killing
equation to leading order. They have the form $\xi^\mu= \chi^\mu
\tilde\xi^{(\mu)}(u,\theta)+o(\chi^\mu)$ for some fall-offs
$\chi^\mu(r)$ to be determined. Here, round brackets around a
single index mean that the summation convention is suspended. For
such vectors, $\cL_\xi \bar g_{\mu\nu}= O(\rho_{\mu\nu})$. Solving
the Killing equation to leading order means finding the highest
orders $\chi^{\mu}(r)$ in $r$ such that equation
\begin{equation}
\cL_\xi \bar g_{\mu\nu} = o(\rho_{\mu\nu}),\label{def_akV22}
\end{equation}
admits non-vanishing $\tilde \xi^{\mu}(u,\theta)$ as solutions.
After a straightforward computation (summarized in
Appendix~\ref{app:proofs}.\ref{app:aKv_known}), one finds
\begin{eqnarray}
\xi^u&=& T(\theta^A)+u\d_{1}Y^1(\theta^A)+o(r^0),\nonumber\\
\xi^r&=& -r\d_{1} Y^1(\theta^A)+o(r),\label{asykvf}\\
\xi^A&=& Y^A(\theta^B)+o(r^0),\qquad A=1\dots n-2.\nonumber
\end{eqnarray}
where $T(\theta^A)$ is an arbitrary function on the $n-2$ sphere,
and $Y^A(\theta^A)$ are the components of the conformal Killing
vectors on the $n-2$ sphere. These vectors form a sub-algebra of
the Lie algebra of vector fields and the bracket induced by the
Lie bracket $\hat \xi=[\xi,\xi^\prime]$ is determined by
\begin{eqnarray}
\hat T &=& Y^A\d_{A} T^\prime + T \d_{1} Y^{\prime 1} - Y^{\prime
A}\d_{A} T -T^\prime \d_{1} Y^{ 1}\,, \\
\hat Y^A &=& Y^B\d_{B} Y^{\prime A} -Y^{\prime B}\d_{B}
Y^{A}.\label{bracket}
\end{eqnarray}
It follows that the gauge transformations with $T=0=Y^A$ form an
ideal in the algebra of infinitesimal diffeomorphisms. As will be
justified in the following section, these transformations can be
considered as proper gauge transformations. The quotient algebra
is defined to be the algebra of asymptotic Killing vectors
$\mathfrak{bms_n}$. It is the semi-direct sum of the conformal
Killing vectors $Y^A$ of Euclidean $n-2$ dimensional space with an
abelian ideal of so-called infinitesimal supertranslations. Note
that the exact Killing vectors of $\bar g$, $\xi_\mu
=a_\mu+b_{[\mu\nu]}x^\nu$ give rise to
\begin{eqnarray}
Y^A_E=\frac{1}{
  s_{(A)}}(b_{[i0]}+b_{[ij]}\frac{x^j}{r})\frac{1}{r}\dd{x^i}{y^{A}},\
T_E=-[a_0+a_i\frac{x^i}{r}],
\end{eqnarray}
and belong to $\mathfrak{bms_n}$, so that $\mathfrak{iso(n-1,1)}$
is a subalgebra of $\mathfrak{bms_n}$.

In order to make contact with conformal methods, we just note that
if $\tilde g_{\mu\nu} = r^{-2}\bar g_{\mu\nu}$ is the metric
induced at the boundary $r$ constant,
\begin{equation}
d \tilde s^2= -\frac{1}{r^2} \, du^2+
\sum_{A=1}^{n-2}s_A(d\theta^A)^2,\label{B}
\end{equation}
one can easily verify that $\mathfrak{bms_n}$ is isomorphic to the
Lie algebra of conformal Killing vectors of the boundary
metric~\eqref{B}, in the limit $r\rightarrow \infty$.

For $n> 4$, the asymptotic algebra contains the infinitesimal
supertranslations parameterized by $T(\theta^A)$ and the
$n(n-1)/2$ dimensional conformal algebra of Euclidean space
$\mathfrak{so(n-1,1)}$ in $n-2$ dimensions, isomorphic to the
Lorentz algebra in $n$ dimensions.

In four dimensions, the conformal algebra of the two-sphere is
infinite-dimensional and contains the Lorentz algebra
$\mathfrak{so(3,1)}$ as a subalgebra. It would of course be
interesting to analyze whether central extensions arise in the
charge algebra representation of $\mathfrak{bms_4}$, but we will
not do so here. Note that in the original discussion
\cite{Sachs:1962aa}, the transformations were required to be
well-defined on the $2$-sphere and $\mathfrak{bms_4}$ was
restricted to the semi-direct sum of $\mathfrak{so(3,1)}$ with the
infinitesimal supertranslations. In this case, there are no non
trivial central extensions, see e.g.~\cite{mccarthy:1978}.

In three dimensions, the conformal Killing equation on the circle
imposes no restrictions on the function $Y(\theta)$. Therefore,
$\mathfrak{bms_3}$ is characterized by $2$ arbitrary functions
$T(\theta),Y(\theta)$ on the circle. These functions can be
Fourier analyzed by defining $P_n \equiv \xi(T=\,\exp{i n
\theta},Y=0)$ and $J_n = \xi(T=0,Y=\exp{in\theta})$. In terms of
these generators, the commutation relations of $\mathfrak{bms_3}$
become
\begin{equation}
i[ J_m,J_n] = (m-n) J_{m+n},\qquad  i[ P_m,P_n] =
0,\label{alg_bms}\qquad  i[J_m,P_n] = (m-n) P_{m+n}.
\end{equation}
In other words, the 6 dimensional Poincar\'e algebra
$\mathfrak{iso(2,1)}$ of 3 dimensional Minkowski spacetime is
enhanced to the semi-direct sum of the infinitesimal
diffeomorphisms on the circle with the infinitesimal
supertranslations.

\subsection{Charge algebra representation of  $\mathfrak{bms_3}$}

In order to determine the Poisson algebra representation of
$\mathfrak{bms_3}$ we need to specify the boundary conditions on
the metric $g_{\mu\nu}$ and also the more precise form of the
subleading terms in the infinitesimal diffeomorphisms.

If we want the infinitesimal diffeomorphism algebra to be the same
for all allowed metrics, we need to require that solving the
Killing equation to leading order for $g$ in place of $\bar g$
will lead to the $\xi^\mu$ given in~(\ref{asykvf}). We will also
need $\cL_\xi g_{\mu\nu} = O(\chi_{\mu\nu})$ so that the
transformation $\delta g_{\mu\nu} = \cL_\xi g_{\mu\nu}$ leaves the
space of allowed metrics invariant. These conditions are satisfied
for metrics of the form
\begin{eqnarray}
g_{uu} = O(1), \qquad g_{ur} = -1+ O(r^{-1}), \qquad g_{u\theta} =
O(1),\nonumber\\
g_{rr} = O(r^{-2}),\qquad g_{r\theta} = O(1), \qquad
g_{\theta\theta} = r^2 + O(r),\label{BC_bms}
\end{eqnarray}
and infinitesimal diffeomorphisms defined by
\begin{eqnarray}
\xi^u&=& T(\theta)+u\d_{\theta}Y(\theta)+O(r^{-1}),\nonumber\\
\xi^r&=&-r\d_{\theta} Y(\theta)+O(r^0),\label{bms_complete} \\
\xi^\theta&=&Y(\theta)-
\frac{u}{r}\d_\theta\d_{\theta}Y^{\theta}(\theta)+
\frac{1}{r}f_{sub}^{\theta}(\theta)+O(r^{-2}),\nonumber
\end{eqnarray}
where $f^\theta_{sub}(\theta)$ is an arbitrary function. In
addition, the charges are finite and integrable off-shell if
\begin{equation}
g_{r\theta} = g_1(\theta) +O(r^{-1}).\label{cond_suppl}
\end{equation}
This latter condition is left invariant under the action of the
infinitesimal diffeomorphisms and is thus consistent with the
preservation of the phase space under gauge
transformations~\eqref{eq:condRf}.

With the boundary conditions~\eqref{BC_bms}-\eqref{cond_suppl},
one can check that
\begin{equation}\eqalign{
&W_{\varQ{\cL^{EH}}{\phi}}^r[\dv \phi,\dv \phi] = O(r^{-1}),\qquad
W^t_{\varQ{\cL^{EH}}{\phi}}[\dv \phi,\dv \phi] = o(r^{-1}), \vspace{2pt}\\
&E_{\mathcal L^{EH}}^{tr}[\dv \phi,\dv \phi]=O(r^{-1}),}
\end{equation}
hold. As a consequence, the charges are conserved
on-shell~\eqref{conserveda}, the condition~\eqref{eq:109} hold and
the surface one-forms~\eqref{ch_grav1} agree with the ones found
in covariant phase space methods~\eqref{kIW}. Finally, the phase
space of asymptotic metrics $\cF$ and the algebra of infinitesimal
diffeomorphism $\cA_g$ are given by~\eqref{BC_bms},
\eqref{cond_suppl} and~\eqref{bms_complete}.

These boundary conditions contain for example the metric
\begin{equation}
ds^2 = -(1-4m)^2du^2-2 du dr - 8J(1-4m)dud\theta -
\frac{8J}{1-4m}drd\theta+(r^2-16J^2)d\theta^2,\label{kerr3}
\end{equation}
which describes a spinning particle in Minkowski spacetime
\cite{Deser:1983tn}. The space of allowed metrics also contains
the dimensional reduction of the Einstein-Rosen waves from four to
three dimensions \cite{Ashtekar:1996cm}, for which the metric at
infinity in a suitable coordinate system is given by $g_{uu} =
O(1)$, $g_{ur}=-1+O(r^{-1})$, $g_{\theta\theta} = r^2$, the others
coefficients zero. The boundary conditions~\eqref{BC_bms} are
larger than the one used in~\cite{Ashtekar:1993ds,Marolf:2006xj}
except for the $g_{rr}$ coefficient which is allowed to fall-off
as $O(r^{-1})$ in their work.

The surface charges~\eqref{def_Q} are given by
\begin{eqnarray}
\cQ_{\xi}[g,\bar g] &=& \cQ_{\xi}[g] - \cQ_{\xi}[\bar g] +
N_{\xi}[\bar g],\\
Q_\xi[g] &=& \frac{1}{16\pi G}\int_{0}^{2\pi}d\theta\,\Big(
(g_{uu}+r^{-1}\d_u g_{\theta\theta})T+\big(2g_{u\theta} +r \d_u
g_{r\theta} \label{eq:chbms}\\
&& \hspace{-25pt}-\d_\theta(r g_{ur}+u g_{uu})+r^{-1}\d_\theta
(g_{\theta\theta}-u\d_u g_{\theta\theta})+2 \d_\theta^2 g_1 -
r^{-1}g_1 \d_u g_{\theta\theta}\big)Y \Big).\nonumber
\end{eqnarray}
We have $\cQ_\xi[\bar g] = 0$ for all $\xi \in \cA_g$ except
$\cQ_{\d_u}[\bar g] = -\frac{1}{8G}$.

From~\eqref{eq:chbms}, it is clear that the infinitesimal
diffeomorphisms~\eqref{bms_complete} admitting $T(\theta) =
Y(\theta) = 0$ are proper gauge transformations, according to
definition~\eqref{eq:107}. The algebra of asymptotic Killing
vectors in thus correctly given by $\mathfrak{bms}_3$, as assumed
in section~\ref{sec:bmsalg}.

We can also see from~\eqref{eq:chbms} that if we impose the
additional condition
\begin{equation}
g_{\theta\theta}=r^2+g_2(\theta)r+O(r^0)
\end{equation}
on the phase space, which is compatible with the solutions of
interest expressed in~\eqref{kerr3} and the paragraph
below~\eqref{kerr3} and with~\eqref{eq:condRf}, the surface charge
$\cQ_\xi[g,\bar g]$ then equals to $\cQ_\xi[g,\bar g] =
\cQ_\xi[g-\bar g]+N_\xi[\bar g]$. The surface charge
$\cQ_\xi[g,\bar g]$ thus become linear in the metric deviation
$h_{\mu\nu} = g_{\mu\nu}-\bar g_{\mu\nu}$, which is the
simplification encountered in~(\ref{AL}). This case was considered
 in~\cite{Barnich:2006av}.

The expression \eqref{eq:chbms} allows us to compute the central
extension of the Poisson algebra representation of
$\mathfrak{bms_3}$ by first deriving $k_\xi[\dv g]=\dv \cQ_\xi[g]$
and then replacing $\dv g_{\mu\nu}$ by $\cL_{\xi^\prime} \bar
g_{\mu\nu}$ with $\xi^\prime$ given in (\ref{bms_complete}). The
result is
\begin{eqnarray}
\cK_{\xi,\xi^\prime} &=& \frac{1}{8\pi G}\int_0^{2\pi} d\theta
 \Big[\d_\theta Y^\theta (\d_\theta\d_\theta T^\prime + T^\prime) -
\d_\theta Y^{\prime\theta} (\d_\theta\d_\theta T + T)
 \Big].
\end{eqnarray}
Let us choose the normalization~$N_\xi[\bar g] = 0$ for all $\xi
\in \cA_g$ except $N_{\d_u}[\bar g]$ that we leave unspecified. In
terms of the generators $\cQ_{P_n} = \cP_n, \cQ_{J_n} = \cJ_n$, we
get the centrally extended algebra
\begin{eqnarray}
i\{ \cJ_m,\cJ_n \} &=& (m-n) \cJ_{m+n},\nonumber\\
i\{ \cP_m,\cP_n \} &=& 0, \label{bms_charge}\\
i\{ \cJ_m,\cP_n \}& = &(m-n) \cP_{m+n}+\frac{1}{4G}
m(m^2-1-8GN_{\d_u}[\bar g] )\delta_{n+m}.\nonumber
\end{eqnarray}
It can easily be shown to be non-trivial in the sense that it
cannot be absorbed into a redefinition of the generators. Only the
commutators of generators involving either $\cJ_0,\cJ_1, \cJ_{-1}$
or $\cP_0,\cP_1,\cP_{-1}$ corresponding to the exact Killing
vectors of the Poincar\'e algebra $\mathfrak{iso(2,1)}$ are free
of central extensions.

The algebra~\eqref{bms_charge} has many features in common with
the anti-de Sitter case: it has the same number of generators, and
a Virasoro type central charge. In fact, these algebras are
related in the same way than their exact counterparts
\cite{Witten:1988hc}: if one introduces the negative cosmological
constant $\Lambda = -\frac{1}{l^2}$ and considers
\begin{eqnarray}
i[ {J_m},{J_n}] &=& (m-n) J_{m+n},\nonumber\\
i[ {P_m},{P_n} ] &=& \frac{1}{l^2} (m-n)J_{m+n},
\label{alg_bms2}\\
i[{J_m},{P_n} ] &=& (m-n) P_{m+n},\nonumber
\end{eqnarray}
the $\mathfrak{bms}_3$ algebra~\eqref{alg_bms} corresponds to the
case $l\rightarrow \infty$. For finite $l$, the charges
$\cL_m^\pm$ corresponding to the generators $L_m^{\pm} =
\frac{1}{2} (\, l P_{\pm m} \pm J_{\pm m})$ form two copies of the
Virasoro algebra,
\begin{eqnarray}
{i} \{ \cL^\pm_m,\cL^\pm_n \} &=& (m-n) \cL^\pm_{m+n}
+\frac{c}{12}m(m^2-1-8G N_{\d_t}[\bar
g])\delta_{n+m},\label{eq:vir}\\
\{ \cL^\pm_m,\cL^\mp_n \} &=& \, 0,
\end{eqnarray}
where $c= \frac{3l}{2G}$ is the central charge for the anti-de
Sitter case, and $\d_t = \d_u$.

In the classical theory of charges developed in this thesis, only
charge differences can be computed. The normalization of the
background are thus left totally arbitrary. One can however invoke
additional arguments in order to fix these normalizations.

Supersymmetry arguments~\cite{Coussaert:1993jp} and results on the
microscopic origin of the entropy of the BTZ black
hole~\cite{Strominger:1997eq} suggest to define the normalization
of the anti-de Sitter spacetime as $N_{\d_t}[\bar g] = -1/8G$. On
the one hand, following the link between~\eqref{eq:vir}
and~\eqref{bms_charge}, one can be given in to temptation to
define the vacuum energy of $2+1$ Minkowski spacetime also as
$-1/8G$ by a continuity argument. On the other hand, covariant
counterterm methods~\cite{Marolf:2006xj}  propose the different
normalization $N_{\d_t}[\bar g] = -1/4G$. This issue deserves
further attention but needs tools that go beyond the scope of this
thesis.

\section{Asymptotically G\"odel spacetimes}
\label{sec:Godel}

A surprising feature of Einstein's general relativity is the fact
that this theory exhibits closed time-like curves. Such
pathological spacetimes include the G\"odel universe
\cite{Godel:1949ga}, the Gott time-machine \cite{Gott:1990zr} and
the region behind the inner horizon of Kerr black holes. Since the
presence of closed time-like curves signals a strong breakdown of
causality, Hawking advocated through his chronology protection
conjecture that ultraviolet processes should prevent such
geometries from forming \cite{Hawking:1991nk}.

The implications of this proposition have been addressed in the
context of string theory in a series of works (see e.g.
\cite{Dyson:2003zn,Israel:2003cx,Johnson:2004zq}, and also
\cite{Costa:2005ej} for an extensive list of references). Also,
higher-dimensional highly supersymmetric G\"odel-like solutions
were found in supergravity \cite{Gauntlett:2002nw,Harmark:2003ud},
indicating that supersymmetry is not sufficient to discard these
causally pathological solutions. Moreover, a particular issue in
the dual description of gravity theories by gauge theories is the
conjecture linking closed time-like curves on the gravity side and
non-unitarity on the gauge theory side
\cite{Herdeiro:2000ap,Caldarelli:2004mz}. It was indeed shown
\cite{Herdeiro:2000ap} in the context of BMPV black holes
\cite{Breckenridge:1996is} that the regime of parameters in which
there exists naked closed timelike curves is also the regime in
which unitarity is violated in the dual CFT. Also, half BPS
excitations in $adS_5\times S^5$ in IIB sugra can be mapped to
fermions configurations \cite{Lin:2004nb}. Causality violation is
shown to be related to Pauli exclusion principle in the dual
theory \cite{Caldarelli:2004mz}.

In this section, we will work out some properties of the G\"odel
black holes derived in section~\ref{sec:threedbh} of
Chapter~\ref{chap:BHapp} through the representation of their
asymptotic symmetries. The theory of interest here will be
(2+1)-dimensional Einstein-Maxwell-Chern-Simons theory, which can
be viewed as a lower-dimensional toy-model for the bosonic part of
$D=5$ supergravity since the field content and the couplings of
both theories are similar. As a main result, published
in~\cite{Compere:2007in}, we will show that the asymptotic
symmetry algebra contains a Virasoro algebra with \emph{negative}
central charge when the generators are chosen to be bounded from
below for the black hole solutions. It indicates that the
representations of the asymptotic symmetry algebra are
non-unitary, in accordance with the
works~\cite{Herdeiro:2000ap,Caldarelli:2004mz}.

Our analysis will present analogies with the one performed in
$adS_3$ space since there is a close relationship between $adS_3$
and $3d$ G\"odel space. Indeed, the latter can be seen as a
squashed $adS_3$ space, where the original isometry group is
broken from $\SL \times \SL$ to $\SL \times U(1)$, as pointed out
in \cite{Rooman:1998xf}. In the context of string theory, the $3d$
G\"odel metric was shown to be part of the target space of an
exact two-dimensional CFT, obtained as an asymmetric marginal
deformation of the $\SL$ WZW model \cite{Israel:2003cx}. In this
case, the effect of the deformation amounts to break the original
$\widehat{\SL} \times \widehat{\SL}$ symmetry of the model down to
$\widehat{\SL} \times \widehat{U(1)}$. As we will show, a similar
pattern will appear at the level of asymptotic symmetries.

After having briefly recalled in section~\ref{sec:gensetup} our
general setup, we will compute in section~\ref{sec_ask} the
asymptotic symmetry algebra of G\"odel spaces. We then define, in
section~\ref{sec:asymptGodel}, a class of field configurations,
which we will refer to as asymptotically G\"odel space-times in
three dimensions, encompassing the previously mentioned black hole
solutions. In section~\ref{sec:PoissonGoedel}, we represent the
algebra of charges by covariant Poisson brackets and show that the
asymptotic symmetry algebra admits central extensions. We conclude
by discussing some of the results.

\subsection{General setup}
\label{sec:gensetup}

Let us start with the Einstein-Maxwell-Chern Simons theory in
$2+1$ dimensions,
\begin{equation}
I = \frac{1}{16 \pi G} \int \, d^3x \,\left[ \sqrt{-g}\left( R +
\frac{2}{l^2} - \frac{1}{4} F^2 \right) -
\frac{\alpha}{2}\eps^{\mu\nu\rho} A_\mu F_{\nu\rho}
\right].\label{action2}
\end{equation}
The gauge parameters of the theory $(\xi,\lambda)$, where $\xi$
generates infinitesimal diffeomorphisms and $\lambda$ is the
parameter of $U(1)$ gauge transformations are endowed with the Lie
algebra structure
\begin{equation}
[(\xi,\lambda),(\xi^\prime,\lambda^\prime)]_{G} =
([\xi,\xi^\prime],[\lambda,\lambda^\prime]),\label{eq:Lie}
\end{equation}
where the $[\xi,\xi^\prime]$ is the Lie bracket and
$[\lambda,\lambda^\prime]\, \equiv \cL_\xi \lambda^\prime -
\cL_{\xi^\prime}\lambda$. We will denote for compactness the
fields as $\phi^i\equiv (g_{\mu\nu},A_\mu)$ and the gauge
parameters as $f^\alpha = (\xi^\mu,\lambda)$. For a given field
$\phi$, the gauge parameters $f$ satisfying
\begin{equation}
\cL_\xi  g_{\mu\nu} \approx 0, \qquad \cL_\xi A_{\mu} + \d_\mu
\lambda \approx 0, \label{eq:red22}
\end{equation}
where $\approx$ is the on-shell equality, will be called the exact
symmetry parameters of $\phi$. Parameters $(\xi,\lambda) \approx
0$ are called trivial symmetry parameters.

The charge one-form for this theory for exact symmetry parameters
was constructed in section~\ref{sec:EM-chern} of
Chapter~\ref{chap:matter} and was rewritten in
section~\ref{conschar} of Chapter~\ref{chap:BHapp}. In the
asymptotic case, the charge one-form may be written as
\begin{equation}
k_{(\xi,\lambda)}[\dv \phi ; \phi]= k^{exact}_{(\xi,\lambda)}[\dv
\phi ; \phi] + k^{s}_{(\cL_\xi g,\cL_\xi A + d\lambda)}[\dv
\phi;\phi],\label{charge_k}
\end{equation}
where $k^{exact}_{(\xi,\lambda)}$ is given by~\eqref{kCS2}. The
supplementary term
 may be deduced from expressions~\eqref{grav_contrib} and
\eqref{Bcharge}\footnote{In~\cite{Compere:2007in}, there is a
minor sign mistake in (2.5) that does not affect the computation
further on. We recall that we use the mostly plus signature.}. It
is given by
\begin{eqnarray}
k^{s}_{(\cL_\xi g,\cL_\xi A + d\lambda)}[\dv \phi;\phi]& =&
\frac{\sqrt{-g}}{32\pi G}\big(\dv g_{\mu \alpha}(D^\alpha \xi_\nu
+ D_\nu \xi^\alpha)\nonumber \\
&& \qquad +\dv A_\mu (\cL_\xi A_\nu + \d_\nu \lambda) \big)
\epsilon^{\mu\nu}_{\,\,\,\,\,\alpha}dx^\alpha
\end{eqnarray}
and vanishes for exact symmetries. The central
charge~\eqref{eq:16} can be expressed here as
\begin{eqnarray}
\cK_{(\xi,\lambda),(\xi^\prime,\lambda^\prime)}[\bar \phi] =
\int_{S^\infty} k_{(\xi^\prime, \lambda^\prime)}[(\cL_\xi \bar
g_{\mu\nu},\cL_\xi \bar A_{\mu} + \d_\mu \lambda);(\bar g,\bar
A)], \label{eq:cc}
\end{eqnarray}
where $\bar \phi$ is a solution we use as background.

One can define an algebra $\cA$ of asymptotic symmetries
$(\xi,\lambda)$ and then a phase space $\cF$ by following the
algorithm presented in section~\ref{sec:algo} of
Chapter~\ref{chap:asymptcharges}. In summary, the asymptotic
algebra is defined by the three conditions:
\begin{itemize}
\item The leading order of the expressions $\cL_\xi \bar  g_{\mu\nu}$ and
$\cL_\xi \bar A_{\mu} + \d_\mu \lambda$ close to the boundary
$S^\infty$ has to vanish.
\item The expression $\cK_{(\xi,\lambda),(\xi^\prime,\lambda^\prime)}[\bar \phi]$
should be a finite constant.
\item The Lie bracket of two such parameters should also satisfy the two previous conditions.
\end{itemize}

\subsection{G\"odel asymptotic symmetry algebra}
\label{sec_ask}

It was shown in section~\ref{sec:threedbh} of
Chapter~\ref{chap:BHapp} that the equations of motion derived
from~\eqref{action2} admit the solution
\begin{eqnarray}\label{GodelMetric}
\bar{ds}^2  &=& \eps dt^2  - 4\alpha r d t d\varphi + (2r -
\frac{2}{l^2}|1-\alpha^2l^2| r^2)d \varphi^2 +
\frac{1}{-2 \eps r+ \Upsilon^{-1}r^2}dr^2  \nonumber\\
\bar A &=&  \frac{2}{l}\sqrt{|1-\alpha^2l^2|}\, r d\varphi,
\label{Sol_backgr}
\end{eqnarray}
where $\eps = \text{sgn}(1-\alpha^2 l^2)$, $\Upsilon =
\frac{l^2}{2(1+\alpha^2l^2)}$ and $\varphi \in [0,2\pi]$. For
$\eps = -1$, this solution is the $3d$ part of the two parameter
generalization \cite{Reboucas:1983hn} of the G\"odel spacetime
\cite{Godel:1949ga} where the stress-energy tensor of the perfect
fluid supporting the metric is generated by the gauge field. For
$\epsilon = +1$, the solution will be called the tachyonic G\"odel
spacetime because the perfect fluid supporting the metric is
tachyonic. We will use this solution as background in the two
sectors of the theory $\eps = \pm 1$.

For $\eps = -1$, the G\"odel solution (\ref{GodelMetric}) admits 5
non-trivial exact symmetries $(\xi,\lambda)$,
\begin{equation}\eqalign{
(\xi_{(1)},0) &=& (\d_t,0) ,\qquad \qquad (\xi_{(2)},0) = (2\alpha \Upsilon \d_t +\d_\varphi,0),\vspace{5pt} \\
(\xi_{(3)},0) &=&(
\frac{2\alpha\Upsilon}{\sqrt{1+2\Upsilon/r}}\sin\varphi \d_t -
\sqrt{2 \Upsilon r+r^2}\cos\varphi \d_r +
\frac{r+\Upsilon}{\sqrt{2\Upsilon r+r^2}}\sin\varphi
\d_\varphi,0),\vspace{5pt}
\\
(\xi_{(4)},0)
&=&(\frac{2\alpha\Upsilon}{\sqrt{1+2\Upsilon/r}}\cos\varphi \d_t +
\sqrt{2 \Upsilon r+r^2}\sin\varphi \d_r +
\frac{r+\Upsilon}{\sqrt{2\Upsilon r+r^2}}\cos\varphi
\d_\varphi,0),\vspace{5pt} \\
(0,\lambda_{(1)}) &=& (0,1).} \label{ask_exact}
\end{equation}
The four Killing vectors form a $ \mathbb{R} \oplus so(2,1)$
algebra. In the case $\eps = +1$, only the two first vectors are
Killing vectors.

Let us now compute the asymptotic symmetries of this background
solution $\bar \phi$. They are of the form
\begin{eqnarray}
\xi &=& \chi_\xi(r)\tilde \xi(t,\varphi)+o(\chi_\xi(r))\nonumber \\
\lambda &=& \chi_\lambda(r)\tilde
\lambda(t,\varphi)+o(\chi_\lambda(r)),\label{form_aKv}
\end{eqnarray}
for some fall-offs $\chi_\xi(r)$, $\chi_\lambda(r)$ and functions
$\tilde \xi(t,\varphi)$, $\tilde \lambda(t,\varphi)$ to be
determined. For such parameters, one has
\begin{equation}
\cL_\xi \bar  g_{\mu\nu} = O(\rho_{\mu\nu}), \qquad \cL_\xi \bar
A_{\mu} + \d_\mu \lambda = O(\rho_\mu), \label{cond_eps}
\end{equation}
where $\rho_{\mu\nu}$ and $\rho_\mu$ depend on the explicit form
of the parameters~\eqref{form_aKv}. Equations~\eqref{cond_eps} are
satisfied to the leading order in $r$ when one imposes $\cL_\xi
\bar  g_{\mu\nu} = o(\rho_{\mu\nu})$ and $\cL_\xi \bar A_{\mu} +
\d_\mu \lambda = o(\rho_\mu)$. If one solves these equations with
the highest order in $r$ for $\chi_\xi(r)$ and $\chi_\lambda(r)$,
one gets the unique solution
\begin{equation}\eqalign{
\xi &=& (F(t,\varphi) +o(r^0)) \d_t +( - r \d_\varphi
\Phi(\varphi)
+o(r^1))\d_r + (\Phi(\varphi)+o(r^0) ) \d_\varphi, \vspace{5pt}\\
\lambda &=& \lambda(t,\varphi) + o(r^0),}\label{asK}
\end{equation}
where $F(t,\varphi)$ and $\Phi(\varphi)$ are arbitrary functions.
We now require the central extension~\eqref{eq:cc} to be a finite
constant. The term diverging in $r$ in~\eqref{eq:cc} vanishes if
we impose $\xi^\varphi = \Phi(\varphi)+o(r^{-1})$. The central
extension is then constant by requiring
\begin{equation}
F(t,\varphi) = F(\varphi),\qquad \lambda(t,\varphi)=
\lambda(\varphi).
\end{equation}
The resulting expression for~\eqref{eq:cc} is given by
\begin{eqnarray}
K_{(\xi,\lambda),(\xi^\prime,\lambda^\prime)}[\bar \phi] &=&
\frac{1}{16\pi G} \int_0^{2\pi} d\varphi \Big[ 2\alpha \Upsilon
\d_\varphi \Phi^\prime \d_\varphi^2\Phi
-\frac{\eps}{2\alpha\Upsilon}\d_\varphi F F^\prime \nonumber \\
&&+ 2\eps \Phi^\prime \d_\varphi F+\alpha\d_\varphi \lambda
\lambda^\prime - ((\xi,\lambda) \leftrightarrow
(\xi^\prime,\lambda^\prime)) \Big].\label{eq:cc2}
\end{eqnarray}
The asymptotic parameters just found form a subalgebra $\cA$ of
the bracket~\eqref{eq:Lie}. The asymptotic parameters which are of
the form
\begin{equation}
\xi = o(r^{0})\d_t +o(r^1)\d_r + o(r^{-1})\d_\varphi, \qquad
\lambda =o(r^{0}),\label{triv}
\end{equation}
will be considered as trivial because (i) they form an ideal of
the algebra $\cA$, (ii) for any $f$ of the form~\eqref{triv} and
$f^\prime \in \cA$, the associated central charge
$K_{f,f^\prime}[\bar \phi]$ vanishes. An additional justification
will be provided in section~\ref{sec:PoissonGoedel}. We define the
asymptotic symmetry algebra $\mathfrak{Godel_3}$ as the quotient
of $\cA$ by the trivial asymptotic parameters~\eqref{triv}. This
algebra can thus be expressed only in terms of the leading order
functions $F(\varphi)$, $\Phi(\varphi)$ and $\lambda(\varphi)$. By
setting $\hat f=[f,f^\prime]_G$, one can write the
$\mathfrak{Godel_3}$ algebra explicitly as
\begin{equation}
\hat F(\varphi) =  \Phi \d_\varphi F^\prime - \Phi^\prime
\d_\varphi F, \qquad \hat \Phi(\varphi) = \Phi \d_\varphi
\Phi^\prime - \Phi^\prime \d_\varphi \Phi,\qquad  \hat \lambda =
\Phi \d_\varphi \lambda^\prime - \Phi^\prime \d_\varphi
\lambda.\label{eq:algGo}
\end{equation}
A convenient basis for non-trivial asymptotic symmetries consists
in the following generators
\begin{eqnarray}
l_n &=& \{ (\xi,\lambda) \in \cA | F(\varphi) = 2\alpha \Upsilon e^{i n \varphi}, \, \Phi(\varphi) = e^{i n \varphi}, \, \lambda(\varphi) = 0 \}, \nonumber\\
t_n &=& \{ (\xi,\lambda) \in \cA | F(\varphi) = e^{i n \varphi}, \, \Phi(\varphi) = 0, \, \lambda(\varphi) = 0 \},\label{gen_back}\\
j_n &=& \{ (\xi,\lambda) \in \cA | F(\varphi) = 0, \,
\Phi(\varphi) = 0, \, \lambda(\varphi) = e^{i n \varphi} \}.
\nonumber
\end{eqnarray}
In terms of these generators, the $\mathfrak{Godel_3}$ algebra
reads
\begin{eqnarray}
i[ {l_m},{l_n}]_G &= &(m-n) l_{m+n},\nonumber\\
i[ {l_m},{t_n} ]_G &=& -n t_{m+n},\label{alg_Godel}\\
i[{l_m},{j_n} ]_G & =& -n j_{m+n},\nonumber
\end{eqnarray}
while the other commutators are vanishing. One can recognize the
exact symmetry parameters~\eqref{ask_exact} as a subalgebra of
$\mathfrak{Godel_3}$. Indeed, one has $t_0 \sim (\xi_{(1)},0)$,
$l_0 \sim (\xi_{(2)},0)$, $l_{-1} \sim (-i\xi_{(3)}+\xi_{(4)},0)$,
$l_1 \sim (i \xi_{(3)}+\xi_{(4)},0)$ and $j_0 \sim
(0,\lambda_{(1)})$ where $\sim$ denote the belonging to the same
equivalence class of asymptotic symmetries.

In $adS_3$, the exact $so(2,2)$ algebra is enhanced in the
asymptotic context to two copies of the Witt algebra. The G\"odel
metric can be interpreted as a squashed $adS_3$ geometry, which
breaks the original $so(2,2)$ symmetry algebra down to $u(1)
\oplus so(2,1) $ \cite{Rooman:1998xf}. The exact Killing symmetry
algebra is here enhanced to a semi-direct sum of a Witt algebra
with a $\widehat{u(1)}$ loop algebra. Moreover, the gauge sector
$u(1)$ is enhanced to another $\widehat{u(1)}$ loop algebra also
forming an ideal of the $\mathfrak{Godel_3}$ algebra.

\subsection{Asymptotically G\"odel fields}
\label{sec:asymptGodel}

We defined in the previous section the asymptotic symmetry algebra
$\mathfrak{Godel_3}$ by a well-defined procedure starting from the
background $\bar \phi$. One can ask which are the field
configurations $\phi$ such that the preceding analysis leads to
the same algebra~\eqref{eq:algGo} with $\bar \phi$ replaced by
$\phi$. The subset of such field configurations which is preserved
under the action of the asymptotic symmetry algebra will then
provide a natural definition of asymptotically G\"odel fields
$\cF$. A set of fields satisfying these conditions is given by
\begin{eqnarray}
g_{tt} &=& \eps+r^{-1}\overset{(1)}{g_{tt}}+O(r^{-2}), \qquad
g_{tr} = O(r^{-2}), \nonumber\\
g_{t\varphi}&=&-2\alpha  r +
\overset{(1)}{g_{t\varphi}}+O(r^{-1}),\qquad g_{rr} =
\frac{\Upsilon}{r^2}+r^{-3}\overset{(1)}{g_{rr}}+O(r^{-4}),
\nonumber\\
g_{r\varphi}&=&r^{-1}\overset{(1)}{g_{r\varphi}}+O(r^{-2}), \qquad
g_{\varphi\varphi} =
-\frac{2}{l^2}|1-\alpha^2l^2|r^2+r^{1}\overset{(1)}{g_{\varphi\varphi}}+O(r^{0}),\nonumber\\
A_t &=& -\frac{\sqrt{(1-\alpha^2l^2)\eps}}{\alpha l
}+r^{-1}\overset{(1)}{A_{t}}+O(r^{-2}),\qquad A_r =
r^{-2}\overset{(1)}{A_r}+O(r^{-3}), \nonumber\\
&&\qquad A_\varphi =
\frac{2}{l}\sqrt{\abs{1-\alpha^2l^2}}r+\overset{(1)}{A_{\varphi}}+O(r^{-1}),\label{BC1}
\end{eqnarray}
where all functions $\overset{(1)}{g_{tt}}$, \dots$\,$ depend
arbitrarily on $t$ and $\varphi$. In order for these field
configurations be left invariant under the asymptotic symmetries,
one has furthermore to restrict the subleading component of
$\xi^\varphi$ to $\xi^\varphi = \Phi(\varphi)+O(r^{-2})$. The
asymptotic symmetries thus become
\begin{eqnarray}
\xi &=& (F(\varphi) +o(r^0)) \d_t +( - r \d_\varphi \Phi(\varphi)
+o(r^1))\d_r + (\Phi(\varphi)+O(r^{-2}) ) \d_\varphi,\nonumber \\
\lambda &=& \lambda(\varphi) + o(r^{0}),
\end{eqnarray}
and always contain the asymptotic form of the exact symmetries
\eqref{ask_exact}.

However, for the purpose of providing a well-defined
representation of the asymptotic symmetry algebra, one has to
restrict the definition of fields $\cF$ by selecting those
satisfying \eqref{eq:105}, \eqref{eq:92a}, \eqref{conserveda} and
\eqref{eq:109}. These conditions are met if the following
differential equation hold,
\begin{equation}\eqalign{
&\overset{(1)}{g_{\varphi\varphi}}-\eps
\Upsilon^{-2}\overset{(1)}{g_{rr}}+4\alpha \eps
+\frac{\eps}{\alpha \Upsilon}\d_t
\overset{(1)}{g_{r\varphi}}\vspace{2pt}
\overset{(1)}{g_{t\varphi}}+\frac{2(\alpha^2l^2-1)}{l^2}\overset{(1)}{g_{tt}}
\\&\qquad \qquad +\frac{2\eps \sqrt{\eps
(1-\alpha^2l^2)}}{\alpha l \Upsilon}(\d_t
\overset{(1)}{A_r}+\overset{(1)}{A_t}) = 0.}\label{BC2}
\end{equation}
We finally define the set of asymptotically G\"odel fields $\phi =
(g,A)$ as those satisfying the boundary conditions~\eqref{BC1} and
\eqref{BC2}. In general, the asymptotic symmetries are allowed to
depend arbitrarily on the fields, $(\xi[g,A],\lambda[g,A])$. They
should however, by construction, obey the same algebra
$\mathfrak{Godel_3}$. A basis for the asymptotic symmetries of
$\phi$ can be written as
\begin{eqnarray}
l_n &=& \{ (\xi,\lambda) \in \cA | F(\varphi) = 2\alpha \Upsilon
f_l[g,A] e^{i n \varphi}, \, \Phi(\varphi) =
e^{i n \varphi}, \, \lambda(\varphi) = 0 \}, \nonumber\\
t_n &=& \{ (\xi,\lambda) \in \cA | F(\varphi) = f_t[g,A] e^{i n
\varphi}, \, \Phi(\varphi) = 0,
\, \lambda(\varphi) = 0 \},\label{gen_gen}\\
j_n &=& \{ (\xi,\lambda) \in \cA | F(\varphi) = 0, \,
\Phi(\varphi) = 0, \, \lambda(\varphi) =f_j[g,A]  e^{i n \varphi}
\}. \nonumber
\end{eqnarray}
where the solution-dependent multiplicative factors $f_l,\;f_t$
and $f_j$ have been added for convenience. We choose $f_l[\bar
g,\bar A] = f_t[\bar g,\bar A]= f_j[\bar g,\bar A]= 1$ in order to
match the asymptotic symmetries~\eqref{gen_back} defined for the
background. The choice of multiplicative factors for generic
fields $(g,A)$ is restricted by the second integrability condition
of~\eqref{eq:92a}.

Note that besides the background itself the asymptotically G\"odel
fields contain the three parameters ($\nu$, $J$, $Q$) particle
$(\eps= -1)$~\eqref{particles} and black hole $(\eps = +1)$
solutions~\eqref{blackhole}\footnote{The solutions written
in~\eqref{particles},~\eqref{blackhole} differ from the solutions
written here by the change of coordinates $r^{here} =
\frac{r^{there}}{\sqrt{|8G\mu^{there}|}}$, $ t^{here} =
\sqrt{|8G\mu^{there}|} t^{there}$, $\nu = 2 \eps
\sqrt{|8G\mu^{there}|}$.}.

\subsection{Poisson algebra}
\label{sec:PoissonGoedel}

We are now ready to represent the asymptotic algebra
$\mathfrak{Godel_3}$ by associated charges in the space of
configurations defined in~\eqref{BC1}-\eqref{BC2}. An explicit
computation shows that the charges associated with each
generator~\eqref{gen_gen} are in general non-vanishing. We denote
these charges by $L_n \equiv \cQ_{l_n}[\phi,\bar \phi]$, $T_n
\equiv \cQ_{t_n}[\phi,\bar \phi]$ and $J_n \equiv
\cQ_{j_n}[\phi,\bar \phi]$. On the contrary, all trivial
asymptotic parameters are associated with vanishing charges and
thus correspond to proper gauge transformations as it should. This
provides additional justification for the quotient
$\mathfrak{Godel_3}$ taken in section~\ref{sec_ask}.

The central extensions~\eqref{eq:cc2} may be explicitly computed
for any pair of generators of the background~\eqref{gen_back}. The
only non-vanishing terms are
\begin{eqnarray}
i K_{l_m,l_n} &=& \frac{c}{12} m(m^2+\eps)\delta_{n+m},\nonumber\\
i K_{t_m,t_n} &=& \frac{\eps}{8G \alpha \Upsilon} m\delta_{m+n,0}.\label{final_cc}\\
i K_{j_m,j_n} &=& -\frac{\alpha}{4G} m \delta_{m+n}.\nonumber
\end{eqnarray}
where the Virasoro-type central charge $c$ reads
\begin{equation} c = -\frac{6\alpha
\Upsilon}{G}= -\frac{3\alpha l^2}{(1+\alpha^2l^2)G}.
\end{equation}
According to Theorem~\ref{lem5} on page~\pageref{lem5}, the
G\"odel algebra is finally represented at the level of charges by
the following centrally extended Poisson algebra
\begin{eqnarray}
i\{L_m,L_n\} &=& (m-n)(L_{m+n}-\cN_{l_{m+n}}) + \frac{c}{12}m(m^2+\eps)\delta_{m+n},\nonumber\\
i\{L_m,T_n\} &=& -n (T_{m+n}-\cN_{t_{m+n}}), \nonumber\\
i\{T_m,T_n\} &=& \frac{\eps}{8G \alpha \Upsilon} m\delta_{m+n},\label{C1} \\
i\{L_{m},J_{n}\} &=& -n(J_{m+n}-\cN_{j_{m+n}}),\nonumber\\
i\{J_{m},J_{n}\} &=& -\frac{\alpha}{4G}m \delta_{m+n}.\nonumber
\end{eqnarray}
The central extensions~\eqref{final_cc} are non-trivial because
they cannot be absorbed into the (undetermined classically)
normalizations of the generators. The $L_n$ form a Virasoro
algebra while the two loop algebras $\{ t_n \}$, $\{ j_n \}$ are
represented by centrally extended $\widehat{u(1)}$ affine
algebras.

\subsection{Discussion}

In $3d$ asymptotically anti-de Sitter spacetimes, the asymptotic
charge algebra which consists of two copies of the Virasoro
algebra \cite{Brown:1986nw} allows one to compute the entropy of
the BTZ black hole via the Cardy formula \cite{Strominger:1997eq}.
One may wonder if an analogous derivation based on the asymptotic
algebra~\eqref{C1} could be performed.

It turns out that the analysis in G\"odel spacetimes is more
tricky. The G\"odel black holes are given in~\eqref{blackhole}. In
this case, the $r$ coordinate has the range $-\infty < r < \infty$
and $\eps = +1$. The solution~\eqref{blackhole} displays an
horizon and therefore describes a regular black hole only if the
inequality
\begin{equation}
2G \nu^2  \geq \frac{J}{2\alpha \Upsilon}\label{eq:bounds}
\end{equation}
holds. The tachyonic G\"odel solution corresponds to $\nu =
+\frac{1}{4G}$, $J=Q=0$. Because the solutions with $\nu$, $J$ and
$Q$ are related by the change of coordinates $r \rightarrow -r$,
$\varphi \rightarrow -\varphi$ with the solutions $-\nu$, $J$,
$-Q$, the conserved quantity $\nu$ associated with $\d_t$ does not
provide a satisfactory definition of mass. However, one can define
the quantity $\mu = 2\eps G\nu^2$ which is by definition positive
for black holes and which equals $-\frac{1}{8G}$ for the G\"odel
background. In particular, in the anti-de Sitter limit $\alpha^2
l^2 \rightarrow 1$, $\mu$ correctly reproduces the mass gap
between the zero mass BTZ black hole and anti-de Sitter space. It
was shown in section \ref{sec:threedbh} of
Chapter~\ref{chap:BHapp} that this quantity is associated with the
Killing vector $4G \eps \nu \d_t$. Note also that $\d_\varphi$ is
associated with $-J + \frac{Q^2}{4\alpha}$.

Choosing the multiplicative factor $f_{l} = 4G \nu$, the charge
associated with the generator $l_0$ of~\eqref{gen_gen} becomes for
the black holes
\begin{equation}
L_0 = 2\alpha \Upsilon \mu - J + \frac{Q^2}{4\alpha} -
\frac{\alpha \Upsilon}{4G} + Q_{l_0}[\bar \phi].
\end{equation}
When $\alpha > 0$, the inequality~\eqref{eq:bounds} imposes that
the spectrum of $L_0$ is bounded from below. The Virasoro
generators $L_n$ may then be associated with operators acting on a
ground state with minimal $L_0$-eigenvalue. When $\alpha <0$, one
may instead consider the generators $L^\prime_n = - L_{-n}$
satisfying also a Virasoro algebra
\begin{eqnarray}
i\{L^\prime_m,L^\prime_n\} &=&
(m-n)(L^\prime_{m+n}-\cN_{l^\prime_{m+n}}) +
\frac{c^\prime}{12}m(m^2+1)\delta_{m+n},
\end{eqnarray}
with $c^\prime = -c = 6\alpha \Upsilon/G$ and for which $
L^\prime_0 = -L_0$ is also bounded from below. Remark that in any
of these two cases, the classical Virasoro central charge ($c$ for
$\alpha >0$ and $c^\prime$ for $\alpha<0$) is negative, which, in
general, implies that the representations of this algebra are
non-unitary. In the anti-de Sitter limit $\alpha^2 \rightarrow
1/l^2$, the central charge tends to minus the usual adS$_3$
central charge $3l/2G$. This indicates a discontinuity in the
limiting procedure.

The Bekenstein-Hawking entropy associated with the black hole
solutions \eqref{blackhole} is given by
\begin{equation}
S_{BH} = 2\pi \sqrt{\alpha \Upsilon G^{-1}(2\alpha \Upsilon \mu-
J)}+2\pi \sqrt{2\alpha^2\Upsilon^2 G^{-1}\mu}.\label{BH2}
\end{equation}
Let us consider without loss of generality the case $\alpha >0$
and define $\Delta_0$ as the value of $L_0$ for the zero mass
black hole $\mu=J=Q=0$, $\Delta_0 = -\alpha
\Upsilon/(4G)+\cQ_{l_0}[\bar \phi]$. We observe that the first
term in~\eqref{BH2} may be written as $$2\pi
\sqrt{|c-24\Delta_0|L_0/6}$$ for $\Delta_0 = 0$ or $\Delta_0 =
-\alpha \Upsilon/(2G)$ and for $Q=0$ in the large mass $\mu \gg
1/(8G)$ limit. In the semi-classical limit $\alpha \Upsilon \gg
G$, the latter formula is the Cardy formula\footnote{Actually, the
determination of the asymptotic density of states in a conformal
field theory from the Cardy formula (see e.g.
\cite{Carlip:1998qw,Dijkgraaf:2000fq}) seems to be meaningful only
when the \emph{effective} central charge $c_{eff} = c - 24
\Delta_0$ is positive (which may encompass non unitary CFTs),
which is the case for $\Delta_0 = -\alpha \Upsilon/(2G)$, and it
is not obvious to us that using an absolute value is the right way
to proceed when it is negative. We thank Mu-In Park and Steve
Carlip for their comments on this point.}
\cite{Bloete:1986qm,Cardy:1986ie,Dijkgraaf:2000fq} for the
Virasoro algebra with generators $L_n$.

It is possible to reproduce the second part of the
entropy~\eqref{BH2} via the Cardy formula by introducing operators
$\hat T_n$ to each element of the affine algebra $T_n$, applying
the Sugawara procedure to obtain a new Virasoro algebra $\tilde
L_n$ with central charge $\tilde c = 1$ and by appropriately
choosing the lowest value $\tilde \Delta_0$ of $\tilde L_0$. In
this case, the effective central charge $|\tilde c-24\tilde
\Delta_0| = 6\alpha \Upsilon/G$ equals the effective central
charge $|c - 24 \Delta_0|$ in the initial Virasoro sector.
However, this construction \emph{a posteriori} is quite
artificial.

There are several points that deserve further investigations. It
would be interesting to study the supersymmetry properties of
these black holes by embedding the Lagrangian~\eqref{action2} in
some supergravity theory. The extension of the asymptotic symmetry
algebra to a supersymmetric asymptotic symmetry algebra in the
spirit of \cite{Banados:1998pi} would then allow one to fix the
lowest value $\Delta_0$ of $L_0$ undetermined classically and left
ambiguous even after the matching of the entropy with the Cardy
formula. Note that the naive dimensional reduction on a 2-sphere
of the $5d$ minimal supergravity \cite{Gauntlett:2002nw} in which
G\"odel black holes were studied \cite{Gimon:2003ms} does not
admit~\eqref{blackhole} as solutions. There are however other
alternatives. Namely, it turns out that the three-dimensional
G\"odel black holes can be promoted to a part of an exact string
theory background along the lines of
\cite{Israel:2004vv,Detournay:2005fz}, and are in particular
solutions to the low energy effective action for heterotic or type
II superstring theories. It could therefore be instructive to
check if the present asymptotic analysis holds in this latter
theories as well and then study the supersymmetry properties of
these solutions.

%% file: conclusion.tex
\chapter*{Summary and outlook}
\addcontentsline{toc}{part}{Summary and outlook}
\chaptermark{Summary and outlook}

In the first part of this thesis, a theory of exact symmetries in
gauge and gravity theories was formulated using techniques of the
variational calculus. Some very satisfactory results are worth
emphasizing. Each reducibility parameter (e.g. Killing vector for
gravity) is associated with a unique finite and conserved surface
charge one-form in field space. These one-forms form a
representation of the Lie algebra of reducibility parameters. For
reducibility parameters associated with integrable surface charge
one-forms, conserved quantities can be defined for a family of
symmetric solutions. Using the geometric properties of horizons,
we showed how the first law of black hole mechanics is universal
in gravitation theories, regardless of the details of the
dynamics, the number of spacetime dimensions, the horizon topology
or the spacetime asymptotic structure.

In the case of exact symmetries, our definitions were shown to
agree with Hamiltonian methods and with covariant phase space
methods when applied, respectively, to Lagrangians of first order
in time derivatives and to diffeomorphic invariant Lagrangians.
These comparisons between different formalisms complement what can
be found in the literature. The systematic derivation of
expressions for conserved charges in Einstein gravity coupled to
matter fields provided a very useful toolkit for the description
of conservation laws in gravity.

As applications, we recovered the charges of Kerr-anti-de Sitter
black holes in any dimensions and we studied the case of black
rings with dipole charges. By deriving the thermodynamics of black
holes in G\"odel backgrounds and black strings in $pp$-waves
backgrounds, we showed that the analysis of classical charges
associated with exact symmetries can be done independently on the
asymptotic structure of spacetimes.

We also constructed a new class of $3d$ black hole and particle
solutions to the Einstein-Maxwell theory with negative
cosmological constant supplemented by a Chern-Simons coupling.
These solutions were shown to arise from identifications on the
non-trivial $3d$ factor of the G\"odel spacetime. They reduce to
the BTZ solutions for two particular choices of the Chern-Simons
coupling.

In the second part of the thesis, asymptotically conserved charges
were defined on the sphere at infinity by integrating surface
charges one-forms associated with asymptotic reducibility
parameters in a convenient phase space of fields. The resulting
representation theorem of the Lie algebra of asymptotic symmetries
by a possibly centrally extended Lie algebra of charges reproduced
similar theorems in Hamiltonian as well as in covariant phase
space methods.

Some advantages of our formalism are worth mentioning. First, the
technical tools used allow to treat gauge theories with
 higher derivatives. Second, the Lagrangian formalism is
suitable to obtain covariant expressions, e.g. in diffeomorphic
theories. Finally, what makes most covariant phase space methods
ambiguous, namely the dependence of the pre-symplectic form on
boundary terms added to the Lagrangian, is avoided here by the
choice of the invariant pre-symplectic form.

For phase spaces which are asymptotically linear, well-known
expressions as ADM or Abbott-Deser charges can be recovered.
Interestingly, the formalism applies for more general boundary
conditions. In general, charge differences with respect to a given
background become non-linear functionals of the field deviation
with respect to the background.

Phase spaces and asymptotic reducibility parameters were found for
three different asymptotic configurations in three-dimensional
gravity by following an unified algorithm. The previous result in
asymptotically anti-de Sitter spacetimes was expanded to flat and
G\"odel asymptotics where centrally extended representations of
the asymptotic symmetry algebras were found. The following pattern
of asymptotic charge algebra in $3d$ gravity now emerges:
\begin{eqnarray*}
&&\text{adS}_3 \rightarrow \text{Two copies of the Virasoro algebra},\\
&&\text{Mink}_3 \rightarrow \text{Centrally extended $\mathfrak{bms_3}$ algebra},\\
&&\text{G\"odel}_3 \rightarrow \text{Centrally extended
$\mathfrak{Godel_3}$ algebra}.
\end{eqnarray*}
The first result, obtained 20 years ago, became 10 years later a
sign for the AdS/CFT correspondence. One may wonder if the other
results hint at similar correspondences, e.g. flat $3d$ gravity
and a field theory admitting $\mathfrak{bms_3}$ as global symmetry
group or gravity with G\"odel asymptotics and a (probably
non-unitary) field theory admitting $\mathfrak{Godel_3}$ as a
symmetry group. However, the serious consideration of these ideas
goes far beyond the scope of this thesis.

Let us finally mention some directions for the future. A technical
issue yet to be clarified is the role of the supplementary term
$E_\cL$ in the charges. This term does not appear in usual
covariant phase space methods, nor in Hamiltonian formalism where
it is trivially zero and it vanishes in all examples treated in
this thesis. Also, an improved algorithm to define phase spaces
and asymptotic reducibility parameters while avoiding the
non-geometrical resolution of the reducibility equations to first
order is still to be found. Asymptotically flat spacetimes at null
infinity in $n \geq 4$ dimensions require additional
considerations because non-conservation and non-integrability of
the charge one-forms are generic in that case~\cite{Wald:1999wa}.

More generally, it would be of interest to compare our formalism
with spinorial techniques which are crucial in the proof of
positive energy theorems and in stability analyses. The link with
quasi-local methods would also be interesting, especially for
numerical applications. Topological charges, like magnetic charges
or the NUT charge in gravity are not associated with reducibility
parameters in the usual formulation of gauge theories. One can ask
if there exists formulations in which these topological charges
can be treated on an equal setting as charges related to gauge
invariance.

%% file: appendices.tex
\appendix

\addcontentsline{toc}{part}{Appendices}

\cleardoublepage
\newpage
\thispagestyle{empty} \mbox{  }
\begin{center}
{\vspace{180pt} \bf \Huge \mbox{ }  Appendices }
\end{center}

\renewcommand{\thechapter}{\Alph{chapter}}
\renewcommand{\theequation}{\Alph{chapter}\arabic{equation}}

\setcounter{chapter}{0} \setcounter{section}{0}
\setcounter{subsection}{0} \setcounter{equation}{0}
\setcounter{footnote}{0}

\chapter{Elements from the variational bicomplex}
\label{app:basicdef}

The variational bicomplex was first introduced in the mid 1970's
as a way of studying the inverse problem of the calculus of
variations, see~\cite{Anderson1991} for a comprehensive review.
More details on the variational bicomplex can be found for
instance in the textbooks
\cite{Andersonbook,Olver:1993,Saunders:1989,Dickey:1991xa}.

\section{Jet spaces and vector fields}
\label{sec:elem-defin-conv}

Let $M$ be the base space with coordinates $x^\mu$, $\mu=0,\dots
n-1$ which is locally isomorphic to $\mathbb R^n$. Local
coordinates in an open set $U$ of the space of fields are denoted
as $\phi^i$. We assume to make it simple that all fields are
Grassmann even. They constitute the fiber bundle $\pi: E
\rightarrow M$ where $E$ is locally $M\times U$. A section, or
history of fields, is then a mapping from $M$ to $E$, $x^\mu
\rightarrow (x^\mu,\phi^i(x^\mu))$. In general, one may allow for
a non-trivial fiber bundle. However, except when explicitly
mentioned, we will not take in consideration such global properties and we will only
work in local coordinate patches.

The jet space $V^k$ at a point $p\in M$ of coordinates $x^\mu_p$
is the equivalence class of sections at $p$ where two sections are
equivalent if they have the same partial derivatives up to the
order $k$ at $p$. The jet fiber of order $k$, $\cJ^k(E)$ is given
locally by $M \times V^k$. It has as coordinates
\begin{equation*}
(x^\mu,\phi^i,\phi_\mu^i,\phi_{\mu_1\mu_2}^i,\dots
\phi^i_{\mu_1\mu_2\dots \mu_k}).
\end{equation*}
Here, the $k$-th order derivatives $\phi^i_{\mu_1\mu_2\dots \mu_k}
\equiv \frac{\partial^k\phi^i(x)}{\partial x^{\mu_1}\dots\partial
x^{\mu_k}}|_{x^\mu_p}$ of a field $\phi^i(x)$ at $p$ are not all
independent because the derivatives are symmetric under
permutations of the derivative indices $\mu_1,\dots,\mu_k$. One
has $\phi_{\mu_1\mu_2}^i= \phi_{\mu_2\mu_1}^i$, etc. The infinite
jet bundle $\cJ^\infty(E)$ is defined from $\cJ^k(E)$ by a
limiting procedure. A point in $\cJ^\infty(E)$ can be identified
with an equivalence class of local sections around a point in $M$
-- equivalent local sections at $p$ have the same Taylor
coefficients to all orders at $p$. As in the classification of
conservation laws the differential order of the sought-after
quantities is not known \textit{a priori}, the appropriate space to
formulate conservation laws is the infinite jet bundle.

Local functions $f(x,[\phi]) \in Loc(E)$ are smooth functions
depending on the coordinates $x^\mu$ of the base space $M$, the
fields $\phi^i$, and a finite number of the jet-coordinates
$\phi^i_{\mu_1\dots\mu_k}$ denoted collectively as $[\phi]$.

As in \cite{DeDonder1935,Andersonbook}, we define derivatives
$\partial^S/\partial \phi^i_{\mu_1\dots\mu_k}$ that act on the
basic variables through
\begin{eqnarray*} &&\frac{\partial^S
  \phi^j_{\nu_1\dots\nu_k}}{\partial \phi^i_{\mu_1\dots\mu_k}} =\delta^j_i\,
\delta^{\mu_1}_{(\nu_1}\dots\delta^{\mu_k}_{\nu_k)}\ ,\quad
\frac{\partial^S
  \phi^j_{\nu_1\dots\nu_m}}{\partial \phi^i_{\mu_1\dots\mu_k}}
=0\quad\mbox{for $m\neq k$},
\nonumber\\[6pt]
&&\frac{\partial^S x^\mu}{\partial \phi^i_{\mu_1\dots\mu_k}}=0,
\end{eqnarray*}
where the round parentheses denote symmetrization with weight one,
\[
\delta^{\mu_1}_{(\nu_1}\delta^{\mu_2}_{\nu_2)} =\frac 12
(\delta^{\mu_1}_{\nu_1}\delta^{\mu_2}_{\nu_2}
+\delta^{\mu_1}_{\nu_2}\delta^{\mu_2}_{\nu_1})\, ,\quad
\mbox{etc.}
\]
For instance, the definition gives explicitly
\[
\frac{\partial ^S\phi^i_{11}}{\partial \phi^i_{11}}=1\ ,\quad
\frac{\partial ^S\phi^i_{12}}{\partial \phi^i_{12}}=
\frac{\partial ^S\phi^i_{21}}{\partial \phi^i_{12}}=\frac 12\
,\quad \frac{\partial ^S\phi^i_{112}}{\partial \phi^i_{112}}=\frac
13\ ,\quad \frac{\partial ^S\phi^i_{123}}{\partial
\phi^i_{123}}=\frac 16\ .
\]
We note that the use of these operators automatically takes care
of many combinatorial factors which arise in other conventions,
such as those used in \cite{Olver:1993}.

A generalized vector field on $\cJ^\infty(E)$ is given by
\[
c^\mu \Q{}{x^\mu} + \sum_{k \geq 0} b^i_{\mu_1\dots \mu_k} \,
\frac{\partial^S}{\partial \phi^i_{\mu_1\dots\mu_k}},
\]
where $c^\mu, \, b^i_{\mu_1\dots \mu_k} \in Loc(E)$. Here $\sum_{k
\geq 0}$ means the sum over all $k$, from $k=0$ to infinity, with
the summand for $k=0$ is given by $b^i\,\partial /\partial
\phi^i$, i.e., by definition $k=0$ means ``no indices $\mu_i$''.
Furthermore, we are using Einstein's summation convention over
repeated indices: for each $k$ there is a summation over all
tupels $(\mu_1,\dots,\mu_k)$. Hence, for $k=2$, the sum over
$\mu_1$ and $\mu_2$ contains both the tupel $(\mu_1,\mu_2)=(1,2)$
and the tupel $(\mu_1,\mu_2)=(2,1)$. These conventions extend to
all other sums of similar type.

The total derivative is the vector field denoted by $\partial_\nu$
which acts on local functions according to
\begin{equation}
\partial_\nu=\frac{\partial}{\partial x^\nu}+\sum_{k  \geq 0}
\phi^i_{\mu_1\dots\mu_k\nu}\,\frac{\partial^S}{\partial
  \phi^i_{\mu_1\dots\mu_k}}\ .\label{def_dmu}
\end{equation}
This vector field is defined such that for local functions
$f(x,[\phi]) \in Loc(E)$ and for sections $x^\mu \rightarrow
(x^\mu,\phi^i(x))$,
\begin{equation*}
(\d_\mu f)|_{\phi^i(x)} = \frac{d}{dx^\mu}\left( f|_{\phi^i (x)
}\right).
\end{equation*}
It also satisfies
\begin{eqnarray*}
&&[\d_\mu,\d_\nu]=0, \qquad [\Q{}{\phi^i},\d_\mu] = 0,\\
&& [\QS{}{\phi^i_{\mu_1\dots \mu_k}},\d_\nu] =
\delta_{(\nu}^{\mu_1}\delta_{\lambda_1}^{\mu_2}\cdots
\delta_{\lambda_{k-1})}^{\mu_k}\QS{}{\phi^i_{\lambda_1 \dots
\lambda_{k-1}}}.
\end{eqnarray*}

The Euler-Lagrange derivative of a local functional $f$ is defined
by
\[
\varQ{f}{\phi^i} = \sum_{k  \geq 0} (-)^k \d_{\mu_1}\dots
\d_{\mu_k}\QS{f}{\phi^i_{\mu_1 \dots \mu_k}}.
\]
It satisfies the remarkable property that $\varQ{f}{\phi^i} = 0$
if and only if $f = \d_\mu j^\mu$ for $j^\mu \in Loc(E)$.

An infinitesimal transformation is defined by the transformations
$x^\mu \rightarrow x^\mu+\eps \,c^\mu$ and $\phi^i \rightarrow
\phi^i +\epsilon \, b^i$ with $c^\mu(x)$
 and $b^i(x,[\phi]) \in Loc(E)$ with which one associates
the vector field $v = c^\mu \Q{}{x^\mu}+b^i \Q{}{\phi^i}$ (one can
also consider $c^\mu \in Loc(E)$ but this generalization is not
needed here). This vector field can be naturally prolonged onto
$\text{pr}\, v \in \cJ^\infty(E)$, see \cite{Barnich:2000cs}. The
resulting vector on the jet space $\cJ^\infty(E)$ is then the sum
of a total derivative $c^\mu \d_\mu$ and of the vector field
\begin{equation}\delta_Q =\sum_{k=0}
(\partial_{\mu_1}\dots\partial_{\mu_k}Q^i) \,
\frac{\partial^S}{\partial\phi^i_{\mu_1\dots\mu_k}}\
,\label{char_def}
\end{equation}
with $Q^i = b^i - \phi^i_\mu c^\mu$ which is called the
characteristic of the vector. The Lie bracket of characteristics
is defined by $[Q_1,Q_2]^i=\delta_{Q_1}Q^i_2-\delta_{Q_2}Q^i_1$
and satisfies $[\delta_{Q_1},\delta_{Q_2}]=\delta_{[Q_1,Q_2]}$.

For notational convenience, let us introduce the set of
multiindices that is the set of all tupels $(\mu_1,\dots,\mu_k)$,
including (for $k=0$) the empty tupel. The one-element tuple
is denoted by $\mu$ without round parentheses, while a generic
tuple is denoted by $(\mu)$. The length, i.e., the number of
individual indices, of a multiindex $(\mu)$ is denoted by $|\mu|$.
We use Einstein's summation convention as well for repeated
multiindices as in \cite{Andersonbook}. For instance, the total
derivative, the variation with characteristic $Q^i$ and the
Euler-Lagrange derivative may be written compactly as
\[ \d_\mu = \frac{\partial}{\partial x^\nu} +
\phi^i_{\nu(\mu)}\QS{}{\phi^i_{(\mu)}},\qquad \delta_Q =
\d_{(\mu)}Q^i\QS{}{\phi^i_{(\mu)}}, \qquad \varQ{f}{\phi^i} =
(-\partial)_{(\mu)} \QS{f}{\phi^i_{(\mu)}},\] where
$(-\partial)_{(\mu)} \hat = (-)^{|\mu|}\partial_{(\mu)}$.

\section{Horizontal complex}
\label{sec:horzi}

Let us consider the exterior algebra $\Lambda(dx^\mu)$ of
differentials $dx^\mu$ which we treat as anticommuting (Grassmann
odd) variables, $dx^\mu dx^\nu=-dx^\nu dx^\mu$. Local horizontal
forms are elements of $\Omega(E) = Loc(E) \otimes
\Lambda(dx^\mu)$, i.e. forms whose coefficients are local
functions. We define the action of the symmetrized derivative on
$dx^\alpha$ as $\frac{\partial ^S dx^\mu}{\partial
\phi^i_{\mu_1\dots\mu_k}}=0$.

If the space $M$ is endowed with a metric $g_{\mu\nu}$ (which can
be contained in the set of fields), one can define the Hodge dual
of an horizontal $p$-form $\omega^p$ as $\star\; \omega^p =
\sqrt{|g|}\omega^{\mu_1\dots \mu_p}(d^{n-p}x)_{\mu_1\dots\mu_p}$
where indices are raised with the metric and where
\begin{eqnarray*}
  (d^{n-p}x)_{\mu_1\dots\mu_p} \hat =
\frac 1{p!(n-p)!}\, \epsilon_{\mu_1\dots \mu_p \mu_{p+1}\cdots
\mu_n} dx^{\mu_{p+1}}\dots dx^{\mu_n}.
\end{eqnarray*}
Here, $\epsilon_{\mu_1\dots\mu_n}$ is the numerically invariant
tensor with $\epsilon_{01\dots n-1} = 1$. We have the relations
\begin{eqnarray*}
dx^\alpha \; (d^{n-p-1}x)_{\mu_1\cdots \mu_{p+1}}  &=&
 (d^{n-p}x)_{[\mu_1\cdots \mu_{p}}\delta^\alpha_{\mu_{p+1}]},\\
(d^{n-p-1}x)_{\mu_1\cdots \mu_{p+1}}\; dx^\alpha &=& (-)^{n-p-1}
(d^{n-p}x)_{[\mu_1\cdots \mu_{p}}\delta^\alpha_{\mu_{p+1}]}.
\end{eqnarray*}
As a consequence, one has $\star \star\, \omega^p =
(-)^{p(n-p)+s}\omega^p$, where $s$ is the signature of the metric
and if $\alpha^{(p)}$ and $\beta^{(q)}$ are $p$ and $q$ forms with
$q \leq p \leq n$, they obey
\[
\beta^{(q)} \wedge \star\, \alpha^{(p)} =
\frac{1}{q!}\alpha^{(p)\mu_1\cdots \mu_{p-q}\rho_1 \cdots
\rho_q}\beta^{(q)}_{\rho_1 \cdots
\rho_q}(d^{n-(p-q)}x)_{\mu_1\cdots \mu_{p-q}}.
\]
The horizontal differential on horizontal forms is defined by
\begin{equation}
\dH=dx^\nu\partial_\nu.\label{def_dH}
\end{equation}
For example, the derivative of a $n-p$ form $k^{(n-p)} =
k^{[\mu_1\cdots \mu_p]}(d^{n-p}x)_{\mu_1\cdots \mu_p}$ is given by
\[\dH k^{(n-p)} = \d_\rho k^{[\mu_1\cdots
\mu_{p-1}\rho]}(d^{n-(p-1)}x)_{\mu_1\cdots \mu_{p-1}}.\]One has
also $[\delta_Q,\dH] = 0$.

The fundamental theorem on the horizontal complex is the algebraic
Poincar\'e
lemma~\cite{Vinogradov:1977,Takens:1979aa,Tulczyjew:1980aa,Anderson:1980aa,
Wilde:1981aa,Tsujishita:1982aa,Brandt:1990gy,
Wald:1990ic,Dubois-Violette:1991is,Dickey:1992aa}
\begin{theorem}\label{poincarelemma}
The cohomology $H^p(\dH,\Omega(E))$ is isomorphic to $\mathbb R$
in form degree 0, vanishes for form degrees $0 < p < n$ and for
$p=n$ is isomorphic to the equivalence classes of $n$-forms $L
\;d^nx$ that differ by an (horizontal) derivative, or stated
differently the equivalence classes of local $n$-forms that admit
the same Euler-Lagrange derivatives.
\end{theorem}

A Cartan calculus can be defined on the algebra $\Omega(E)$. The
inner product by a vector $c$ is given by $i_c = c^\mu
\Q{}{dx^\mu}$ and the Lie differential is defined by
\begin{equation}
\cL_c = i_c \dH + \dH i_c \label{app:Liec}.
\end{equation}
When acting on horizontal forms, any vector field $v = c^\mu
\Q{}{x^\mu}+b^i \Q{}{\phi^i}$ can be prolonged as $\text{pr}\, v =
c^\mu \d_\mu + \delta_Q + \dH c^\mu \Q{}{dx^\mu} = \delta_Q +
\cL_c$ such that it satisfies $[\text{pr}\,v,\dH]=0$. For example,
a vector field acting on a $n$-form $L \,d^n x$ can be written as
\begin{equation}
\text{pr}\,v\, (L\, d^nx) = \delta_Q L\, d^nx +\dH(c^\mu L
(d^{n-1}x)_\mu).\label{app:deltaL}
\end{equation}

\section{Lie-Euler operators and $T$ form}

Except for a different notation, we follow
closely~\cite{Andersonbook} in this section.

Multiple integrations by parts can be done using the following. If
for a given collection $P_i^{(\mu)}$ of local functions, the
equality
\begin{eqnarray}
  \label{eq:20}
 \partial_{(\mu)}Q^i P_i^{(\mu)} =\partial_{(\mu)}(Q^i R_i^{(\mu)})
\end{eqnarray}
holds for all local functions $Q^i$, then
\begin{eqnarray}
  \label{eq:21}
  R_i^{(\mu)}=\left(\begin{array}{c}
    |\mu|+|\nu| \\ |\mu| \end{array}\right)(-\partial)_{(\nu)}\,
P_i^{((\mu)(\nu))} . \label{intparts}
\end{eqnarray}
Conversely, if \eqref{eq:20} holds for a given collection
$R_i^{(\mu)}$ then
\begin{eqnarray}
  \label{eq:22}
  P_i^{(\mu)}=\left(\begin{array}{c}
    |\mu|+|\nu| \\ |\mu|
    \end{array}\right)\partial_{(\nu)}R_i^{((\mu)(\nu))}.
\end{eqnarray}
By definition, when
$P_i^{(\mu)}=\frac{\partial^Sf}{\partial\phi^i_{(\mu)}}$, the
higher order Euler-Lagrange derivatives $\frac{\delta
f}{\delta\phi^i_{(\mu)}}$ are given by the associated
$R_i^{(\mu)}$,
\begin{eqnarray}
\frac{\delta f}{\delta\phi^i_{(\mu)}} \hat =
\left(\begin{array}{c}
    |\mu|+|\nu| \\ |\mu| \end{array}\right)(-\partial)_{(\nu)}\,
\frac{\partial^Sf}{\partial\phi^i_{((\mu)(\nu))}}.\label{higherLie}
\end{eqnarray}
As as consequence, \bea \delta_Q f=
\partial_{(\mu)}\Big[ Q^i\,\frac{\delta f}{\delta
  \phi^i_{(\mu)}} \Big], \qquad \forall f,\,Q^i \in Loc(E)  \label{fund} \eea
By definition, $\delta/\delta\phi^i$ is the usual Euler-Lagrange
derivative. The crucial property of these operators is that they
``absorb total derivatives'', \bea
&|\mu|=0:& \frac{\delta (\partial_\nu f)}{\delta \phi^i}=0,\label{eA7}\\
&|\mu|>0:& \frac{\delta (\partial_\nu f)}{\delta \phi^i_{(\mu)}}
=\delta^{(\mu}_\nu\frac{\delta f}{\delta \phi^i_{(\mu^\prime))}}\
,\quad (\mu)=(\mu(\mu^\prime)),\label{eA8} \eea where, e.g.,
\[
\delta^{(\mu}_\nu\frac{\delta f}{\delta
\phi^i_{\lambda)}}=\frac{1}{2}\big(\delta^{\mu}_\nu\frac{\delta
f}{\delta \phi^i_{\lambda}}+\delta^{\lambda}_\nu\frac{\delta
f}{\delta \phi^i_{\mu}}\big).
\]
It may be also deduced that
\begin{eqnarray}
\varQ{(\d_\nu f)}{\phi^i_{\rho(\mu)}} =
\frac{1}{|\mu|+1}\delta^\rho_\nu
\varQ{f}{\phi^i_{(\mu)}}+\frac{|\mu|}{|\mu|+1}\delta^{(\mu_1}_\nu
\varQ{f}{\phi^i_{\rho \mu_2 \cdots \mu_{|\mu|}) } }.
\label{higherLieprop}
\end{eqnarray}

By considering the particular case where \eqref{eq:20},
\eqref{eq:21} are used in terms of $Q_2$ with
\begin{eqnarray}
P_i^{(\mu)}[\vddl{\omega^n}{\phi}]=\ddl{^S
  Q^j_1}{\phi^i_{(\mu)}}\vddl{\omega^n}{\phi^j},\label{eq:25}
\end{eqnarray}
we get $\delta_{Q_2}(Q^j_1)\vddl{\omega^n}{\phi^j}=
\partial_{(\mu)}\Big(Q_2^i R^{(\mu)}_i[\vddl{\omega^n}{\phi}]\Big)$.
Splitting the term without derivatives on the r.h.s from the
others and defining
\begin{eqnarray}
T_{Q_1}[Q_2,\vddl{\omega^n}{\phi}]&=&\partial_{(\mu)}\Big(Q_2^i
  R^{(\mu)\nu}_i[\frac{\partial}{\partial
    dx^\nu}\vddl{\omega^n}{\phi}]\Big),
 \label{eq:27} \\& =&\hspace{-2pt}\left(\hspace{-2pt}\begin{array}{c}
    |\mu|+1+|\rho|\\ |\mu|+1 \end{array}\hspace{-2pt}\right)\hspace{-2pt}
\partial_{(\mu)}\Bigg(Q_2^i (-\6)_{(\rho)}
\Big(\ddl{^SQ^j_1}{\phi^i_{((\mu)(\rho)\nu)}}\frac{\partial}{\partial
    dx^\nu}\vddl{\omega^n}{\phi^j}\Big)
\Bigg),\nonumber
\end{eqnarray}
gives
\begin{eqnarray}
\delta_{Q_2}(Q^j_1)\vddl{\omega^n}{\phi^j}=Q^i_2 R_i+\dH
T_{Q_1}[Q_2,\vddl{\omega^n}{\phi}], \quad R_i=(-\6)_{(\nu)}\Big(
\ddl{^SQ_1^j}{\phi^i_{(\nu)}}\vddl{\omega^n}{\phi^j}\Big).\label{eq:28}
\end{eqnarray}
Note also that the variation of the $T$ form can be written as
\begin{multline}
  \label{eq:34a}
  \delta_{Q_3}T_{Q_1}[Q_2,\vddl{\omega^n}{\phi}]=T_{Q_1}[\delta_{Q_3}Q_2,
\vddl{\omega^n}{\phi}]+
T_{Q_1}[Q_2,\delta_{Q_3}\vddl{\omega^n}{\phi}]+
\\+T_{\delta_{Q_3}Q_1}[Q_2,\vddl{\omega^n}{\phi}]-Y_{Q_1,Q_3}[Q_2,
\vddl{\omega^n}{\phi}],
\end{multline}
where
\begin{multline}
  \label{eq:16bis}
Y_{Q_1,Q_3}[Q_2,\vddl{\omega^n}{\phi}]=
\left(\begin{array}{c}|\mu|+|\rho|+1\\ |\mu|+1\end{array}\right)
\6_{(\mu)} \Big(Q_2^i(-\6)_{(\rho)}\\\big(\frac{\partial}{\partial
    dx^\nu}\vddl{\omega^n}{\phi^j}\frac{\6^S \partial_{(\sigma)} Q^k_3}
{\6\phi^i_{((\mu)(\rho)\nu)}}\frac{\6^S Q_1^j}
{\6\phi^k_{(\sigma)}}\big) \Big).
\end{multline}

\section{Horizontal and vertical bicomplex}
\label{sec:vert_bicompl}

Let us denote by $\Omega^p(J^\infty(E))$ the ring of differential
$p$-forms on $J^\infty(E)$ and $\Omega(J^\infty(E))$ the ring of
all differential forms on $J^\infty(E)$. A differential form
$\omega$ on $J^\infty(E)$ is called a contact form if for every
equivalence class of local sections of $E$, the pull-back of
$\omega$ on $M$ is zero. The set of contact forms on $J^\infty(E)$
defines an ideal in $\Omega(J^\infty(E))$ which is generated
locally by the so-called vertical one forms $\dv\phi^i_{\mu_1\dots
\mu_k} = \text{d}\phi^i_{\mu_1\dots \mu_k}-\phi^i_{\mu_1\dots
\mu_k\mu_{k+1}}dx^{\mu_{k+1}} $ which are Grassmann odd. Remember
that $\phi^i_{\mu_1\dots \mu_k}$ are not all independent because
its derivatives are symmetric under permutations of the indices. A
local basis of the full exterior algebra $\Omega(J^\infty(E))$ is
thus given by the forms
\begin{equation}
dx^\mu, \; \dv\phi^i, \; \dv\phi^i_\mu,\; \dv\phi^i_{\mu\nu},\dots
\end{equation}
We can now distinguish the forms $\omega \in
\Omega^{p,s}(J^\infty(E))$ of type $(p,s)$ as the forms that can
be written as
\begin{equation}
\omega = f_{i_1 \cdots i_s}^{(\nu_1)\cdots
(\nu_1)}(x^\mu,[\phi^i])dx^{\mu_1}\wedge \cdots \wedge
dx^{\mu_p}\wedge \dv\phi^{i_1}_{(\nu_1)}\wedge \cdots \wedge
\dv\phi^{i_s}_{(\nu_s)}.
\end{equation}
The forms of type $(p,0)$ constitute the horizontal forms
described in section~\ref{sec:horzi}. Note for future purposes
that the inner product of the form $\star\, \omega^{(p,s)}$ dual
to $\omega^{(p,s)}$ with the vector $c$ is given explicitly by
\begin{equation}
i_c \star \omega^{(p,s)} = (-)^s (p+1)
\sqrt{|g|}c^{[\mu_{p+1}}\omega^{\mu_1 \cdots
\mu_p]}(d^{n-(p+1)}x)_{\mu_1 \cdots \mu_p}.\label{inner_prod}
\end{equation}
The exterior derivative $\text{d}: \Omega^{p}(J^\infty(E))
\rightarrow \Omega^{p+1}(J^\infty(E))$ can be decomposed into
horizontal and vertical differentials
\begin{equation}
\text{d} = \dH + \dv.
\end{equation}
The horizontal differential has been defined on horizontal forms
in~\eqref{def_dH}. It is extended to the vertical generators in
such a way that $\{\dH,\dv \}=0$. Acting on $(p,s)$ forms, it is
given explicitly by $\dH = dx^\nu \d_\nu$ with
\begin{equation}
\d_\nu = \frac{\partial}{\partial x^\nu} +
\phi^i_{\nu(\mu)}\QS{}{\phi^i_{(\mu)}}+\dv\phi^i_{\nu(\mu)}\QS{}{\dv\phi^i_{(\mu)}}.\label{def_dH2}
\end{equation}
The vertical differential is given by
\begin{equation}
\dv=\sum_{k \geq 0} \dv\phi^i_{\mu_1\dots\mu_k}
\frac{\partial^S}{\partial \phi^i_{\mu_1\dots\mu_k}}\ .
\label{vertdiff}
\end{equation}
It satisfies $\dv (\dv\phi^i_{(\mu)})=0$ and $(\dv)^2 = 0$. The
variational bicomplex for the fiber bundle $\pi:\,E \rightarrow M$
is the double complex $(\Omega^{*,*}(J^\infty(E)),\dH,\dv)$.

For any vector field $v$ of $E$, there is a unique vector field
$\text{pr}\, v \in \cJ^\infty(E)$ called the prolongation of $v$
such that $v$ and $\text{pr}\, v$ agree on functions on $E$ and
such that the contact ideal is preserved under the Lie derivative
with respect to $\text{pr}\, v$. The prolonged vector field
differs from the one defined before equation~\eqref{char_def} by
vertical generators. Using the defining relation $[\text{pr}\,
v,\dH]= 0 = [\text{pr}\, v,\dv]$, any vector field $v = c^\mu
\Q{}{x^\mu}+b^i \Q{}{\phi^i}$ can be prolonged as $\text{pr}\, v =
c^\mu \d_\mu + \delta_Q + \dH c^\mu \Q{}{dx^\mu}$ with $Q^i = b^i
- c^\mu \phi^i_\mu$. The vector field under characteristic form
$\delta_Q$ is now given by
\begin{eqnarray}
\delta_Q &=& \d_{(\mu)}Q^i\QS{}{\phi^i_{(\mu)}}+ \d_{(\mu)}\dv
Q^i\QS{}{\dv\phi^i_{(\mu)}}.\label{app:delta_Q}
\end{eqnarray}
in place of~\eqref{char_def} and satisfies
$[\delta_Q,\dH]=0=[\delta_Q,\dv]$. We have still
$[\delta_{Q_1},\delta_{Q_2}]=\delta_{[Q_1,Q_2]}$.

The inner product of a form $\omega \in \Omega(J^\infty(E))$ with
respect to a vector field $Q^i$ is defined as $i_Q \omega =
\d_{(\mu)}Q^i \QS{}{\dv\phi^i_{(\mu)}}\omega$. It satisfies
\begin{eqnarray}
  \{i_{Q},\dv\}=\delta_{Q},\qquad  [i_{Q_1},\delta_{Q_2}]=i_{[Q_1,Q_2]}.\label{eq:comm2}
\end{eqnarray}

\paragraph{Augmented variational bicomplex} In the context of gauge
theories, we also consider the augmented bicomplex whose basic
variables are the original set of fields $\phi^i$ and several
copies $f^\alpha_{a}$, $a=1,2,3\dots$ of the gauge parameters. The
whole set of fields is denoted as
$\Phi^\Delta_a=(\phi^i,f^\alpha_a)$ and the variational bicomplex
is extended to this complete set, e.g. $\dv^\Phi$ is defined in
terms of $\Phi^\Delta_a$,
\begin{eqnarray}
\dv^\Phi&=&\sum_{k=0} \d_{(\mu)}\dv\phi_{}^i
\QS{}{\phi_{(\mu)}^i}+\d_{(\mu)}\dv f^\alpha_a \QS{}{
f^\alpha_{a(\mu)}}.\label{vertdiff2}
\end{eqnarray}
When $\dv^\Phi$ is restricted to act on the fields $\phi^i$ and
their derivatives alone we denote it by $\dv$ which is given
by~\eqref{vertdiff}.

%

\section{Horizontal homotopy operators}
\label{app:horizhomo}

The horizontal homotopy
operator~\cite{Tulczyjew:1980aa,Andersonbook}
\begin{equation}
I^p_{\dv\phi}\,:\,\Omega^{p,s} \longrightarrow \Omega^{p-1,s+1}
\end{equation}
is defined by
\bea I^p_{\dv\phi}\omega^{p,s}=
\frac{|\mu|+1}{n-p+|\mu|+1}\
\partial_{(\mu)}\Big( \dv\phi^i
\frac{\delta}{\delta\phi^i_{((\mu)\nu)}}\,\frac{\partial
\omega^{p,s}}{\partial dx^\nu}\Big) \label{phihomotopy} \eea for
$\omega^{p,s}$ a $(p,s)$-form, $p,\, s \geq 0$. Note that there is
a summation over $(\mu)$ by Einstein's summation convention. The
following result (see e.g.~\cite{Andersonbook}) is the key for
showing local exactness of the horizontal part of the variational
bicomplex: \bea &0 \leq p< n:&\dv \omega^{p,s}=
I^{p+1}_{\dv\phi}(\dH
\omega^{p,s})-\dH(I^p_{\dv\phi}\omega^{p,s});
\label{cc1a}\\
&p=n:& \dv \omega^{n,s} =\dv\phi^i
\frac{\delta\omega^{n,s}}{\delta\phi^i} -\dH
(I^n_{\dv\phi}\omega^{n,s}).\label{cc2a} \eea The last relation is
sometimes called the ``first variation formula''. Note that the
homotopy~\eqref{phihomotopy} enjoys the property
\begin{equation}
[\dv,I^p_{\dv\phi}]=0. \label{app:propdv}
\end{equation}
Similarly, one can define the homotopy $I_Q^p$ obtained by
replacing $\dv\phi^i$ in~\eqref{phihomotopy} by $Q^i$. It also
obeys
 \bea &0\leq  p<
n:&\delta_Q \omega^{p,s}= I^{p+1}_{Q}(\dH \omega^{p,s})+\dH
(I^p_{Q}\omega^{p,s}),
\label{cc1}\\
&p=n:& \delta_Q \omega^{n,s} =Q^i
\frac{\delta\omega^{n,s}}{\delta\phi^i} +\dH
(I^n_{Q}\omega^{n,s}).\label{cc2} \eea

In the context of the extended jet-bundle of gauge theories, we
will also use the following homotopy operators that only involve
the gauge parameters: for local functions $g_a^\alpha$,
\begin{eqnarray}
I^{p}_{g}\,:\,\Omega^{p,s} \longrightarrow \Omega^{p-1,s}
\end{eqnarray}
is defined by
\begin{eqnarray}
  \label{eq:45}
I^{p}_{g}\omega^{p,s}= \frac{|\lambda|+1}{n-p+|\lambda|+1}
\partial_{(\lambda)}  \Big(g_a^\alpha
  \vddl{}{f^\alpha_{a(\lambda)\rho}} \Q{\omega^{p,s}}{dx^\rho}\Big).
\end{eqnarray}
For a form $\omega_f^{p,s}$ linear in $f^\alpha$ and its
derivatives $0 \leq  p < n$, the following relation holds,
\begin{eqnarray}
 I^{p+1}_{g}\dH \omega_f^{p,s} + \dH  I^{p}_{g} \omega_f^{p,s} =
 \omega_g^{p,s},
\end{eqnarray}
where $\omega_g^{p,s}$ is the form $\omega_f^{p,s}$ with
$f_a^\alpha$ and their derivatives replaced by $g_a^\alpha$ and
their derivatives.

In the augmented variational bicomplex, one can consider the
augmented homotopy operator $I_{\dv\Phi}^p \hat = I_{\dv \phi}^p +
I_{\dv f}^p$ where $I_{\dv \phi}^p$ is given
by~\eqref{phihomotopy} and $I_{\dv f}^p$ by~\eqref{eq:45}. It
obeys $I^{p+1}_{\dv\Phi}\dH \omega_f^{p,s} - \dH I^{p}_{\dv\Phi}
\omega_f^{p,s} = \dv^\Phi \omega^{p,s}$ where $\dv^\Phi$ is the
augmented vertical generator~\eqref{vertdiff2}.

\section{Commutation relations} \label{sec:comm-relat}

Starting from
$\delta_{Q_1}\delta_{Q_2}\omega^n-\delta_{Q_2}\delta_{Q_1}\omega^n=
\delta_{[Q_1,Q_2]}\omega^n$ and using \eqref{cc2} both on the
inner terms of the l.h.s and on the r.h.s gives
\begin{eqnarray}
  \label{eq:50}
  Q_2^i\delta_{Q_1}\vddl{\omega^n}{\phi^i}-Q_1^i\delta_{Q_2}
\vddl{\omega^n}{\phi^i} = \dH( I^n_{[Q_1,Q_2]}\omega^n-
\delta_{Q_1}I^n_{Q_2}\omega^n+\delta_{Q_2}I^n_{Q_1}\omega^n ).
\end{eqnarray}

Starting from $\dv( \delta_Q\omega^n)=\delta_Q(\dv\omega^n)$ and
using~\eqref{fund}, we get $\partial_{(\mu)}(\dv \phi^i
\vddl{\delta_Q\omega^n}{\phi^i_{(\mu)}})$
$=\partial_{(\mu)}(\delta_Q(\dv
\phi^i\vddl{\omega^n}{\phi^i_{(\mu)}}))$, which can be written as
\[\partial_{(\mu)}(\dv \phi^i
[\vddl{}{\phi^i_{(\mu)}},\delta_Q]{\omega^n})=
\partial_{(\mu)}(\dv Q^i\vddl{\omega^n}{\phi^i_{(\mu)}}).\]Applying
$\vddl{}{\dv\phi^i_{\mu_1\dots\mu_k}}$ gives
\begin{eqnarray}
    \label{eq:1}
[\vdl{\phi^i_{\mu_1\dots\mu_k}},\delta_Q]\omega^n\hspace{-5pt}&=&\hspace{-6pt}\sum_{l\leq
k} \hspace{-2pt}\left(\begin{array}{c}
    l+|\nu| \\ l \end{array}\right)\hspace{-2pt}(-\partial)_{(\nu)}\Big(
\frac{\6^S
 Q^j}{\6\phi^i_{((\nu)\mu_1\dots\mu_l}}
\vddl{\omega^n}{\phi^j_{\mu_{l+1}\dots\mu_k)}} \Big).
\end{eqnarray}
In particular,
\begin{eqnarray}
  \label{eq:3}
  [\vdl{\phi^i},\delta_Q]\omega^n=(-\partial)_{(\nu)}\Big(
\frac{\6^S
 Q^j}{\6\phi^i_{(\nu)}}
\vddl{\omega^n}{\phi^j} \Big).
\end{eqnarray}
When combined with \eqref{eq:28}, we get
\begin{eqnarray}
  \label{eq:32}
  Q^i_2[\delta_{Q_1},\vdl{\phi^i}]\omega^n=
-\delta_{Q_2}Q^j_1\vddl{\omega^n}{\phi^j} +\dH
T_{Q_1}[Q_2,\vddl{\omega^n}{\phi}].
\end{eqnarray}

Similarly, applying $\vddl{}{\dv\phi^i_{\mu_1\dots\mu_k}}$ to
$\partial_{(\mu)}(\dv \phi^i
\vddl{(\delta_Q\omega)}{\phi^i_{(\mu)}})=\partial_{(\mu)}(\dv
(Q^i\vddl{\omega}{\phi^i_{(\mu)}}))$, gives
\begin{eqnarray}
\vdl{\phi^i_{\mu_1\dots\mu_k}}(\delta_Q\omega) &=&\sum_{l\leq
  k}\vdl{\phi^i_{(\mu_1\dots\mu_l}}\Big(Q^j\vddl{\omega}{\phi^j_{\mu_{l+1}\dots
  \mu_k)}}\Big).
\end{eqnarray}
Applying $\vddl{}{\dv\phi^i_{(\lambda)}}$ to
$\dv\vddl{\omega^n}{\phi^j}=\vddl{}{\phi^j}(\dv\phi^i\vddl{\omega^n}{\phi^i})$,
we also get
\begin{eqnarray}
  \label{eq:39}
  \vddl{}{\phi^i_{(\lambda)}}\vddl{\omega^n}{\phi^j}=(-)^{|\lambda|}\frac{\6^S}
{\6\phi^j_{(\lambda)}}\vddl{\omega^n}{\phi^i}.
\end{eqnarray}

Starting from $\dH ([\delta_{Q_1},I^n_{Q_2}]\omega^n)=
\delta_{[Q_1,Q_2]}\omega^n-\delta_{Q_1}Q^i_2\vddl{\omega^n}{\phi^i}-
Q^i_2\delta_{Q_1}\vddl{\omega^n}{\phi^i}+
Q^i_2\vddl{(\delta_{Q_1}\omega^n)}{\phi^i}$ and using
\eqref{eq:32} to compute the last two terms, we find
\begin{eqnarray}
  \label{eq:19}
  \dH ([\delta_{Q_1},I^n_{Q_2}]\omega^n)=\dH
  (I^n_{[Q_1,Q_2]}\omega^n)-\dH T_{Q_1}[Q_2,\vddl{\omega^n}{\phi}].
\end{eqnarray}
Similarly, for $p<n$, by evaluating $\dH
([\delta_{Q_1},I^p_{Q_2}]\omega^p)$ one finds
\begin{eqnarray}
  \label{eq:24}
  \dH
([\delta_{Q_1},I^p_{Q_2}]\omega^p)=\dH (I^p_{[Q_1,Q_2]}\omega^p)+
(I^{p+1}_{[Q_1,Q_2]}-[\delta_{Q_1},I^{p+1}_{Q_2}])(\dH\omega^p).
\end{eqnarray}
By the same type of arguments, one shows
\begin{multline}
  \label{eq:29}
  \dH\Big(
\delta_{Q_1}(I^n_{Q_2}\omega^n)-(1\leftrightarrow 2)\Big)=\\=
\dH\Big(I^n_{[Q_1,Q_2]}\omega^n-I^n_{Q_1}(\delta_{Q_2}\omega^n)-
  T_{Q_1}[Q_2,\vddl{\omega^n}{\phi}]-(1\leftrightarrow 2)\Big),
\end{multline}
\begin{multline}
  \label{eq:29bis}
  \dH\Big(
\delta_{Q_1}(I^p_{Q_2}\omega^p)-(1\leftrightarrow 2)\Big)=\\=
\dH\Big(I^p_{[Q_1,Q_2]}\omega^p\Big)+(I^{p+1}_{[Q_1,Q_2]}
-\delta_{Q_1}I^{p+1}_{Q_2}+\delta_{Q_2}I^{p+1}_{Q_1})(\dH
\omega^p).
\end{multline}

\section{Presymplectic $(n-1,2)$ forms} \label{sec:prop-w_vddl-1}

Let us define the $(n-1,2)$-forms
\begin{eqnarray}
  \label{eq:66a}
W_{{\delta\omega^n}/{\delta\phi}}=- \half
I^n_{\dv\phi}\big(\dv\phi^i
  \vddl{\omega^n}{\phi^i}\big),  \qquad
\Omega_{\omega^n}=\dv I^n_{\dv\phi}\omega^n,
\end{eqnarray}
and the $(n-2,2)$-form
\begin{eqnarray}
E_{\omega^n}=\half
I^{n-1}_{\dv\phi}I^n_{\dv\phi}\omega^n\label{eq:26},
\end{eqnarray}
where the horizontal homotopy is given in~\eqref{phihomotopy}

Using~\eqref{cc1a}, \eqref{cc2a} and \eqref{app:propdv} we obtain
\begin{eqnarray}
  \label{eq:80}
  \half I^n_{\dv\phi}\big(\dv\phi^i
  \vddl{\omega^n}{\phi^i}\big)=\dv I^n_{\dv\phi}\omega^n +\half \dH
  (I^{n-1}_{\dv\phi}I^n_{\dv\phi}\omega^n),
\end{eqnarray}
so that
\begin{eqnarray}
  \label{eq:79}
  -W_{{\delta\omega^n}/{\delta\phi}}=\Omega_{\omega^n}+\dH
E_{\omega^n},\qquad\dv \Omega_{\omega^n}=0.
\end{eqnarray}
$\Omega_{\omega^n}$ is the presymplectic $(n-1,2)$ form usually
used in the context of covariant phase space methods. Contrary to
$\Omega_{\omega^n}$, $W_{{\delta\omega^n}/{\delta\phi}}$ involves
only the Euler-Lagrange derivatives of $\omega^n$ and is thus
independent of $\dH$ exact $n$-forms that are added to $\omega^n$.
For this reason, we call $W_{{\delta\omega^n}/{\delta\phi}}$ the
invariant presymplectic $(n-1,2)$ form.

For first order theories, the invariant presymplectic $(n-1,2)$
form coincides with the ``symplectic'' density $\hat \omega$
considered in~\cite{Julia:2002df},
\begin{equation}
W_{{\delta\omega^n}/{\delta\phi}} = \half \dv\phi^i \wedge
\dv\phi^j \QS{}{\phi^i_\nu}\left( \Q{}{dx^\nu}
\varQ{\omega^n}{\phi^j}\right).
\end{equation}
The ``second variational formula'', obtained by applying $\dv$ to
\eqref{cc2a}, can be combined with \eqref{eq:79} to give
\begin{eqnarray}
  \label{eq:13}
  \dv\phi^i\dv \vddl{\omega^n}{\phi^i}=\dH \Omega_{\omega^n}=-\dH
  W_{{\delta\omega^n}/{\delta\phi}}.
\end{eqnarray}
Our surface charges are related to
$W_{{\delta\omega^n}/{\delta\phi}}$, which is $\dv$-closed only up
to a $\dH$ exact term,
\begin{eqnarray}
\dv W_{{\delta\omega^n}/{\delta\phi}}=\dH\dv
E_{\omega^n}.\label{eq:57}
\end{eqnarray}
When $\omega^n = \dH \omega^{n-1}$, we have
\begin{equation}
E_{\dH \omega^{n-1}} = \dv I^{n-1}_{\dv\phi}\omega^{n-1}+\dH
I_{\dv\phi}^{n-2}I_{\dv\phi}^{n-1}\omega^{n-1}.
\end{equation}
Therefore, the quantity $\dv E_{\omega^n}$ do not depend on exact
terms added to $\omega^n$ up to $\dH$-exact terms,
\begin{equation}\label{app:invE}
\dv E_{\omega^n+\dH \omega^{n-1}} = \dv E_{\omega^n}+\dH(\cdot).
\end{equation}
Using the definitions of the homotopy operator
\eqref{phihomotopy}, the higher order Euler-Lagrange derivatives
and the $T$ form~\eqref{eq:27}, the invariant symplectic
form~\eqref{eq:66a} smeared with two vectors fields,
$i_{Q_2}i_{Q_1}W_{{\delta\omega^n}/{\delta\phi}} =
W_{{\delta\omega^n}/{\delta\phi}}[Q_1,Q_2]$ is given by
\begin{equation}
W_{{\delta\omega^n}/{\delta\phi}}[Q_1,Q_2] = \half
\left(I^n_{Q_1}(Q_2^i \varQ{\omega^n}{\phi^i})+
T_{Q_1}[Q_2,\varQ{\omega^n}{\phi^i}]- (Q_1 \leftrightarrow
Q_2)\right),
\end{equation}
which is manifestly antisymmetric in its arguments. The following
proposition provides a crucial alternative formula for the
invariant symplectic form:
\begin{prop}\label{la}
\begin{eqnarray}
 \label{eq:4}  \label{eq:36}
W_{\vddl{\omega^n}{\phi}}[Q_1,Q_2] &= &
I^n_{Q_1}(Q_2^i\vddl{\omega^n}{\phi^i})-T_{Q_2}[Q_1,\vddl{\omega^n}{\phi}] , \\
&=&\hspace*{-3pt}
\left(\hspace*{-5pt}\begin{array}{c}|\mu|+|\rho|+1\\
|\mu|+1\end{array}\hspace*{-5pt}\right) \6_{(\mu)}
\Big(Q_1^i(-\6)_{(\rho)}(Q^j_2\frac{\6^S }
{\6\phi^i_{((\mu)(\rho)\nu)}}\frac{\partial}{\partial
    dx^\nu}\vddl{\omega^n}{\phi^j} )
\hspace*{-2pt}\Big).\nonumber
\end{eqnarray}
\end{prop}
\noindent The equality of the two right-hand sides of the first
and second line is a direct consequence of
definitions~\eqref{higherLie}, \eqref{phihomotopy} and
\eqref{eq:27}. The equality in the first line is proven
in~Appendix~\ref{app:proofs}.\ref{proofofla}.

For later purposes, let us also write
\begin{multline}
  \label{eq:34bis}
  \delta_{Q_3}W_{{\delta\omega^n}/{\delta\phi}}[Q_1,Q_2]=
W_{{\delta\omega^n}/{\delta\phi}}[\delta_{Q_3}Q_1,Q_2]+
W_{{\delta\omega^n}/{\delta\phi}}[Q_1,\delta_{Q_3}Q_2]+\\
+ Z_{{\delta\omega^n}/{\delta\phi}}[Q_1,Q_2,Q_3]
\end{multline}
where
\begin{multline}
  \label{eq:16ter}
Z_{{\delta\omega^n}/{\delta\phi}}[Q_1,Q_2,Q_3]=
\left(\begin{array}{c}|\mu|+|\rho|+1\\ |\mu|+1\end{array}\right)
\6_{(\mu)} \Big(Q_1^i(-\6)_{(\rho)}\\\big(Q^j_2\partial_{(\sigma)}
Q^k_3 \frac{\6^S } {\6\phi^k_{(\sigma)}}\frac{\6^S }
{\6\phi^i_{((\mu)(\rho)\nu)}}\frac{\partial}{\partial
    dx^\nu}\vddl{\omega^n}{\phi^j}\big)
\Big).
\end{multline}
Starting from $[\delta_{Q_1},\delta_{Q_2}]\omega^n=
\delta_{[Q_1,Q_2]}\omega^n$ and using \eqref{cc2} on the outer
terms of the l.h.s, \eqref{eq:36} and \eqref{eq:32} gives
\begin{multline}
  \label{eq:65a}
  Q_1^i\delta_{Q_2}\vddl{\omega^n}{\phi^i}-Q_2^i\delta_{Q_1}
\vddl{\omega^n}{\phi^i} = \dH( I^n_{[Q_1,Q_2]}\omega^n-2
W_{{\delta\omega^n}/{\delta\phi}}[Q_1,Q_2]\\-
\delta_{Q_1}I^n_{Q_2}\omega^n+\delta_{Q_2}I^n_{Q_1}\omega^n ).
\end{multline}
Adding to \eqref{eq:50} gives in particular
\begin{eqnarray}
  \label{eq:65b}
  \dH W_{{\delta\omega^n}/{\delta\phi}}[Q_1,Q_2]=\dH( I^n_{[Q_1,Q_2]}\omega^n-
\delta_{Q_1}I^n_{Q_2}\omega^n+\delta_{Q_2}I^n_{Q_1}\omega^n ).
\end{eqnarray}

\vspace{1cm}

\chapter{Elements from Lagrangian gauge field theories}
\label{app:fieldtheories}

\section{The Lagrangian}

In field theories, the action is a local functional
\[
I[L,\phi^i(x)] = \int_M d^n x L(x,[\phi])|_{\phi^i(x)}
\]
whose equations of motion are used to define the dynamics of the
theory. The Lagrangian $L(x,[\phi])$ is required to depend only on
a finite number $N_D$ of derivatives of $\phi^i$. In the jet space
$J^\infty(E)$, the surface defined by the equations
\[
\d_{(\mu)}\varQ{L}{\phi^i} = 0, \qquad |\mu|=0,\dots,\, N_D\, ,
\]
is called the stationary surface. Local functions pulled back onto
the stationary surface will be called ``on-shell'' and an equality only
valid on the stationary surface will be called ``weakly
vanishing''. Under appropriate regularity conditions on the
Euler-Lagrange equations of motion
\cite{Henneaux:1992ig,Barnich:2000zw}, which we always assume to
be fulfilled, we have $f \approx g$ if and only if the local
functions $f$ and $g$ differ by terms involving local functions
that are linear and homogeneous in $\vddl{L}{\phi^i}$ and their
derivatives.

\section{Symmetries}

A symmetry of the action is a vector field that leaves the
Lagrangian invariant up to a total derivative, $\text{pr}\, v \,(
L\, d^n x) = \dH (j^\mu (d^{n-1}x)_{\mu})$. As a consequence
of~\eqref{app:deltaL}, any symmetry of the action is equivalent to
a symmetry in characteristic form. As a consequence
of~\eqref{eq:3}, any symmetry $\text{pr}\, v$ of the action is a
symmetry of the equations of motion, $\text{pr}\,
v(\varQ{L}{\phi^i}) \approx 0$. However, a symmetry of the
equations of motion is not necessarily a symmetry of the action.

A Noether identity is an identity among the functions
$\d_{(\mu)}\varQ{L}{\phi^i}$ defining the stationary surface,
\begin{equation*}
N^{i(\mu)}\d_{(\mu)}\varQ{L}{\phi^i} = 0 ,\qquad \forall
\phi^i_{(\mu)}\in \cJ^\infty(E),
\end{equation*}
with $N^{i(\mu)} \in Loc(E)$. The Noether operators are defined by
$N^i \hat = N^{i(\mu)}\d_{(\mu)}$ and satisfy
$N^i(\varQ{L}{\phi^i}) \equiv 0$. The Noether identities that
vanish on-shell, e.g. $N^i = \mu^{[ ij]}\varQ{L}{\phi^j}$, are
called trivial.

For $Z=Z^{(\mu)}\partial_{(\mu)}$ a differential operator, one
defines the adjoint operator by $Z^+ \hat =\, (-\partial)_{(\nu)}[
Z^{(\nu)}\cdot \;]$. The adjoint operator can be decomposed in
components $Z^{+(\mu)}$ as $Z^+=Z^{+(\mu)}\partial_{(\mu)}$. One
has $Z^{++} = Z$.

Gauge theories are Lagrangian theories admitting non-trivial
Noether identities. Gauge transformations are linear mappings from
the space of local functions $Loc(E)$ to the vector space of
symmetries of the action.

For $f \in Loc(E)$, we denote by
\[ \delta_f \phi^i = R^i_f = R^i_\alpha (f^\alpha) \]
a generating set of gauge symmetries of $L$. Here, the operators
$R^i_\alpha$ are defined as $\sum_{k\geq 0} R_\alpha^{i(\mu_1
\dots \mu_k)} \d_{\mu_1}\cdots \d_{\mu_k}$ and act on local
functions $f^\alpha$. Gauge transformations of the form
\[
\delta_M \phi^i = M^{+i}[\vddl{L}{\phi}],
\]
with $M^{+i}[\vddl{L}{\phi}]= (-\partial)_{(\mu)}
\Big(M^{[j(\nu)i(\mu)]} \6_{(\nu)}\vddl{L}{\phi^j}\Big)$ are
called trivial gauge transformations.

The generating property means that every symmetry $\delta_g \phi^i
= X^i_g$ that depends linearly and homogeneously on an arbitrary
gauge parameter $g$ is given by a combination of the gauge
transformations $R_f^i$ and trivial transformations,
\[
X^i_g=R^{i}_\alpha (
Z^{\alpha(\nu)}\d_{(\nu)}g)+M^{+i}_g[\vddl{L}{\phi}],
\]
with $M^{+i}_g[\vddl{L}{\phi}]= (-\partial)_{(\mu)}\Big(g\,
M^{[j(\nu)i(\mu)]} \6_{(\nu)}\vddl{L}{\phi^j} \Big)$.

Noether's second theorem proves that
\begin{theorem}
There is a bijection between the gauge transformations and the
Noether identities. The Noether operator $N^i$ corresponds to the
gauge transformation of characteristic $N^{+i}(f)$ for $f \in
Loc(E)$ and, conversely, the gauge transformation of
characteristic $R^i_f$ are associated with the Noether operator
$R^{+i}$.
\end{theorem}
This theorem can be used to integrate by parts the expression
$R^i_f\vddl{\cL}{\phi^i}$ as
\begin{equation}
R^i_f\vddl{\cL}{\phi^i}= f^\alpha
R^{+i}_\alpha(\vddl{\cL}{\phi^i})+\dH S_f \label{app:gaugerel}
\end{equation}
where $R^{+i}_\alpha(\vddl{\cL}{\phi^i}) = 0$ are the Noether
identities and $S_f = S^{\mu
\,i}_\alpha(\varQ{L}{\phi^i},f^\alpha) (d^{n-1}x)_\mu$ is the
Noether current vanishing on-shell.

Let us finally define reducibility parameters or symmetry
parameters as sets of local functions $f^\alpha \in Loc(E)$ that
satisfy
\[ R_\alpha^i[\phi](f^\alpha) \approx 0, \qquad \forall
\phi^i_{(\mu)}\in \cJ^\infty(E). \] A trivial reducibility or
symmetry parameter $f^\alpha$ is a set of weakly vanishing local
functions $f^\alpha \approx 0$. The equivalence class of
reducibility parameters modulo trivial ones is called the set of
non-trivial reducibility parameters.

\section{Linearized theory}
\label{app:linearized}

For $\bar \phi^i(x)$ a solution of the Euler-Lagrange equations of
motion, one can expand the Lagrangian around $\phi^i = \bar
\phi^i(x) + \eps \varphi^i$ as (see \cite{Barnich:2001jy} for
details)
\begin{equation*}
L[\bar \phi^i(x) + \eps \varphi^i] = L[\bar
\phi^i(x)]+\varphi^i_{(\mu)}\frac{\d^{S}L}{\d^S
\phi^i_{(\mu)}}|_{\bar \phi(x)} \eps + L^{free}[\varphi] \eps^2 +
O(\eps^3),
\end{equation*}
where the first term is a constant that can be dropped
classically, the second term is a total divergence according
to~\eqref{cc2a} because $\bar \phi^i(x)$ is a solution and the
term in $\eps^2$ is the relevant term with
\begin{equation*}
L^{free}[\varphi] = \half
\varphi^i_{(\mu)}\varphi^j_{(\nu)}\frac{\d^{S^2}L}{\d^S
\phi^i_{(\mu)}\d^S \phi^j_{(\nu)}}|_{\bar \phi(x)}.
\end{equation*}
The equations of motion of the linearized theory around $\bar
\phi^i(x)$ are given by $\varQ{L^{free}}{\varphi^i}=0$. The
fundamental relation~\eqref{app:gaugerel} can also be expanded in
powers of $\eps$ and is given to lowest non-trivial order by
\begin{equation*}
R^i_{f[\bar \phi]}[\bar \phi]\vddl{L^{free}}{\varphi^i}= \d_\mu
S^{\mu \,i}_\alpha(\varQ{L^{free}}{\varphi^i},f^\alpha[\bar
\phi]),
\end{equation*}
where the characteristic $R^i_\alpha$ and the possibly field
dependent parameter $f^\alpha$ have been evaluated at $\bar
\phi^i(x)$. This relation expresses the gauge invariance of
$L^{free}$ under the transformation $\delta \varphi^i =
R^i_{f[\bar \phi]}[\bar \phi]$. Assuming that the theory is
linearizable, cfr~\cite{Barnich:1995db}, these gauge
transformations provides a generating set of gauge transformations
for the Lagrangian $L^{free}\,d^n x$.

The reducibility equations for the linearized theory are given by
\[ R_\alpha^i[\bar \phi](g^\alpha) \approx 0, \qquad \forall
\varphi^i_{(\mu)}\in \cJ^\infty(E), \] with
$g^\alpha(x,[\varphi])$ and where $\approx$ means here ``up to
terms vanishing when the linearized equations of motion are
satisfied''. In particular, if the reference solution $\bar
\phi^i$ admits solutions $f^\alpha[\bar \phi]$ to
$R^i_{\alpha}[\bar \phi](f[\bar \phi]) = 0$, $g^\alpha =
f^\alpha[\bar \phi]$ are reducibility parameters of the linearized
theory.

\chapter{Technical proofs}
\label{app:proofs}

\section{Proof of Proposition \bref{lem11}}
\label{appb}

For compactness, let us define a generalized gauge transformation
through
$$\delta^T_{f_1}\Phi^\Delta_2=(R^i_{f_1},[f_1,f_2]^\alpha).$$
In the variational bicomplex, the operator $\delta^T_f$ is defined
by
\begin{equation*}
\delta_{f}^T = \delta_{R_{f}} + \d_{(\mu)}[f, f_a]
\QS{}{f_{a(\mu)}},
\end{equation*}
with $\delta_{R_{f}}$ given in~\eqref{app:delta_Q}. It satisfies
$[\delta_{f}^T,\dH]=0=[\delta_{f}^T,\dv]$.

According to the same reasoning that led to \eqref{eq:alg_curr},
combined with \eqref{1.18} and the definitions
\eqref{sec1cur}-\eqref{defM} of Noether currents for gauge
symmetries, we get
\begin{eqnarray}
  \dH \Big(\delta^T_{f_1}S_{f_2}
  -M_{f_1,f_2}[\vddl{L}{\phi},\vddl{L}{\phi}]-T_{R_{f_1}}[R_{f_2},\vddl{\cL}{\phi}]\Big) =0.
\end{eqnarray}
Applying the contracting homotopy~\eqref{eq:45} with respect to
the gauge parameters $f^\alpha_1$ now gives
\begin{eqnarray}
  \label{eq:10f}
\delta^T_{f_1}S_{f_2}
=M_{f_1,f_2}+T_{R_{f_1}}[R_{f_2},\vddl{\cL}{\phi}]+ \dH
N_{f_1,f_2},
\end{eqnarray}
where
\begin{eqnarray}
  N_{f_1,f_2}[\vddl{L}{\phi}]=I^{n-1}_{f_1}\big(
\delta^T_{f_1}S_{f_2}
-M_{f_1,f_2}-T_{R_{f_1}}[R_{f_2},\vddl{\cL}{\phi}]\big).
\end{eqnarray}
By applying $1=\{I_{f_1},\dH\}$ to $\delta^T_{f_1} k_{f_2}$ and
using $\dH k_{f_2}= -\dv S_f+ I^{n}_{\dv\phi}(\dH S_f)$, we get
\begin{eqnarray}
  \label{eq:13b}
\delta^T_{f_1}  k_{f_2}
  = I^{n-1}_{f_1}\Big(-\delta^T_{f_1} \dv
 S_{f_2}+\delta^T_{f_1}( I^n_{\dv\phi}
 (\dH S_{f_2}))\Big)
+\dH (\cdot).
\end{eqnarray}
Using the property~\eqref{cc1a} of the homotopy operators, the
expression inside the parenthesis of r.h.s of \eqref{eq:13b}
becomes
\begin{multline}
-\delta^T_{f_1} \dv  S_{f_2}+\delta^T_{f_1}( I^n_{\dv\phi}
 (\dH S_{f_2})) =
[\delta^T_{f_1}, I^n_{\dv\phi}](\dH S_{f_2})  +\dH I^{n-1}_{\dv
\phi}(\delta^T_{f_1}S_{f_2}).
\end{multline}
{}From equation \eqref{eq:10f}, we get
\begin{multline}
\delta^T_{f_1}  k_{f_2}=I^{n-1}_{f_1}(
 [\delta^T_{f_1},
I^n_{\dv\phi}](\dH S_{f_2}))+\dv N_{f_1,f_2}+\\+
I^{n-1}_{\dv\phi}(M_{f_1,f_2}+
  T_{R_{f_1}}[R_{f_2},\vddl{\cL}{\phi}])
+ \dH (\cdot).
\end{multline}
Using \eqref{eq:40}, \eqref{eq:34a}, and \eqref{eq:34bis}, the
direct computation of $[\delta^T_{f_1}, I^n_{\dv\phi}](\dH
S_{f_2})$ gives
 \begin{multline}
   \label{eq:38}
   [\delta^T_{f_1}, I^n_{\dv\phi}](\dH S_{f_2})=
W_{{\delta\cL}/{\delta\phi}}[R_{f_2},\dv R_{f_1}]+T_{R_{f_2}}[\dv
R_{f_1},\vddl{\cL}{\phi}]\\-
Y_{R_{f_2},R_{f_1}}[\dv\phi,\vddl{\cL}{\phi}]
+Z_{{\delta\cL}/{\delta\phi}}[R_{f_2},\dv\phi,R_{f_1}] -
W_{\delta_{R_{f_1}}\varQ{\cL}{\phi}}[R_{f_2},\dv\phi].
 \end{multline}
If
\begin{multline}
\cT_{f_1,f_2}[\dv\phi]\, \hat =\;
 \Bigg[
I^{n-1}_{f_1}\Bigg( W_{{\delta\cL}/{\delta\phi}}[\dv
R_{f_1},R_{f_2}]+T_{R_{f_2}}[\dv
R_{f_1},\vddl{\cL}{\phi}]\\-Y_{R_{f_2},R_{f_1}}
[\dv\phi,\vddl{\cL}{\phi}]
+Z_{{\delta\cL}/{\delta\phi}}[\dv\phi,R_{f_2},R_{f_1}] -
W_{\delta_{R_{f_1}}\varQ{\cL}{\phi}}[\dv\phi,R_{f_2}] \Bigg) \\
+\dv N_{f_1,f_2}+ I^{n-1}_{\dv\phi}(M_{f_1,f_2}+
  T_{R_{f_1}}[R_{f_2},\vddl{\cL}{\phi}]) \Bigg],\label{c2}
\end{multline}
we finally have
\begin{eqnarray}
 \delta_{R_{f_1}} k_{f_2}[\dv \phi] = -k_{[f_1,f_2]}[\dv \phi]
+ \cT_{f_1,f_2}[\dv\phi] +\dH (\cdot).\label{main1}
\end{eqnarray}
Now $\cT_{f_1,f_2}[\dv\phi]=0$ if (i) $\phi^s$ is a solution to
the Euler-Lagrange equations of motion, (ii) $R_{f_2}|_{\phi^s}=0$
and, (iii) $\dv\phi$ is tangent to the space of solutions at
$\phi^s$. This proves Proposition \bref{lem11}. \qed

\section{Proof of Proposition~\bref{prop_hamilt}}
\label{proof_hamilt}

Let us denote the set of fields collectively by $\phi^i = \{z^A,
\lambda^a\}$. In order to construct the surface charges, we first
have to compute the current $S_f$ defined according to
\eqref{1.3}-\eqref{sec1cur} as
\begin{eqnarray}
R^i_\alpha (f^\alpha) \varQ{\cL_H}{\phi^i} &=&
\Big[\sigma^{AB}\frac{\delta(\gamma_a f^a )}{\delta
z^B}(\sigma_{AC} \dot z^C -
\varQ{h}{z^A} - \varQ{\lambda^b \gamma_b}{z^A})\nonumber\\
&& \hspace{-2cm}+ \,\,(\frac{D f^a}{Dt} + \{f^a,\hat h_E\}_{alt}+
\cC_{bc}^a(f^b,\lambda^c)-V_b^a(f^b))\, (-\gamma_a)\Big]d^nx.
\end{eqnarray}
Note that for any function $g$ not involving time derivatives of
$z^A$, one has
\begin{eqnarray}
\partial_{(k)}Q^A\frac{\partial g}{\partial z^A_{(k)}} =
Q^A\vddl{g }{z^A} +\d_i V^i_A(Q^A,g),\label{eq:99}
\end{eqnarray}
with $V^i_A(Q^A,g)=\d_{(j)} (Q^A\varQ{g}{z^A_{(j)i}} )$. In other
words, $V^i_A(Q^A,g)$ coincides with the components of the
$(n-2)$-form $I^{n-1}_Q(g d^{n-1}x)$ as defined in
\eqref{phihomotopy}, \eqref{cc2} with $\phi^i$ replaced by $z^A$
and $n$ replaced by $n-1$, i.e., for spatial forms with no time
derivatives on $z^A$. One has
\begin{eqnarray}
 R^i_\alpha (f^\alpha) \varQ{\cL_H}{\phi^i} &=&
 \Big[- \frac{d}{dt}(\gamma_a f^a ) \nonumber\\
&+&\d_k\big(V^k_B[\dot z^B - \sigma^{BA}\varQ{h_E}{z^A},\gamma_a
f^a ] +j^{ka}_b(\gamma_a, f^b)\big)\Big]d^nx,\label{eqApp:5}
\end{eqnarray}
where the total derivative is defined by $\frac{d}{dt} =
\frac{D}{Dt}+\d_{(i)}\dot z^A \QS{}{z^A_{(i)}}$ while the current
$j^{ka}_b(\gamma_a,f^b)$ is determined in terms of the Hamiltonian
structure operators through the formula
\begin{eqnarray}
\d_k j^{ka}_b(\gamma_a,f^b) &=& \gamma_a \cV^a_b(f^b) - f^b
\cV^{+a}_b(\gamma_a) \nonumber\\
&&\quad - \gamma_c \cC^c_{ab}(f^a,\lambda^b) +
f^a\cC^{+c}_{ab}(\gamma_c,\lambda^b).
\end{eqnarray}
The weakly vanishing Noether currents $S^\mu_f$ are thus given by
\begin{eqnarray}
S^0_f &=& -  \gamma_a f^a,\label{sham0}\\
S^k_f &=& V^k_B[\dot z^B - \sigma^{BA}\varQ{h_E}{z^A},\gamma_a f^a
]+j^{ka}_b(\gamma_a, f^b).\label{shami}
\end{eqnarray}

Note that $k^{[0i]}_f[\dv\phi,\phi]$, which is the relevant part
of the surface one-form $k_f$ at constant time, only involves the
canonical variables $\dv z^A,z^A$ and the gauge parameters $f^a$,
but not the Lagrange multipliers $\lambda^a$ nor their variations,
$\dv \lambda^a$. This is so because $S^0_f$ does not involve
$\lambda^a$ while the terms in $S^k_f$ with time derivatives
involve only time derivatives of $z^a$ and no Lagrange
multipliers:
\begin{eqnarray}
k^{[0i]}_f[\dv\phi;\phi]=k^{[0i]}_f[\dv z;z]\label{eq:53}.
\end{eqnarray}
More precisely,
using~\eqref{phihomotopy}-\eqref{sham0}-\eqref{shami}, the $0i$
components of the $(n-2)$-form~\eqref{def} can be written as
\begin{eqnarray}
  \label{eq:14}
  k^{[0i]}_f[\dv z;z]=\frac{|k|+1}{|k|+2}
\partial_{(k)}[\dv z^A \varQ{(-  \gamma_a
f^a)}{z^A_{(k)i}}-\dv z^A\varQ{V^i_B[\dot z^B,\gamma_a f^a
]}{z^A_{(k)0}}].
\end{eqnarray}
Equation \eqref{higherLieprop} then allows one to show that
$\varQ{V^i_B[\dot z^B,\gamma_a f^a
]}{z^A_{(k)0}}=\frac{1}{|k|+1}\varQ{(\gamma_a f^a) }{z^A_{(k)i}}$
 so that the terms nicely combine to give
\begin{eqnarray}
  \label{eq:37}
  k^{[0i]}_f[\dv z;z]=-V^i_A[\dv z^A,\gamma_a f^a].
\end{eqnarray}
Taking into account \eqref{eq:99}, we thus have proved the
theorem.\qed

\section{Proof of Proposition~\bref{prop:GEO}}
\label{GEO:app_firstlaw}

Let us prove the relation~\eqref{GEO:to_prove}. Using the
decomposition~\eqref{GEO:surf_form}, the left-hand side of
equation~\eqref{GEO:to_prove} can written explicitly as
\begin{eqnarray}
\oint_H k^K_{\cL^{EH},\delta \xi } -\oint_H i_\xi I^{n}_{\delta
g}\cL_{EH}& =& \oint_{H} \frac{d\mathcal{A}}{16\pi G} \Big(
-\delta\xi^{\mu;\nu}(\xi_\mu n_\nu - \xi_\nu n_\mu) \nonumber\\
&&+ \xi^\mu (\delta
g_{\mu\nu}^{\,\,\,\,\,\,\,;\nu}-g^{\alpha\beta}\delta
g_{\alpha\beta;\mu}) \Big).\label{GEO:topr}
\end{eqnarray}
We have to relate this expression to the variation of the surface
gravity $\kappa$. This is merely an exercise of differential
geometry.

Since the variation is chosen to commute with the total
derivative, the coordinates are left unchanged $\delta x^\mu = 0$
and the horizon $S(x) = 0$ stay at the same location in $x^\mu$.
The covariant vector normal to the horizon $\xi_\mu = f
\partial_\mu S$, where $f$ is a $\kappa$-dependent normalization
function, satisfies
\begin{equation}
\delta \xi_\mu \overset{\mathcal{H}}{=} \delta \ln f\,
\xi_\mu,\label{GEO:dxi}
\end{equation}
where $\delta \xi_\mu \equiv \delta (g_{\mu\nu}\xi^\nu)$. From the
variation of \eqref{GEO:xi0} and of the second normalization
condition~\eqref{GEO:norm}, one obtains
\begin{equation}
\delta \xi^\mu \xi_\mu \overset{\mathcal H}{=} 0, \qquad \delta
n^\mu \xi_\mu \overset{\mathcal H}{=} \delta \ln
f,\label{GEO:rel_delta}
\end{equation}
which shows that $\delta \xi^\mu$ has no component along $n^\mu$
and $\delta n^\mu$ has a component along $n^\mu$ which equals
$-\delta \ln f$\footnote{Note also the following property that is
useful in order to prove the first law in the way of
\cite{Iyer:1994ys}. Using~\eqref{GEO:defq} and \eqref{GEO:dxi}, we
have
\begin{equation}
\half (\delta \xi_{\mu,\nu} - \delta \xi_{\nu,\mu})
\overset{\mathcal H}{=} \xi_\mu (\delta \ln f q_\nu + \delta
q_\nu) - \xi_\nu (\delta \ln f q_\mu + \delta q_\mu).
\end{equation}
It implies in particular that the expression $\delta
\xi_{[\mu;\nu]}$ has no tangential-tangential component, $\delta
\xi_{[\mu;\nu]}\eta^\mu \tilde \eta^\nu \overset{\mathcal H}{=}
0$, $\forall \eta, \tilde \eta$ orthogonal to $\mathcal H$.}.

Let us develop the variation of $\kappa$ starting from the
definition~\eqref{GEO:kappa}. One has
\begin{eqnarray}
\delta \kappa &=& \frac{1}{2}(\xi^\mu \xi_\mu)_{;\nu} \delta n^\nu
+ \frac{1}{2} (\delta \xi_\mu \xi^\mu + \xi_\mu \delta \xi^\mu
)_{;\nu}n^\nu, \\
&=& \frac{1}{2} \delta \xi_{\mu ; \nu}(\xi^\mu n^\nu + \xi^\nu
n^\mu) + \xi^\mu_{\,\,\, ;\nu}(\delta \xi_\mu n^\nu + \xi_\mu
\delta
n^\nu)\nonumber \\
&& + \frac{1}{2}n^\nu  (\xi_\mu \delta \xi^\mu)_{;\nu}-\frac{1}{2}
n^\nu \mathcal{L}_\xi \delta \xi_\nu, \label{GEO:eq_kappa}
\end{eqnarray}
where all expressions are implicitly pulled-back on the horizon.
The first term in~\eqref{GEO:eq_kappa} is recognized as
$-\frac{1}{2} \delta \xi_{\mu ; \nu}g^{\mu\nu}$ after
using~\eqref{GEO:metric}, \eqref{GEO:dxi} and \eqref{GEO:ortho}.
According to \eqref{GEO:dxi}-\eqref{GEO:rel_delta}, the second
term can be written as
\begin{equation}
\xi^\mu_{\,\,\, ;\nu}(\delta \xi_\mu n^\nu + \xi_\mu \delta n^\nu)
= \xi_{\mu ; \nu} \xi^\mu \delta\eta^\nu,
\end{equation}
for some $\delta \eta^\nu$ tangent to $\mathcal{H}$. This term
vanishes thanks to~\eqref{GEO:ortho}. The third term can be
written as
\begin{eqnarray}
\frac{1}{2}n^\nu (\xi_\mu \delta \xi^\mu)_{;\nu} = -\frac{1}{2}
n^\nu \mathcal{L}_{\delta \xi}\xi_\nu + n_\nu \xi_\mu \delta
\xi^{\mu ; \nu}.
\end{eqnarray}
Now, the Lie derivative of $\delta \xi_\mu$ along $\xi$ can be
expressed as
\begin{equation}
\mathcal{L}_\xi \delta \xi_\mu = -\mathcal{L}_{\delta
\xi}\xi_\mu,\label{GEO:Lie_diff}
\end{equation}
by using the Killing equation~\eqref{GEO:Kill} and its
variation~\eqref{GEO:varKill}. The fourth term can then be written
as
\begin{equation}
-\frac{1}{2} n^\nu \mathcal{L}_\xi \delta \xi_\nu= \frac{1}{2}
n^\nu \mathcal{L}_{\delta \xi} \xi_\nu.
\end{equation}
Adding all the terms, the variation of the surface gravity becomes
\begin{eqnarray}
\delta \kappa &=& -\frac{1}{2} (\delta \xi_\mu)^{;\mu} + \delta
\xi^{\mu;\nu}\xi_\mu n_\nu ,\nonumber\\
&=& -\frac{1}{2} \delta g_{\mu\nu}^{\,\,\,\,\,\,\,;\mu}\xi^\nu -
\frac{1}{2} \delta \xi^\mu_{\,\,\, ;\mu} + \frac{1}{2}
\delta\xi^{\mu;\nu}(\xi_\mu n_\nu - \xi_\nu n_\mu) + \frac{1}{2}
\delta\xi^{\mu;\nu}(\xi_\mu n_\nu + \xi_\nu
n_\mu),\nonumber\\
&=& -\frac{1}{2} \delta g_{\mu\nu}^{\,\,\,\,\,\,\,;\mu}\xi^\nu -
\delta \xi^\mu_{\,\,\, ;\mu} + \frac{1}{2}
\delta\xi^{\mu;\nu}(\xi_\mu n_\nu - \xi_\nu n_\mu) + \frac{1}{2}
\delta \xi^{\mu;\nu}\gamma_{\mu\nu}.\label{GEO:lastt}
\end{eqnarray}
The last line is a consequence of \eqref{GEO:metric}. Contracting
\eqref{GEO:varKill} with $g^{\mu\nu}$ we also have
\begin{equation}
\delta \xi^\mu_{\;\,;\mu} = -\frac{1}{2} \xi^\mu g^{\alpha
\beta}\delta g_{\alpha\beta ;\mu}.
\end{equation}
Finally, the last term in \eqref{GEO:lastt} reduces to
$\frac{1}{2} \delta t^{\alpha}_{\,\,\, |\alpha} $ where $|\alpha$
denotes the covariant derivative with respect to the $n-2$ metric
$\gamma_{\mu\nu}$ and $\delta t^{\mu} = \gamma^\mu_{\,\,\, \nu
}\delta \xi^\nu$ is the pull-back of $\delta \xi^\mu$ on
$\mathcal{H}$. Indeed, one has
\begin{eqnarray}
\frac{1}{2} \delta \xi^{\mu;\nu}\gamma_{\mu\nu} &=& \frac{1}{2}
\delta
t^{\mu;\nu}\gamma_{\mu\nu},\\
&=& \frac{1}{2} (\delta t^{\mu}_{\;\, ,\nu}\gamma_\mu^{\;\,\nu} +
\Gamma_{\mu; \nu
\alpha}\gamma^{\mu\nu}\delta t^\alpha),\\
&=& \frac{1}{2} \delta t^\mu_{\,\,\, |\mu},
\end{eqnarray}
where $|\mu$ denotes the covariant derivative with respect to the
$n-2$ metric $\gamma_{\mu\nu}$. The first line
uses~\eqref{GEO:Kill}-\eqref{GEO:rel_delta} and
$\gamma_{\mu\nu}\xi^\nu=0=\gamma_{\mu\nu}n^\nu$. The last line
uses the decomposition~\eqref{GEO:metric} and $\delta t^\alpha
n_\alpha= 0 =\delta t^\alpha \xi_\alpha$. We have finally the
result
\begin{equation}
\delta \kappa =  -\frac{1}{2} \delta
g_{\mu\nu}^{\,\,\,\,\,\,\,;\mu}\xi^\nu + \frac{1}{2} \xi^\mu
g^{\alpha \beta}\delta g_{\alpha\beta ;\mu} + \frac{1}{2}
\delta\xi^{\mu;\nu}(\xi_\mu n_\nu - \xi_\nu n_\mu) + \frac{1}{2}
(\gamma^\mu_{\,\,\,
\nu}\delta\xi^\nu)_{|\mu}.\label{GEO:deltakappa}
\end{equation}
Expression~\eqref{GEO:topr} therefore equals to
\begin{eqnarray}
\oint_H k^K_{\cL^{EH},\delta \xi } -\oint_H i_\xi I^{n}_{\delta
g}\cL_{EH}& =& -\oint_{H} \frac{d\mathcal{A}}{8\pi G} \delta
\kappa,
\end{eqnarray}
and the result~\eqref{GEO:to_prove} follows because $\delta
\kappa$ is constant on the horizon.\qed

Remark that in classical derivations
\cite{Bardeen:1973gs,Carter1973}, it is assumed that the Killing
vectors $\partial_t$ and $\partial_\varphi$ have the same
components before and after the variation,
\begin{equation*}
\delta (\partial_t)^\mu = \delta (\partial_{\varphi^a})^\mu = 0.
\end{equation*}
One then has $\delta \xi^\mu = \delta \Omega^a
(\partial_{\varphi^a})^\mu$ and the variation of
$\kappa$~\eqref{GEO:lastt} reduces to the well-known expression
\begin{equation*}
\delta \kappa =  -\frac{1}{2} \delta
g_{\mu\nu}^{\,\,\,\,\,\,\,;\mu}\xi^\nu + \delta \Omega^a
(\partial_{\varphi^a})^{\mu;\nu}\xi_\mu n_\nu.
\end{equation*}

\section{Proof of Proposition~\bref{lem4}}
\label{appc} Contracting the vertical one-forms of~\eqref{eq:57}
with the tangent vectors $R_{f_a}$ and $R_{f_b}$ to $\cF^s$, one
gets after applying the homotopy $I_{f_a}$ and integrating on
$S^{\infty,t}$
\begin{equation}
\oint_{S^{\infty,t}} I_{f_a} \big( i_{R_{f_b}}i_{R_{f_a}}\dv
W_{{\delta \cL}/{\delta\phi}}[\dv\phi,\dv\phi]\big) = 0,
\end{equation}
as a consequence of~\eqref{eq:109}. Using~\eqref{eq:comm2}
and~\eqref{1.18}, we get on solutions $\phi^s$,
\begin{eqnarray}
&& \oint_{S^{\infty,t}} I_{f_a} \bigg( \delta_{R_{f_b}}W_{{\delta
\cL}/{\delta\phi}}[\dv\phi,R_{f_a}]|_{\phi^s}
-\delta_{R_{f_a}}W_{{\delta
\cL}/{\delta\phi}}[\dv\phi,R_{f_b}]|_{\phi^s}\nonumber \\
&&+ \dv W_{{\delta \cL}/{\delta\phi}}[R_{f_a},R_{f_b}]|_{\phi^s} -
W_{{\delta \cL}/{\delta\phi}}[\dv\phi,R_{[f_a,f_b]}]|_{\phi^s}
\bigg) = 0.\label{eq:4.15}
\end{eqnarray}
Now, the integrability conditions~\eqref{eq:92a} imply
\begin{eqnarray}
\label{eq:92b} \oint_{S^{\infty,t}} \delta_{R_{f_b}} k_{f_a}[\dv
\phi]=\oint_{S^{\infty,t}} \dv
  k_{f_a}[R_{f_b}].
\end{eqnarray}
Owing to~\eqref{4.13}, one gets
\begin{eqnarray}\label{eq:4.14}
\oint_{S^{\infty,t}} I_{f_a} \delta_{R_{f_b}}W_{{\delta
\cL}/{\delta\phi}} [\dv \phi,R_{f_a}] = \oint_{S^{\infty,t}}
I_{f_a} \dv W_{{\delta \cL}/{\delta\phi}}[R_{f_b},R_{f_a}].
\end{eqnarray}
Note that because~\eqref{eq:29abis}, one has for solutions
$\phi^s$ and one-forms $\dv^s\phi$ tangent to $\cF^s$,
\begin{eqnarray}\label{eq:note1}
I_{f_a} \delta_{R_{f_a}}W_{{\delta \cL}/{\delta\phi}} [\dv^s
\phi,R_{f_b}]|_{\phi^s} =I_{f_b} \delta_{R_{f_a}}W_{{\delta
\cL}/{\delta\phi}} [\dv^s \phi,R_{f_b}]|_{\phi^s} +\dH(\cdot).
\end{eqnarray}
Using the note~\eqref{eq:note1} one can plug~\eqref{eq:4.14} two
times into~\eqref{eq:4.15} for $\dv^s\phi$ tangent to $\cF^s$ to
get
\begin{eqnarray}
\oint_{S^{\infty,t}} I_{f_a}W_{{\delta \cL}/{\delta\phi}} [\dv^s
\phi,R_{[f_a,f_b]}]|_{\phi^s}  = \oint_{S^{\infty,t}} I_{f_b}\dv
W_{{\delta \cL}/{\delta\phi}} [R_{f_b},R_{f_a}]|_{\phi^s}.
\end{eqnarray}
Using~\eqref{eq:29abis}, we then get the result~\eqref{eq:87} and
the proposition is demonstrated. \qed.

\section{Proof of Proposition \bref{la}}
\label{proofofla}

Let $R[Q_1,Q_2]$ be the r.h.s of~\eqref{eq:4}. Proposition
\bref{la} amounts to showing
\begin{eqnarray}
  \label{eq:67}
  R[Q_1,Q_2]=-R[Q_2,Q_1].
\end{eqnarray}
Splitting the derivatives $(\mu)$ in those acting on $Q^i_1$,
denoted by $(\alpha)$, and in those acting on the remaining
expression, denoted by $(\mu^\prime)$ and regrouping the indices
$((\mu^\prime)(\rho)) \equiv (\sigma)$, we get,
\begin{multline}
R[Q_1,Q_2]=\sum_{|\alpha|\geq 0}\sum_{|\sigma| \geq
|\mu^\prime|\geq
  0}\left(\begin{array}{c}|\sigma|+|\alpha|+1\\
|\mu^\prime|+|\alpha|+1\end{array}\right)
\left(\begin{array}{c}|\mu^\prime|+|\alpha|\\
|\alpha|\end{array}\right)(-)^{|\mu^\prime|}\\
\d_{(\alpha)}Q^i_1(-\d)_{(\sigma)}
\Big(Q_2^j\frac{\partial^S}{\d\phi^i_{((\sigma)(\alpha)\nu)}}
\frac{\partial}{\partial  dx^\nu}\vddl{\omega^n}{\phi^j}\Big).
\end{multline}
We now evaluate $\sum_{|\sigma| \geq |\mu^\prime|\geq 0}$ as
$\sum_{|\sigma| \geq 0}\sum_{|\mu^\prime|=0}^{|\sigma|}$ and use
the fact that
\begin{eqnarray}
  \label{eq:33}
  \sum_{|\mu^\prime|=0}^{|\sigma|} \left(\begin{array}{c}|\sigma|+|\alpha|+1\\
|\mu^\prime|+|\alpha|+1\end{array}\right)
\left(\begin{array}{c}|\mu^\prime|+|\alpha|\\
|\alpha|\end{array}\right)(-)^{|\mu^\prime|}=1,
\end{eqnarray}
for all $|\alpha|,|\sigma|$, so that
\begin{equation}
R[Q_1,Q_2] =\6_{(\alpha)}Q_1^i
(-\6)_{(\sigma)}\Big(Q^j_2\frac{\6^S}
{\6\phi^i_{((\alpha)(\sigma)\nu)}} \frac{\partial}{\partial
dx^\nu}\vddl{\omega^n}{\phi^j} \Big).\label{eq:app1}
\end{equation}
Expanding the $\sigma$ derivatives,
\begin{equation}
R[Q_1,Q_2] =
 \d_{(\alpha)}Q^i_1\; \d_{(\beta)}Q^j_2 \; C_{ij}^{(\alpha)(\beta)},
\end{equation}
where
\begin{equation}
C_{ij}^{(\alpha)(\beta)} = (-)^{|\beta|} \left(\begin{array}{c}|\rho|+|\beta|\\
|\beta|\end{array}\right)(-\d)_{(\rho)}
\frac{\d^S}{\d\phi^i_{(\alpha)(\beta)(\rho)\nu}}\varQ{}{\phi^j}\frac{\d}{\d
dx^\nu }\omega^n.\label{defCij}
\end{equation}
Antisymmetry~\eqref{eq:67} amounts to prove that
\begin{equation}
C_{ij}^{(\alpha)(\beta)} =
-C_{ji}^{(\beta)(\alpha)}.\label{eq:antisym}
\end{equation}
From equation \eqref{eq:39}, we get
\begin{eqnarray}
  \label{eq:75}
&&  C_{ij}^{(\alpha)(\beta)} = -(-)^{|\alpha|} \left(\begin{array}{c}|\rho|+|\beta|\\
|\beta|\end{array}\right)\d_{(\rho)}
\vddl{}{\phi^i_{(\alpha)(\beta)(\rho)\nu}}\varQ{}{\phi^i}\frac{\d}{\d
dx^\nu }\omega^n
\end{eqnarray}
Using the definition of higher order Lie
operators~\eqref{higherLie} we get
\begin{eqnarray}
 C_{ij}^{(\alpha)(\beta)} &=& - \sum_{|\sigma^\prime| \geq |\rho| \geq 0} (-)^{|\alpha|+|\sigma^\prime|+|\rho|} \left(\begin{array}{c}|\rho|+|\beta|\\
|\beta|\end{array}\right) \nonumber \\
&&\left(\begin{array}{c}|\alpha|+|\beta|+|\sigma^\prime|+1\\
|\alpha|+|\beta|+|\rho|+1\end{array}\right) \d_{(\sigma^\prime)}
\frac{\d^S}{\phi^j_{(\alpha)(\beta)(\sigma^\prime)\nu}}\varQ{}{\phi^i}\frac{\d}{\d
dx^\nu }\omega^n.
\end{eqnarray}
Evaluating $\sum_{|\sigma^\prime| \geq |\rho| \geq 0}$ as
$\sum_{|\sigma^\prime| \geq 0}\sum_{|\rho|=0}^{|\sigma^\prime|}$
and using the equality
\begin{eqnarray}
  \label{eq:81}
  \sum_{|\rho|=0}^{|\sigma^\prime|}(-)^{|\rho|}\left(\begin{array}{c}|\rho|+|\beta|\\
|\beta|\end{array}\right)\left(\begin{array}{c}|\alpha|+|\beta|+|\sigma^\prime|+1\\
|\alpha|+|\beta|+|\rho|+1\end{array}\right)=
\left(\begin{array}{c}|\sigma^\prime|+|\alpha|\\
|\alpha|\end{array}\right),
\end{eqnarray}
we finally obtain
\begin{eqnarray}
 C_{ij}^{(\alpha)(\beta)}\hspace{-6pt} &=&\hspace{-6pt} -(-)^{|\alpha|}\left(\hspace{-3pt}\begin{array}{c}|\sigma^\prime|+|\alpha|\\
|\alpha|\end{array}\hspace{-3pt}\right)(-\d)_{(\sigma^\prime)}
\frac{\d^S}{\d\phi^j_{(\alpha)(\beta)(\sigma^\prime)\nu}}\varQ{}{\phi^i}\frac{\d}{\d
dx^\nu }\omega^n.
\end{eqnarray}
Comparing with~\eqref{defCij}, we have~\eqref{eq:antisym} as it
should. \qed

An explicitly antisymmetric expression for
$C_{ij}^{(\alpha)(\beta)}$ can also be found along the following
lines. Using the definition of the Euler-Lagrange derivatives, it
is straightforward to show that for any local function $f$,
\begin{equation}
\QS{}{\phi^i_{(\alpha)}}\varQ{f}{\phi^j} =
\sum_{m=0}^{|\alpha|}(-)^{m+|\alpha|}\left(\begin{array}{c}|\tau|+|\alpha|-m\\
|\alpha|-m\end{array}\right) (-\d)_{(\tau)}\QS{}{\phi^i_{(\alpha_1
\dots \alpha_{m} }}\QS{f}{\phi^j_{   \alpha_{m+1}\dots
\alpha_{|\alpha|})(\tau)}}\label{eq:app2}
\end{equation}
Using then~\eqref{eq:app2} where one replace $(\alpha)$ by
 $(\alpha)(\beta)$ and taking into account all combinatorial
 factors, one obtains that
\begin{lemma}\label{lemma2}
For all local functions $f$, one has
\begin{eqnarray}
\QS{}{\phi^i_{(\alpha)(\beta)}}\varQ{f}{\phi^j} &=&
\sum_{m=0}^{|\alpha|}(-)^{m+|\alpha|}\sum_{n=0}^{|\beta|}(-)^{n+|\beta|} \left(\begin{array}{c}|\tau|+|\alpha|-m+|\beta|-n\\
|\alpha|-m\end{array}\right)
\times \nonumber\\
&&\hspace{-2cm}\left(\begin{array}{c}|\tau|+|\beta|-n\\
|\beta|-n\end{array}\right)\left(\begin{array}{c}|\alpha|+|\beta|\\
|\beta|\end{array}\right)^{-1}
\left(\begin{array}{c}m+n\\m\end{array}\right) \times \nonumber\\
&&\hspace{-2cm}(-\d)_{(\tau)}\QS{}{\phi^i_{(\alpha_{1}\dots
\alpha_{m})(\beta_1\dots \beta_n) }}\QS{f}{\phi^j_{(\alpha_{m+1}
\dots \alpha_{|\alpha|})(\beta_{n+1}\dots\beta_{|\beta|})(\tau)}},
\end{eqnarray}
where the indices are totally symmetrized, $(\alpha) =
((\alpha_{1}\dots \alpha_{m})(\alpha_{m+1} \dots
\alpha_{|\alpha|}))$ and $(\beta) = ((\beta_{1}\dots
\beta_{m})(\beta_{m+1} \dots \beta_{|\beta|}))$.
\end{lemma}
Splitting further the indices $(\alpha)$ in Lemma~\bref{lemma2}
into $(\alpha)\nu$ and posing $n^* = |\tau|+|\beta|-n$ and $m^* =
|\alpha|-m$, we obtain
\begin{eqnarray*}
\QS{}{\phi^i_{(\alpha)(\beta)\nu}}\varQ{f}{\phi^j}\hspace{-6pt}
&=&\hspace{-6pt}
\sum_{m=0}^{|\alpha|}\sum_{n=0}^{|\beta|}(-)^{m^*+n^*} \left(\hspace{-3pt}\begin{array}{c}m^*+n^*\\
m^*\end{array}\hspace{-3pt}\right)\hspace{-4pt}
\left(\hspace{-3pt}\begin{array}{c}n^*\\
|\tau|\end{array}\hspace{-3pt}\right)\hspace{-4pt}\left(\hspace{-3pt}\begin{array}{c}|\alpha|+|\beta|\\
|\beta|\end{array}\hspace{-3pt}\right)^{-1}\hspace{-4pt}\left(\hspace{-3pt}\begin{array}{c}m+n\\m\end{array}\hspace{-3pt}\right)  \nonumber\\
&&\hspace{-3cm} \d_{(\tau)} \left(
\frac{m+n+1}{|\alpha|+|\beta|+1} \QS{}{\phi^i_{\nu
(m)(n)}}\QS{f}{\phi^j_{(m^*)(n^*)}} -
\frac{m^*+n^*+1}{|\alpha|+|\beta|+1}
\QS{}{\phi^i_{(m)(n)}}\QS{f}{\phi^j_{(m^*)(n^*)\nu}} \right).
\end{eqnarray*}
We can now develop \eqref{defCij} by using the last expression
with $(\alpha)$ replaced by $(\alpha)(\beta)$ and $(\beta)$
replaced by $(\rho)$. The resulting expression involves the
following summation in $\tau$, $m$, $n$ and $\rho$ (with now $m^*
= |\alpha|+|\beta|-m$ and $n^*= |\tau|+|\rho|-n$)
\begin{equation}
\sum_{|\tau|\geq 0}\sum_{|\rho|\geq 0}\sum_{n=
0}^{|\rho|}\sum_{m=0}^{|\alpha|+|\beta|}\Big( \cdot\Big)
\end{equation}
which can be rewritten as
\begin{equation}
\sum_{n \geq 0} \sum_{n^* \geq 0}\sum_{m=0}^{|\alpha|+|\beta|}
\sum_{k=0}^{n^*}\Big( \cdot\Big)
\end{equation}
by expressing $\tau$ in terms of $n^*$ and posing $k = |\rho|-n$.
Using then
\begin{eqnarray}
&&\sum_{k=0}^{n^*} (-)^k \frac{(|\alpha|+|\beta|)!(k+n)!}{(k+n+|\alpha|+|\beta|+1)!} \left(\begin{array}{c}n^*\\
k\end{array}\right)\left(\begin{array}{c}k+n+|\beta|\\
|\beta|\end{array}\right) \nonumber\\
&&\qquad =  \frac{(|\beta|+n)!(|\alpha|+n^*)!}{(|\alpha|+|\beta|+n+n^*+1)!} \left(\begin{array}{c}|\alpha|+|\beta|\\
|\alpha|\end{array}\right),
\end{eqnarray}
we obtain
\begin{eqnarray}
C_{ij}^{(\alpha)(\beta)}\hspace{-6pt}&=&\hspace{-9pt}\sum_{n,n^*\geq
0} \hspace{-3pt}\sum_{m=0}^{|\alpha|+|\beta|} (-)^{|\alpha|+m}
\frac{(|\beta|+n)!(|\alpha|+n^*)!}{(|\alpha|+|\beta|+n+n^*+1)!} \left(\hspace{-3pt}\begin{array}{c}m+n\\
n\end{array}\hspace{-3pt}\right)\hspace{-2pt}\left(\hspace{-3pt}\begin{array}{c}m^*+n^*\\
n^*\end{array}\hspace{-3pt}\right) \nonumber\\
&&(-\d)_{(n)(n^*)} \bigg( (n+m+1)
\QS{}{\phi^i_{\nu (m)(n)}}\QS{f}{\phi^j_{(m^*)(n^*)}}\nonumber \\
&& \hspace{1.5cm} - (n^*+m^*+1) \QS{}{\phi^i_{
(m)(n)}}\QS{f}{\phi^j_{(m^*)(n^*)\nu}}\bigg).
\end{eqnarray}
Exchanging the role of $n$ and $n^*$ and of $m$ and $m^*$ in the
second term in the parenthesis, we finally obtain an expression
explicitly antisymmetric under the exchange of $i \leftrightarrow
j $, $(\alpha) \leftrightarrow (\beta)$.

\section{Explicit computation of the $\mathfrak{bms}_n$ algebra}
\label{app:aKv_known}

Introducing the notation $\tilde\xi^u=U(u,\theta^A)$,
$\tilde\xi^r=R(u,\theta^A)$, $\tilde \xi^A=Y^A(u,\theta^B)$, the
$rr$-component of equation~\eqref{def_akV22} reduces to
\begin{equation}
-2 U \d_r\chi^u+\d_r o(\chi^u) =o(\rho_{rr}).
\end{equation}
The equation requires $\chi^u=r^0$. Since $\d_r o(r^0) =
o(r^{-1})$, we have $\rho_{rr} = r^{-1}$. The $ur$-component of
equation~\eqref{def_akV22} then gives
\begin{equation}
-\d_u U +o(r^{0}) -\d_r\chi^r R + \d_r o(\chi^r) =o(\rho_{ur}).
\end{equation}
This leads to $\chi^r = r$, $R + \d_u U = 0$ and $\rho_{ur}=r^0$.
The $uu$-component of equation~\eqref{def_akV22} reduces to
\begin{equation}
- 2 r \d_u R + o(r) =o(\rho_{uu}).
\end{equation}
It imposes $\d_u R = 0$ and gives $\rho_{uu} = r$. From the $rA$
component,
\begin{eqnarray}
  -\partial_{A} U+o(r^0)+ \partial_r \chi^A Y^A r^2 s_A
+r^2\partial_r  o(\chi^A)=o(\rho_{rA}),
\end{eqnarray}
we get $\chi^A=r^0$ and $\rho_{rA}=r$. The $uA$-component of
equation~\eqref{def_akV22} is
\begin{eqnarray}
r^2 \d_u Y^A + r^2 o(r^0) +r\d_A R + o(r^1) &=&o(\rho_{uA}),
\end{eqnarray}
implying $\d_u Y^A = 0$, and $\rho_{uA}=r^2$. Finally, the $AA$
and $AB$ with $A\neq B$ components of equation~\eqref{def_akV22}
are given by
\begin{eqnarray}
2 r^2 R s_{A} + 2r^2 \d_{(A)} Y^{(A)} s_{(A)} + r^2
Y^C \d_C s_{A} +o(r^2)&=& o(\rho_{AA}),\\
 r^2\d_B Y^{(A)} s_{(A)} + r^2 \d_A Y^{(B)}s_{(B)}+o(r^2) &=&
o(\rho_{AB}).
\end{eqnarray}
One finds the following conditions
\begin{eqnarray}
&& d_u Y^A = 0, \quad R + \d_A Y^{(A)} + \sum_{C<A}Y^C
\cot{\theta^C}=0,\\
&& \d_B Y^{(A)} s_{(A)} + \d_A Y^{(B)}s_{(B)} = 0,
\end{eqnarray}
with $\rho_{AA}=r^2=\rho_{AB}$. The constraints imposed by
\eqref{def_akV22} on $U$, $R$ and $Y^A$ are summarized by
\begin{eqnarray}
&R = -\d_1 Y^1, \label{R},\qquad \d_u U=\d_{1}Y^1,\qquad\d_u \d_u U = 0,\qquad \d_u Y^A = 0,\label {U}\\
 & \d_{1} Y^1=\d_{{(A)}}Y^A+\sum_{B<A}\cot \theta^B Y^B,\quad \forall A,
\label{eq:conf1}\\
&\d_{A} Y^{B}s_{(B)}+\d_{B} Y^{A}s_{(A)}=0,\quad A\neq B,\quad
A,B=1,\dots,n-2. \label{eq:conf2}
\end{eqnarray}
The last two equations allow one to identify $Y^A(\theta^B)$ with
the conformal Killing vectors of the sphere in $n-2$ dimensions
with metric $g^{(n-2)}_{AB} = \delta_{AB}s_{(A)}$.

%% file: bibliothesis.tex
\providecommand{\href}[2]{#2}\begingroup\raggedright\endgroup